\documentclass[
11pt, 
english, 
doublespacing, 
headsepline, 
]{MastersDoctoralThesis} 

\usepackage[utf8]{inputenc} 
\usepackage[T1]{fontenc}
\usepackage{palatino,mathpazo}

\usepackage[autostyle=true]{csquotes} 
\usepackage[utf8]{inputenc}
\usepackage{amssymb,amsmath}
\DeclareMathOperator{\sgn}{sgn}
\usepackage{enumitem}
\usepackage{bookmark}
\usepackage{mathtools}

\usepackage[nottoc,notlot,notlof]{tocbibind}
\usepackage{afterpage}
\usepackage{needspace}
\usepackage{slashed}
\usepackage{hyperref}
\usepackage{graphicx}
\usepackage{subcaption}

\usepackage{verbatim}
\usepackage{bold-extra}
\usepackage{enumitem}
\usepackage[activate={true,nocompatibility},final=true,kerning=true,spacing=true,tracking=true,shrink=30,stretch=30,factor=0]{microtype}
\microtypecontext{spacing=nonfrench}
\usepackage{titlesec}
\definecolor{chapternumbergray}{rgb}{0.83, 0.83, 0.83}

\newfont{\chapterNumber}{eurb10 scaled 7000}

\titleformat{\chapter}[display]%
{\relax}{\mbox{}\marginpar{\vspace*{-\baselineskip}\color{chapternumbergray}\chapterNumber\thechapter}}{0pt}%
{\LARGE\itshape}[\normalsize\vspace*{.8\baselineskip}\titlerule]%

\titlespacing*{\chapter}{0pt}{0cm}{1cm}
\titleformat{\section}{\Large}{\makebox[0cm][r]{\thesection\hspace{1em}}}{0em}{\scshape\lowercase}
\titlespacing*{\section}{0pt}{\baselineskip}{\baselineskip}
\titleformat{\subsection}{\large}{\thesubsection}{.6em}{\itshape}
\titlespacing*{\subsection}{0pt}{\baselineskip}{\baselineskip}
\titleformat{\subsubsection}{\bfseries}{}{}{}
\titlespacing*{\subsubsection}{0pt}{\baselineskip}{\baselineskip}

\makeatletter
\def\cleardoublepage{\clearpage\if@twoside%
	\ifodd\c@page\else
	\vspace*{\fill}
	\hfill
	\begin{center}
		\it 
	\end{center}
	\vspace{\fill}
	\thispagestyle{empty}
	\newpage
	\if@twocolumn\hbox{}\newpage\fi\fi\fi
}
\makeatother

\let\oldfootnote\footnote
\def\footnote{\ifhmode\unskip\fi\oldfootnote}
\makeatletter
\RenewDocumentCommand{\thesistitle} { O{#2} m }{%
   \def\shorttitle{#1}%
   \def\@title{#2}%
   \def\ttitle{#2}%
}
\makeatother

\usepackage[utf8]{inputenc}
\usepackage{xcolor}
\usepackage{physics}

\usepackage{pgf} 

\usepackage{amsthm}

\usepackage{cancel}

\usepackage{wrapfig}
\usepackage{tabularx,}
\usepackage{comment}
\usepackage{latexsym}
\usepackage{mathtools}
\usepackage{dsfont}
\usepackage{mathrsfs}
\usepackage{adjustbox}
\usepackage{float}

\usepackage{framed}
\usepackage{amsfonts}
\usepackage{ytableau}

\def\e{\epsilon}

\newcommand{\be}{\begin{equation}}
\newcommand{\ee}{\end{equation}}

\newcommand{\beq}{\begin{equation}}
\newcommand{\eeq}{\end{equation}}
\newcommand{\bea}{\begin{eqnarray}}
\newcommand{\eea}{\end{eqnarray}}
\newcommand{\bi}{\begin{itemize}}
\newcommand{\ei}{\end{itemize}}

\newcommand{\M}{\mathcal{M}}

\renewcommand{\thesection}{\arabic{section}}

\newcommand{\bs}{\begin{split}}
\newcommand{\es}{\end{split}}

\newcommand{\commie}[1]{}

\definecolor{mred}{rgb}{0.5, 0.0, 0.0}

\usepackage[bottom]{footmisc}
\makeatletter
\def\blfootnote{\xdef\@thefnmark{}\@footnotetext}
\makeatother

\usepackage[framemethod=default]{mdframed}
\newmdenv[skipabove=10pt,
skipbelow=7pt,
rightline=false,
leftline=true,
topline=false,
bottomline=false,
linecolor=mred,
backgroundcolor=mred!5,
innerleftmargin=4pt,
innerrightmargin=10pt,
innertopmargin=0pt,
leftmargin=2pt,
rightmargin=0pt,
linewidth=2pt,
innerbottommargin=-10pt]{lbBox}

\thesistitle{\textit{Dark~Matter,~Rotation~Curves,\\and~the~Morphology~of~Galaxies}\\ {\fontsize{12}{12}}} 
\supervisor{Prof. Augusto \textsc{Sagnotti}} 
\examiner{} 
\degree{Doctoral Thesis} 
\author{Kirill \textsc{Zatrimaylov}} 
\addresses{kirill.zatrimaylov@sns.it} 

\subject{Theoretical Physics} 
\university{\href{https://www.sns.it/}{Scuola Normale Superiore}} 
\department{\href{https://www.sns.it/it/classe-scienze}{Classe di Scienze}} 
\group{\href{https://www.sns.it/it/didattica/phd-ordinamento-degli-studi/corsi-studio}{Corso di Perfezionamento in Fisica}} 
\faculty{\href{https://www.sns.it/}{Scuola Normale Superiore}} 

\begin{document}
\begin{titlepage}
	\begin{center}
				\vspace*{.018\textheight}
				\begin{figure}[th]
					\centering
					\includegraphics[scale=0.5]{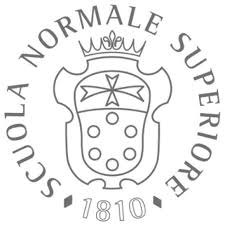}
					\label{fig:logo}
				\end{figure}
		{\scshape\LARGE \univname\par}\vspace{0.7cm} 
		\textsc{\Large Tesi di Perfezionamento in Fisica}\\[0.5cm] 
		
		\HRule \\[0.2cm] 
		{\huge \bfseries \ttitle\par}\vspace{0.4cm} 
		\HRule \\[1.0cm] 
		
		\begin{minipage}[t]{0.4\textwidth}
			\begin{flushleft} \large
				\emph{Candidato:}\\
				\href{kirill.zatrimaylov@sns.it}{\authorname} 
			\end{flushleft}
		\end{minipage}
		\begin{minipage}[t]{0.4\textwidth}
			\begin{flushright} \large
				\emph{Relatore:} \\
				\href{augusto.sagnotti@sns.it}{\supname} 
			\end{flushright}
		\end{minipage}\\[1.5cm]
		
		
		\groupname\\\deptname\\[1cm] 
		
		
		{\large XXXIII Ciclo \\ Anno Accademico 2020-2021} 
		\vfill
	\end{center}
\end{titlepage}

\begin{abstract}
\thispagestyle{empty}
In this thesis, we investigate some aspects of dark matter phenomenology and its predictive power in explaining the flattening of galaxy rotation curves at large distances. After outlining the Standard Model of particle physics, its symmetries and possible extensions in Chapter~\ref{C1}, we review key facts about dark matter and various types of dark matter models in Chapter~\ref{C2}. In Chapter~\ref{C3} we discuss some alternatives to cold dark matter, which include modified Newtonian dynamics (MOND), superfluid dark matter and emergent gravity, and highlight the difficulties that are encountered in attempts to extend these frameworks to full-fledged relativistic settings. In Chapter~\ref{C4} we turn to explore a completely different option, namely that flattened rotation curves reflect the presence of prolate dark--matter bulges or string--like objects around galaxies, without the need for any infrared modification of gravity. To test this model, we fit a number of galaxy rotation curves and find that the presence of a string--like filament yields improvement in fit quality of about 40--70\% in some cases, while the deformation of a dark halo yields only modest improvement by about 6--7\%. In Chapter~\ref{C5} we collect some concluding remarks.
\end{abstract}

	\baselineskip=20pt
	
	\vskip 12pt
	\newpage
\begin{acknowledgements}
	\thispagestyle{empty}
	First and foremost, I would like to express my gratitude to my advisor, Prof. Augusto Sagnotti, for guiding me on my PhD project and proposing a key concept behind it, for introducing me to the basics of String Theory, for his patience in helping me become a qualified theoretical physicist, and for the interesting and fruitful conversations that we've had.
	
	I would like to thank Prof. Andrea Ferrara for his many useful comments on the manuscript, Andrea Pallotini for providing technical assistance and access to the SNS workstation, Prof. Wilfried Buchmüller for his kind hospitality during my stay at DESY, as well as Prof. Emilian Dudas, Prof. Mariano Cadoni, Prof. Stacy McGaugh, Federico Lelli, and Pengfei Li for their comments and clarifications, and to Prof. Pier Stefano Corasaniti for his helpful and instructive criticism as the referee of one of my papers. Besides, I am grateful to Prof. Ugo Moschella, Prof. Sergio Cacciatore, and Prof. Vittorio Gorini for the kindness extended to me during my stay at the 2018 Como Lake School ``Waves on the Lake``, and to Prof. Arthur Hebecker for making my stay at the 2018 W.E. Heraeus Summer School ``Saalburg`` pleasant and comfortable.
	
	I would like to express my gratitude to the professors at Novosibirsk State University, particularly Prof. Alexander D. Dolgov, Prof. Ilya F. Ginzburg, Prof. Simon I. Eidelman, Prof. Alexander I. Milstein, Prof. Victor S. Fadin, Prof. Valery G. Serbo, Prof. Leonid V. Il'ichev, Andrey V. Grabovsky, Prof. David A. Shapiro, Prof. Konstantin V. Lotov, and to many others, as well as to Prof. Vitaly Vanchurin, with whom I collaborated on side projects, to Prof. Alberto Guglielmi, to my physics tutor Irina Akhmetyanova, my math school teacher Boris Tanygin, and my primary school teacher Alina Maslova.
	
	I would also like to thank my friends and collaborators, first and foremost Ivano Basile for helping me to understand various aspects of String Theory, for emotional support, a lot of enjoyable discussions, and help with Cython codes. In addition, I would like to express my gratitude to Riccardo Antonelli, Eftychia Sagkrioti, Harman Deep Kaur, Denis Bitnii, Karapet Mkrtchyan, Ehsan Hatefi, Antonio Camerlengo, Marco Costa, Alice Balbi, and many others. 
	
	Last but not least, I'm grateful to my family, my parents and grandparents.
\end{acknowledgements}	
    \newpage
	\tableofcontents
	\newpage
	
\begin{chapter}{Introduction}\label{C0}
At the most fundamental level reached by Science today, Nature is described by two theories: the Standard Model of Particle Physics (SM)~\cite{SM}, which is an account of small-scale physics down to $10^{-18}$ m based on Quantum Mechanics and Special Relativity, and General Relativity (GR)~\cite{GR}, which describes astrophysical and cosmological phenomena at scales up to $10^{26}$ m.

The Standard Model is a Quantum Field Theory~\cite{QFT}, in which each type of elementary particle is described as an excitation of the corresponding field. The known particles are divided into two classes, fermions with spin $\frac{1}{2}$, which are the building blocks of matter, and bosons, which comprise the carriers of the strong and electroweak interactions. The spin--1 bosons lie in the adjoint representation of the group $SU(3)\otimes SU(2)\otimes U(1)$, which defines the gauge symmetries of the theory. The $SU(3)$ sector contains 8 bosons known as gluons that mediate the strong interaction, the $SU(2)$ sector includes 3 bosons $W^\pm$ and $W^0$, and the $U(1)$ sector contains a single boson $B$ (the $W^{\pm, 0}$ and $B$ bosons mediate the electroweak interaction). The SM also includes a boson doublet $(H^+ H^0)$ with spin 0, which includes yet another type of particle, known as the Higgs boson. After a phase transition in the early Universe, the $H^0$ component is postulated to have acquired a non--zero vacuum expectation value, and as a result the $W^\pm$ bosons and a superposition of $W^0$ and $B$ known as the $Z$--boson, all of which are coupled to $H^0$, become massive, while the orthogonal superposition of $W^0$ and $B$, which is not coupled to $H^0$, remains massless and is the widely known photon $\gamma$. The electroweak interaction thus splits into a short--range weak force that is responsible for $\beta$--decays of atomic nuclei, and the long--range electromagnetic force whose macroscopic manifestations permeate our life and our technology.

The fermionic content of SM comprises three generations, each of which contains four types of particles. Two of these particles are known as quarks: they interact with both strong and electroweak sectors. Due to the confining nature of the strong interaction, quarks are only observed in bound three--quark or quark--antiquark states called baryons and mesons respectively (collectively, they are known as hadrons). The other two particles are called leptons: one of them is charged with respect to the electromagnetic and weak sectors, while the other, known as the neutrino, only manifests itself in weak interactions. The first generation includes the $u$ and $d$ quarks, the electron $e$, and the electronic neutrino $\nu_e$, the second one contains the $s$ and $c$ quarks, the muon $\mu$, and the muon neutrino $\nu_\mu$, and finally the third generation includes the $t$ and $b$ quarks, the $\tau$--lepton, and the $\tau$--neutrino $\nu_\tau$. The three generations are completely identical to each other, with the only exception of particle masses, which become larger going from one generation to the next. Just like the gauge bosons of the weak interaction, fermions acquire their masses through couplings to the Higgs field, known as the Yukawa couplings.

In addition to gauge symmetries, the strong and electromagnetic sectors of SM are invariant under discrete symmetries, the $C$, $P$, and $T$ transformations. The first one is charge conjugation, which inverts the sign of the charge, the second is the parity transformation, which interchanges left--handed and right--handed particles, and the third one is time reversal. The weak sector breaks the $C$ and $P$ symmetries (and, in some Yukawa couplings, even the composite $CP$ symmetry) because it only interacts with left--handed particles and right--handed antiparticles. However, it respects the combined $CPT$ symmetry, consistently with a classic theorem due to Lüders, Pauli and Zumino: every theory that is local, Lorentz invariant, and bounded from below has to be invariant under $CPT$~\cite{PauliLuders}.

Finally, the SM has some accidental symmetries that are not purposely built into the Lagrangian, but are the consequences of gauge invariance or renormalizability. One of them is baryonic symmetry, namely the conservation of a special baryonic charge ascribed to three--quark bound states, and the other two are the custodial symmetry and the flavour symmetry. The former is just the $SO(4)$ rotation symmetry of the Higgs doublet, broken by the Higgs vacuum value and by the gauge and Yukawa couplings, and the latter is the rotation symmetry between the three generation of fermions that is also broken by the Yukawa couplings.

The fourth interaction, gravity, does not fit into the SM, and is described by general relativity as the curvature of spacetime induced by the presence of matter. Still, gravity is also somehow a gauge theory, and actually its study led to the basic tools to quantize gauge theories. General Relativity accounts for the cosmological evolution of the Universe~\cite{Cosmology}, which is permeated by a relic radiation, the CMB. This radiation records remarkably subtle details on an early inflationary epoch~\cite{Inflation} that provides a rationale for the high degree of homogeneity that we observe today. Ideas drawn from Particle Physics, along the lines that led to the Higgs field, can account for the transition from an early epoch of accelerated expansion to a subsequent epoch of deceleration, while adding Quantum Mechanics into the picture gives a rationale to the inception of structure formation~\cite{Chib_Mukh}. Yet, the attempts to quantize gravity in the same fashion as the other three forces encounter theoretical, phenomenological, and conceptual obstacles~\cite{gs}.

However, a number of phenomena cannot be explained within either SM or GR. One of the most prominent among them is dark matter (DM). Relying on gravitational effects in galaxies by Fritz Zwicky in the 1930s~\cite{Zwicky:1933gu,Zwicky:1937zza} and subsequent studies by Sinclair Smith~\cite{Smith:1936mlg} and Vera Rubin~\cite{Rubin:1978kmz,Rubin:1980zd}, it has been possible to conclude that the total amount of gravitating matter within a galaxy is far greater than the amount of luminous matter. This result has been confirmed by several independent lines of investigation, and most notably by studies of primordial structure formation, observations of gravitational lensing~\cite{Natarajan:2017sbo,Tyson:1998vp,Markevitch:2003at,Mahdavi:2007yp}, hot gas in galaxy clusters~\cite{Buote:2003tw}, Kaiser's effect (the distortion of galactic redshift due to their peculiar motion)~\cite{Peacock:2001gs}, and the spectrum of cosmic microwave background (CMB). According to the most accurate CMB spectrum measurements by the Planck collaboration, dark matter amounts to about $85\%$ of the matter content of the Universe~\cite{PLANCK}.

It has been established that DM cannot consist of already known elementary particles or astrophysical objects. Therefore, explain it origin requires either an extension of the SM, or new types of astrophysical objects, or possibly a modification of GR. The first category of DM candidates involves various types of particles proposed to resolve other issues of the Standard Model (such as the sterile neutrino~\cite{Minkowski:1977sc}, the axion~\cite{Peccei:1977hh,Kim:1979if,Shifman:1979if}, and the supersymmetric partners of SM particles~\cite{Gervais:1971ji,Volkov:1972jx,Golfand:1971iw}), while the second category includes primordial black holes (PBHs) that could have formed in the early Universe in post--inflationary epochs~\cite{Zeldovich:1967lct,Hawking:1971ei}. Finally, the latter kind of proposals involves a modification of the laws of gravity at large distances, and a prominent example to this effect is known as MOND (modified Newtonian dynamics)~\cite{Milgrom:1983ca}. Despite its relative obscurity, this option is sometimes seen as a very economical one, since it explains a wealth of galaxy--scale phenomenology at the cost of introducing just one new parameter. Unlike other models, however, MOND can only account for one specific class of observations (gravitational dynamics of galaxies), but there have been numerous attempts to embed it within a broader framework. The two most notable examples are emergent gravity, which draws some inspiration from the AdS/CFT correspondence~\cite{ADSCFT} and models spacetime as an elastic medium that responds to the presence of baryonic matter~\cite{Verlinde:2016toy}, and superfluid dark matter, which proposes that dark matter undergoes a phase transition at galactic scales, producing MOND--type effects~\cite{Berezhiani:2015bqa}.

In this Thesis, we study several models that incorporate MOND and formulate a ``no--go`` statement regarding them that considerably limits their explanatory power. This result, in our opinion, points toward alternative interpretations of the phenomenology attributed to MOND. We then propose one possible alternative interpretation for the surprising shape of galactic rotation curves, ascribing their origin to the presence of elongated string--like objects at the centers of galaxies, and discover some observational evidence for their presence in a number of galaxies of the SPARC catalogue~\cite{Lelli:2016zqa}.

\section{Synopsis}
The contents of this Thesis are as follows.

After an overview of the Standard Model of particle physics (SM), its field content and symmetries in the Section~\ref{C1S1} of Chapter~\ref{C1}, we move on to discuss some of its unresolved issues and possible extensions of the SM that would accommodate them in Section~\ref{C1S2}. In particular, we bring up the problem of neutrino masses that may point to the existence of a fourth flavor of neutrinos (known as sterile neutrinos~\cite{Minkowski:1977sc}) in Subsection~\ref{C1S2S1}, the strong CP problem that may necessitate the introduction of a scalar boson known as the axion~\cite{Peccei:1977hh,Kim:1979if,Shifman:1979if} in Subsection~\ref{C1S2S2}, and the discrepancies between the Standard Model (which describes the electromagnetic, strong and weak forces) and general relativity (which is the theory of the fourth interaction, gravity) in Subsection~\ref{C1S2S3}. One of these discrepancies arises when one attempts to quantize gravity in the same fashion as the three other forces, due to the fact that canonical quantum gravity is non--renormalizable and diverges at two loops~\cite{gs}. Another issue has to do with the vacuum energy that is not directly observed in QFT, but becomes a physical observable in GR: its value measured from cosmological observations (the so--called ``dark energy``) is smaller than the value expected from the Standard Model by 120 orders of magnitude~\cite{PLANCK}. 

In Subsection~\ref{C1S2S4}, we discuss a possible solution to these issues in the form of supersymmetry: namely, if one assumes that each bosonic field in the Standard Model has a fermionic counterpart (and vice versa), their contributions to the vacuum energy would exactly cancel, and the observed nonzero value of VE can result from supersymmetry breaking~\cite{Gervais:1971ji,Volkov:1972jx,Golfand:1971iw}. The simplest supersymmetric theory that incorporates SM is known as the minimal supersymmetric Standard Model (MSSM). Moreover, if one includes the graviton and its superpartner with spin $\frac{3}{2}$ known as the gravitino into the model (this extended model is known as supergravity~\cite{SUGRA}), it removes the two--loop divergence. Supergravity is still expected to diverge at seven loops, but, as we note in Subsection~\ref{C1S2S5}, it may be the low--energy limit of a different framework known as String Theory~\cite{Scherk:1974ca,Yoneya:1974jg,stringtheory}. Within String Theory, particle states are understood as excitations of one--dimensional objects of finite length (``strings``). The presence of a finite string length scale effectively ``cures`` the theory of UV divergences. However, the supersymmetric String Theory can only be consistently quantized in ten spacetime dimensions, which means that the extra six dimensions have to be compactified, possibly on Ricci--flat manifolds known as the Calabi--Yau manifolds. The exact type of compactification, and thus the choice of internal manifold, determines the spectrum of the resulting four--dimensional theory. At this point it is not clear which choice, if any, would reproduce the Standard Model (this is known as the ``landscape problem``), also because the theory is relatively under control only in the presence of supersymmetry. In addition, there exist not one but five supersymmetric string theories (type I, type IIA and IIB, and heterotic strings based on the $SO(32)$ and $E8\otimes E8$ gauge groups), which, along with the eleven--dimensional supergravity, are linked via transformations known as dualities, and are believed to be various limits of a single model known as the M--theory~\cite{Witten:1995ex}. The exact formulation of M--theory remains unknown so far, as does the behavior of the theory in the presence of supersymmetry breaking, which is expected to impinge in an important way on vacuum stability.

Finally, in Subsection~\ref{C1S2S6} we overview a number of problems from the cosmology of the early Universe (namely, the flatness problem, the horizon problem, and the structure formation problem) that may be resolved if the early Universe underwent a phase of accelerated expansion~\cite{Inflation}. Physically, this regime of accelerated expansion can be realized through a scalar field called the inflaton. The inflaton is slowly rolling down a potential, producing a near--constant value of vacuum energy that drives the expansion. Moreover, the quantum fluctuations of the inflaton and of the metric tensor classicalize during inflation, producing the seeds for structure formation. In the context of String Theory, the inflationary scenario may alleviate the vacuum stability issues. Namely, a specific scenario of supersymmetry breaking, known as brane supersymmetry breaking, can trigger the onset of inflationary phase, and would also leave an imprint on the spectrum of primordial perturbations~\cite{climbing}.

In Chapter~\ref{C2}, we overview the problem of dark matter that also indicates the presence of physics beyond the Standard Model. In Section~\ref{C2S1}, we touch upon the observational evidence for an additional matter component in the Universe, including the galaxy rotation curves~\cite{Zwicky:1933gu,Zwicky:1937zza,Smith:1936mlg,Rubin:1978kmz,Rubin:1980zd}, the CMB spectrum~\cite{PLANCK}, the gravitational lensing~\cite{Natarajan:2017sbo,Tyson:1998vp,Markevitch:2003at,Mahdavi:2007yp}, and the primordial structure formation, and in Section~\ref{C2S2}, we consider several discrepancies between the prevailing cold dark matter (CDM) paradigm and observational data, namely the issues of core--cusp, missing satellites~\cite{Klypin:1999uc}, ``too big to fail``, and the unexpectedly tight correlations between luminous and dark matter distributions encoded in the radial acceleration relation (RAR)~\cite{McGaugh:2016leg} and the universal rotation curve (URC)~\cite{Karukes:2016eiz,Persic:1995ru,Salucci:2007tm}. In Section~\ref{C2S3}, we overview the primary dark matter candidates, most of which are related to the extensions of the Standard Model that we discussed earlier. Namely, the Subsection~\ref{C2S3S1} is dedicated to weakly interacting massive particles (WIMPs) that naturally appear in the spectrum of supersymmetric theories~\cite{Goodman:1984dc}, the Subsection~\ref{C2S3S2} covers primordial black holes (PBHs) that could have formed in post--inflationary epoch~\cite{Zeldovich:1967lct,Hawking:1971ei}, and in the Subsection~\ref{C2S3S3}, we bring up the sterile neutrinos~\cite{Minkowski:1977sc}. In~\ref{C2S3S4}, we consider the so--called ``fuzzy`` dark matter, namely ultralight scalar bosons that form a Bose--Einstein condensate~\cite{Hui:2016ltb}. This type of particles, known as axion--like particles (ALPs), can emerge in String Theory from compactification of gauge p--forms~\cite{Svrcek:2006yi}. In~\ref{C2S3S5}, we discuss the possibility that dark matter is comprised of a hidden sector of particles, rather than just one new particle; this sector may interact with the Standard Model particles via mediators like the dark photon. Such hidden sectors are also expected to exist in models motivated by String Theory~\cite{Acharya:2016fge}. Finally, in~\ref{C2S3S6} we review dynamical dark matter that can decay into Standard Model particles~\cite{Dienes:2011ja,Dienes:2011sa}.

In Chapter~\ref{C3}, we give consideration to a drastically different option, namely that the phenomenology ascribed to dark matter instead arises from infrared modifications of gravity that give rise to the so--called MOND (modified Newtonian dynamics) and explain the radial acceleration relation and the flattening of rotation curves at large distances~\cite{Milgrom:1983ca}. There are numerous relativistic extensions and physical motivations for MOND, including emergent gravity that models spacetime as an elastic medium and the MOND force as an elastic force~\cite{Verlinde:2016toy} and superfluid dark matter that behaves like CDM on intergalactic scales, but undergoes a phase transition within galaxies, producing a long--range phonon force~\cite{Berezhiani:2015bqa}. However, we demonstrate that the covariant formulation of emergent gravity, introduced in~\cite{Hossenfelder:2017eoh}, is unbounded from below, while superfluid dark matter provides no clear mechanism for the phase transition and the coupling of phonons to baryonic matter. We also touch upon several other possible relativistic completions of MOND, such as $f(R)$ gravity, which, as we point out, cannot be tailored to produce Newtonian behavior at smaller distances and MONDian behavior at large distances (depending on the exact form of $f(R)$, one would obtain either only the Newtonian solution or an infinitely large family of solutions). Then, drawing upon an earlier work~\cite{Boran:2017rdn} that analyzes the LIGO--Virgo results, we rule out a broader class of theories, which puts severe constraints on incorporating MOND within general relativity.

We conclude with Chapter~\ref{C4}, where we present a possible alternative model for the flattening of rotation curves, based on geometric intuition from classical electrostatics. Our key idea is that the presence of a long, infinitely thin filament at the center of a galaxy would produce a quasi--logarithmic potential, resulting in near--constant rotation velocity at large radii. To test this idea, we fitted 83 galaxy rotation curves from the SPARC sample~\cite{Lelli:2016zqa}, and found the evidence for the presence of a string--like filament in about 24 of them, with varying degree of conclusiveness, but giving fit quality improvement of about 40--70\% in best cases. We overview different candidates for the filament--like objects, from purely astrophysical one like black hole jets to the more exotic ones like cosmic strings that touch upon theoretical high--energy physics. For comparison, we also fitted two of these galaxies with a different setup that incorporates a longitudinal deformation of the dark halo itself instead of the filament. However, in this case the improvement in agreement with the observational data was marginal at best, at around 6--7\%.

\section{Publications}
The results that I shall present in this Thesis are based upon the following published articles:

	\begin{itemize}
		\item {K. Zatrimaylov,\newline ``\textbf{On Filaments, Prolate Halos and Rotation Curves}'', In: \emph{Journal of Cosmology and Astroparticle Physics}, 2021.04 (2021): 056.}
		
		\vspace{0.5em}
	
		\item {K. Zatrimaylov,\newline ``\textbf{A Critique of Covariant Emergent Gravity}'', In: \textit{Journal of Cosmology and Astroparticle Physics}, 2020.08 (2020): 024.}
		
	\end{itemize}

\end{chapter}

\chapter{The Standard Model and Beyond}\label{C1}
\label{Chapter1}
\section{The Standard Model and its symmetries}\label{C1S1}
\subsection{Gauge symmetries and the field content of SM}
The Standard Model (SM)~\cite{SM} is the 3+1-dimensional Quantum Field Theory~\cite{QFT} which describes all currently known elementary particles and interactions. The gauge symmetry group of the SM is $SU(3)\otimes SU(2)\otimes U(1)$, and it defines the bosonic content of the theory: $U(1)$ has one gauge field $B_\mu$, $SU(2)$ has $2^2-1=3$ fields ($W^a$, with $a$ running from 1 to 3), while SU(3) has $3^2-1=8$ of them, known as gluons ($G^A_\mu$, with $A$ running from 1 to 8). Upon quantization, they give us the corresponding vector bosons. The fields change in the following way under the gauge transformations:\\
\beq
B_\mu \rightarrow B_\mu + \frac{1}{g'}\partial_\mu\xi \ ,
\eeq
\beq
W^a_\mu \ \rightarrow \ W^a_\mu + \frac{1}{g}\partial_\mu\xi^a + \epsilon^{abc}W_\mu^b\xi^c \ ,
\eeq
\beq
G^A_{\mu} \ \rightarrow \ G^A_{\mu} + \frac{1}{g_s}\partial_\mu\xi^A + f^{ABC}G^B_\mu\xi^C \ .
\eeq
In addition, we have two complex scalar Higgs fields:
\beq
H \ = \ \frac{1}{\sqrt{2}}\begin{pmatrix}
	H^+ \\
	H^0
	\end{pmatrix} \ = \ \begin{pmatrix}
	\phi_1+i\phi_2 \\
	\phi_3+i\phi_4
\end{pmatrix} \ ,
\eeq
which transform as a doublet in the fundamental representation of $SU(2)$:
\beq
H \ \rightarrow \ \exp\left(i\frac{\sigma^a}{2}\xi^a(x)\right) H \ ,
\eeq
where $\sigma^a$ are the Pauli matrices, which are the generators of $SU(2)$.\\
Given the requirement of gauge invariance, the Lagrangians of gauge fields should be built not out of the fields themselves, but of their gauge invariant field strengths:\\
\beq
F_{\mu\nu}  \ = \  \partial_\mu \ B_\nu \ - \ \partial_\nu \ B_\mu \ ,
\eeq
\beq
G^A_{\mu\nu}  \ = \  \partial_\mu G^A_\nu - \partial_\nu G^A_\mu - g_sf^{ABC}G^B_\mu G^C_\nu \ ,
\eeq
\beq
W^a_{\mu\nu}  \ = \  \partial_\mu W^a\nu - \partial_\nu W^a_\mu + g\epsilon^{abc}W^b_\mu\, W^c_\nu \ ,
\eeq
and the Higgs Lagrangian should be a functional only of $H^\dagger H$. The kinetic term would be expected to have the form\\
\beq
\partial_\mu H^\dagger\partial^\mu H \ ,
\eeq
but to make it gauge invariant, we have to replace the ordinary derivative with the covariant derivative:
\beq
\partial_\mu \rightarrow D_\mu  \ = \  \partial_\mu - \frac{i}{2}g\vec{\sigma}\vec{W_\mu} - \frac{i}{2}g'B_\mu \ .
\eeq
In total, the bosonic part of the SM Lagrangian is given by:
\beq
\begin{gathered}
\mathcal{L}_b  \ = \  - \frac{1}{4}B_{\mu\nu}B^{\mu\nu} - \frac{1}{2}W^a_{\mu\nu}W^{a\mu\nu} - \frac{1}{2}G_{\mu\nu}^aG^{a\mu\nu} + \frac{1}{2}(D_\mu H^\dagger\,D^\mu H^\dagger) - V(H^\dagger H) \ ,
\end{gathered}
\eeq
where 
\beq
V (H^\dagger H)  \ = \  - \mu^2(H^\dagger H) + \frac{\lambda}{2}(H^\dagger H)^2 \ ,
\eeq
so that the Higgs field has a tachyonic-type negative mass and a quartic self-interaction term. The extrema of the potential are given by the condition
\beq
\frac{\delta V}{\delta H^\dagger}  \ = \  H\left( -\mu^2+\lambda H^\dagger H \right)  \ = \  0 \ ,
\eeq
\emph{i.e.} either $H \ = \ 0$ or $|H| \ = \ \frac{\mu}{\sqrt{\lambda}}$. The first extremum is a local maximum, which means it corresponds to a false vacuum, while the second one is a minimum (or, more precisely, a valley of minima), corresponding to the true vacuum. The transition from the former to the latter, which occurs in the early Universe when $T$ drops below about $10^{15}$ K, breaks the electroweak symmetry, since the kinetic Higgs term generates the effective mass terms for electroweak bosons (due to charge conservation, we assume that the nonzero vacuum expectation value $v \ = \ \frac{\sqrt{2}\mu}{\sqrt{\lambda}}$ is acquired by $H^0$):\\
\beq
\begin{gathered}
\langle (D_\mu \phi)^2 \rangle  \ = \  \frac{g^2v^2}{4}W^+_\mu W^{-\mu} + \frac{v^2(g^2+g'^2)}{8}Z_\mu Z^\mu \ ,
\end{gathered}
\eeq
where
\beq
W^\pm  \ = \  \frac{W^1 \mp iW^2}{\sqrt{2}} \ ,
\eeq
\beq
Z  \ = \  W^3\cos\theta_w - B\sin\theta_w \ ,
\eeq
\beq
\cos\theta_w \ = \ \frac{g}{\sqrt{g^2+g'^2}} \ , \qquad \sin\theta_w \ = \ \frac{g'}{\sqrt{g^2+g'^2}} \ .
\eeq
The masses of the resulting combinations are therefore
\beq
M_W  \ = \  \frac{vg}{2} \ , \qquad M_Z  \ = \  \frac{v\sqrt{g^2 + g'^2}}{2} \ ,
\eeq
so that $M_Z\geq M_W$. The combination orthogonal to Z,
\beq
A  \ = \  B\cos\theta_w + W^3\sin\theta_w \ ,
\eeq
remains massless, and is interpreted as the photon field. The original bosonic states can be rewritten in terms of these, as
\beq
W^1  \ = \  \frac{W^+ + W^-}{\sqrt{2}} \ ,
\eeq
\beq
W^2  \ = \  i\frac{W^+ - W^-}{\sqrt{2}} \ ,
\eeq
\beq
W^3  \ = \  A\sin\theta_w + Z\cos\theta_w \ ,
\eeq
\beq
B  \ = \  A\cos\theta_w - Z\sin\theta_w \ ,
\eeq
and the covariant derivative therefore loos like
\beq
\begin{gathered}
D_\mu  \ = \  \partial_\mu - \frac{i}{\sqrt{2}}gW^+\left(\frac{\sigma_1 - i\sigma_2}{2}\right) - \frac{i}{\sqrt{2}}gW^-\left(\frac{\sigma_1 + i\sigma_2}{2}\right) - \\
iA_\mu g\sin\theta_w\left(\frac{\sigma_3}{2} + \frac{Y}{2}\right) - iZ_\mu\left(g\cos\theta_w\frac{\sigma_3}{2} - g'\sin\theta_w\frac{Y}{2}\right) \ ,
\end{gathered}
\eeq
or alternatively,
\beq
\begin{gathered}
D_\mu  \ = \  \partial_\mu - ieQA_\mu - i\frac{e}{\cos\theta_w\sin\theta_w}\left(T^3 - \sin^2\theta_wQ\right)Z_\mu - i\frac{g}{\sqrt{2}}\left(W_\mu^+T^+ + W_\mu^-T^-\right) \ ,
\end{gathered}
\eeq
where
\beq
T^+  \ = \  \frac{\sigma_1 + i\sigma_2}{2}  \ = \  \begin{pmatrix} 0 & 1 \\
	0 & 0 \end{pmatrix} \ ,
\eeq
and
\beq
T^-  \ = \  \frac{\sigma_1 - i\sigma_2}{2}  \ = \  \begin{pmatrix} 0 & 0\\
	1 & 0 \end{pmatrix}
\eeq
are the isospin-raising and lowering operators,
\beq
e  \ = \  g\sin\theta_w  \ = \  \frac{gg'}{\sqrt{g^2 + g'^2}}
\eeq
is the elementary electric charge (equal in natural units to the square root of electromagnetic fine--structure constant: $e = \sqrt{\alpha}$), and
\beq
Q  \ = \  \frac{\sigma_3}{2} + \frac{Y}{2}
\eeq
is the conserved electromagnetic charge (the generator which remains unbroken when $SU(2)\otimes U(1)_Y$ breaks down to $U(1)_Q$).
The reason why exactly this combination of generators corresponds to the unbroken symmetry is that when $H^0$ acquires nonzero vacuum expectation value, it is no longer symmetric under U(1), and neither is H symmetric under the transformations that mix $H^+$ and $H^0$. The only symmetry which remains unbroken is the $U(1)$ symmetry of $H^+$; it corresponds to the generator
\beq
Q=T_3 + \frac{1}{2}Y \ ,
\eeq
which in matrix form looks like\\
\beq
\begin{pmatrix}
\frac{1}{2} & 0\\
0 & -\frac{1}{2}
\end{pmatrix} + \frac{1}{2}\begin{pmatrix}
1 & 0\\
0 & 1
\end{pmatrix}  \ = \  \begin{pmatrix}
1 & 0\\
0 & 0
\end{pmatrix} \ .
\eeq
This generator is known as the electric charge, and corresponds to the massless vector boson $A_\mu$.

In addition, the SM includes three generations of fermions, represented by spinor fields; each generation contains two leptons (particles uncharged under SU(3), i.e. SU(3) singlets) and two quarks (charged under SU(3), i.e. SU(3) color triplets). 

The first generation consists of a doublet of an electron ($e$) and electron neutrino ($\nu_e$), and another doublet consisting of an up and a down quark ($u$ and $d$), as well as their right-handed singlet copies:
\beq
\begin{pmatrix}
	\nu_e \\
	e
\end{pmatrix}_L \ , \begin{pmatrix}
	u \\
	d
\end{pmatrix}_L \ , (e)_R \ , (u)_R \ , (d)_R \ .
\eeq
The second generation contains the muon $\mu$ and muon neutrino $\nu_\mu$, and the strange and charm quarks ($s$ and $c$):
\beq
\begin{pmatrix}
	\nu_\mu \\
	\mu
\end{pmatrix}_L \ , \begin{pmatrix}
	c \\
	s
\end{pmatrix}_L \ , (\mu)_R \ , (c)_R \ , (s)_R \ .
\eeq
In the third generation, we have the tau lepton $\tau$ and tau neutrino $\nu_\tau$, as well as the top and bottom quarks ($b$ and $t$):
\beq
\begin{pmatrix}
	\nu_\tau \\
	\tau
\end{pmatrix}_L \ , \begin{pmatrix}
	t \\
	b
\end{pmatrix}_L \ , (\tau)_R \ , (t)_R \ , (b)_R \ .
\eeq
The particles in each generation have the same charges as their counterparts from the previous generation, but larger masses.

The quarks transform under $SU(3)$ as:\\
\beq
\psi \rightarrow \exp\left(iT^a\xi^a\right) \psi \ ,
\eeq
with $a$ running from 1 to 8.\\
The spinors can also be decomposed into their left-handed and right-handed parts using the projector $\gamma_5$:
\beq
\psi \ = \ \psi_L+\psi_R \ ,
\eeq
where
\beq
\psi_L \ = \ \frac{I-\gamma_5}{2}\psi \ , \\
\psi_R \ = \ \frac{I+\gamma_5}{2}\psi \ .
\eeq
These projectors have the properties\\
\beq
\begin{gathered}
\bar{\psi}_R\psi_R \ = \  0
\end{gathered}
\eeq
and
\beq
\begin{gathered}
\bar{\psi}_R\gamma^\mu\psi_L \ = \  0
\end{gathered}
\eeq
due to the fact that $\gamma_5$ anticommutes with all $\gamma$ matrices, and
\beq
(1 + \gamma_5)(1 - \gamma_5) \ = \ 0 \ ;
\eeq
likewise, $\bar{\psi}_L\psi_L$ and $\bar{\psi}_L\gamma^\mu\psi_R$ would also be zero. Therefore the Lagrangian for a fermionic field with spin $\frac{1}{2}$, which has the form
\beq
\mathcal{L}  \ = \  \bar{\psi} \left(i\slashed{D} + m\right) \psi \ ,
\eeq
can be rewritten as
\beq
\begin{gathered}
\left(\bar{\psi}_L + \bar{\psi}_R\right)\left(i\slashed{D} + m\right)\left(\psi_L + \psi_R\right)  \ = \  \bar{\psi}_Li\slashed{D}\psi_L + \bar{\psi}_Ri\slashed{D}\psi_R + m\left(\bar{\psi}_L\psi_R + \bar{\psi}_R\psi_L\right) \ .
\end{gathered}
\eeq
However, according to observational data, only the left-handed components of fermions participate in the weak interaction, i.e. they form doublets under $SU(2)$, while the right-handed components are singlets. It means that both left- and right-handed fermions transform under the $U(1)$:\\
\beq
\psi_{L,R} \rightarrow\exp\left(iY\xi\right)\psi_{L,R} \ ,
\eeq
but only the left-handed ones transform under $SU(2)$:\\
\beq
\psi_L \rightarrow\exp\left(i\frac{\vec{\sigma}}{2}\vec{\xi}\right)\psi_L \ .
\eeq
Within each generation, the left-handed lepton doublets have the weak hypercharge $Y=-1$; as a result, components with $T_3=\frac{1}{2}$ (neutrinos) have zero electric charge and components with $T_3=-\frac{1}{2}$ (electrons, muons and $\tau$-leptons) have electric charge -1. Right-handed leptons have $Y=-2$ and $\tau_3=0$, so the electric charge is also -1. Left-handed quarks have $Y=\frac{1}{3}$ and electric charges $\frac{2}{3}$ (for $u, c, t$ with isospin $+\frac{1}{2}$) and $-\frac{1}{3}$ (for $d, s, b$ with isospin $-\frac{1}{2}$), while right-handed $u, c, t$ quarks have $Y=\frac{4}{3}$, and right-handed $d, s, b$ quarks have $Y=-\frac{2}{3}$.

For simplicity, we shall adopt the notation $L_i$ for the leptons ($e, \mu, \tau$), $\nu_i$ for the corresponding leptonic neutrinos ($\nu_e, \nu_\mu, \nu_\tau$), $U_i$ for the quarks with $T_3=+\frac{1}{2}$ and their right-handed counterparts ($u, c, t$), and $D_i$ for the quarks with $T_3=-\frac{1}{2}$ and their right-handed counterparts ($d, s, b$). The index $i=1,2,3$ labels the generation.

This means that the left-handed kinetic term for each lepton generation would have the form\\
\beq
\bar{\nu}_{iL}i\slashed{\partial}\nu_{iL} + \bar{L}_{iL}i\slashed{\partial}L_{iL} + eA_\mu J_{eiL}^\mu + \frac{e}{\cos\theta_w}Z_\mu\left(J_{3i}^\mu - \sin^2\theta_wJ_{eiL}^\mu\right) + \frac{g}{\sqrt{2}}\left(W^{+ \mu}J^+_{i\mu} + W^{-\mu}J^-_{i\mu}\right) \ ,
\eeq
where:
\beq
\begin{gathered}
J^\mu_{eiL}  \ = \  -\bar{L}_{iL}\gamma^\mu L_{iL} \ ,\\
J^\mu_{3i}  \ = \  \frac{1}{2}\left(\bar{\nu}_{iL}\gamma^\mu\nu_{iL} - \bar{L}_{iL}\gamma^\mu L_{iL}\right) \ , \\
J^{+\mu}_i  \ = \  \bar{\nu}_{iL}\gamma^\mu L_{iL} \ ,\\
J^{-\mu}_i  \ = \  \bar{L}_{iL}\gamma^\mu\nu_{iL} \ .
\end{gathered}
\eeq
For the right-handed copies, we don't have the terms with three W-bosons, so the kinetic term is
\beq
\begin{gathered}
\bar{L}_{iR}i\slashed{\partial}L_{iR} + eA_\mu J^\mu_{eiR} - \frac{e}{\cos\theta_w}\sin^2\theta_wZ_\mu J^\mu_{eiR} \ ,
\end{gathered}
\eeq
with
\beq
J^\mu_{eiR}  \ = \  -\bar{L}_{iR}\gamma^\mu L_{iR} \ .
\eeq
The right-handed neutrinos are not coupled to the gauge bosons, so they are not directly detectable, and are assumed not to exist within the Standard Model. In total, the leptonic part of the Lagrangian has the form\\
\beq
\begin{gathered}
\sum^3_{i=1}\left(\bar{\nu}_{iL}i\slashed{\partial}\nu_{iL} + \bar{L}_{i}i\slashed{\partial}L_{i} + eA_\mu J_{ei}^\mu + \frac{e}{\cos\theta_w}Z_\mu(J_{3i}^\mu - \sin^2\theta_wJ_{ei}^\mu) + \frac{g}{\sqrt{2}}(W^{+ \mu}J^+_{i\mu} + W^{-\mu}J^-_{i\mu})\right) \ ,
\end{gathered}
\eeq
where
\beq
\begin{gathered}
L_i  \ = \  L_{iL} + L_{iR} \ ,\\
J^\mu_{ei}  \ = \  J^\mu_{eiL} + J^\mu_{eiR}  \ = \  -\bar{L}_i\gamma^\mu L_i \ .
\end{gathered}
\eeq
For the quarks, the kinetic and the interaction terms are the same, but \textit{both} components of the left-handed isospin doublets have right-handed singlet counterparts; in addition, quarks are triplets under the $SU(3)$, so they have an additional interaction term with the SU(3) gauge field $G^a_\mu$:
\beq
\begin{gathered}
\sum^3_{i=1}\left(\bar{U}_{i}i\slashed{\partial}U_i + \bar{D}_{i}i\slashed{\partial}D_{i} + eA_\mu J_{ei}^\mu + \frac{e}{\cos\theta_w}Z_\mu(J_{3i}^\mu-\right.\\
\left.\sin^2\theta_wJ_{ei}^\mu) + \frac{g}{\sqrt{2}}(W^{+ \mu}J^+_{\mu i} + W^{-\mu}J^-_{\mu i}) + g_sG^a_\mu J^{a\mu}_i\right) \ ,
\end{gathered}
\eeq
where
\beq
\begin{gathered}
U_i  \ = \  U_{iL} + U_{iR} \ ,\\ 
D_i \ = \ D_{iL}+D_{iR} \ ,
\end{gathered}
\eeq
and the currents are defined as
\beq
\begin{gathered}
J_{ei}^\mu  \ = \  \frac{2}{3}\bar{U}_i\gamma^\mu U_i - \frac{1}{3}\bar{D}_i\gamma^\mu D_i \ ,\\
J_{3i}^\mu \ = \ \frac{1}{2}\left(\bar{U}_{iL}\gamma^\mu U_{iL} - \bar{D}_{iL}\gamma^\mu D_{iL}\right) \ ,\\ J^{+\mu}_i \ = \ \bar{U}_{iL}\gamma^\mu D_{iL} \ ,\\
J^{-\mu}_i \ = \ \bar{D}_{iL}\gamma^\mu U_{iL} \ ,\\
J^{a\mu}_i \ = \ \bar{U}_iT^a\gamma^\mu U_i + \bar{D}_iT^a\gamma^\mu D_i \ .
\end{gathered}
\eeq
In total, the isospin-flipping part of the Lagrangian can be written as
\beq
\mathcal{L}_{\pm}  \ = \  \frac{g}{\sqrt{2}}\left(W^{+\mu}J^+_{\mu} + W^{-\mu}J^-_\mu\right) \ .
\eeq
At second order in perturbation theory, it gives
\beq
\left(\frac{g}{\sqrt{2}}\right)^2J^+_\mu J^-_\nu<W^{+\mu}W^{-\nu}> \ .
\eeq
Here
\beq
<W^{+\mu}W^{-\nu}>  \ = \  \frac{g^{\mu\nu} - \frac{k^\mu k^\nu}{M_W^2}}{M_W^2 - k^2}
\eeq
is the propagator of the W-boson; at energies below $M_W$, we can neglect terms of order $\frac{k^2}{M_W^2}$, so the effective theory would be just two-current interaction:
\beq
\mathcal{L}_\pm  \ = \  \left(\frac{g}{\sqrt{2}M_W}\right)^2J^+_\mu J^{-\mu} \ .
\eeq
For example, the $\beta$-decay, which converts a neutron into a proton:
\beq
n \rightarrow p + e^- + \bar{\nu}_e
\eeq
is a quark-level weak process:
\beq
d \rightarrow u + W^- \rightarrow u + e^- + \bar{\nu}_e \ , 
\eeq
given by
\beq
\begin{gathered}
\mathcal{L}_\beta  \ = \  \frac{G_F}{\sqrt{2}}\left(\bar{e}\gamma^\mu(1-\gamma_5)\nu_e\right)\left(\bar{u}\gamma_\mu(1 - \gamma_5)d\right) \ ,
\end{gathered}
\eeq
where
\beq
\frac{G_F}{\sqrt{2}} \ = \ \left(\frac{g}{2\sqrt{2}M_W}\right)^2
\eeq
is the so-called Fermi constant.

However, we cannot write the mass terms for either quarks or leptons because a term of the form $\bar{\psi}_L\psi_R$ or vice versa would have a free $SU(2)$ index. The Higgs field solves this problem via chirality--flipping interactions with fermions:
\beq
\mathcal{L}_{int}  \ = \  f_{ij}\bar{\psi}^a_{Li}H^a\psi_{Rj} \ .
\eeq
When the VEV of $H^2$ changes from 0 to $\frac{v}{\sqrt{2}}$, we get a term of the form
\beq
\frac{v}{\sqrt{2}}f_{ij}\bar{\psi}^2_{Li}\psi_{Rj} \ ,
\eeq
which is basically the mass term with the mass $\frac{v}{\sqrt{2}}f_{ij}$ if $\psi^2_{iL}$ and $\psi_{jR}$ are left- and right-handed copies of the same fermion. Now, if we want the mass term for the $\psi^1_{iL}$ component, we need to introduce another type of interaction term:
\beq
\tilde{\mathcal{L}}_{int}  \ = \  \tilde{f}_{ij}\epsilon_{ab} \bar{\psi}^a_{Li}H^b\psi_{Rj} \ ,
\eeq
where $\epsilon_{ab}$ is the invariant Levi--Civita tensor:
\beq
\begin{gathered}
\epsilon_{12}  \ = \  -\epsilon_{21} \ = \  1 \ ,\\ \epsilon_{11}  \ = \  \epsilon_{22}  \ = \  0 \ .
\end{gathered}
\eeq
When the Higgs VEV becomes non--zero, it would be
\beq
\tilde{\mathcal{L}}_{int}  \ = \  \frac{v}{\sqrt{2}}\tilde{f}_{ij}\bar{\psi}_{Li}\psi_{Rj} \ ,
\eeq
which is the mass term with the mass $\frac{v}{\sqrt{2}}\tilde{f}_{ij}$, where $\psi_{Rj}$ is the right-handed copy of $\psi_{Li}$.\\

For quarks, the matrices $f^u_{ij}$ and $f^d_{ij}$ are not diagonal, but they can be diagonalized by changing the flavour basis:
\beq
\begin{gathered}
\begin{pmatrix}
	u \\
	c \\
	t \\
\end{pmatrix}_L  \ = \  T^L_U\begin{pmatrix}
u' \\
c' \\
t' \\
\end{pmatrix}_L \ ,\\
$ $\\
\begin{pmatrix}
d \\
s \\
b \\
\end{pmatrix}_L  \ = \  T^L_D\begin{pmatrix}
d' \\
s' \\
b' \\
\end{pmatrix}_L \ , 
\end{gathered}
\eeq
where $Q_{iL}'$ are mass eigenstates, $Q_{iL}$ are the eigenstates of weak interaction, and $T^{L}_{U,D}$ are the transition matrices. In mass eigenstate basis the isospin-changing weak interaction terms would look like
\beq
\mathcal{L}_\pm  \ = \  \frac{g}{\sqrt{2}}\left(W^+_\mu \bar{U}'_{Li}V_{ij}\gamma^\mu D'_{Lj} + W^-_\mu \bar{D}'_{Li}V^\dagger_{ij}\gamma^\mu U'_{Lj}\right) \ ,
\eeq
where
\beq
V_{ij}  \ = \  (T^L_U)^\dagger_{ik}(T^L_D)_{kj}
\eeq
is the so-called Cabibbo-Kobayashi-Maskawa (CKM) matrix. Since the basis-changing transformations are unitary, this matrix also has the property of unitarity.
\beq
V_{ik}V^\dagger_{kj}  \ = \  \delta_{ij} \ ;
\eeq
Up to redefinitions of the fields, one needs only four quantities to parametrize the CKM matrix. The standard parametrization is
\beq
V_{ij}  \ = \  \begin{pmatrix}
	c_{12}c_{13} & s_{12}c_{13} & s_{13}e^{-i\delta_{13}} \\
	-s_{12}c_{23}-c_{12}s_{23}s_{13}e^{i\delta_{13}} & c_{12}c_{23}-s_{12}s_{23}s_{13}e^{i\delta_{13}} & s_{23}c_{13} \\
	s_{12}s_{23}-c_{12}c_{23}s_{13}e^{i\delta_{13}} & -c_{12}s_{23} & c_{23}c_{13}
\end{pmatrix} \ ,
\eeq
with the parameters $\theta_{12}$ (known as the Cabibbo angle), $\theta_{23}$, $\theta_{13}$, $\delta_{13}$ (here we use the notation $c_{ij}=\cos\theta_{ij}$, $s_{ij}=\sin\theta_{ij}$).

\subsection{Discrete symmetries}
In addition to the continuous gauge symmetries, the SM Lagrangian manifests discrete symmetries, namely C, P, and T transformations, in the electromagnetic and strong sectors; however, they are broken by weak interactions, which conserve only the combined CPT symmetry.

The C-transformation is charge conjugation; it means that if we have the Lagrangian for a spinor $\psi$
\beq
\mathcal{L}  \ = \  \psi^\dagger\gamma^0\left(\gamma^\mu (i\partial_\mu-eA_\mu)-m\right)\psi \ ,
\eeq
then for a charge-conjugate spinor $\psi^c$ we would have
\beq
\mathcal{L}  \ = \  \psi^{c\dagger}\gamma^0\left(\gamma^\mu(i\partial_\mu + eA_\mu) - m\right)\psi^c \ . \label{F}
\eeq
Since the spinor transforms under $U(1)$ as
\beq
\psi \rightarrow\exp\left(ie\xi\right)\psi \ ,
\eeq
we can take its complex conjugate, which would transform as
\beq
\psi^c \rightarrow\exp\left(-ie\xi\right)\psi^c \ .
\eeq
Since the Lagrangian is real, we can write it as its complex conjugate:
\beq
\mathcal{L}^*  \ = \  \mathcal{L}  \ = \  (\psi^*)^\dagger\gamma^{0*}\left(\gamma^{\mu *} (-i\partial_\mu + eA_\mu) - m\right)\psi^* \ ,
\eeq
which in turn can be rewritten as:
\beq
\mathcal{L}  \ = \  \psi^TC^TC^{-1^T}\gamma^{0*}\left(\gamma^{\mu*} (-i\partial_\mu + eA_\mu) - m\right) C^{-1}C\psi^* \ ,
\eeq
where C is an arbitrary matrix. Now, if
\beq
-C^{-1^T}\gamma^{0*}\gamma^{\mu*}C^{-1}  \ = \  \gamma^0\gamma^\mu \ , \label{M}
\eeq
then $C\psi^*$ would be the ``charge conjugate`` spinor $\psi^c$. In Weyl basis, where $\gamma^2$ is purely imaginary, and all the other $\gamma$'s are real, it can be chosen as
\beq
C  \ = \  \gamma^2 \ ,
\eeq
which means that $C^T=-C, C^*=-C, C^{-1}=C$, and \eqref{M} reads:
\beq
-C\gamma^{\mu*}C  \ = \  \gamma^\mu \ ,
\eeq
so for $\mu=0,1,3$ $\gamma^{\mu*}=\gamma^\mu$, and C anticommutes with $\gamma^\mu$, while for $\mu=2$ $\gamma^{2*}=-\gamma^2$, and $\gamma^2$ commutes with itself. Since $\gamma^2$ anticommutes with $\gamma^5$, C transforms $\psi_L$ into $\psi^*_R$ and $\psi_R$ into $\psi^*_L$.

The P-transformation is the reflection of the spatial coordinates:
\beq
\vec{r} \rightarrow -\vec{r} \ ,
\eeq
so it should act on the Lagrangian in the following way:
\beq
i\bar{\psi}\left(\gamma^0\partial_t - \vec{\gamma}\vec{\nabla}\right)\psi\rightarrow i\bar{\psi}\left(\gamma^0\partial_t+\vec{\gamma}\vec{\nabla}\right)\psi \ .
\eeq
For the Lagrangian to be invariant, the spinor also has to transform as:
\beq
\psi \rightarrow P\psi \ ,
\eeq
and P can be chosen as $\gamma^0$, since it commutes with itself and anticommutes with $\gamma^i$'s. Since $\gamma^0$ anticommutes with $\gamma^5$, the P-transformation interchanges right-handed and left-handed fermions.

Finally, the T-transformation reverses time:
\beq
i\bar{\psi}\left(\gamma^0\partial_t - \vec{\gamma}\vec{\nabla}\right)\psi\rightarrow i\bar{\psi}\left(-\gamma^0\partial_t - \vec{\gamma}\vec{\nabla}\right)\psi \ .
\eeq
The T-operation is antilinear, which means we have to complex conjugate the coefficients:
\beq
i\gamma^\mu\rightarrow -i(\gamma^\mu)^* \ .
\eeq
In the Weyl representation, in which $\gamma^2$ is imaginary and the other three matrices are real, $i\gamma^2$ and $i\gamma^0$ would not change (the latter due to time reversal), and $i\gamma^1$ and $i\gamma^3$ would change sign. To compensate for this, the spinors should be multiplied by the T-matrix:
\beq
i\psi^\dagger T^\dagger \gamma^0\left(\gamma^0\partial_t + \gamma^1\partial_1 - \gamma^2\partial_2 + \gamma^3\partial_3\right)T\psi \ .
\eeq
This matrix can be chosen as $\gamma^1\gamma^3$, so that it would anticommute with $\gamma^1$ and $\gamma^3$, and commute with $\gamma^0$ and $\gamma^2$.

However, due to the chirality of the weak interaction, it violates the C and P symmetries, i.e. the $W^\pm$-bosons do not interact with right-handed particles and left-handed antiparticles. After the combined CP transformation the particles would still couple to the $W^\pm$-bosons, since $\gamma^5$ commutes with $\gamma^0\gamma^2$, but the spinors would be complex conjugated. I.e. the isospin-flipping interaction Lagrangian for right-handed antiparticles is:
\beq
\begin{gathered}
\mathcal{L}_{\pm}^{CP}  \ = \  \frac{g}{\sqrt{2}}W^-_\mu \left(U'^T_{Li}V_{ij}\gamma^\mu D'^*_{Lj} + \nu^T_{iL}\gamma^\mu L^*_{iL}\right) + \frac{g}{\sqrt{2}}W^+_\mu \left(D'^T_{Li}V^\dagger_{ij}\gamma^\mu U'^*_{Lj} + L^T_{iL}\gamma^\mu\nu^*_{iL}\right) \ ,
\end{gathered}
\eeq
and since the Lagrangian is real, we can complex conjugate it to obtain:
\beq
\begin{gathered}
\mathcal{L}_{\pm}^{CP}  \ = \  \frac{g}{\sqrt{2}}W^+_\mu \left(\bar{U}'_{Li}V^*_{ij}\gamma^\mu D'_{Lj} + \bar{\nu}_{iL}\gamma^\mu L_{iL}\right) + \frac{g}{\sqrt{2}}W^-_\mu \left(\bar{D}'_{Li}V^T_{ij}\gamma^\mu U'_{Lj} + \bar{L}_{iL}\gamma^\mu\nu_{iL}\right) \ .
\end{gathered}
\eeq
The leptonic interactions are invariant under CP, but quark interactions are not, due to the fact that the CKM matrix is complex, and $V_{ij}$ is substituted for $V^*_{ij}$. Therefore, if we have two interfering amplitudes for a weak process with left-handed particles:
\beq
M  \ = \  |M_1|e^{i\phi_1}e^{i\psi_1} + |M_2|e^{i\phi_2}e^{i\psi_2} \ ,
\eeq
where $\psi_1$ and $\psi_2$ are the CKM contributions to the phases, and $\phi_1$ and $\phi_2$ are other contributions. The amplitude for right-handed antiparticles would be:
\beq
M^{CP}  \ = \  |M_1|e^{i\phi_1}e^{-i\psi_1} + |M_2|e^{i\phi_2}e^{-i\psi_2} \ .
\eeq
Therefore the transition probability for left-handed particles is proportional to
\beq
\begin{gathered}
|M|^2  \ = \  |M_1|^2+|M_2|^2 + 2|M_1||M_2|\cos\left((\phi_1 - \phi_2) + (\psi_1 - \psi_2)\right) \ ,
\end{gathered}
\eeq
while for the right-handed antiparticles it is proportional to
\beq
\begin{gathered}
|M^{CP}|^2  \ = \  |M_1|^2 + |M_2|^2 + 2|M_1||M_2|\cos\left((\phi_1 - \phi_2) - (\psi_1 - \psi_2)\right) \ ,
\end{gathered}12
\eeq
so the rates of the two processes are different:
\beq
\begin{gathered}
|M|^2 - |M^{CP}|^2  \ = \  -4|M_1||M_2|\sin(\phi_1 - \phi_2)\sin(\psi_1 - \psi_2) \ .
\end{gathered}
\eeq
The combined CPT symmetry is however conserved due to the Pauli-Lüders theorem, which states that any local Lorentz-invariant QFT with energy bounded from below is CPT invariant~\cite{PauliLuders}.

\subsection{Accidental symmetries}

In addition, the Standard Model manifests so-called accidental symmetries: these are the symmetries which are not specifically imposed, but emerge as a consequence of gauge invariance or renormalizability (symmetry-breaking terms would have mass dimension larger than 4).

One of these symmetries is the baryon--number conservation: each quark can be assigned the quantum number B, with $B=\frac{1}{3}$ for quarks and $B=-\frac{1}{3}$ for antiquarks (each baryon would therefore have $B=+1$, each antibaryon $B=-1$, and each meson $B=0$), corresponding to the phase rotation symmetry when all the quark spinors are rotated simultaneously: 
\beq
\begin{gathered}
q \ \rightarrow \ \exp\left(i\frac{1}{3}\xi\right)q \ ,\\
\bar{q} \ \rightarrow \ \exp\left(-i\frac{1}{3}\xi\right)\bar{q} \ ,
\end{gathered}
\eeq
where
\beq
q \ = \ \begin{pmatrix}
	u \\
	d \\
	s \\
	c \\
	b \\
	t
\end{pmatrix} \ .
\eeq
Likewise, for leptons we have the lepton flavour number conservation: electrons and electron neutrinos are assigned the conserved number $L_e=+1$ (-1 for their antiparticles), muons and muon neutrinos are assigned $L_\mu=+1$, and $\tau$-leptons and $\tau$-neutrinos have $L_\tau=+1$. Each of them corresponds to the phase rotation of a given lepton generation:
\beq
\begin{gathered}
\nu_i \rightarrow e^{i\xi}\nu_i \ ,\\
L_i \rightarrow e^{i\xi}L_i \ ,\\
\bar{\nu}_i \rightarrow e^{-i\xi}\bar{\nu}_i \ ,\\
\bar{L}_i \rightarrow e^{-i\xi}\bar{L}_i \ ,
\end{gathered}
\eeq
where $i$ is the lepton generation number. The conservation of these three numbers is violated by the so-called neutrino oscillations, which are an observed phenomenon beyond the Standard Model in its original form addressed here; however, so far there are no observations of any processes violating the conservation of total lepton number:
\beq
L  \ = \  L_e + L_\mu + L_\tau \ .
\eeq
Other accidental symmetries include the custodial symmetry and the flavour symmetry. The former has to do with the fact that
\beq
H^\dagger H  \ = \  \phi_1^2 + \phi_2^2 + \phi_3^2 + \phi_4^2
\eeq
is invariant under the SO(4) rotation group, which is isomorphic to $SU(2)\otimes SU(2)$. When $\phi_3$ acquires a nonzero VEV, this symmetry breaks down to SO(3), isomorphic to $SU(2)$:
\beq
\phi_1^2 + \phi_2^2 + \phi_4^2 \ .
\eeq
Under this symmetry the fields $W^a$ would transform as a triplet, and the right-handed fermions as doublets, but it is broken to $U(1)_Q$ by the gauge coupling g' and Yukawa couplings of fermions to Higgs.

The flavour symmetry is the rotation of three fermion generations; we can independently rotate left- and right-handed states:
\beq
\begin{gathered}
Q_L\rightarrow U^{ij}_{Q_L}Q_L \ ,\\
U_R\rightarrow U^{ij}_{U_R}U_R \ ,\\
D_R\rightarrow U^{ij}_{D_R}D_R \ ,\\
L_L\rightarrow U^{ij}_{L_L}L_L \ ,\\
e_R\rightarrow U^{ij}_{e_R}e_R \ .
\end{gathered}
\eeq
Each of those is a $U(3)$ symmetry, so the total flavour symmetry group is $(U(3))^5$. It is broken by the Yukawa quark couplings to baryonic and leptonic symmetries.

The masses and charges of all Standard Model particles are given in the Table~\ref{TabSM}.\\
\begin{table}
\centering
	\begin{tabular}{ | l | l | l | l | l |}
		\hline
		Particle & Mass & $Y$ & $T_3$ & $Q$ \\ \hline
		$(\nu_e)_L$ & ~--- & -1 & 1/2 & 0 \\ \hline
		$(e)_L$ & 0.51 MeV & -1 & -1/2 & -1 \\ \hline
		$(u)_L$ & 2.2$^{+0.5}_{-0.4}$ MeV & 1/3 & 1/2 & 2/3 \\ \hline
		$(d)_L$ & 4.7$^{+0.5}_{-0.3}$ MeV & 1/3 & -1/2 & -1/3 \\ \hline
		$(e)_R$ & 0.51 MeV & -2 & 0 & -1 \\ \hline
		$(u)_R$ & 2.2$^{+0.5}_{-0.4}$ MeV & 4/3 & 0 & 2/3 \\ \hline
		$(d)_R$ & 4.7$^{+0.5}_{-0.3}$ MeV & -2/3 & 0 & -1/3 \\ \hline
		$(\nu_\mu)_L$ & ~--- & -1 & 1/2 & 0 \\ \hline
		$(\mu)_L$ & 105.66 MeV & -1 & -1/2 & -1 \\ \hline
		$(c)_L$ & 1.275$^{+0.025}_{-0.035}$ GeV & 1/3 & 1/2 & 2/3 \\ \hline
		$(s)_L$ & 95$^{+9}_{-3}$ MeV & 1/3 & -1/2 & -1/3 \\ \hline
		$(\mu)_R$ & 0.51 MeV & -2 & 0 & -1 \\ \hline
		$(c)_R$ & 1.275$^{+0.025}_{-0.035}$ GeV & 4/3 & 0 & 2/3 \\ \hline
		$(s)_R$ & 95$^{+9}_{-3}$ MeV & -2/3 & 0 & -1/3 \\ \hline
		$(\nu_\tau)_L$ & ~--- & -1 & 1/2 & 0 \\ \hline
		$(\tau)_L$ & 1776.86$\pm$0.12 MeV & -1 & -1/2 & -1 \\ \hline
		$(t)_L$ & 173$\pm$0.4 GeV & 1/3 & 1/2 & 2/3 \\ \hline
		$(b)_L$ & 4.18$^{+0.4}_{-0.3}$ GeV & 1/3 & -1/2 & -1/3 \\ \hline
		$(\tau)_R$ & 1776.86$\pm$0.12 MeV & -2 & 0 & -1 \\ \hline
		$(t)_R$ & 173$\pm$0.4 GeV & 4/3 & 0 & 2/3 \\ \hline
		$(b)_R$ & 4.18$^{+0.4}_{-0.3}$ GeV & -2/3 & 0 & -1/3 \\ \hline
		$W^\pm$ & 80.379$\pm$0.012 GeV & 0 & $\pm$1 & $\pm$1 \\ \hline
		$Z$ & 91.1876$\pm$0.0021 GeV & 0 & 0 & 0 \\ \hline
        $\gamma$ & 0 & 0 & 0 & 0 \\ \hline
        g & 0 & 0 & 0 & 0 \\ \hline
        $H^0$ & 125.18$\pm$0.16 GeV & 1 & -1/2 & 0 \\ \hline
	\end{tabular}
	\caption{Masses and charges of the Standard Model particles.}
	\label{TabSM}
\end{table}

\section{Phenomena beyond the Standard Model}\label{C1S2}
\subsection{Neutrino oscillations and the seesaw mechanism}\label{C1S2S1}
Numerous observations (solar neutrinos, neutrino beams, etc.) indicate that neutrino fluxes of a given flavour decrease over long distances, implying that neutrinos can change their flavour~\cite{Fukuda:1998mi}. This means that the weakly interacting flavour states of neutrinos (``eigenstates of detection``) do not coincide with the propagating mass states (``eigenstates of time evolution``), and can be represented as superpositions of them:
\beq
|\nu_i>  \ = \  \sum_jU_{ij}|\nu'_j> \ ,
\eeq
where $\nu_i$ are the flavour eigenstates, $\nu'_i$ are the mass eigenstates, and $U_{ij}$ is the transition matrix, the analogue of CKM matrix for neutrinos. It evolves in time as:
\beq
|\nu_i(t)>  \ = \  \sum_jU_{ij}e^{-i(E_i-p)t} |\nu'_j> \ .
\eeq
Given that at small masses
\beq
E_i  \ = \  \sqrt{m_i^2 + p^2}\approx p + \frac{m_i^2}{2p} \ ,
\eeq
and assuming that all superposition components propagate with the same momentum (i.e. the flavour eigenstate is also the momentum eigenstate), we obtain
\beq
|\nu_i>  \ = \  \sum_jU_{ij} e^{-i\frac{m_j^2t}{2p}}|\nu'_j> \ .
\eeq
Therefore, if a neutrino is emitted with flavour i, the probability of its detection with flavour j after a time period $\tau$ is
\beq
|<\nu_j|\nu_i>|^2  \ = \  |\sum_kU^*_{jk}U_{ik}e^{-i\frac{m_k^2t}{2p}}|^2 \ .
\eeq
However, within the Standard Model the neutrinos are massless because there are no right-handed neutrino states. One way to  introduce neutrino masses is the so-called seesaw mechanism. First, we introduce the right--handed neutrino states and couple them to the Higgs field:
\beq
f^\nu\epsilon_{ab}\bar{L}^aH^b\nu_R \ ,
\eeq
which gives rise to a standard (Dirac) mass term of the form
\beq
M_D\left(\bar{\nu}_L\nu_R + \bar{\nu}_R\nu_L\right) \ .
\eeq
However, we can also add the so--called Majorana mass terms (they can only be added for neutral particles, since they violate the $U(1)$ symmetry): 
\beq
M_{M1}\bar{\nu}_L\nu^c_L+M_{M2}\bar{\nu}^c_R\nu_R+ c. c.
\eeq
The total mass term is therefore given by
\beq
\begin{pmatrix}
	\bar{\nu}_L\\
	\bar{\nu}^c_R
\end{pmatrix}\begin{pmatrix}
M_{M1} & M_D \\
M_D & M_{M2}
\end{pmatrix}\begin{pmatrix}
\nu^c_L & \nu_R
\end{pmatrix} \ .
\eeq
Diagonalizing this matrix, we derive the equation on its eigenvalues:
\beq
\lambda^2-(M_{M1}+M_{M2})\lambda+(M_{M1}M_{M2}-M_D^2) \ = \ 0
\eeq
with the solutions
\beq
\lambda \ = \ \frac{M_{M1}+M_{M2}\pm\sqrt{(M_{M1}-M_{M2})^2+4M_D^2}}{2} \ .
\eeq
Now, one can see that when one eigenvalue increases, the other decreases, which gives rise to the name ``seesaw``. For specificity, if we consider the case $M_{M1}=0, M_{M2}\gg M_D$, we obtain
\beq
\lambda \ = \ M_{M2} \left(\frac{1\pm\sqrt{1+4(\frac{M_D}{M_{M2}})^2}}{2}\right) \ ,
\eeq
i.e.
\beq
\begin{gathered}
\lambda_1 \ \approx \ M_{M2} \ , \ \lambda_2 \ \approx \ -\frac{M_D^2}{M_{M2}} \ .
\end{gathered}
\eeq
Therefore, if the Standard Model neutrinos are very light ($M_D\approx M_M$), the right-handed sterile neutrinos would naturally be much heavier~\cite{Minkowski:1977sc}.

\subsection{The Strong CP Problem}\label{C1S2S2}
It is possible to add the following term to the Standard Model Lagrangian:
\beq
\mathcal{L}_\theta  \ = \  \theta\frac{g_s^2}{32\pi^2} G_{\mu\nu}G_{\alpha\beta}\epsilon^{\mu\nu\alpha\beta} \ .
\eeq
It is known as the topological term: in the presence of gravity it does not involve the metric, and may be rewritten as the divergence of the so-called Chern--Simons (Bardeen) current:
\beq
\begin{gathered}
\mathcal{L}_\theta  \ = \  \theta\frac{g_s^2}{32\pi^2}\partial_\mu J^{\mu}_{CS} \ ,
\end{gathered}
\eeq
defined as
\beq
\begin{gathered}
J^\mu_{CS}  \ = \  2\epsilon^{\mu\nu\alpha\beta}\left(A^a_\nu F^a_{\alpha\beta} + \frac{2}{3}g_s\epsilon^{abc}A^a_\nu A^b_\alpha A_\beta^c\right) \ .
\end{gathered}
\eeq
This means that the contribution of $\mathcal{L}_\theta$ to the Standard Model action is just a surface integral at infinity:
\beq
S_\theta  \ = \  \theta\frac{g_s^2}{16\pi^2}\int dS_\mu \ \epsilon^{\mu\nu\alpha\beta} \ \left(A_\nu^aF^a_{\alpha\beta} + \frac{2}{3}g_s\epsilon^{abc}A^a_\nu A^b_\alpha A^c_\beta\right) \ .
\eeq
The first term identically vanishes, since one focuses naturally on configurations of fields for which at infinity
\beq
F^a_{\mu\nu} \ = \ 0 \ .
\eeq

Therefore for an abelian theory like QED, in which the second term is not present, the topological term is identically zero and does not contribute to the action. For a non--abelian theory, however, the second term can be non--zero due to the fact that the vector potential does not have to vanish. Instead, it is given by pure gauge at infinity:
\beq
A_\mu  \ = \  A_\mu^aT^a  \ = \  \frac{i}{g_s}U^{-1}\partial_\mu U \ .
\eeq
Due to the property of the generators
\beq
Tr[T^aT^bT^c] \ = \ i\epsilon^{abc} \ ,
\eeq
the action would be given by
\beq
\begin{gathered}
S_\theta  \ = \  n\theta \ ,
\end{gathered}
\eeq
where
\beq
n \ = \ -\frac{1}{24\pi^2}\int dS_\mu \ \epsilon^{\mu\nu\alpha\beta} \ Tr\left[(U^{-1}\partial_\nu U)(U^{-1}\partial_\alpha U)(U^{-1}\partial_\beta U)\right]
\eeq
is the so-called winding number. In the general case, two different $U$-matrices cannot be smoothly transformed into each other, which means two different ``pure gauge`` configurations of $A_\mu$ are also topologically non--equivalent. To transform one into the other, we have to pass through configurations with
\beq
F^a_{\mu\nu} \neq 0 \ ,
\eeq
which means they correspond to different vacua with different values of $n$.

Besides the QCD vacua, the topological term can also emerge from the so-called chiral, or axial anomaly, generally known as the Adler--Bell--Jackiw (ABJ) anomaly. Namely, we can observe that the Lagrangian of a massless fermion
\beq
\mathcal{L}  \ = \  i\bar{\psi}\slashed{D}\psi
\eeq
is also invariant under the transformation
\beq
\begin{gathered}
\psi\rightarrow e^{i\xi\gamma_5}\psi \ ,\\
\bar{\psi}\rightarrow\bar{\psi}e^{i\xi\gamma_5} \ ,
\end{gathered}
\eeq
known as the axial symmetry, due to the fact that all the $\gamma_\mu$ matrices anticommute with $\gamma_5$. However, if we write the partition function for this Lagrangian:
\beq
Z  \ = \  \int\mathcal{D}\bar{\psi} \ \mathcal{D}\psi \ \exp\left[i\int d^4 x \ \bar{\psi} i\slashed{D} \psi\right] \ ,
\eeq
we would see that its measure is not invariant under the axial transformation, if one demands that it be invariant under the usual phase transformations related to electromagnetic interactions. To understand this, we decompose the fermionic fields into eigenvectors of $\slashed{D}$:
\beq
\begin{gathered}
\psi  \ = \  \sum_na_n\phi_n \ , \\
\bar{\psi}  \ = \  \sum_nb_n\bar{\phi}_n \ ,
\end{gathered}
\eeq
with
\beq
\begin{gathered}
(i\slashed{D})\phi_n  \ = \  \lambda_n\phi_n \ ,\\
\bar{\phi}_n(i\slashed{D})  \ = \  -i\partial_\mu\phi_n\gamma^\mu  \ = \  \lambda_n\bar{\phi}_n \ ,
\end{gathered}
\eeq
and the eigenvectors normalized as follows:
\beq
\begin{gathered}
\int d^4 x \ \phi^\dagger_n\phi_m  \ = \  \int d^4 x \ \bar{\phi}_n\bar{\phi}^\dagger_m  \ = \  \delta_{nm} \ .
\end{gathered}
\eeq
In this basis, the measure would be
\beq
\mathcal{D}\bar{\psi} \ \mathcal{D}\psi  \ = \  \Pi_n(a_nb_n) \ .
\eeq
Now, after the transformation, the new fields
\beq
\psi'  \ = \  (1+i\xi\gamma_5)\psi \ ,\\
\bar{\psi}' \ = \ \bar{\psi}(1+i\xi\gamma_5)
\eeq
may also be decomposed:
\beq
\begin{gathered}
\psi'  \ = \  \sum_n a'_n\phi_n \ ,\\
\bar{\psi}'  \ = \  \sum_nb'_n\bar{\phi}_n \ ,
\end{gathered}
\eeq
with the new coefficients given by
\beq
\begin{gathered}
a'_n \ = \ \sum_m\left(\delta_{mn}+i\int d^4x\xi(x)(\phi^\dagger_n\gamma_5\phi_m)\right)a_m \ ,\\
b'_n \ = \ \sum_m\left(\delta_{mn}+i\int d^4x\xi(x)(\phi^\dagger_n\gamma_5\phi_m)\right)b_m \ .
\end{gathered}
\eeq
The determinant of
\beq
J_{nm}  \ = \  \delta_{nm}+i\int d^4x\xi(x)(\phi^\dagger_n\gamma_5\phi_m) 
\eeq
can be calculated with the use of the identity
\beq
\ln\left(\det(M)\right)  \ = \  \tr\left(\ln(M)\right) \ ,
\eeq
which in this case gives us
\beq
\ln(J) \ = \ i\sum_n\int d^4 x \ \xi(x) \ (\phi^\dagger_n\gamma_5\phi_n)
\eeq
($\xi$ is infinitesimal, so we neglect $O(\xi^2)$).

This expression should be regulated, and it's natural to introduce a cutoff at large momenta:
\beq
\ln(J) \ = \ i\int d^4 x \ \xi(x) \ \lim\limits_{\Lambda\rightarrow\infty}\sum_n\int d^4x\xi(x)(\phi^\dagger_n\gamma_5\phi_n)\exp\left(\frac{\lambda_n^2}{\Lambda^2}\right) \ ,
\eeq
which results in
\beq
i\int d^4 x \ \xi(x) \ \lim\limits_{\Lambda\rightarrow\infty}<x|\tr[\gamma_5\exp\left(-\frac{\slashed{D}^2}{\Lambda^2}\right)]|x> \ .
\eeq
Then we use another identity
\beq
\begin{gathered}
(\slashed{D})^2  \ = \  D_\mu D^\mu + \frac{g_s}{2}\sigma^{\mu\nu}T^aG^a_{\mu\nu} \ ,
\end{gathered}
\eeq
where
\beq
\sigma^{\mu\nu}  \ = \  \frac{i}{2}[\gamma^\mu, \gamma^\nu] \ ,
\eeq
and obtain
\beq
\begin{gathered}
\ln(J)  \ = \  i\int d^4x\xi(x)\lim\limits_{\Lambda\rightarrow\infty}<x|Tr\left[\gamma_5\exp\left(-\frac{-D_\mu D^\mu-\frac{g_s}{2}\sigma^{\mu\nu}T^aG^a_{\mu\nu}}{\Lambda^2}\right)\right]|x> \ .
\end{gathered}
\eeq
Only the term with four $\gamma$-matrices would be nonzero, so we can write
\beq
\begin{gathered}
\ln(J) \ = \ i\int d^4 x \ \xi(x) \ G^a_{\mu\nu}G^a_{\alpha\beta} \ \tr\left[\gamma^5\sigma^{\mu\nu}\sigma^{\alpha\beta}\right] \ \lim\limits_{\Lambda\rightarrow\infty}\frac{g_s^2}{8\Lambda^4}<x|\exp\left(-\frac{-D_\mu D^\mu}{\Lambda^2}\right)|x> \ ,
\end{gathered}
\eeq
where the trace is given by
\beq
Tr[\gamma_5\sigma_{\mu\nu}\sigma_{\alpha\beta}] \ = \ -Tr[\gamma_5\gamma_\mu\gamma_\nu\gamma_\alpha\gamma_\beta] \ = \ 4i\epsilon_{\mu\nu\alpha\beta} \ ,
\eeq
and the matrix element is dominated by
\beq
<x|\exp\left(-\frac{-D_\mu D^\mu}{\Lambda^2}\right)|x> \ = \ \int\frac{d^4k}{(2\pi)^4}\exp(\frac{k^2}{\Lambda^2}) \ .
\eeq
After a Wick rotation, it gives us
\beq
i\frac{\Lambda^4}{16\pi^2} \ ,
\eeq
so finally
\beq
\ln(J) \ = \ -i\int d^4 x \ \xi(x) \ \frac{g_s^2}{32\pi^2}G^a_{\mu\nu}G^a_{\alpha\beta} \ \epsilon^{\mu\nu\alpha\beta} \ .
\eeq
Since for fermionic variables the measure transforms as
\beq
\mathcal{D}\bar{\psi}\mathcal{D}\psi \rightarrow  J^{-2}\mathcal{D}\bar{\psi}\mathcal{D}\psi \ ,
\eeq
the axial transformation would give us an effective extra contribution to the Lagrangian of the form
\beq
\mathcal{L}_\xi  \ = \  \xi(x)\frac{g_s^2}{16\pi^2}G^a_{\mu\nu}G^a_{\alpha\beta}\epsilon^{\mu\nu\alpha\beta} \ ,
\eeq
effectively changing the value of $\theta$:
\beq
\theta \rightarrow \theta+2\xi \ .
\eeq
Since $\xi$ can be arbitrary, it means that the physics should not depend on it, i.e. the QCD partition function must not be a function of $\theta$, which solves the strong CP problem. However, this solution would be feasible only if we had at least one massless quark (namely, the up quark), which we could use to ``rotate away`` the $\theta$ parameter. This can also be seen from the value of the neutron electric dipole moment, calculated with an effective hadronic Lagrangian:
\beq
d_n\propto\theta\frac{m_um_d}{m_u+m_d} \ ,
\eeq
which vanishes for $m_u=0$.

However, lattice simulations indicate that this is not the case; at the same time, experimental measurements of the dipole moment have constrained $\theta$ to be not larger than $10^{-10}$, which makes it a ``fine-tuning`` problem~\cite{Baker:2006ts}.

One solution, known as the Peccei-Quinn mechanism~\cite{Peccei:1977hh}, is to promote $\theta$ to a dynamical field called axion, which is coupled to the topological term:
\beq
\mathcal{L}_a  \ = \  \frac{1}{2}(\partial_\mu a)^2 + \frac{a}{f_{a}}\frac{g_s^2}{32\pi^2}G^a_{\mu\nu}G^a_{\alpha\beta}\epsilon^{\mu\nu\alpha\beta} \ .
\eeq

Now, we can combine $\theta$ and $a$ into
\beq
\Theta  \ = \  \theta+\frac{a}{f_a} \ ,
\eeq
and the total Lagrangian would therefore be only a function of $\Theta$, and not of $\theta$ and $a$ separately:
\beq
\mathcal{L}_\theta+\mathcal{L}_a  \ = \  \frac{1}{2}(\partial_\mu\Theta)^2 + \frac{g_s^2}{32\pi^2}\Theta G^a_{\mu\nu}G^a_{\alpha\beta}\epsilon^{\mu\nu\alpha\beta} \ .
\eeq
Therefore, since the whole QCD Lagrangian except the topological term preserves P and CP, it means that the effective potential, calculated from quantum corrections, would have to be even in $\Theta$ and have a stationary point at $\Theta=0$.

Initially the constant $f_a$ was thought to be around the electroweak scale, which contradicted the observational bounds on the axion-photon interactions; however, Kim, Shifman, Vainshtein, and Zakharov (KSVZ) showed that $f_a$ can be made arbitrarily large, bypassing the observational constraints (the so-called ``invisible axion``)~\cite{Kim:1979if,Shifman:1979if}.
\subsection{Gravity and Dark Energy}\label{C1S2S3}

The fourth interaction, gravity, is described by the Einstein-Hilbert action~\cite{GR}:
\beq
\label{EH}
S \ = \ \frac{1}{16\pi G} \ \int d^4 x \ \sqrt{-g} \ R \ ,
\eeq
where
\beq
R  \ = \  R_{\mu\nu}g^{\mu\nu}
\eeq
is the Ricci scalar,
\beq
R_{\mu\nu}  \ = \  \partial_\alpha\Gamma^\alpha_{\mu\nu} - \partial_\nu\Gamma^{\alpha}_{\mu\alpha} + \Gamma^\alpha_{\mu\nu}\Gamma^\beta_{\alpha\beta} - \Gamma^\beta_{\mu\alpha}\Gamma^\alpha_{\nu\beta}
\eeq
is the Ricci tensor,
\beq
\Gamma^\gamma_{\alpha\beta}  \ = \  \frac{1}{2}g^{\gamma\rho}\left(\partial_\alpha g_{\beta\rho} + \partial_\beta g_{\alpha\rho} - \partial_\rho g_{\alpha\beta}\right)
\eeq
are the Christoffel symbols, and
\beq
g  \ = \  \det(g_{\mu\nu})
\eeq
is the metric determinant. Now, if we vary the sum of the Einstein-Hilbert action and the matter action over the metric $g_{\mu\nu}$, we obtain the Einstein equations:
\beq
R_{\mu\nu} - \frac{1}{2}g_{\mu\nu}R  \ = \  8\pi GT_{\mu\nu} \ ,
\eeq
where
\beq
T_{\mu\nu}  \ = \  -\frac{2}{\sqrt{-g}}\frac{\delta}{\delta g^{\mu\nu}}(\sqrt{-g}\mathcal{L}_m) \ ,
\eeq
equivalent to
\beq
T_{\mu\nu}  \ = \  -2\frac{\delta\mathcal{L}_m}{\delta g^{\mu\nu}}+g_{\mu\nu}\mathcal{L}_m \ ,
\eeq
is the stress-energy tensor, and $\mathcal{L}_m$ is the Lagrangian of matter (it can be either the Standard Model Lagrangian or the Lagrangian of some effective theory). In particular,
\beq
T_{00}  \ = \  \rho
\eeq
is the energy density, and
\beq
T_{ii}  \ = \  p
\eeq
is the pressure in isotropic solutions where $T_{ii}$ is the same for all three spatial coordinates.
\begin{figure}
\centering
\includegraphics[scale=0.5]{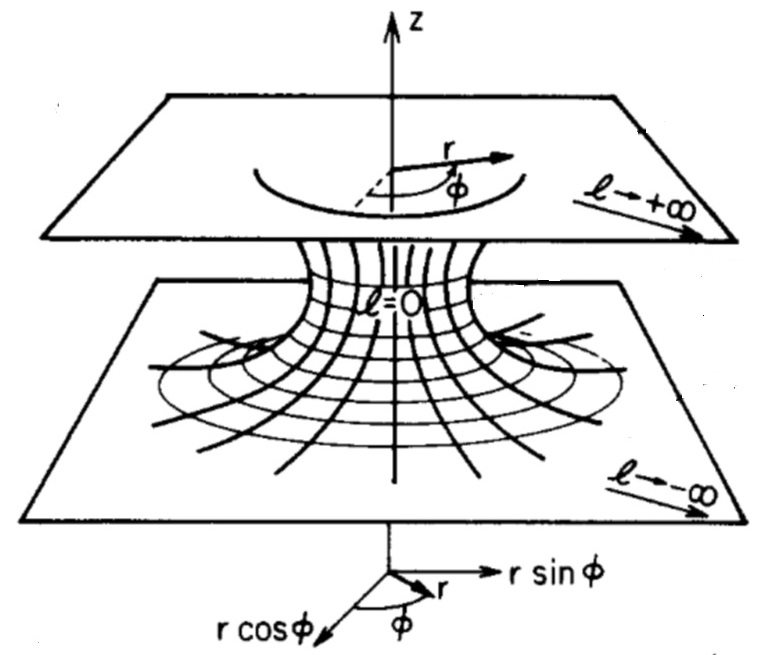}
\caption{Wormhole throat embedding}
\label{Wormhole}
\end{figure}

General relativity does not impose a priori any constraints on $T_{\mu\nu}$; therefore, by choosing matter with an ``exotic`` stress-energy tensor, it is possible to construct solutions that are usually considered pathological. One such solution is the so-called Ellis (or Morris--Thorne) wormhole which connects two disconnected regions of space, potentially allowing for closed timelike curves and violations of causality~\cite{Ellis:1973yv,Bronnikov:1973fh,Morris:1988cz}. Without loss of generality, any spherically symmetric metric can be written in the form:\\
\beq
ds^2  \ = \  -e^{2\Phi(r)}dt^2 +\left(1 - \frac{b(r)}{r}\right)^{-1}dr^2 + r^2\left(d\theta^2+\sin^2\theta d\phi^2\right) \ ,
\eeq
where $b(r)$ and $\Phi(r)$ are arbitrary functions. However, if we require that the radius varies non-monotonously, i.e. it goes from infinity to some minimal value $a$, and then back to infinity, with the intervals $(+\infty,a)$ and $(a,+\infty)$ not identified with each other, it would be the wormhole metric, connecting either two disconnected points in spacetime or two different Universes.

Now, we can take a slice of spacetime at constant time and $\theta=\frac{\pi}{2}$ and obtain a two-dimensional surface:
\beq
ds_2^2  \ = \  \left(1 - \frac{b}{r}\right)^{-1}dr^2 + r^2d\phi^2 \ ,
\eeq
which may be embedded into three-dimensional cylindrical coordinates:
\beq
ds_3^2  \ = \  dr^2 + dz^2 + r^2d\phi^2
\eeq
via the condition
\beq
\frac{dz}{dr}  \ = \  \pm\left(\frac{r}{b} - 1\right)^{-1/2} \ .
\eeq
The ``wormhole`` connects the + and - solutions, i.e. those with $z>0$ and those with $z<0$ (fig.~\ref{Wormhole}). For it to be traversable, it should connect to the flat spacetime smoothly, which means 
\beq
\frac{dr}{dz}  \ = \  0\Leftrightarrow b(a) \ = \ a \ ,
\eeq
where $a$ is the minimal size of the radius. Besides, it has to satisfy the flare-out condition:
\beq
\frac{d^2r}{dz^2}  \ = \  \frac{b - b'r}{2b^2}>0 \equiv b(r) > b'(r)r \label{W}
\eeq
for $r>a$.

Now, if we solve the Einstein equations for this metric, we see that the 00-component of the stress-energy tensor, i.e. the energy density, is given by:
\beq
\rho  \ = \  \frac{b'(r)}{8\pi r^2} \ , \label{rho}
\eeq
while the radial pressure (the rr-component of the stress--energy tensor) is:
\beq
p_r  \ = \  \frac{b/r-2(r-b)\Phi'}{8\pi r^2} \ . \label{p}
\eeq
Assuming $\Phi$ is non-singular near $r=a$, we can rewrite the flare-out condition \eqref{W} as:
\beq
\rho-p_r < 0 \ .
\eeq
To avoid these kinds of solutions, it has been proposed to supplement general relativity with the so-called null energy condition (NEC). In its simplest form, it says:\\
\beq
T_{00}  \ = \  \rho > 0 \ ,
\eeq
i.e. negative masses cannot exist. Its covariant formulation is\\
\beq
T_{\mu\nu}n^\mu n^\nu > 0 \ ,
\eeq
where $n$ is an arbitrary null vector, i.e. a vector satisfying the condition:
\beq
g_{\mu\nu}n^\mu n^\nu  \ = \  0 \ .
\eeq
Now, if we align $n$ in the radial direction:
\beq
n \ = \ (1,1,0,0) \ ,
\eeq
NEC would basically state that the flare-out condition \eqref{W} is wrong, and Morris-Thorne wormholes cannot be traversable.

For a homogeneous isotropic Universe, we would have a Friedmann-Robertson-Walker (FRW) metric of the form
\beq \label{FRW}
ds^2  \ = \  -dt^2+a^2(t)d\vec{r}^2 \ ,
\eeq
and the stress-energy tensor components would be
\beq
T_{00}  \ = \  \rho \ ,
\eeq
\beq
T_{ij}  \ = \  a^2g_{ij}p \ .
\eeq
Now, if we choose
\beq
n^0 \ = \ 1 \ ,
\eeq
the spatial components should have the form
\beq
n_i  \ = \  a^{-1}\nu_i \ ,
\eeq
with
\beq
g^{ij}\nu_i\nu_j  \ = \  1 \ ,
\eeq
and the null energy condition for this vector would be
\beq
\rho+p > 0 \ .
\eeq
The Einstein equations for the FRW metric, known as the Friedmann equations, have the form
\beq \label{FE1}
H^2  \ = \  \frac{8\pi G}{3}\rho - \frac{\kappa}{a^2} \ ,
\eeq
\beq \label{FE2}
2\dot{H}+3H^2  \ = \  -8\pi Gp-\frac{\kappa}{a^2} \ ,
\eeq
where
\beq
H \ = \ \frac{\dot{a}}{a}
\eeq
is the Hubble parameter, and $\kappa$ is the curvature of the Universe (if it is nonzero, it can be normalized to $\pm1$; the cases of $k=+1,0,-1$ are known as the closed, flat, and open universe respectively). Eq.~\eqref{FE1} can also be rewritten in the form
\beq
\Omega+\Omega_k \ = \ 1 \ ,
\eeq
where
\beq \label{Omega}
\Omega \ = \ \frac{\rho}{\rho_c} \ , \ \Omega_k \ = \ - \frac{\kappa}{a^2H^2}
\eeq
are the contributions of the matter density (both non--relativistic matter and radiation) and of the curvature to the critical density of the Universe:
\beq
\rho_c \ = \ \frac{3H^2}{8\pi G} \ .
\eeq
Substituting~\eqref{FE1} into~\eqref{FE2}, we obtain
\beq
\dot{H}  \ = \  -4\pi G(\rho + p) + \frac{\kappa}{a^2} \ .
\eeq
From NEC, it follows that the first term on the right-hand side is always negative, and therefore in most cases (for zero and negative curvature, and sometimes even for positive curvature) the Hubble rate can only decrease during the evolution of the Universe, which rules out the so-called Big Bounce models and time-symmetric solutions, in which the Universe starts expanding after a phase of contraction.

In addition, the covariant stress-energy conservation law is
\beq
\nabla_\mu T^{\mu\nu}  \ = \  \dot{\rho} + 3H(\rho + p) \ = \ 0 \ ,
\eeq
and if the second term is always positive, this means that $\dot{\rho}$ must always be negative, i.e. the density of the Universe always decreases during its evolution.\\
Now, assuming Minkowski metric with a small perturbation:
\beq
g_{\mu\nu}  \ = \  \eta_{\mu\nu} + h_{\mu\nu} \ ,
\eeq
we can derive the Einstein equations to the first order (the so-called linearized gravity):
\beq
\begin{gathered}
\frac{1}{2}\left(\partial_\gamma\partial_\mu h^\gamma_\nu+\partial_\gamma\partial_\nu h^\gamma_\mu-\square h_{\mu\nu}-\partial_\mu\partial_\nu h-\eta_{\mu\nu}\partial_\alpha\partial_\beta h^{\alpha\beta}+\eta_{\mu\nu}\square h\right) \ = \ 8\pi GT_{\mu\nu} \ .
\end{gathered}
\eeq
However, the solutions are not unique because the gauge transformation
\beq\label{DA}
h_{\mu\nu}\rightarrow h_{\mu\nu}-\partial_\mu\xi_\nu-\partial_\nu\xi_\mu \ , 
\eeq
leaves the equations invariant at first order. A more generic form of this expression for non--flat backgrounds may be derived by considering the coordinate transformation
\beq\label{CoordTransform}
x^\mu \ \rightarrow \ x^\mu+\xi^\mu \ .
\eeq
In general, a coordinate transformation of the metric is given by
\beq
g'_{\mu\nu}(x') \ = \ \frac{\partial x^\alpha}{\partial x'^\mu}\frac{\partial x^\beta}{\partial x'^\nu}g_{\alpha\beta}(x) \ ,
\eeq
and substituting~\eqref{CoordTransform}, one obtains at first order in $\xi$:
\beq
g'_{\mu\nu} \ = \ g_{\mu\nu}-g_{\alpha\nu}\partial_\mu\xi^\alpha-g_{\mu\beta}\partial_\nu\xi^\beta-\xi^\alpha\partial_\alpha g_{\mu\nu} \ ,
\eeq
or, equivalently,
\beq
g'_{\mu\nu} \ = \ g_{\mu\nu}-\nabla_\mu\xi_\nu-\nabla_\nu\xi_\mu \ .
\eeq
For Minkowski metric, this is exactly~\eqref{DA}. 

We can choose a gauge (the so-called de Donder gauge) such that
\beq
\partial_\mu h^\mu_\nu-\frac{1}{2}\partial_\nu h  \ = \  \partial_\mu(\bar{h}^\mu_\nu) \ = \ 0 \ ,
\label{Q}
\eeq
where
\beq
\bar{h}_{\mu\nu}  \ = \  h_{\mu\nu} - \frac{1}{2}\eta_{\mu\nu}h \ ,
\eeq
or, equivalently,
\beq
h_{\mu\nu}  \ = \  \bar{h}_{\mu\nu} - \frac{1}{2}\eta_{\mu\nu}\bar{h} \ .
\eeq
This is achieved by choosing the vector $\xi$ which satisfies:
\beq
\square\xi_\mu  \ = \  \partial_\mu(\bar{h}_\nu^\mu) \ .
\eeq
In this gauge, the Einstein equations simplify to:
\beq
-\square\bar{h}_{\mu\nu} \ = \ 16\pi GT_{\mu\nu} \ .
\eeq
In absence of source, the equation is just the wave equation:
\beq
\square\bar{h}_{\mu\nu} \ = \ 0 \ ,
\eeq
with solutions of the form
\beq
\bar{h}_{\mu\nu} \ = \ \epsilon_{\mu\nu}e^{ik_\mu x^\mu} \ ,
\eeq
with $k_\mu k^\mu=0$, i.e. gravitational waves propagate at the speed of light. As a symmetric matrix, $\epsilon_{\mu\nu}$ has 10 independent components. The Lorentz gauge condition gives the constraints
\beq
k^\mu\epsilon_{\mu\nu} \ = \ 0 \ .
\eeq
$\nu$ runs from 1 to 4, so we have 4 constraints, and therefore 6 degrees of freedom left. Finally, there is the residual gauge freedom due to the fact that we can shift $h_{\mu\nu}$ by $\xi$ without breaking the Lorentz gauge condition~\eqref{Q} if $\xi$ satisfies:
\beq
\square\xi_\mu \ = \ 0 \ .
\eeq
Then, if $\xi$ is chosen as
\beq
\xi_\mu \ = \ C_\mu e^{ik_\mu x^\mu} \ ,
\eeq
$A_{\mu\nu}$ can be made transverse to the time vector:
\beq
k^\mu A_{\mu0}\rightarrow k^\mu\left(A_{\mu0}+ik_\mu C_0+ik_0C_\mu\right) \ = \ 0 
\eeq
by requiring
\beq
(C_\mu k^\mu) \ = \ \frac{i}{2}\epsilon_\gamma^\gamma \ .
\eeq
This condition comprises four equations, which can be satisfied by choosing the four components of $C_\mu$. Therefore, if $h_{0\mu}=0$, the 0-component of the Lorentz condition gives us
\beq
h \ = \ 0 \ ,
\eeq
and therefore
\beq
h_{\mu\nu} \ = \ \bar{h}_{\mu\nu} \ .
\eeq
This gauge is known as the transverse-traceless (TT) gauge. Now, given the two gauge conditions, we have $10-4-4=2$ degrees of freedom. If we orient $\vec{k}$ alongside the z axis:
\beq
k \ = \ \left(w,0,0,w\right) \ ,
\eeq
these degrees would be $h_{xx}=-h_{yy}=h_+$ and $h_{xy}=h_{yx}=h_\times$, i.e. 
\beq
\epsilon_{\mu\nu} \ = \ \begin{pmatrix}
	0 & 0 & 0 & 0 \\
	0 & h_+ & h_\times & 0 \\
	0 & h_\times & -h_+ & 0 \\
	0 & 0 & 0 & 0
\end{pmatrix} \ .
\eeq
Using analogy with electromagnetism, one is led to expect that gravitational waves result from streams of spin-2 massless particles called gravitons, and the Minkowski metric would therefore be the vacuum expectation value of the gravitational field. Now, if we absorb the Planck mass into the metric:
\beq
h_{\mu\nu}\rightarrow M_{Pl}h_{\mu\nu} \ ,
\eeq
it is possible to write the effective action
\beq\label{EffGR}
\begin{gathered}
S_{eff} \ = \ \frac{1}{64\pi}\int d^4 x \ \left[-(\partial_\alpha h_{\mu\nu})^2+O(\frac{1}{M_P}h^3)\right]+\frac{1}{M_P}\int d^4x \ h^{\mu\nu}T_{\mu\nu} \ ,
\end{gathered}
\eeq
and quantize it using covariant quantization techniques. However, this theory is nonrenormalizable. Just like the effective Fermi theory, it has a dimensionful coupling constant $M^{-1}_{P}$, but its symmetries eliminate completely the divergences at one loop~\cite{tv}. However, at two loops a divergence shows up~\cite{gs}. In addition, the Lagrangian is non-polynomial, so it contains an infinite series of terms with growing powers of $M^{-1}_P$, and would require infinitely many counterterms of the form $R^2, R_{\mu\nu}R^{\mu\nu}$, etc.

Another discrepancy between General Relativity and the Standard Model is the breaking of the vacuum energy shift symmetry. The SM fields have the zero-point energy equal to
\beq \label{VE}
\rho \ = \ \sum_i\int\frac{d^3k}{(2\pi)^3} \ \frac{1}{2}\hbar w_i \sim M_{Pl}L^{-3}_{Pl} \ .
\eeq
The existence of vacuum energy is evidenced by the Casimir effect, but since in quantum theory only transition energies between states are observable, the zero-point energy is not physically relevant, and we can shift the ground state energy by an arbitrary constant. In General Relativity, however, this is not the case. Any vacuum energy gives a contribution to the matter action of the form:
\beq
\mathcal{L}_\Lambda  \ = \  \Lambda\int d^4x \ \sqrt{-g} \ .
\eeq
Because of the multiplier $\sqrt{-g}$, it would yield a nonzero contribution to the stress-energy tensor:
\beq
T_{\mu\nu}  \ = \  \Lambda \ g_{\mu\nu} \ ,
\eeq
and the Einstein equations
\beq
R_{\mu\nu}-\frac{1}{2}g_{\mu\nu}R  \ = \  8\pi G\Lambda g_{\mu\nu}
\eeq
no longer have a Minkowski solution with $R_{\mu\nu}=0$. Instead, the spherically symmetric solution for $\Lambda>0$ would be \emph{the de Sitter (dS) universe}: its metric has the FRW form~\eqref{FRW}, with
\beq\label{dSmetric}
a(t) \ \propto \ e^{\sqrt{\frac{8\pi G}{3}\Lambda}t} \ .
\eeq
Likewise, for a negative $\Lambda$ we would obtain a solution known as \emph{the anti-de Sitter (AdS) universe}, with the metric:
\beq\label{AdSmetric}
-(1+\frac{r^2}{L^2})dt^2 \ + \ (1+\frac{r^2}{L^2})^{-1}dr^2 \ + \ r^2d\Omega^2 ,
\eeq
where $L$ is the AdS radius, yielded by
\beq
L \ = \ \sqrt{-\frac{3}{8\pi G\Lambda}} \ .
\eeq
Technically, D--dimensional dS and AdS spacetimes can be seen as hypersurfaces embedded in a (D+1)--dimensional spacetime whose metric is given by:
\beq
ds^2 \ = \ -d\tau^2+d\vec{x}^2\pm dz^2 \ ,
\eeq
and the embedding condition is
\beq
-\tau^2+\vec{x}^2\pm z^2 \ = \ -L^2 \ .
\eeq
The plus sign in front of $z^2$ corresponds to positive $\Lambda$ and negative $L^2$ (dS), and the minus sign corresponds to negative $\Lambda$ and positive $L^2$ (AdS). For dS, the parametrization
\beq
\begin{gathered}
\tau \ = \ \sqrt{-L^2}\sinh\left(\frac{t}{\sqrt{-L^2}}\right)+\frac{r^2}{2\sqrt{-L^2}}e^{\frac{t}{\sqrt{-L^2}}} \ ,\\ 
\ z \ = \ \sqrt{-L^2}\cosh\left(\frac{t}{\sqrt{-L^2}}\right)-\frac{r^2}{2\sqrt{-L^2}}e^{\frac{t}{\sqrt{-L^2}}} \ , \ x_i \ = \ e^{\frac{t}{\sqrt{-L^2}}}\xi_i \ , 
\end{gathered}
\eeq
with $\sum_i\xi_i^2=r^2$, yields the FRW metric with the scale factor given by~\eqref{dSmetric}, while for AdS, the parametrization
\beq
\tau \ = \ \sqrt{L^2-r^2}\cos\left(\frac{t}{L}\right) \ , \ z \ = \ \sqrt{L^2-r^2}\sin\left(\frac{t}{L}\right) \ , \ x_i \ = \ r\xi_i \ , 
\eeq
with $\sum_i\xi_i^2=1$, produces the metric~\eqref{AdSmetric}.

An even more puzzling fact is that the value of $\Lambda$, estimated from Planck data, is actually positive (as one would expect in the vacuum energy scenario), but its value is around 70\% of the critical density of the Universe, i.e. about 120 orders smaller than~\eqref{VE} (it is known as ``dark energy``)~\cite{PLANCK}. If taken at face value, this result would imply extreme fine--tuning: the (negative) cosmological constant, which should be added by hand to the Einstein equations to cancel the vacuum energy contribution, would have to be \emph{almost} equal to the vacuum energy, but have a difference at order $10^{-120}$.

We can also consider a more generic class of actions for GR, known as $f(R)$ gravity:
\beq
\label{fR}
\mathcal{S} \ = \ -\frac{1}{16\pi G} \ \int \ d^4x \ \sqrt{-g} \ f(R) \ 
\eeq
(other scalars constructed from the Riemann and Ricci tensors, such as $R_{\mu\nu}R^{\mu\nu},$ should not be present in the action because they would produce the Ostrogradsky instability~\cite{Ostrogradsky:1850fid,Woodard:2006nt,Stelle:1977ry}). Performing the conformal rescaling of the metric:
\beq
\tilde{g}_{\mu\nu} \ = \ f'(R)g_{\mu\nu} \ = \ e^{\sqrt{\frac{2}{3}\kappa}\phi}g_{\mu\nu} \ ,
\eeq
with $\kappa=8\pi G$, we can rewrite the action~\eqref{fR} in the form
\beq
\mathcal{S} \ = \ - \int \ d^4x \ \sqrt{-\tilde{g}} \ \left(\frac{1}{2\kappa}\tilde{R}\ + \ \frac{1}{2}\tilde{g}^{\mu\nu}\partial_\mu\phi\partial_\nu\phi \ + \ U(\phi) \right) \ ,
\eeq
where
\beq
U(\phi) \ = \ \frac{Rf' \ - \ f}{2\kappa f'^2} \ .
\eeq
Since the matter Lagrangian would still couple to the ``old`` metric $g_{\mu\nu}$, it would effectively interact with $\phi$:
\beq
\mathcal{S}_m \ = \ \int \ d^4x \ \sqrt{-e^{-\sqrt{\frac{2}{3}\kappa}\phi}\tilde{g}} \ \mathcal{L}_m(e^{-\sqrt{\frac{2}{3}\kappa}\phi}\tilde{g}) \ .
\eeq
This means that $f(R)$ gravity is dual to standard Einstein gravity with an additional scalar field~\cite{Sotiriou:2008rp}.
\subsection{Supersymmetry, Minimal Supersymmetric Standard Model (MSSM), and Supergravity}\label{C1S2S4}
\label{SUSY}
One way to resolve the vacuum energy problem is to supplement the Standard Model with an additional symmetry, known as supersymmetry (SUSY), which interchanges bosonic and fermionic fields~\cite{Gervais:1971ji,Volkov:1972jx,Golfand:1971iw}. According to the Coleman--Mandula ``no-go`` theorem, a quantum field theory can have no symmetry group with a Lie algebra larger than $ISO(1,3)\otimes G$, where $ISO(1,3)$ is the Poincare group, and $G$ is the internal symmetry group ($SU(3)\otimes SU(2)\otimes U(1)$ in the case of the Standard Model). This means that spacetime symmetries and internal symmetries cannot be combined in any way~\cite{Coleman:1967ad} by ordinary generators. However, this restriction may be bypassed if the Lie algebra is substituted by a Lie superalgebra, which includes both commutation and anticommutation relations. Namely, one can introduce a number of generators $(Q^I)^i_\alpha$, known as supercharges, and their Dirac conjugates $(\bar{Q}^I)^i_\alpha$, where $i$ is the spinor index, $\alpha$ is the index taking values 1 and 2, and $I$ is the supercharge number (a theory can have more than one supersymmetry). Since these generators are spinors, it means that by acting on bosonic (i.e. scalar, vector or tensor) states, they would transform them into fermionic (i.e. spinor) states, and vice versa:
\beq
\begin{gathered}
Q^I_\alpha|B>\rightarrow|F> \ , \\
\bar{Q}^I_\alpha|F>\rightarrow|B> \ .
\end{gathered}
\eeq
In four dimensions, the supercharges satisfy the following anticommutation properties with each other:
\beq
\{Q_\alpha^I,\bar{Q}_\beta^J\} \ = \ 2\delta^{IJ}\sigma^\mu_{\alpha\beta}P_\mu \ , \label{D}
\eeq
\beq
\{Q^I_\alpha,Q^J_\beta\} \ = \ \epsilon_{\alpha\beta}Z^{IJ} \ ,
\eeq
\beq
\{\bar{Q}^I_\alpha,\bar{Q}^J_\beta\} \ = \ \epsilon_{\alpha\beta}(Z^{IJ})^* \ ,
\eeq
and the following commutation properties with the Poincare group generators:
\beq
[Q^I_\alpha,P_\mu] \ = \ 0 \ ,
\eeq
\beq
[\bar{Q}^I_\alpha,P_\mu] \ = \ 0 \ ,
\eeq
\beq
[M_{\mu\nu},Q^I_\alpha] \ = \ i(\sigma_{\mu\nu})_\alpha^\beta Q_\beta^I \ ,
\eeq
\beq
[M_{\mu\nu},\bar{Q}^I_\alpha] \ = \ i(\bar{\sigma}_{\mu\nu})_\alpha^\beta \bar{Q}_\beta^I \ .
\eeq
In these relations, $\sigma_{\mu\nu}$ is defined as
\beq
\sigma_{\mu\nu} \ = \ \frac{1}{4}\left[\sigma_\mu,\sigma_\nu\right] \ , \ \bar{\sigma}_{\mu\nu} \ = \ \frac{1}{4}\left[\bar{\sigma}_\mu,\bar{\sigma}_\nu\right] \ ,
\eeq
and $Z^{IJ}$ is the so-called central charge with the antisymmetry property:
\beq
Z^{JI} \ = \ -Z^{IJ} \ .
\eeq
In the simplest variant of SUSY, the so-called $N=1$ SUSY (N denotes the number of supercharges), there is only one supercharge $Q_\mu$ and its Dirac conjugate $\bar{Q}_\alpha$, and the central charge is zero. We can also use the four--component spinor notation, in which:
\beq
\{Q,\bar{Q}\} \ = \ 2\gamma^\mu P_\mu \ , \ \left[Q,M_{\mu\nu}\right] \ = \ \frac{i}{2}\gamma_{\mu\nu}Q \ ,
\eeq
with
\beq
\gamma_{\mu\nu} \ = \ \frac{1}{2}\left[\gamma_\mu,\gamma_\nu\right] \ .
\eeq
The vacuum state is annihilated by both supercharges:
\beq
Q^I_\alpha|0> \ = \ \bar{Q}^I_\alpha|0> \ = \ 0 \ ; \label{E}
\eeq
therefore, if we take the vacuum average of both sides of \eqref{D}, we obtain
\beq
<0|P_\mu|0> \ = \ 0 \ .
\eeq
This means that in a supersymmetric theory, the energy and momentum of the ground state should be zero, if computed with the prescription determined by the algebra; this is explained by the fact that for any bosonic loop diagram contributing to the vacuum energy, there is a fermionic diagram of the same magnitude but of the opposite sign.

The extension of Standard Model with $N=1$ SUSY is known as minimal supersymmetric Standard Model (MSSM)~\cite{Fayet:1977yc}. It contains scalar superpartners of leptons and quarks known as sleptons and squarks (due to the chirality of the weak interaction, left-handed and right-handed states have separate superpartners) and spin-$\frac{1}{2}$ counterparts of gauge fields called gluinos (for gluons), winos, and bino (for W- and B-bosons). Besides, instead of a single Higgs doublet, we should have two of them, and therefore two superpartners (Higgsinos). There are two reasons for this: the first one is gauge anomaly cancellation, and the second is that in supersymmetric models, we cannot have Yukawa couplings to $\epsilon^{ab}H^b$, so we need two separate Higgs doublets to give masses to up- and down-type quarks.

The simplest way to write the Lagrangian of MSSM is to introduce superspace, i.e. to supplement the four spacetime coordinates with the so-called Grassmann coordinates $\theta_\alpha$ and $\bar{\theta}_{\dot{\alpha}}$, with $\alpha$ and $\dot{\alpha}$ taking values 1 and 2:
\beq
(x_\mu,\theta_\alpha,\bar{\theta}_{\dot{\alpha}}) \ .
\eeq
All Grassmann variables anticommute with each other:
\beq
\begin{gathered}
\{\theta_\alpha,\theta_\beta\} \ = \ \{\theta_\alpha,\bar{\theta}_{\dot{\alpha}}\} \ = \{\bar{\theta}_{\dot{\alpha}},\bar{\theta}_{\dot{\beta}}\} \ = \ 0 \ ,
\end{gathered}
\eeq
and as a result they are nilpotent:
\beq
\begin{gathered}
\theta_\alpha\theta_\alpha \ = \ \bar{\theta}_{\dot{\alpha}}\bar{\theta}_{\dot{\alpha}} \ = \ 0
\end{gathered}
\eeq
for each $\alpha$ and $\dot{\alpha}$ (here we are not summing over them).

In the following, we consider an algebra of four Grassmann variables ($\theta^{1,2}$, $\bar{\theta}^{1,2}$). In this case, products of more than two variables ($\theta_\alpha\theta_\beta\theta_\gamma...$ and $\bar{\theta}_{\dot{\alpha}}\bar{\theta}_{\dot{\beta}}\bar{\theta}_{\dot{\gamma}}...$) would vanish, since at least two of the indices $\alpha, \beta, \gamma$ has to have the same value.

This means that an arbitrary function of superspace coordinates (a superfield) may be written as:
\beq
\begin{gathered}
F(x,\theta,\bar{\theta}) \ = \ f(x)+\theta\psi(x)+\bar{\theta}\bar{\chi}(x)+\theta^2\alpha(x)+\bar{\theta}^2\beta(x)+\\
\theta\sigma^\mu\bar{\theta}v_\mu(x)+\theta^2\bar{\theta}\bar{\phi}(x)+\bar{\theta}^2\theta\xi(x)+\theta^2\bar{\theta}^2\rho(x) \ ,
\end{gathered}
\eeq
where $f(x), \alpha(x), \beta(x), \rho(x)$ are scalars, $\psi(x), \bar{\chi}(x), \bar{\phi}(x), \xi(x)$ are spinors, and $v_\mu(x)$ is a vector, and
\beq
\begin{gathered}
\theta\theta \ = \ \epsilon^{\alpha\beta}\theta_\alpha\theta_\beta \ = \ \theta^\alpha\theta_\alpha \ , \\
\bar{\theta}\bar{\theta} \ = \ \epsilon^{\dot{\beta}\dot{\alpha}}\bar{\theta}_{\dot{\alpha}}\bar{\theta}_{\dot{\beta}} \ = \ \bar{\theta}_{\dot{\alpha}}\bar{\theta}^{\dot{\alpha}}
\end{gathered}
\eeq
are the scalar products.

The integrals over Grassmann variables, known as the Berezin integrals, are determined demanding translational invariance, and are given by
\beq
\begin{gathered}
\int d^2\theta \ = \ \int d^2\bar{\theta} \ = \ 0 \ ,\\
\int d^2\theta\theta\theta \ = \ \int d^2\bar{\theta}\bar{\theta}\bar{\theta} \ = \ 1 \ . \\
\end{gathered}
\eeq
The supercharges can be defined as differential operators in superspace, just like the momentum is the translation operator in ordinary space:
\beq
\begin{gathered}
Q_\alpha \ = \ i\partial_\alpha+\sigma^\mu_{\alpha\dot{\alpha}}\bar{\theta}^{\dot{\alpha}}\partial_\mu \ ,\\
\bar{Q}_{\dot{\alpha}} \ = \ -i\bar{\partial}_{\dot{\alpha}}-\theta^\alpha\sigma^\mu_{\alpha\dot{\alpha}}\partial_\mu \ .
\end{gathered}
\eeq
However, a derivative of a superfield over Grassmann variables would not be a superfield itself, since the supersymmetry transformations do not commute with derivation:
\beq
\left[\delta_{\epsilon,\bar{\epsilon}},\partial_\alpha\right]\neq0\neq\left[\delta_{\epsilon,\bar{\epsilon}},\partial_{\dot{\alpha}}\right] \ .
\eeq
This means we should introduce the ``SUSY covariant derivatives``:
\beq
\begin{gathered}
D_\alpha \ = \ \partial_\alpha+i\sigma^\mu_{\alpha\dot{\alpha}}\bar{\theta}^{\dot{\alpha}}\partial_\mu \ ,\\
\bar{D}_{\dot{\alpha}} \ = \ \bar{\partial}_{\dot{\alpha}}+i\theta^\alpha\sigma^\mu_{\alpha\dot{\alpha}}\partial_\mu \ ,
\end{gathered}
\eeq
whose anticommutator is given by
\beq
\begin{gathered}
\{D_{\alpha},\bar{D}_{\dot{\beta}}\} \ = \ 2i\sigma^\mu_{\alpha\dot{\beta}}\partial_\mu \ ,
\end{gathered}
\eeq
and which anticommute with the supercharges:
\beq
\begin{gathered}
\{D_\alpha,Q_\beta\} \ = \ \{D_\alpha,\bar{Q}_{\dot{\beta}}\} \ = \ 
\{\bar{D}_{\dot{\alpha}},Q_\beta\} \ = \ \{\bar{D}_{\dot{\alpha}},\bar{Q}_{\dot{\beta}}\} \ = \ 0 \ .
\end{gathered}
\eeq
Therefore, since $\epsilon$ and $\bar{\epsilon}$ also anticommute with $D$ and $\bar{D}$ (because they anticommute with $\theta, \bar{\theta}, \partial, \bar{\partial}$), $\epsilon Q$ and $\bar{\epsilon}\bar{Q}$ would commute with $D$ and $\bar{D}$. This means that 
\beq
[D_\alpha,\delta_{\epsilon,\bar{\epsilon}}] \ = \ [\bar{D}_{\dot{\alpha}},\delta_{\epsilon,\bar{\epsilon}}] \ = \ 0 \ ,
\eeq
and $D_\alpha Y$ and $\bar{D}_{\dot{\alpha}}Y$ are also superfields.

Now, if we integrate Y over superspace, we get a function $\mathcal{L}$ in the real space:
\beq
\mathcal{L}(x) \ = \ \int Y(x,\theta,\bar{\theta})d^2\theta d^2\bar{\theta} \ ,
\eeq
which transforms under SUSY as:
\beq
\delta_{\epsilon,\bar{\epsilon}}\mathcal{L} \ = \ \int\left(i(\theta\sigma\bar{\epsilon}-\epsilon\sigma\bar{\theta})^\mu\partial_\mu+\epsilon^\alpha\partial_\alpha+\bar{\epsilon}^\beta\bar{\partial}_\beta\right)Yd^2\theta d^2\bar{\theta} \ .
\eeq
The last two terms vanish because they would have no more than three $\theta$ and $\bar{\theta}$ (i.e. the highest-order terms would be $\theta^2\bar{\theta}$ and $\theta\bar{\theta}^2$), while the nonvanishing terms would need to have $\theta^2\bar{\theta}^2$. The first term is a total derivative and vanishes upon integration over $d^4x$; therefore $\mathcal{L}$ is supersymmetric by construction. This means that if we define Y as a ``superfield Lagrangian``, i.e. a functional of superfields: 
\beq
Y\left(\phi(x,\theta,\bar{\theta}),\psi(x,\theta,\bar{\theta}),V_\mu(x,\theta,\bar{\theta})\right) \ ,
\eeq
then $\mathcal{L}$ would be a supersymmetric Lagrangian.

One particular type of superfield is a chiral superfield, defined by the condition
\beq
\bar{D}_{\dot{\alpha}}\Phi \ = \ 0 \ .
\eeq
It is convenient to introduce new coordinates in superspace: 
\beq
y^\mu_{\pm} \ = \ x^\mu\pm i\theta^\alpha\sigma^\mu_{\alpha\dot{\alpha}}\bar{\theta}^{\dot{\alpha}} \ , \label{R}
\eeq
and since
\beq
\begin{gathered}
\bar{D}_{\dot{\alpha}}y^\mu_+ \ = \ \bar{D}_{\dot{\alpha}}\theta^\alpha \ = \ 0 \ ,
\end{gathered}
\eeq
a chiral superfield can be defined as an arbitrary function of $y_+$ and $\theta$: 
\beq
\begin{gathered}
\Phi \ = \ \phi(y_+)+\sqrt{2}\theta^\alpha\psi_\alpha(y_+)+\theta\theta F(y_+) \ ,
\end{gathered}
\eeq
which gives us:
\beq
\begin{gathered}
\Phi \ = \ \phi(x)+i\theta^\alpha\sigma^\mu_{\alpha\dot{\alpha}}\bar{\theta}^{\dot{\alpha}}\partial_\mu\phi(x)-\frac{1}{4}\theta\theta\bar{\theta}\bar{\theta}\square\phi+\sqrt{2}\theta\psi(x)-\frac{i}{\sqrt{2}}\theta^2\partial_\mu\psi\sigma^\mu\bar{\theta}+\theta\theta F(x) \ . \label{Ch}
\end{gathered}
\eeq
Likewise, we can define an anti-chiral field $\bar{\Phi}$, which satisfies the condition
\beq
D_\alpha\bar{\Phi} \ = \ 0 \ ,
\eeq
and is therefore a function of $y^-$ and $\bar{\theta}$:
\beq
\begin{gathered}
\bar{\Phi} \ = \ \bar{\phi}(y^-)+\sqrt{2}\bar{\theta}_{\dot{\alpha}}\bar{\psi}^{\dot{\alpha}}(y^-)+\bar{\theta}\bar{\theta}\bar{F}(y^-) \ ,
\end{gathered}
\eeq
i.e.
\beq
\bar{\Phi} \ = \ \bar{\phi}(x)-i\theta^\alpha\sigma^\mu_{\alpha\dot{\alpha}}\bar{\theta}^{\dot{\alpha}}\partial_\mu\bar{\phi}(x)-\frac{1}{4}\theta\theta\bar{\theta}\bar{\theta}\square\bar{\phi}+\sqrt{2}\bar{\theta}\bar{\psi}(x)+\frac{i}{\sqrt{2}}\bar{\theta}^2\theta\sigma^\mu\partial_\mu\bar{\psi}(x)+\bar{\theta}\bar{\theta}\bar{F}(x) \ .
\eeq
Due to the chain rule for derivatives, if $\Phi$ is a chiral superfield, then an arbitrary function $\Lambda(\Phi)$, which is holomorphic: 
\beq
\frac{\partial\Lambda}{\partial\bar{\Phi}} \ = \ 0 \ ,
\eeq
is also chiral:
\beq
\bar{D}_{\dot{\alpha}}\Lambda \ = \ 0 \ .
\eeq
The kinetic Lagrangian for a number of chiral superfields $\Phi^i$ is given by the so-called Kähler potential $K(\Phi^i,\bar{\Phi}^i)$:
\beq
\mathcal{L} \ = \ \int d^2\theta d^2\bar{\theta}K\left(\Phi^i,\bar{\Phi}^i\right) \ .
\eeq
Since $\mathcal{L}$ corresponds to the $\theta\theta\bar{\theta}\bar{\theta}$-part of K, K should not depend on $D_\alpha\Phi$ and $\bar{D}_{\dot{\alpha}}\bar{\Phi}$ to avoid higher derivatives of fields in the Lagrangian and resulting nonlocality. In addition, it should be real, to guarantee the reality of the Lagrangian: 
\beq
\bar{K}(\bar{\Phi}^i,\Phi^i) \ = \ K(\Phi^i,\bar{\Phi}^i) \ .
\eeq
In the simplest model with just one chiral field $\Phi$, these conditions imply that $K$ can be expressed as a polynomial of the form:
\beq
K(\Phi,\bar{\Phi}) \ = \ \sum^\infty_{m,n=1}c_{mn}\bar{\Phi}^m\Phi^n \ ,
\eeq
with 
\beq
c^*_{mn} \ = \ c_{nm} \ .
\eeq
The Lagrangian is invariant under the following transformation of the Kähler potential:
\beq\label{KT}
K(\Phi,\bar{\Phi})\rightarrow K(\Phi,\bar{\Phi})+\Lambda(\Phi)+\bar{\Lambda}(\bar{\Phi}) \ ,
\eeq
where $\Lambda(\Phi)$ is a holomorphic function of $\Phi$. This symmetry is due to the fact that $\Lambda$ is itself a chiral field (as was previously proven), and therefore its $\theta^2\bar{\theta}^2$ component is a total derivative; therefore polynomial terms with $m=0$ or $n=0$ are irrelevant. 

Now, from \eqref{R} we see that $\theta$ and $\bar{\theta}$ have mass dimension $[M]=-\frac{1}{2}$ (since $x^\mu$ has $[M]=-1$). Since the Lagrangian has $[M]=4$, it means that the Kähler potential $K\propto\theta^2\bar{\theta}^2\mathcal{L}$ would have $[M]=4-4*\frac{1}{2}=2$. At the same time, $\Phi$ has $[M]=1$, since $\phi$, as a scalar, has $[M]=1$, and $\psi$, as a spinor, has $[M]=\frac{3}{2}$; this means that the coefficients $c_{mn}$ have $[M]=2-m-n$. Renormalizability requires that the coefficients should not have negative mass dimension ($[M]\ge0$), which means that the polynomial has only one term with $m=n=1$:
\beq
K(\Phi,\bar{\Phi}) \ = \ \bar{\Phi}\Phi \ .
\eeq
Upon integration, the $\theta^2\bar{\theta}^2$-part of K gives us the kinetic terms plus the $F^2$-term:
\beq
\begin{gathered}
\mathcal{L}_K \ = \ \partial_\mu\phi\partial^\mu\phi^*+\frac{i}{2}\left(\partial_\mu\psi\sigma^\mu\bar{\psi}-\psi\sigma^\mu\partial_\mu\bar{\psi}\right)+\bar{F}F \ .
\end{gathered}
\eeq
However, we do not get the self-interactions and Yukawa couplings for $\phi$ and $\psi$; to obtain them, we needs to introduce the additional term $W(\Phi)$, known as the superpotential. Since $W$ is holomorphic, it should be integrated only over half superspace and supplemented with a Hermitian conjugate to satisfy the reality condition:
\beq
\mathcal{L}_W \ = \ \int d^2\theta W(\Phi)+\int d^2\bar{\theta}\bar{W}(\bar{\Phi}) \ .
\eeq
From the chiral superfield decomposition \eqref{Ch}, we see that the $\theta\theta$-component of the superpotential is given by:
\beq
W(\Phi)|_{\theta\theta} \ = \ \left(\frac{\delta W}{\delta\phi}F-\frac{1}{2}\frac{\delta^2W}{\delta\phi^2}\psi\psi\right)\theta\theta \ ,
\eeq
and the complete Lagrangian would therefore be given by:
\beq
\begin{gathered}
\mathcal{L} \ = \ \partial_\mu\phi\partial^\mu\phi^*+\frac{i}{2}\left(\partial_\mu\psi\sigma^\mu\bar{\psi}-\psi\sigma^\mu\partial_\mu\bar{\psi}\right)+\bar{F}F+\\
\frac{\delta W}{\delta\phi}F-\frac{1}{2}\frac{\delta^2W}{\delta\phi^2}\psi\psi+\frac{\delta \bar{W}}{\delta\phi^*}\bar{F}-\frac{1}{2}\frac{\delta^2\bar{W}}{\delta\phi^{*2}}\bar{\psi}\bar{\psi} \ .
\label{Z}
\end{gathered}
\eeq
Since $F$ and $\bar{F}$ have no kinetic term, they do not propagate, and can be integrated out (they are known as the auxiliary fields), so we finally obtain:
\beq
\begin{gathered}
\mathcal{L} \ = \ \partial_\mu\phi\partial^\mu\phi^*+\frac{i}{2}\left(\partial_\mu\psi\sigma^\mu\bar{\psi}-\psi\sigma^\mu\partial_\mu\bar{\psi}\right)-\\
|\frac{\delta W}{\delta\phi}|^2-\frac{1}{2}\left(\frac{\delta^2W}{\delta\phi^2}\psi\psi+\frac{\delta^2\bar{W}}{\delta\phi^{*2}}\bar{\psi}\bar{\psi}\right) \ .
\end{gathered}
\eeq
Since $d\theta d\theta$ has mass dimension 1, the superpotential should have $[M]=3$, which means that it should be at most cubic in $\Phi$ to be renormalizable: 
\beq
W(\Phi) \ = \ \sum^3_{n=1}c_n\Phi^n \ .
\eeq
In particular, the choice
\beq
W(\Phi) \ = \ \frac{m}{2}\Phi^2+\frac{g}{3}\Phi^3
\eeq
corresponds to a theory known as the Wess-Zumino model:
\beq\label{WessZumino}
\begin{gathered}
\mathcal{L}_{WZ} \ = \ \partial_\mu\phi\partial^\mu\phi^*+\frac{i}{2}\left(\partial_\mu\psi\sigma^\mu\bar{\psi}-\psi\sigma^\mu\partial_\mu\bar{\psi}\right)-\\
m^2|\phi|^2-mg|\phi|^2\left(\phi+\phi^*\right)-g^2|\phi|^4-g\left(\phi\psi\psi+\phi^*\bar{\psi}\bar{\psi}\right) \ .
\end{gathered}
\eeq
This can be easily generalized to models with several chiral multiplets $\Phi^i$. In this case, we would have
\beq
\begin{gathered}
\label{KW}
K(\Phi^i,\bar{\Phi}^i) \ = \ \bar{\Phi}^i\Phi^i \ ,\\
W(\Phi^i) \ = \ c_i\Phi^i+\frac{1}{2}m_{ij}\Phi^i\Phi^j+\frac{1}{3}g_{ijk}\Phi^i\Phi^j\Phi^k \ ,
\end{gathered}
\eeq
with the bosonic Lagrangian given by
\beq\label{BL}
\mathcal{L}_b \ = \ g_{i\bar{j}}\partial_\mu\phi^i\partial^\mu\phi^{*\bar{j}} - g^{i\bar{j}}\partial_iW\partial_{\bar{j}}\bar{W} \ ,
\eeq
where
\beq
g_{i\bar{j}} \ = \ \frac{\delta^2K}{\delta\phi^i\delta\phi^{*\bar{j}}}
\eeq
is known as the Kähler metric.

The Kähler potential is invariant under the global transformation
\beq
\begin{gathered}
\Phi\rightarrow\exp\left(i\Lambda\right)\Phi \ , \ \bar{\Phi}\rightarrow\bar{\Phi}\exp\left(-i\Lambda\right) \ .
\end{gathered}
\eeq
However, if we want to promote $\Lambda$ to a local parameter $\Lambda(x,\theta,\bar{\theta})$, we would need
\beq
\bar{D}_{\dot{\alpha}}(e^{i\Lambda}\Phi) \ = \ 0 \ ,
\eeq
and therefore $\Lambda$ should also be a chiral superfield. However, this means that the action would no longer be invariant:
\beq
\bar{\Phi}e^{-i\bar{\Lambda}}e^{i\Lambda}\Phi\neq\bar{\Phi}\Phi \ .
\eeq
As in ordinary field theory, the solution is to introduce a new field $V$, known as the vector superfield:
\beq
\mathcal{L} \ = \ \int d^2\theta\bar{\Phi}e^{V}\Phi \ .
\eeq
Due to the reality of the Lagrangian, $V$ should be real and transform as
\beq\label{VTL}
e^V\rightarrow e^{i\bar{\Lambda}}e^Ve^{-i\Lambda} \ \Leftrightarrow \ V\rightarrow V+i\left(\bar{\Lambda}-\Lambda\right) \ .
\eeq
Choosing an appropriate gauge, known as the Wess-Zumino gauge, we can cast $V$ in the form
\beq
V \ = \ \theta\sigma^\mu\bar{\theta}v_\mu+i\theta^2\bar{\theta}\bar{\lambda}(x)-i\bar{\theta}^2\theta\lambda(x)+\frac{1}{2}\theta^2\bar{\theta}^2D(x) \ .
\eeq

In this gauge, all powers of V higher than 2 vanish, so the Kähler potential with interaction can be easily calculated (here we performed the rescaling $V\rightarrow2gV$):
\beq
\begin{gathered}
\label{VecInt}
\int d^2\theta d^2\bar{\theta}\left(\bar{\Phi}\Phi+2g\bar{\Phi}V\Phi+2g^2\bar{\Phi}V^2\Phi\right) \ = \  \frac{i}{2}\left(\bar{\phi}\partial_\mu\phi-\partial_\mu\bar{\phi}\phi\right)v^\mu+\\
\frac{i}{\sqrt{2}}\left(\bar{\phi}(\lambda\psi)-\phi(\bar{\psi}\bar{\lambda})\right)+\frac{1}{2}\psi\sigma^\mu\bar{\psi}v_\mu+\frac{1}{2}\bar{\phi}\phi D+\frac{1}{2}\bar{\phi}\phi(v_\mu v^\mu)
\end{gathered}
\eeq
Finally, combining \eqref{VecInt} and \eqref{Z}, we obtain:
\beq
\begin{gathered}
\mathcal{L}_\phi \ = \ D_\mu\phi D^\mu\bar{\phi}+\frac{i}{2}(D_\mu\psi\sigma^\mu\bar{\psi}-\psi\sigma^\mu D_\mu\bar{\psi})+i\sqrt{2}g(\bar{\phi}(\lambda\psi)-\phi(\bar{\psi}\bar{\lambda}))+\\
\bar{F}F+g\bar{\phi}\phi D+\frac{\delta W}{\delta\phi}F-\frac{1}{2}\frac{\delta^2W}{\delta\phi^2}\psi\psi+\frac{\delta \bar{W}}{\delta\phi^*}\bar{F}-\frac{1}{2}\frac{\delta^2\bar{W}}{\delta\phi^{*2}}\bar{\psi}\bar{\psi} \ ,
\label{CL}
\end{gathered}
\eeq
where
\beq
D_\mu \ = \ \partial_\mu-igv_\mu
\eeq
is the covariant derivative.

Now we have to add the kinetic terms for the vector field itself. These are given by
\beq
\mathcal{L}_W \ = \ \frac{1}{4}\int d^2\theta (W^\alpha W_\alpha)+\frac{1}{4}\int d^2\bar{\theta}(\bar{W}^{\dot{\alpha}}\bar{W}_{\dot{\alpha}}) \ , \label{er}
\eeq
where $W$ is the superfield strength:
\beq
\begin{gathered}
W_\alpha \ = \ -\frac{1}{4}\bar{D}\bar{D}D_\alpha V \ , \ \bar{W}_{\dot{\alpha}} \ = \ -\frac{1}{4}DD\bar{D}_{\dot{\alpha}}V \ ,
\end{gathered}
\eeq
which is invariant under the gauge transformation:
\beq
\begin{gathered}
W_\alpha\rightarrow W_\alpha-\frac{1}{4}\bar{D}\bar{D}D_\alpha(\Phi+\bar{\Phi}) \ = \ W_\alpha \ .
\end{gathered}
\eeq
To calculate $W_\alpha$, it would be more convenient to express $V$ in terms of $y_+$, since $\bar{D}_{\dot{\alpha}}y_+=0$:
\beq
\begin{gathered}
V_{WZ} \ = \ \theta\sigma^\mu\bar{\theta}v_\mu(y_+)+i\theta^2\bar{\theta}\bar{\lambda}(y_+)-i\bar{\theta}^2\theta\lambda(y_+)+\frac{1}{2}\theta^2\bar{\theta}^2\left(D(y_+)-i\partial_\mu v^\mu(y_+)\right) \ .
\end{gathered}
\eeq
Now,
\beq
\begin{gathered}
D_\alpha V_{WZ} \ = \ \sigma^\mu_{\alpha\dot{\alpha}}\bar{\theta}^{\dot{\alpha}}v_\mu+2i\theta_\alpha\bar{\theta}\bar{\lambda}+\theta^2\bar{\theta}^2\sigma^\mu_{\alpha\dot{\alpha}}\partial_\mu\bar{\lambda}^{\dot{\alpha}}-i\bar{\theta}^2\lambda_\alpha+\theta_\alpha\bar{\theta}^2D+2i\bar{\theta}^2\sigma^{\mu\nu}_{\alpha\beta}\theta^\beta\partial_\mu v_\nu \ ,
\end{gathered}
\eeq
and 
\beq
\begin{gathered}
W_\alpha \ = \ \theta^2\sigma^\mu_{\alpha\dot{\alpha}}\partial_\mu\bar{\lambda}^{\dot{\alpha}}-i\lambda_\alpha+\theta_\alpha D+2i\sigma^{\mu\nu}_{\alpha\beta}\theta^\beta\partial_\mu v_\nu \ .
\end{gathered}
\eeq
The last term can be rewritten as 
\beq
i\sigma^{\mu\nu}_{\alpha\beta}\theta^\beta F_{\mu\nu} \ ,
\eeq
with
\beq
F_{\mu\nu} \ = \ \partial_\mu v_\nu-\partial_\nu v_\mu \ .
\eeq
And the Lagrangian \eqref{er} would be
\beq
\begin{gathered}
\mathcal{L}_W \ = \ -i\lambda\sigma^\mu\partial_\mu\bar{\lambda}+\frac{1}{2}D^2-\frac{1}{4}F_{\mu\nu}F^{\mu\nu} \ .
\end{gathered}
\eeq
Finally, we can add the term
\beq
\mathcal{L}_{FI} \ = \ 2g\xi\int d^2\theta d^2\bar{\theta}V \ = \ g\xi D(x) \ ,
\eeq
known as the Fayet-Iliopoulos term. It is SUSY invariant by definition, and gauge invariant due to the fact that
\beq
\int d^2\theta d^2\bar{\theta}(\Psi+\bar{\Psi})
\eeq
is a total derivative. Now, if we integrate out the auxiliary field $D$, we obtain
\beq
D \ = \ -g\bar{\phi}\phi-g\xi \ ,
\eeq
which generates the potential term
\beq
V \ = \ \frac{g^2}{2}(\bar{\phi}\phi+\xi)^2 \ .
\eeq
This theory is the $N=1$ supersymmetric QED, i.e. the abelian U(1) model. However, we can also consider a non--abelian theory, N=1 super-Yang-Mills (SYM), by making V a matrix:
\beq
V\rightarrow V^aT^a \ .
\eeq
This means that $\Lambda^a$ would also be a matrix:
\beq
(e^{V^aT^a})_{ij}\rightarrow (e^{i\bar{\Lambda}^aT^a})_{ik}(e^{V^aT^a})_{kl}(e^{-i\Lambda^aT^a})_{lj}\ ,
\eeq
the covariant derivative would have the form
\beq
D_\mu \ = \ \partial_\mu-\frac{i}{2}v_\mu^aT^a \ ,
\eeq
and the field strength would instead be defined as:
\beq
W_\alpha \ = \ -\frac{1}{4}\bar{D}\bar{D}e^{-V}D_\alpha e^V \ ,
\eeq
which gives us
\beq
-\frac{1}{4}\bar{D}\bar{D}\left(D_\alpha V+\frac{1}{2}[D_\alpha V, V]\right) \ .
\eeq
The first term has already been calculated, and the second one is
\beq
\begin{gathered}
\frac{1}{8}\bar{D}\bar{D}[V, D_\alpha V] \ = \ \frac{1}{2}[v_\mu,v_\mu]\sigma^{\mu\nu}_{\alpha\beta}\theta^\beta-\frac{i}{2}\theta^2\sigma^\mu_{\alpha\dot{\alpha}}[v_\mu,\bar{\lambda}^{\dot{\alpha}}] \ ,
\end{gathered}
\eeq
so eventually we get the field strength
\beq
W_\alpha \ = \ \theta^2\sigma^\mu_{\alpha\dot{\alpha}}D_\mu\bar{\lambda}^{\dot{\alpha}}-i\lambda_\alpha+\theta_\alpha D+i\sigma^{\mu\nu}_{\alpha\beta}\theta^\beta F_{\mu\nu} \ ,
\eeq
with
\beq
\begin{gathered}
D_\mu\bar{\lambda}^{\dot{\alpha}} \ = \ \partial_\mu\bar{\lambda}^{\dot{\alpha}}-\frac{i}{2}[v_\mu,\bar{\lambda}^{\dot{\alpha}}] \ ,\\
F^a_{\mu\nu} \ = \ \partial_\mu v_\nu-\partial_\nu v_\mu-\frac{i}{2}[v_\mu,v_\nu] \ .
\end{gathered}
\eeq
Besides, after the substitution $V\rightarrow2gV$, the kinetic term may be written in the form:
\beq
\frac{1}{32\pi}Im\left[\tau\int d^2\theta Tr[W^\alpha W_\alpha]\right] \ ,
\eeq
where
\beq
\tau \ = \ \frac{\theta_{YM}}{2\pi}+i\frac{4\pi}{g^2}
\eeq
is the complexified gauge coupling. This yields the standard kinetic term plus the topological term:
\beq
\begin{gathered}
-iTr[\lambda\sigma^\mu\partial_\mu\bar{\lambda}]+\frac{1}{2}Tr[D^2]-\frac{1}{4}Tr[F_{\mu\nu}F^{\mu\nu}]+g^2\frac{\theta_{YM}}{32\pi^2}Tr[F_{\mu\nu}\tilde{F}^{\mu\nu}] \ .
\end{gathered}
\eeq
The D--term may once again be integrated out, and since we have no Fayet--Iliopoulos contribution for non--abelian gauge groups, we get just
\beq
V \ = \ \frac{g^2}{2}(\bar{\phi}T^a\phi)^2 \ .
\eeq
More generically, one can derive the form of the D--term from the requirement that the gauge transformations of the fields:
\beq
\delta_I\phi^a \ = \ \xi^a_I
\eeq
leave the Kähler potential invariant up to a gauge transformation:
\beq
K(\phi,\bar{\phi}) \ = \ K(\phi,\bar{\phi})+\xi^a_I\partial_aK+\bar{\xi}^{\bar{a}}_I\partial_{\bar{a}}K+\Lambda(\phi)+\bar{\Lambda}(\bar{\phi}) \ .
\eeq
Therefore, given the transformation law of the vector superfield~\eqref{VTL}, the D--term would be given by
\beq
D_I \ = \ \frac{i}{2}\left(\xi^a_I\partial_aK-\bar{\xi}^{\bar{a}}_I\partial_{\bar{a}}K+\Lambda(\phi)-\bar{\Lambda}(\bar{\phi})\right) \ .
\eeq
In particular, if we require the Kähler potential to be gauge invariant, $\Lambda$ would have to be an imaginary constant, producing a Fayet--Iliopoulos term.

One particular case of $N=1$ supersymmetric model is MSSM, which has the same field content as the Standard Model, but with the fields replaced by superfields, up to a doubling in the Higgs sector that we have anticipated. Namely, it contains electroweak and gluon vector superfields, along with fermionic and Higgs chiral superfields. The superpotential of MSSM reads:
\beq
W \ = \ \mu H_uH_d+y_u\bar{u}H_uQ+y_d\bar{d}H_dQ+y_e\bar{e}H_dL \ .
\eeq
In principle, one can also add lepton- and baryon-number violating terms
\beq
W_{\Delta L=1} \ = \ \alpha^{ijk}L_iL_j\bar{e}_k+\beta^{ijk}L_iQ_j\bar{d}_k+\gamma^iL_iH_u \ 
\eeq
and
\beq
W_{\Delta B=1} \ = \ \lambda^{ijk}\bar{u}_i\bar{d}_j\bar{d}_k \ .
\eeq
However, given the tight observational constraints on B- and L-violating processes (other than neutrino oscillations), $\alpha, \beta, \gamma$, and $\lambda$ would have to be extremely small. One natural way to set them to zero is to introduce a new symmetry, the so-called matter parity:
\beq
P_M \ = \ (-1)^{3(B-L)} \ .
\eeq
Fermionic supermultiplets would have $P_M=-1$, while Higgs and gauge boson supermultiplets, which do not carry a baryon or lepton number, would have $P_M=+1$. Therefore, requiring that the Lagrangian be $P_M$-even eliminates these terms. 

Unlike B and L, which are known to be violated by sphaleron processes, $P_M$ can in principle be an exact symmetry of the theory.
It may be recast in terms of the so-called R-parity:
\beq
P_R \ = \ (-1)^{3(B-L)+2s} \ ,
\eeq
but unlike matter parity, the R-parity cannot be defined for the whole supermultiplet because it would be different for particles and their superpartners. The R-parity can be understood to emerge from the continuous R-symmetry of supersymmetric theories, which corresponds to the $U(1)$ transformation of Grassmann variables: 
\beq
\theta\rightarrow e^{i\alpha}\theta \ , \ \bar{\theta}\rightarrow e^{-i\alpha}\bar{\theta} \ 
\eeq
(the discrete R-parity corresponds to the mirror reflection: $\theta\rightarrow-\theta, \bar{\theta}\rightarrow-\bar{\theta}$). To preserve the invariance of Berezin integrals, we would also have to demand
\beq
d\theta\rightarrow e^{-i\alpha}d\theta \ , \ d\bar{\theta}\rightarrow e^{i\alpha}d\bar{\theta} \ .
\eeq
Now, if a chiral field $\Phi$ has the R-charge $n$, then, per the decomposition~\eqref{Ch}, the scalar component $\phi$ should also have the charge $n$, the spinor component $\psi$ has to have the charge $(n-1)$, and $F$ would have $(n-2)$. The vector superfields would have to remain uncharged due to the form of the interaction terms~\eqref{VecInt}, which means that the vector field $v_\mu$ would also be uncharged, and the gaugino $\lambda$ would have the charge +1.

From these considerations, one can see that the Standard Model particles would not be charged under the R-symmetry, while their superpartners (gauginos, higgsinos, squarks, and sleptons) would have $R=\pm1$. Now, since the term $\mu H_uH_d$ breaks the R-symmetry, one would have to demote it to the discrete R-parity, defined as $P_R=(-1)^R$, making the SM particles R-even and their superpartners R-odd.

However, in SUSY models, the masses of particles and their superpartners are exactly equal, which means that superpartners should have already been discovered in collider experiments. The fact that we do not observe them means that, if supersymmetry has a role in Nature, it should be broken in such a way that makes superpartners heavier than the SM particles.

There are two ways of breaking SUSY: spontaneous (the Lagrangian is supersymmetric, but the system is in a non-supersymmetric vacuum state), which is similar to electroweak symmetry breaking, and explicit (the Lagrangian contains SUSY-breaking terms, and some of them can have a spontaneous origin in supergravity, the local version of supersymmetry that we shall discuss shortly). The SUSY ground state has zero energy due to \eqref{E}, which means
\beq
V(\phi,\bar{\phi}) \ = \ \bar{F}F+\frac{1}{2}D^2=0
\eeq
if SUSY is unbroken (here $\xi=0$ for non--abelian groups, while $T^a$ is just $1$ for abelian ones). Field configurations satisfying
\beq
F^i \ = \ \frac{\partial W}{\partial\phi^i} \ = \ 0
\eeq
are known as F-flat directions, while those which satisfy
\beq
D^i\propto\bar{\phi}^iT^a_{ij}\phi^j \ = \ 0
\eeq
for non--abelian case or
\beq
D\propto\bar{\phi}\phi+\xi \ = \ 0
\eeq
for abelian case are known as D-flat directions. The violations of these conditions result in F- and D- spontaneous SUSY breaking, respectively.

The generic conditions that define a vacuum (supersymmetric or not) are
\beq
\frac{\delta V}{\delta\phi} \ = \ 0 \ , \ \frac{\delta V}{\delta\bar{\phi}} \ = \ 0 \ .
\eeq
The first of these equations is equivalent to
\beq
- \ \left<F^j\right>\left<\frac{\partial W}{\partial\phi^i\partial\phi^j}\right> \ + \ g\left<D^a\right>\left<\bar{\phi}_j\right>(T^a)^j_i \ = \ 0 \ ,
\eeq
and can be written in the matrix form:
\beq
\begin{pmatrix}
\left<\frac{\partial W}{\partial\phi^i\partial\phi^j}\right> & -g\left<\bar{\phi}_j\right>(T^a)^j_i \\
	-g\left<\bar{\phi}_i\right>(T^a)^i_j & 0
\end{pmatrix}\begin{pmatrix} \left<F^j\right> \\
\left<D^a\right>
\end{pmatrix} \ = \ 0
\eeq
(the second line turns to zero due to the gauge invariance of the superpotential: $\left<F^j\right>\left<\bar{\phi}_j\right>(T^a)^j_i\propto\delta\left<W\right>=0$).

Therefore, for a SUSY-breaking vacuum in which either $\left<F^j\right>$ or $\left<D^a\right>$ are nonzero, the matrix has a nontrivial eigenvector corresponding to the eigenvalue $\lambda=0$. However, as can be seen from the Lagrangian~\ref{CL}, this matrix is also the mass matrix for the fermions $\psi$ and $\lambda$:
\beq
\frac{1}{2}\begin{pmatrix} \psi^i & i\sqrt{2}\lambda^a
\end{pmatrix}\begin{pmatrix}
\left<\frac{\partial W}{\partial\phi^i\partial\phi^j}\right> & -g\left<\bar{\phi}_j\right>(T^a)^j_i \\
	-g\left<\bar{\phi}_i\right>(T^a)^i_j & 0
\end{pmatrix}\begin{pmatrix} \psi^j \\
i\sqrt{2}\lambda^a
\end{pmatrix} \ ,
\eeq
which means that the fermion state
\beq
\chi\propto\left<F^j\right>\psi^j+i\sqrt{2}\left<D^a\right>\lambda^a
\eeq
would be massless. It is known as \emph{the goldstino}, the analogue of a Goldstone boson for broken supersymmetry. The presence of a massless particle in the spectrum is potentially problematic, since such a particle has never been detected. We shall touch upon one possible solution to this problem later on, when we discuss supergravity: just as Goldstone bosons can be eaten by vectors making them massive, so Goldstinos can be eaten by spinor--vectors, gravitino fields, making them massive.

For a generic cubic superpotential of the form~\eqref{KW}, the F-term is yielded by
\beq
F_i \ = \ c_i+m_{ij}\Phi^j+g_{ijk}\Phi^j\Phi^k \ ,
\eeq
and therefore we can see that the necessary condition for F-term SUSY breaking should be $c_i\neq0$ (otherwise there is always the vacuum $\Phi^i=0$, for which SUSY is unbroken). For a case of three fields $\Phi_{1,2,3}$ with the superpotential
\beq
W=c\Phi_1+m\Phi_2\Phi_3+g\Phi_1\Phi_3^2 \ ,
\eeq
known as the O'Raifeartaigh model, the F-terms would be
\beq
F_1 \ = \ c+g\phi_3^2 \ , \ F_2 \ = \ m\phi_3 \ , \ F_3 \ = \ 2g\phi_1\phi_3 \ .
\eeq
For $c\neq0$, they cannot be put to zero simultaneously, resulting in F-term SUSY breaking.

For the same reason, the D-term breaking needs the Fayet-Iliopoulos term, which means that the gauge group of the theory should not be semisimple (i.e. it should include at least one abelian $U(1)$ subgroup). 

The simplest example of D-term SUSY breaking involves two scalar fields $\Phi_1$ and $\Phi_2$ charged under $U(1)$ with opposite charges $\pm e$, with the Kähler potential and superpotential given by
\beq
\begin{gathered}
K \ = \ \bar{\Phi}_1e^{2eV}\Phi_1 \ + \ \bar{\Phi}_2e^{-2eV}\Phi_2 \ , \\ 
W \ = \ m(\Phi_1\Phi_2 \ + \ \bar{\Phi}_1\bar{\Phi}_2) \ ,
\end{gathered}
\eeq
corresponding to the potential
\beq
\begin{gathered}
V \ = \ -\frac{1}{2}D^2-|F_1|^2-|F_2|^2-eD(|\phi_1|^2-|\phi_2|^2)-\\
\xi D-m(F_1\phi_2+F_2\phi_1+\bar{F}_1\bar{\phi}_2+\bar{F}_2\bar{\phi}_1)
\end{gathered}
\eeq
(in the FI term, we absorbed $e, \xi_1,$ and $\xi_2$ into a single constant $\xi$ for convenience).

The equations of motion
\beq
\begin{gathered}
D \ = \ -\xi-e(|\phi_1|^2-|\phi_2|^2) \ ,\\
F_1 \ = \ -m\bar{\phi}_2 \ , \ F_2 \ = \ -m\bar{\phi}_1 
\end{gathered}
\eeq
do not have a solution with $D=F_1=F_2=0$ for nonzero $\xi$. After integrating out the auxiliary fields, one obtains
\beq
V \ = \ \frac{1}{2}\xi^2 +\frac{1}{2}e^2\left(|\phi_1|^2-|\phi_2|^2\right)^2+(m^2+e\xi)|\phi_1|^2+(m^2-e\xi)|\phi_2|^2 \ .
\eeq
For $m^2>e\xi$, all terms are positive, and the potential has a minimum at $\phi_1=\phi_2=0$, corresponding to D-breaking of SUSY. If, on the contrary, $e\xi$ is larger than $m^2$, the term in front of $\phi_2$ would be negative, and the minimum would be $\phi_1=0, \ |\phi_2|^2=\frac{e\xi-m^2}{e^2}$, resulting in both $D$ and $F_1$ distinct from zero.

So far we have not discussed the option of explicit supersymmetry breaking. For a theory like MSSM that includes scalar and spinor fields, one could introduce SUSY-breaking mass and interaction terms for these fields (the so-called ``soft`` terms):
\beq
\mathcal{L}_{br} \ = \ \mu\bar{\psi}\psi-m_i\phi_i^2+g_{ij}\phi_i\phi_j+\lambda_{ijk}\phi_i\phi_j\phi_k \ .
\eeq
However, one can effectively describe this mechanism in terms of spontaneous SUSY breaking by taking the coupling constants in the Kähler potential and superpotential and promoting them to superfields (such fields are known as spurions). For instance, it is possible to obtain the scalar mass term by modifying the Kähler potential:
\beq
K \ = \ Z\bar{\Phi}\Phi \ , \ Z \ = \ 1+\theta^2\alpha+\bar{\theta}^2\bar{\alpha}+\theta^2\bar{\theta}^2\rho \ .
\eeq
This modification yields the following additional Lagrangian terms:
\beq
\Delta\mathcal{L} \ = \ \rho\bar{\phi}\phi+\bar{\alpha}F\bar{\phi}+\alpha\bar{F}\phi \ ,
\eeq
which, upon integrating out $F$, produce the scalar mass term:
\beq
(\rho-|\alpha|^2)\bar{\phi}\phi \ .
\eeq
Performing the same operation on the superpotential couplings:
\beq
\begin{gathered}
\frac{m}{2}\Phi^2 \ , \ m^2\rightarrow m^2+2\kappa\theta^2 \ ,\\
\frac{g}{3}\Phi^3 \ , \ g\rightarrow g+3\sigma\theta^2 \ ,
\end{gathered}
\eeq
we get
\beq
\Delta\mathcal{L} \ = \ (\kappa-m\alpha)\phi^2+(\bar{\kappa}-\bar{m}\bar{\alpha})\bar{\phi}^2+(\sigma-g\alpha)\phi^3+(\bar{\sigma}-\bar{g}\bar{\alpha})\bar{\phi}^3 \ .
\eeq
The spurion terms can appear within the so-called \emph{messenger paradigm}: according to it, supersymmetry breaking occurs spontaneously within some hidden sector of the model, and is transferred to the visible sector via a mediator field which interacts with particles from both sectors. A scenario of this kind would have the fields from the visible sector interact with the hidden sector field $X$ via effective higher-order operators, similarly to the Fermi interaction:
\beq
\mathcal{L}_{int}=\int d^2\theta d^2\bar{\theta}\frac{c}{M^2}\bar{X}X\bar{\Phi}\Phi \ ,
\eeq
where $M$ is the characteristic energy scale of the mediator. Now, if X acquires a nonzero F-term:
\beq
X \ = \ X_0+\theta^2F \ ,
\eeq
it would produce mass terms for the supermultiplet $\Phi$ of order $m\sim|F|/M$. The SUSY breaking scale would therefore be around
\beq
\sqrt{|F|}\sim\sqrt{mM} \ .
\eeq
For $M=M_P$ (this scenario is known as gravity mediated SUSY breaking, since the mediator is likely to be the gravitational field), if we expect the superpartners' masses to be around the electroweak scale (100-1000 GeV), the breaking would occur at around $10^{10}$--$10^{11}$ GeV. Another feasible scenario is gauge mediated SUSY breaking, involving messengers charged under the gauge groups of the Standard Model. This interaction can be parametrized in terms of the superpotential:
\beq
W=\alpha X\bar{\Phi}\Phi \ ,
\eeq
and the masses would be of order $m\sim\alpha|F|/M$, where $\alpha$ is the gauge coupling constant. Again, assuming EW scale for $m$ and taking $\alpha\sim10^{-2}$, one obtains
\beq
\frac{|F|}{M} \ \sim \ 10^5 GeV \ ,
\eeq
and, given that the energy scale of the mediator cannot be smaller than the energy scale of SUSY breaking ($M^2\ge|F|$), one obtains the lower bound on this scale:
\beq
\sqrt{|F|}\ge10^5 GeV \ .
\eeq
The upper bound can be obtained from the consideration that the contribution from gauge interactions should be much larger than the contributions from the Planck scale:
\beq
\alpha\frac{|F|}{M}\ge\frac{|F|}{M_P} \ .
\eeq
Assuming that the difference is at least one-two orders, we obtain $M\le10^{15}-10^{16}$ GeV, and, substituting this number into $|F|\sim(Mm/\alpha)$, we see that the SUSY breaking scale should be below $10^{10}$ GeV.

Up until this point, we have worked on flat Minkowski background, neglecting gravitational effects. However, as we shall see, we would need to take them into account if we want to promote SUSY to a local symmetry (this framework is known as supergravity)~\cite{SUGRA}~\footnote{Some preliminary steps in that direction were made in~\cite{VolkovSoroka}.}. To be specific, let us consider the Wess--Zumino model with $m=g=0$. The Lagrangian~\eqref{WessZumino} is invariant under the supersymmetry transformation
\beq
\begin{gathered}
\phi\rightarrow\phi+\epsilon\chi \ , \ \chi\rightarrow\chi-i\sigma^\mu\bar{\epsilon}\partial_\mu\phi \ ,\\
\phi^*\rightarrow\phi^*+\bar{\epsilon}\bar{\chi} \ , \ \bar{\chi}\rightarrow\bar{\chi}+i\epsilon\sigma^\mu\partial_\mu\phi^*
\end{gathered}
\eeq
due to the identity
\beq
\sigma^\mu_{\alpha\beta}\sigma^\nu_{\gamma\delta} \ = \ -\frac{1}{2}g^{\mu\nu}\epsilon_{\alpha\gamma}\epsilon_{\beta\delta} \ .
\eeq
However, if the transformation parameter $\epsilon$ is not constant, the symmetry no longer holds, and the transformed Lagrangian acquires a term of the form
\beq
\delta\mathcal{L} \ = \ \partial_\mu\epsilon^\alpha K^\mu_\alpha \ + \ \partial_\mu\bar{\epsilon}_{\dot{\alpha}}\bar{K}^{\mu\dot{\alpha}} \ ,
\eeq
with
\beq
K^\mu_\alpha \ \propto \ \chi_\alpha\partial^\mu\phi^* \ .
\eeq
To cancel this term, we would need to introduce a new gauge field $\Psi_\mu^\alpha$ that couples to $K_\alpha^\mu$ and transforms as
\beq
\Psi_\mu^\alpha \ \rightarrow \ \Psi_\mu^\alpha-\partial_\mu\epsilon^\alpha \ .
\eeq
Due to having both a Lorentz index and a spinor index, this field would have spin $\frac{3}{2}$. However, this new term would also vary under the supersymmetry transformation: namely, if we perform the SUSY transformation on $\chi$, we obtain an additional term of the form
\beq
g^{\alpha\mu}\psi_\alpha\sigma^\nu\bar{\epsilon}\partial_\mu\phi^*\partial_\nu\phi
\eeq
The metric tensor $g^{\mu\nu}$ can be recast in the form
\beq
\frac{1}{2}g^{\mu\nu} \ - \ \frac{1}{4}\epsilon^{\alpha\gamma}\epsilon^{\beta\delta}\sigma^\mu_{\alpha\beta}\sigma^\nu_{\gamma\delta} \ ,
\eeq
leading up to
\beq
\psi^\mu\sigma^\nu\bar{\epsilon}T_{\mu\nu} \ ,
\eeq
where $T_{\mu\nu}$ is the stress--energy tensor of the scalar field. 

To compensate for this term, we would need to introduce one more field with two Lorentz indices (spin 2) that couples to the stress--energy tensor of matter and transforms as $\propto\psi^\mu\sigma^\nu\bar{\epsilon}$, but this field is just the metric tensor. Therefore the spin-$\frac{3}{2}$ field has to be the graviton's superpartner, known as the gravitino. This extension of $N=1$ SUSY is known as minimal supergravity (mSUGRA)~\cite{Chamseddine:1982jx}.

The kinetic term and the mass term for a field with spin $\frac{3}{2}$ are given by the Rarita--Schwinger action~\cite{Rarita:1941mf}:
\beq
\mathcal{L}_{RS} \ = \ -\frac{1}{2}\bar{\Psi}_\mu\gamma^{\mu\nu\rho}\partial_\nu\Psi_\rho+\frac{1}{2}m\bar{\Psi}_\mu\gamma^{\mu\nu}\Psi_\nu \ ,
\eeq
with the equation of motion
\beq
\gamma^{\mu\nu\rho}\partial_\nu\Psi_\rho \ = \ m\gamma^{\mu\nu}\Psi_\nu \ .
\eeq
Now, if we act on both sides of the EOM with the derivative $\partial_\mu$, we obtain the condition
\beq\label{RSC1}
\gamma^{\mu\nu}\partial_\mu\Psi_\nu \ = \ 0
\eeq
(the left--hand side vanishes due to the antisymmetry of $\gamma^{\mu\nu\rho}$). Then, we can multiply both sides of the EOM by $\gamma_\mu$ to find
\beq\label{RSC2}
3m\gamma^\nu\Psi_\nu \ = \ 2\gamma^{\nu\rho}\Psi_\rho \ = \ 0 \ ,
\eeq
where we used~\eqref{RSC1}. Finally, substituting the identities
\beq
\gamma^{\mu\nu\rho} \ = \ \gamma^\mu\gamma^{\nu\rho}-g^{\mu\nu}\gamma^\rho+g^{\mu\rho}\gamma^\nu
\eeq
and
\beq
\gamma^{\mu\nu} \ = \ \gamma^{\mu}\gamma^{\nu}-g^{\mu\nu}
\eeq
into the Rarita--Schwinger equation and using the constraints~\eqref{RSC1} and~\eqref{RSC2}, we obtain
\beq
\gamma^\nu\partial_\nu\Psi^\mu+m\Psi^\mu \ = \ 0 \ ,
\eeq
which is simply the Dirac equation. Therefore the Rarita-Schwinger equation is equivalent to the Dirac equation plus the two constraints~\eqref{RSC1} and~\eqref{RSC2}.

Supergravity can be more conveniently formulated in the so--called Cartan (or tetrad) formalism, namely by expressing the curved metric in the form:
\beq
g_{\mu\nu} \ = \ e^a_\mu e^b_\nu \eta_{ab} \ ,
\eeq
where $\eta_{ab}$ is the flat Minkowski metric, and $e^a_\mu$ is an object known as the vielbein. In this formalism, the $N=1$ supergravity action would be given by
\beq\label{SGA}
\mathcal{S} \ = \ \frac{1}{2\kappa^2} \ \int \ d^Dx \ e \ e^{a\mu}e^{b\nu}R_{\mu\nu ab}-\frac{1}{2\kappa^2} \ \int \ d^Dx \ e \ \bar{\psi}_\mu\gamma^{\mu\nu\rho}D_\nu\psi_\rho
\eeq
with $e=\sqrt{-g}$ and
\beq
D_\nu\psi_\rho \ = \ \partial_\nu\psi_\rho+\frac{1}{4}w_{\nu ab}\gamma^{ab}\psi_\rho  \ .
\eeq
(the supergravity action has no torsion term there, although the $\Gamma$'s acquire torsion).

One can show that the action~\eqref{SGA} is invariant under the SUSY transformations
\beq
\begin{gathered}
\delta\psi_\mu \ = \ D_\mu\epsilon \ = \ \partial_\mu\epsilon+\frac{1}{4}w_{\mu ab}\gamma^{ab}\epsilon \ ,\\
\delta e_\mu^a \ = \ \frac{1}{2}\bar{\psi}\gamma^a\psi_\mu \ ,
\end{gathered}
\eeq
provided the connection acquires a peculiar torsion term bilinear in the gravitino field.

We should also consider the possibility of a non--zero cosmological constant. The commutator of two Lorentz transformation generators is given by
\beq
[M_{AB},M_{CD}] \ = \ ig_{BC}M_{AD}+ig_{AD}M_{BC}-ig_{AC}M_{BD}-ig_{BD}M_{AC} \ ,
\eeq
and, as we have seen in Chapter~\ref{C1S2S3}, the four--dimensional dS and AdS spacetimes can be seen as embeddings within a five--dimensional spacetime with signature $(-,+,+,+,-)$ for AdS, and $(-,+,+,+,+)$ for dS.

The Lorentz generators can be split into two groups: $M_{\mu\nu}$ and $P_\mu=\frac{1}{L}M_{\mu5}$, with the commutation relations
\beq\label{PoincAlg}
[P_{\mu},P_{\nu}] \ = \ \pm \frac{i}{L^2}M_{\mu\nu} \ , \ [Q,P_\mu] \ = \ \frac{i}{2L}\gamma_\mu Q
\eeq
with the plus sign for the AdS case, and minus for dS. For the theory to be supersymmetric, the supercharges have to satisfy the Jacobi identity:
\beq
\left[P_\mu,[P_\nu,Q]\right] \ + \ \left[Q,[P_\mu,P_\nu]\right] \ + \ \left[P_\nu,[Q,P_\mu]\right] \ = \ 0 \ .
\eeq
Now, substituting~\eqref{PoincAlg}, we obtain
\beq
-\frac{1}{2L^2}\gamma_{\mu\nu}Q\pm\frac{1}{2L^2}\gamma_{\mu\nu}Q \ = \ 0 \ .
\eeq
This condition is satisfied only for the plus sign, i.e. AdS space. This means that in the presence of a positive cosmological constant supersymmetry is always broken.

The gravitino mass term makes the gravitino effectively massless in AdS, consistently with the fact that a gauge symmetry, local supersymmetry, is present.

Besides, in AdS space, it is possible to have a massless graviton accompanied by a gravitino superpartner bearing a mass term. The reason is that the operator $P^2$, which is usually used to define mass, does not commute with $Q$, which means it does not have the same value for all the states in a supermultiplet. However, it is possible to introduce another definition of mass, via the operator
\beq
-\frac{1}{2}M_{AB}M^{AB} \ = \ H^2+J^2-\frac{1}{2}L^+_iL^-_i-\frac{1}{2}L^-_iL^+_i \ ,
\eeq
where
\beq
H \ = \ iP_0
\eeq
corresponds to energy,
\beq
J^2 \ = \ -\frac{1}{2}M_{ij}M^{ij} \
\eeq
is the spin, and
\beq
L^{\pm}_i \ = \ -iM_{0i}\pm M_{5i} \ 
\eeq
are the raising and lowering operators. The operator $M_{AB}M^{AB}$ commutes with $Q$, and using the commutation relation
\beq
\left[J^+_i,J^-_i\right] \ = \ -2H\delta_{ij} \ ,
\eeq
we can rewrite it as
\beq
H(H-3)+J^2-L_i^+L_i^- \ . 
\eeq
For the ground state, it is given by
\beq\label{GroundState}
E(E-3)+S(S+1) \ ,
\eeq
but since the mass is invariant, it has to be the same for the ground state and for the excited states belonging to the same irreducible representation. For an excited state with the energy $E+1$ and spin $s-1$, the mass would be given by
\beq
(E+1)(E-2)+S(S-1)-<E+1,S-1|L_i^+L_i^-|E+1,S-1> \ .
\eeq
This expression has to be exactly equal to~\eqref{GroundState}, but since the last term is the square norm of the state vector $|L_i^-|E+1,S-1>|^2$, it must be non--negative. Therefore, we have to require
\beq
E \ \ge \ S+1 \ .
\eeq
When this bound is saturated, the value of the mass operator is $2(S^2-1)$, so if we want the states that saturate the bound to be massless and the operator to reduce to $P^2$ in the Minkowski limit ($L\rightarrow\infty$), we can define the mass operator as
\beq
m^2 \ = \ -\frac{1}{2L^2}M_{AB}M^{AB}-\frac{2}{L^2}(S^2-1) \ .
\eeq
For the ground states, it is given by
\beq
m^2L^2 \ = \ E(E-3)-S(S-1)+2 \ ,
\eeq
which means that the energy levels are:
\beq
E \ = \ \frac{3}{2}\pm\sqrt{(S-\frac{1}{2})^2+L^2m^2} \ .
\eeq
If the energy is real, the mass should satisfy the condition
\beq
m^2 \ \ge \ -\frac{1}{4L^2} \ , 
\eeq
the so--called Breitenlohner--Freedman bound. It implies the possibility of stable AdS vacua even in the presence of tachyonic states.

As is the case with global SUSY, the matter sector of SUGRA is determined by a Kähler potential and a superpotential, which however are not restricted by renormalizability, since the gravity portion does not respect this condition. However, the ordinary derivatives in~\eqref{BL} are replaced with covariant derivatives, and the F--potential $|\partial W|^2$ is substituted for
\beq
V_F \ = \ e^{K/M_P^2}\left[g^{i\bar{j}}D_iWD_{\bar{j}}\bar{W}-3\frac{|W|^2}{M_P^2}\right] \ ,
\eeq
with the ''covariant Kähler derivative'' $D_i$ given by
\beq
D_i \ = \ \partial_i+\frac{\partial_iK}{M_P^2} \ .
\eeq
To preserve the invariance of the scalar potential under the transformation of the Kähler potential~\eqref{KT}, we would also have to transform the superpotential:
\beq
W \ \rightarrow \ e^{-\Lambda/M_P^2}W \ .
\eeq
It is also possible to define a single gauge invariant combination of $K$ and $W$:
\beq
G \ = \ K+M_P^2\ln\left(\frac{|W|^2}{M_P^6}\right) \ ,
\eeq
so that the scalar potential would be a function only of $G$:
\beq
V_F \ = \ e^{G/M_P^2}\left[g^{i\bar{j}}\partial_iG\partial_{\bar{j}}G-3M_P^4\right] \ .
\eeq
The D-terms are given by
\beq
D_I \ = \ iM_P^2\xi^a_I\frac{D_aW}{W} \ .
\eeq

When supersymmetry is broken, the gravitino field ``eats`` the goldstino field and acquires mass, which can explain why the massless goldstino has not been observed. In the scenario of F--term SUSY breaking, when the D--terms are negligible, the mass of the gravitino would be given by
\beq
m_g \ = \ e^{K/2M_P^2}\frac{|W|}{M_P^2} \ .
\eeq

The theory described above is known as matter--coupled $N=1$ supergravity. However, just like global SUSY theories, SUGRA can have a larger number of supersymmetries, although these extensions are not chiral. In 4 dimensions, the largest possible number of supersymmetries is 8 (for $N>8$, the theory would contain higher--spin states with spins larger than 2, whose long--range interactions have long been fraught with difficulties), with a field content of 8 gravitini, 28 vectors, 56 fermions with spin $\frac{1}{2}$, and 70 scalars~\cite{Cremmer:1978ds}. The maximum number of spacetime dimensions that SUGRA can live in without containing higher--spin states is 11~\cite{Cremmer:1978km}. This model contains only one supermultiplet, namely the graviton $g_{MN}$, the gravitino $\psi^a_M$, and the 3--form $A_{MNP}$. Cremmer and Julia first obtained the $N=8$ theory as a low-energy limit of the 11--dimensional SUGRA if seven spatial dimensions are compactified on a torus of small size.

\subsection{String Theory}\label{C1S2S5}
The $N=8$ supergravity, which is believed to be the least divergent of all supergravity theories~\cite{Bern:2011qn}, share with Einstein gravity the presence of a dimensional parameter, the Planck mass, which determines the strength of its interactions. The most sophisticated symmetry arguments that are currently available cannot exclude that ultraviolet divergences show up at seven loops~\cite{Banks:2012dp,Bern:2018jmv}, and the current view, following the original proposals of Scherk and Schwarz and Yoneya of the 1970's~\cite{Scherk:1974ca,Yoneya:1974jg}, it to regard supergravity models as low--energy limits of a different, far more intricate, theory, known as String Theory, or, more precisely as we shall see shortly, M--theory~\cite{stringtheory}.

String Theory emerged, in a different context, in 1968, when Gabriele Veneziano discovered a peculiar $2\rightarrow2$ scattering amplitude, initially associated to scalar mesons and given by the formula
\beq
A \ = \ \frac{\Gamma\left(-1-\alpha's\right)\Gamma\left(-1-\alpha't\right)}{\Gamma\left(-2-\alpha'(s+t)\right)} \ ,
\eeq
subsequently known as the Veneziano amplitude, where
\beq
s \ = \ -(p_1+p_2)^2 \ , \ t \ = \ -(p_1-p_3)^2
\eeq
are the Mandelstam variables, and $\alpha'$ is a parameter with the dimension of length squared~\cite{Veneziano:1968yb}.

It was soon recognized that the consistency of the theory demands that the poles of this expression are
\beq
\alpha's \ = \ n-1 \ ,
\eeq
with $n=0, 1, 2, ...$, corresponding to an infinite tower of states ($\alpha'$ is known as the Regge slope). Notice that lowest-mass state of this tower is a tachyon, which signals an instability of the vacuum! However, at large $s$ and $t$ we also have the asymptotic $|A|\propto |t|^{\alpha's+1}$; given that $t$ is proportional to $\sin^2\theta$, this means that the corresponding tree--level diagram includes virtual particles with spin $J=n$ that is arbitrarily high. Since $s=M^2$ in the rest frame, one can thus see signs of the relation $M^2\propto J$, which is the natural property of a rotating relativistic string. Therefore, modeling hadrons as vibrating string-like objects, the parameter $\sqrt{\alpha'}=l_s$ was initially associated to the characteristic length scale of hadronic physics, of the order of nuclear sizes. The reinterpretation of the mid 1970's raised it considerably, bringing it typically close to the Planck scale.

We know that the action of a relativistic point--like particle is, up to a normalization factor, the integral of the worldline traversed by the particle:
\beq\label{RPA}
\mathcal{S} \ = \ -m \ \int \ d\tau \ \sqrt{-\eta_{\mu\nu}\partial_\tau X^\mu\partial_\tau X^\nu} \ ,
\eeq
where $\tau$ is the proper time. Likewise, the trajectory of a string would be a two--dimensional surface known as the worldsheet, so the natural generalization of~\eqref{RPA} is an integral over the surface of the worldsheet:
\beq
\mathcal{S} \ = \ -\frac{1}{2\pi\alpha'} \ \int \ d\tau d\sigma \ \sqrt{-\det(\partial_\alpha X\partial_\beta X)} \ .
\eeq
This expression is known as the Nambu--Goto action (the second variable $\sigma$ characterizes the spatial configuration of the string). One can show that the Nambu--Goto action is equivalent to the simpler Polyakov action
\beq
\mathcal{S} \ = \ -\frac{1}{4\pi\alpha'} \ \int \ d\tau d\sigma \ \sqrt{-g} \ g^{\alpha\beta} \eta_{\mu\nu}\partial_\alpha X^\mu\partial_\beta X^\nu \ ,
\eeq
where we have introduced an auxiliary worldsheet metric $g_{\alpha\beta}$.

The theory described by this action is just a conformal field theory of a D--component scalar multiplet in a 2--dimensional spacetime, and therefore it can be canonically quantized. By properly choosing the gauge, it is possible to reduce the worldsheet metric to two--dimensional Minkowski metric, up to a conformal factor that does not contribute to the action, and the equation of motion is then the familiar wave equation
\beq
\partial^\alpha\partial_\alpha X^\mu \ = \ 0 \ ,
\eeq
with the generic solution
\beq
X(\tau,\sigma) \ = \ X^+(\tau+\sigma) \ + \ X^-(\tau-\sigma) \ .
\eeq
It admits two types of solutions, namely closed strings, with $\sigma$ going from $0$ to $2\pi$ and the boundary condition
\beq
X^\mu(\tau,\sigma+2\pi) \ = \ X^\mu(\tau,\sigma) \ ,
\eeq
and open strings, with $\sigma$ going from $0$ to $\pi$, and the boundary conditions given either by
\beq
\partial_\sigma X^\mu(\tau,\sigma)\Big|_{\sigma=0,\pi} \ = \ 0 \ ,
\eeq
known as the Neumann boundary condition, or by
\beq
X^\mu(\tau,\sigma)\Big|_{\sigma=0,\pi} \ = \ c^\mu \ ,
\eeq
known as the Dirichlet boundary condition. The Dirichlet boundary condition implies that the string's ends are fixed on some higher--dimensional object; such objects are known as D--branes~\cite{Polchinski:1995mt}. Unlike open strings, the states of closed strings are constrained by the so--called level matching condition, according to which the quantum states should have an equal number of left--moving and right--moving modes.

However, upon quantization the open--string sector yields a massless spin-1 field, and the closed--string sector yields a massless spin-2 field, corresponding to long--range forces. This picture is strictly at odds with hadronic physics, but makes perfect sense in the context of particle physics and quantum gravity: the spin-1 state can be identified with the photon (and its non--abelian counterparts with Yang--Mills fields), and the spin-2 state with the graviton. This makes String Theory a candidate theory for quantum gravity, and one can easily argue that the replacement of point-like particles with strings can resolve the problem of UV divergences. Namely, as we have seen in Section~\ref{C1S2S3}, the effective coupling constant for gravity is
\beq
\alpha_G \ \sim \ \left(\frac{E}{M_P}\right)^2 \ ,
\eeq
which means that it becomes larger than 1 at energies around the Planck scale. But for two interacting strings, the interaction energy would be limited to the area of their intersection $(\Delta x)^2$, i.e.
\beq
\alpha_G \ \sim \ \left(\frac{E}{M_P}\right)^2\left(\frac{\Delta x}{l_s}\right)^2 \ .
\eeq
Finally, using the Heisenberg uncertainty principle, we can replace $\Delta x$ with $E^{-1}$, which means that near the Planck scale, $\alpha_G$ tends to a constant value of order $(M_Pl_s)^{-2}$.

Nonetheless, the theory described by the Polyakov action can only be consistently quantized in 26--dimensional spacetime. In addition, it contains a source of instability in the form of a tachyonic state with negative mass, but does not contain fermions because one can only obtain bosonic states from products of $X^\mu$. The solution is to introduce worldsheet fermions:
\beq
\mathcal{S} \ = \ -\frac{1}{4\pi\alpha'} \ \int \ d\tau d\sigma \ \sqrt{-g} \ \left(\partial_\alpha X^\mu\partial^\alpha X_\mu \ + \ \bar{\psi}^\mu\rho^\alpha\partial_\alpha\psi_\mu \right) \ ,
\eeq
making the theory supersymmetric ($\rho^\alpha$ are two--dimensional Dirac matrices). A theory described by this action requires 10 spacetime dimensions for consistent quantization. 

For the closed string in ten--dimensional spacetime, the physical observables (which are bilinears of fermions) should be periodic under $\sigma\rightarrow\sigma+2\pi$, which means that the fermions themselves have to be either periodic:
\beq
\psi(\tau,\sigma+2\pi) \ = \ \psi(\tau,\sigma) \ ,
\eeq
i.e. the so--called Ramond (R) fermions, or antiperiodic:
\beq
\psi(\tau,\sigma+2\pi) \ = \ -\psi(\tau,\sigma) \ ,
\eeq
i.e. the so--called Neveu--Schwarz (NS) fermions. Together with the level matching condition, it gives us four types of states: NS-NS, NS-R, R-NS, and R-R. The NS worldsheet fermions are spacetime bosons, while the R fermions are also fermionic in regards to spacetime. Since the Ramond vacuum state is fermionic, we actually have two vacua $|L>$ and $|{R}>$ with opposite chiralities. It should also be noted that the massless state in the R sector corresponds to the vacuum (i.e. it carries one spacetime spinor index and no Lorentz indices), but in the NS sector, it corresponds to the first excited state (i.e. it carries one Lorentz index and no spinor indices), while the NS vacuum is tachyonic.

The tachyon can be eliminated, while also ensuring other consistency conditions in the form of modular invariance (when closed strings only are present) and the absence of gauge and gravitational anomalies, via a procedure known as the Gliozzi--Scherk--Olive (GSO) projection~\cite{Gliozzi:1976qd}. One possible choice of the GSO projection removes the ground NS state and leaves only one Ramond vacuum for both right-moving and left-moving fermions. Choosing $|L>$ for the left--movers and $|{R}>$ for the right--movers yields the sectors $|NS>\otimes|NS>$, $|L>\otimes|NS>$, $|NS>\otimes|{R}>$, and $|L>\otimes|{R}>$: this theory is known as type IIA String Theory. At the massless level, the $NS-NS$ states bear two Lorentz indices, and consequently can be decomposed into the modes of a symmetric tensor $g_{\mu\nu}$ (graviton), of an antisymmetric tensor $B_{\mu\nu}$ (Kalb--Ramond field), and of a scalar trace $\phi$ (dilaton). The $L-NS$ state bears one Lorentz index and one spinor index, and therefore can be decomposed into the modes of a spin $\frac{3}{2}$-field $\Psi^a_\mu$ (a left-handed gravitino) and of a spin $\frac{1}{2}$-field $\lambda^a$ (a right-handed dilatino). The $NS-{R}$ sector yields the opposite--chirality states of a right-handed gravitino and of a left-handed dilatino. Finally, the RR states correspond to a 1-form $A_\mu$ and an antisymmetric 3-form $A_{\mu\nu\rho}$. The forms $B_{\mu\nu}$ and $A_{\mu\nu\rho}$ are generalizations of Maxwell's electromagnetic potential. While Maxwell's potential, which bears one Lorentz index, is produced by pointlike electric sources (zero--dimensional particles), a 2--form like the Kalb--Ramond field is associated with string-like electric sources (one--dimensional objects), and in general, a (p+1)--form indicates the presence of p--dimensional electric sources, known as p--branes. The analogue of electric-magnetic duality for branes implies that p--branes have (D-p-4)--dimensional magnetic duals, corresponding to (D-p-3)--dimensional forms.

Likewise, choosing the same Ramond vacuum for both left--movers and right--movers yields the type IIB String Theory with the sectors $|NS>\otimes|NS>$, $|R>\otimes|NS>$, $|NS>\otimes|R>$, and $|R>\otimes|R>$. The first two sectors give us the same field content as in the type IIA case, namely the bosons $g_{\mu\nu}, B_{\mu\nu}, \phi$, and the fermions $\Psi^a_\mu$ and $\lambda^a$. However, the $NS-R$ sector yields copies of the gravitino and the dilatino with the same chirality ($\Psi'^a_\mu$ and $\lambda'^a$), and the $R-R$ sector produces a scalar (0-form) $A$, a 2--form $A_{\mu\nu}$, and a self--dual 4-form $A_{\mu\nu\rho\sigma}$. Therefore, unlike type IIA, the type IIB theory is chiral. The low--energy limits of these two types of String Theory are known as type IIA and IIB supergravity. 

Type IIA and type IIB string theories are linked via a symmetry known as the T--duality~\cite{Dai:1989ua,Dine:1989vu}. Namely, if a string is compactified on a circle of radius $R$, its momentum would be quantized ($p=\frac{n}{R}$), and there would be an additional quantum number determining the contributions to the masses of the states due to the winding of the string around the circle. If the string is wrapped around the circle $m$ times, its mass spectrum would be:
\beq
M^2 \ = \ \frac{n^2}{R^2}+\frac{m^2R^2}{\alpha'^2} + ...
\eeq
One can easily see that this spectrum is equivalent to the one that would result from a string wrapped $n$ times around a circle of radius $\frac{\alpha'}{R}$, and with corresponding momenta. However, the expression for the bosonic string coordinate reads:
\beq
X(\tau,\sigma) \ = \ x+2\alpha'\frac{n}{R}\tau+2mR\sigma+... \ ,
\eeq
or, equivalently,
\beq
X(\tau,\sigma) \ = \ x+(\alpha'\frac{n}{R}+mR)(\tau+\sigma)+(n\frac{\alpha'}{R}-mR)(\tau-\sigma)+... \ .
\eeq
The T--duality transformation changes the sign of the third term, which corresponds to the right--moving mode, and therefore flips the sign of the right-moving portion of X. Due to supersymmetry, this would also affect the chirality of fermions, interchanging the $|R>$ and $|\bar{R}>$ vacuum states in the right--moving sector, so that the type IIA theory would become type IIB and vice versa.

Nonetheless, the type IIA/B theories are not the only option, as we can also combine the left--moving superstring modes and right--moving bosonic string modes or vice versa (these models are known as heterotic string theories)~\cite{Gross:1984dd}. Since superstrings live in 10 dimensions and bosonic strings only exist in 26 dimensions, the extra 16 dimensions can be compactified on special tori that grant modular invariance. This crucial property of string amplitudes grants, for example, that the one-loop diagram for closed strings, does not depend on a special choice of time on it, precisely as suggested by the shape of this surface. To this end, or equivalently in order to cancel the Lorentz and Yang--Mills chiral anomalies, the lattice defining the special tori has to be related to the weight lattices of $SO(32)$ or $E8\otimes E8$. The compactified bosonic coordinates can be equivalently presented as 32 fermions $\xi^A$. Both heterotic theories contain the graviton, the Kalb--Ramond field, the dilaton, the gravitino and the dilatino (the supergravity sector), together with the Yang--Mills vector fields in the adjoint representation of $SO(32)$ or $E8\otimes E8$ and their spinor superpartners (the super-Yang-Mills sector). One can show that the two internal tori (and therefore the two heterotic theories) are also linked by T--duality, after one space-time dimension is compactified on a circle.

Finally, there is the type I String Theory. It can be derived~\cite{Sagnotti:1987tw} acting on the type IIB spectrum with the projection operator
\beq
P \ = \ \frac{1}{2}\left(1+\Omega\right) \ ,
\eeq
where $\Omega$ reverses the orientation of the string: $\sigma\rightarrow-\sigma$ (this is known as the orientifold projection). This projection eliminates the Kalb--Ramond field, the antisymmetric combinations of the two gravitinos and dilatinos, the scalar $A$, and the 4--form $A_{\mu\nu\rho\sigma}$; as a result, we are left with the graviton, the dilaton, one gravitino, one dilatino, and the 2--form $A_{\mu\nu}$. However, this theory is anomalous, but one can also include open strings. Open strings correspond to super--Yang--Mills states, and in our case, choosing the $SO(32)$ gauge group cancels all anomalies. In the spacetime picture, in which the gauge forms live on special p-branes called D-branes, the orientifold projection is ascribed to the so--called orientifold plane $O_-$ that carries -16 units of brane charge and brane energy density, and requires 16 D9--branes to cancel it, leading to the emergence of the $SO(32)$ group~\cite{Polchinski:1995mt}. This theory has the same low--energy field content as the heterotic $SO(32)$ theory, and its effective action can be mapped to the one of the heterotic theory via a duality known as the S--duality~\cite{Sen:1994fa}. Namely, the low--energy action of type I theory is given by
\beq
S_I \ = \ \frac{1}{2\kappa^2_{10}} \ \int \ d^{10}x \ \sqrt{-G} \ \left[e^{-2\phi}\left(R+4\partial_\mu\phi\partial^\mu\phi\right)-\frac{1}{4}\left[e^{-\phi}|F|^2+...\right]\right] \ ,
\eeq
where $F$ is the field strength of the two--form $A_{\mu\nu}$. On the other hand, the effective action for the heterotic theory reads
\beq
S_{HE} \ = \ \frac{1}{2\kappa^2_{10}} \ \int \ d^{10}x \ \sqrt{-G} \ e^{-2\phi}\left[R+4\partial_\mu\phi\partial^\mu\phi-\frac{1}{4}|F|^2+...\right] \ .
\eeq
Making in these expressions the replacement $G_{\mu\nu}\rightarrow e^{\phi}G_{\mu\nu}$ one can see that they are mapped into one another substituting $\phi$ for $-\phi$, which indicates a weak-strong coupling duality link between them, since $e^\phi$ is the string coupling.

Finally, starting from the 11--dimensional SUGRA and compactifying it on a circle, one obtain the ten--dimensional fields $g_{\mu\nu}, A_{\mu\nu\rho}$ and $\Psi^a_\mu$, as well as the vector $A_\mu=g_{\mu,11}$, the scalar $\phi=g_{11,11}$, the antisymmetric 2-form $B_{\mu\nu}=A_{\mu\nu,11}$, and the spinor $\lambda^a=\Psi^a_{11}$, i.e. the field content of type IIA supergravity~\cite{Townsend:1995kk}. Likewise, compactifying it on an interval, one can obtain the massless modes of the heterotic $E8\otimes E8$ String Theory, with two $E8$ groups corresponding to the two ends of the interval~\cite{Horava:1996ma,Horava:1995qa}. This indicates that in both cases string coupling opens up an additional space--time dimension.

\begin{figure}
\centering
	\includegraphics[scale=0.7]{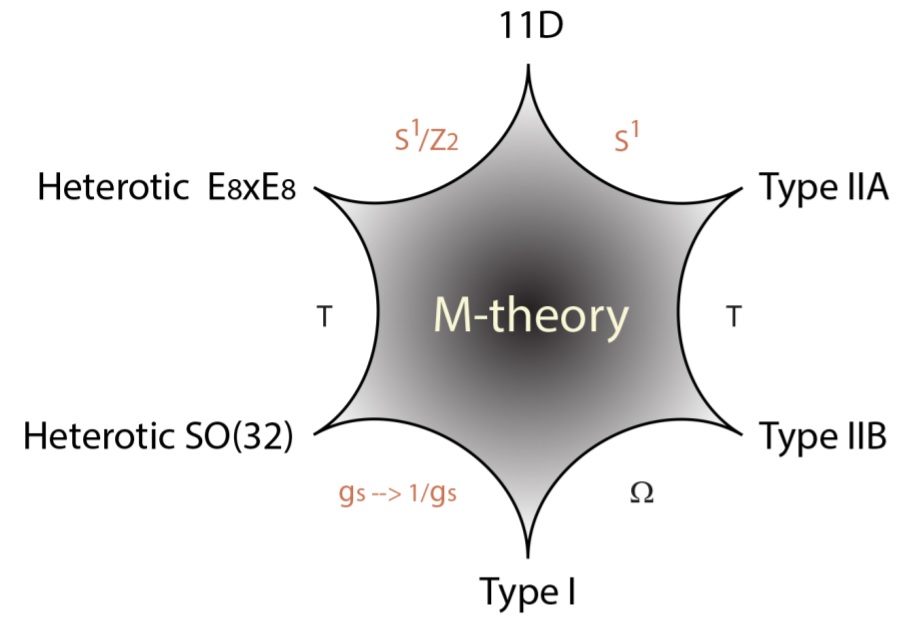}
	\caption{The dualities between different string theories (and 11--dimensional supergravity), all of which are believed to be various low--energy limits of the underlying M--theory.}
\label{Mtheory}
\end{figure}

Drawing upon these links, Edward Witten conjectured in 1995 that all of the aforementioned theories (type I strings, type IIA/B strings, two heterotic string theories, and 11--dimensional SUGRA) are various limits of a single 11--dimensional framework known as the M--theory (fig.~\ref{Mtheory})~\cite{Witten:1995ex}. Remarkably, M--theory is not a theory of strings, as it does not include the antisymmetric 2--form; instead, it is believed to contain higher--dimensional branes that effectively produce strings in 10 dimensions via a compactification on a circle. Nonetheless, the exact formulation of M--theory remains a major conundrum in String Theory, as does establishing quantitative links with the Standard Model: to obtain the four--dimensional (supersymmetric) SM from ten--dimensional strings one can to compactify six extra dimensions on a Ricci--flat manifold known as the Calabi--Yau manifold. The properties of the resulting four--dimensional fields depend on the type of compactification and of the Calabi--Yau itself (these characteristics are called moduli), and in total, String Theory is believed to yield around $10^{500}$ four--dimensional vacua. This is known as the landscape problem; however, it is not unique to String Theory, and emerges due to the fact that general relativity lacks a global energy minimum principle. As a result, one has no clear criteria for preferring one spacetime configuration over another.

The key issue in String Theory today is the breaking of supersymmetry, which is only partly realized via compactification on Calabi--Yau manifolds, since they reduce SUSY to $N=1$. Breaking SUSY completely impinges on vacuum stability in ways that are currently not under control.

There are different scenarios to break supersymmetry in String Theory, and a particularly enticing one, called brane supersymmetry breaking, rests on a variant of the type I String Theory. This involves the $O_+$ orientifold plane, with \emph{positive} charge and energy density. To cancel the charge, one needs 16 D9--antibranes, and this setting gives rise to the $USp(32)$ gauge group, but both branes and antibranes have positive energy density, so that the energy density is not cancelled as in the type-I superstring. This simplest manifestation of brane supersymmetry breaking is known as the Sugimoto model~\cite{bsb}. This produces an energy level shift between the bosonic and fermionic states, leading to supersymmetry breaking, with also a specific term in the Einstein-frame dilaton potential:
\beq\label{DilPot}
V \ = \ V_0e^{\gamma\,\phi} \ ,
\eeq
with $\gamma=\frac{3}{2}$. We shall discuss its potential role in Cosmology in the next chapter.
\subsection{Inflation}\label{C1S2S6}
Yet another possible extension of the Standard Model is motivated by a number of problems from the cosmology of the early Universe. The first of them is the flatness problem: according to the recent results of the Planck collaboration, the contribution of the curvature of the Universe to its critical density, given by $\Omega_k$ in~\eqref{Omega}, should be smaller than 0.01~\cite{PLANCK}. Then, given that the Hubble parameter depends on the temperature as $\frac{T^2}{M_P}$, and the scale factor as $T^{-1}$, we can estimate that the curvature at Planck time would have to be
\beq
\Omega_c\frac{a_0H_0}{a(t_P)H(t_P)}\propto\Omega_c\frac{H_0^2}{T_0^2}<10^{-60} \ ,
\eeq
which requires an extreme amount of fine--tuning.

The second issue is known as the horizon problem: at present, the size of the observable Universe is $\sim10^{62}$ $l_P$. Near the Planck time, it was $(T_0/M_P)*10^{62}$ $l_P\sim10^{30}$ $l_P$, which means that at that time, the Universe consisted of around $10^{90}$ casually separated regions, and there is no obvious reason why it ought to appear highly homogeneous from cosmological observations, as it does.

Finally, there ought to be a mechanism for the formation of primordial inhomogeneities that give rise to cosmological structures. One way to resolve all these issues is be to assume that soon after the Planck epoch, the Universe underwent a period of exponential expansion, with $a\propto e^{Ht}$ (this period is known as inflation)~\cite{Inflation}. Assuming that the expansion continued for a sufficiently long period $\tau$, a single region of size $l_P$ can be stretched by
\beq
e^{H\tau}\int_0^\tau \ dt \ e^{-Ht} \ \approx \ H^{-1}e^{H\tau} \ ,
\eeq
which, for an appropriate choice of parameters, can be larger than the size of the observable Universe.
Likewise, since the curvature depends on the scale factor as $a^{-2}$, during the inflationary period it would be diluted by a factor $\propto e^{-2H\tau}$, which resolves the flatness problem.

The simplest models of inflation rest on a simple scalar Lagrangian of the type
\beq
\mathcal{L}_\phi \ = \ -\frac{1}{2}(\partial_\mu\phi\partial^\mu\phi)-V(\phi) \ ,
\eeq
minimally coupled to Einstein gravity. If the field VEV were spatially homogeneous, its density and pressure would be given by
\beq
\rho \ = \ \frac{1}{2}\dot{\phi}^2+V(\phi) \ , \ p \ = \ \frac{1}{2}\dot{\phi}^2-V(\phi) \ ,
\eeq
and the Hubble equation would be
\beq
\label{Heq}
H^2=\frac{8\pi G}{3}\left(\frac{1}{2}\dot{\phi}^2+V(\phi)\right) \ .
\eeq
Under the condition
\beq
\label{Co1}
\frac{\dot{\phi}^2}{2}\ll V(\phi) \ , 
\eeq
one thus obtains a near-constant value of $H$ and $\frac{p}{\rho}=-1$, corresponding to $a(t)\propto\e^{Ht}$.

The time evolution equation of $\phi$ in the FRW metric would be:
\beq
\label{KG}
\ddot{\phi}+3H\dot{\phi}+V'(\phi)=0 \ .
\eeq
To prolong the period of inflation, we can also impose a ``slow--roll`` condition:
\beq
\label{Co2}
|\ddot{\phi}|\ll3H|\dot{\phi}| \ .
\eeq
In an analogy to classical mechanics, when the slow-roll condition is satisfied, it means that the ``force`` created by the potential is compensated by the ``friction``, and the field does not ``accelerate``. Therefore, combining eqs.~\eqref{Heq} and~\eqref{KG} and using the conditions~\eqref{Co1} and~\eqref{Co2} gives 
\beq\label{Slowroll}
H^2 \ = \ \frac{8\pi G}{3}V \ , \ \dot{\phi}^2 \ = \ \frac{V'^2}{24\pi G V} \ .
\eeq
Therefore, one can reformulate the conditions~\eqref{Co1} and~\eqref{Co2} as
\beq
\epsilon \ \ll \ 1 \ , \ \eta \ \ll \ 1 \ ,
\eeq
with
\beq\label{Epsiloneta}
\epsilon \ = \ \frac{V'^2}{16\pi G V^2} \ , \ \eta \ = \ \frac{|V''|}{8\pi GV} \ .
\eeq
Under these approximations, one can estimate the total number of e--foldings during inflation:
\beq
N \ = \ \int^{t_f}_{t_i}Hdt \ = \ \int^{\phi_f}_{\phi_i}\frac{H}{\dot{\phi}}d\phi \ \approx \ 8\pi G\int^{\phi_f}_{\phi_i}\frac{V}{V'}d\phi \ \approx \ 2\sqrt{\pi G}\Delta\phi  \ .
\eeq
A common choice of $V$ is the power-law
\beq
V(\phi) \ = \ \frac{\phi^\lambda}{M^{\lambda-4}} \ ,
\eeq
corresponding to
\beq
\epsilon \ \propto \ \frac{M_P^2}{\phi^2} \ .
\eeq
This model is known as ``large-field inflation``, since we need values of $\phi$ above the Planck scale to satisfy the condition. Alternatively, one could consider ``small-field inflation`` with a potential of the form
\beq
V(\phi) \ = \ V_0(1-\frac{\phi^2}{M^2}) \ , 
\eeq
\beq
\epsilon \ \propto \ \frac{M_P^2\phi^2}{M^4} \ .
\eeq
In this case, the value of $\phi$ should be smaller than $M^2/M_P$.

A model proposed by Alexei Starobinsky in 1979~\cite{Starobinsky:1979ty} involves a particular type of $f(R)$ gravity:
\beq
S \ = \ \frac{1}{16\pi G} \ \int d^4 x \ \sqrt{-g} \ \left(R \ + \ \frac{R^2}{6M^2} \right) \ .
\eeq
As we have seen before, these types of modes are dual to gravity coupled to a scalar field, and in this case the potential has the interesting shape
\beq
V(\phi) \ = \ \frac{3M^2}{4\kappa}(1-e^{-\sqrt{\frac{2}{3}\kappa}\phi})^2 \ ,
\eeq
which combines a flat region where inflation would take place and a dip where it would end eventually.

Let us address the mechanism of structure formation during inflation, which has to do with the perturbations of the inflaton:
\beq
\phi(t) \ \rightarrow \ \phi(t)+\delta\phi(t,\vec{r}) \ ,
\eeq
and of the metric tensor~\cite{Chib_Mukh}. The metric perturbations can be divided into three classes depending on how they transform under the Lorentz group: the scalar, the vector, and the tensor ones. The scalar perturbations have the form
\beq
(\delta g_{\mu\nu})_S \ = \ a^2(\tau) \begin{pmatrix}
-2\Phi & \partial_iB \\
\partial_iB & 2\Psi\delta_{ij}+2\left(\partial_i\partial_j-\frac{1}{3}\delta_{ij}\triangle\right)E
\end{pmatrix} \ ,
\eeq
while the vector ones are
\beq
(\delta g_{\mu\nu})_V \ = \ a^2(\tau) \begin{pmatrix}
0 & S_i \\
S_i & \partial_iF_j+\partial_jF_i
\end{pmatrix} \ .
\eeq
Finally, the tensor perturbations are given by
\beq
(\delta g_{\mu\nu})_T \ = \ a^2(\tau) \begin{pmatrix}
0 & 0 \\
0 & h_{ij}
\end{pmatrix} \ .
\eeq
Here we have changed the time variable $t$ to the conformal time variable $\tau$, defined as
\beq
\tau \ = \ \int_t^\infty \ \frac{dt'}{a(t')} \ , 
\eeq
such that the unperturbed metric is
\beq
ds^2  \ = \  a^2(\tau)(-d\tau^2+d\vec{r}^2) \ .
\eeq
At linear order, these perturbations propagate independently from each other. Due to the gauge transformations of the metric, outlined in Section~\ref{C1S2S3}, one can eliminate some of these degrees of freedom. Namely, if we shift the coordinates by the four--vector:
\beq
\delta\tau \ = \ \xi(x^\mu) \ , \ \delta x^i \ = \ \partial^i\beta(x^\mu) \ + \ v^i(x^\mu) \ ,
\eeq
where $\partial_iv_i \ = \ 0$, the inflaton perturbation would transform as:
\beq
\delta\phi(\tau,\vec{r}) \rightarrow \delta\phi(\tau,\vec{r}) \ - \ \xi\phi'(\tau) \ ,
\eeq
and the scalar metric perturbations as
\beq
\begin{gathered}
\Phi \ \rightarrow \ \Phi \ - \ \xi' \ - \ \frac{a'}{a}\xi \ , \\
B \ \rightarrow \ B+\xi+\beta' \ , \\
\Psi \ \rightarrow \ \Psi-\frac{1}{3}\triangle\beta-\frac{a'}{a}\xi \ , \\
E \ \rightarrow \ E-\beta \ .
\end{gathered}
\eeq
By choosing $\beta=E, \xi=-E'-B$, we can remove $B$ and $E$, so that we would be left with only two degrees of freedom, $\Phi$ and $\Psi$:
\beq
a^2(\tau)\left(-(1+2\Phi)d\tau^2+(1+2\Psi)d\vec{r}^2\right) \ .
\eeq
The non--zero Christoffel symbols would be:
\beq
\begin{gathered}
\Gamma^0_{00} \ = \ \mathcal{H}+\Phi' \ , \ \Gamma^0_{ii} \ = \ \mathcal{H}\left(1+2(\Psi-\Phi)\right)+\Psi' \ , \ \Gamma^i_{i0} \ = \ \mathcal{H}+\Psi' \ , \\
\\
\Gamma^i_{00} \ = \ \Gamma^0_{i0} \ = \ \partial_i\Phi \ , \ \Gamma^k_{ij} \ = \ \partial_i\Psi\delta_{jk}+\partial_j\Psi\delta_{ik}-\partial_k\Psi\delta_{ij} \ ,
\end{gathered}
\eeq
and the Ricci tensor components:
\beq
\begin{gathered}
R_{00} \ = \ \triangle\Phi-3\mathcal{H}'-3\Psi''+3\mathcal{H}(\Phi'-\Psi') \ ,\\
R_{ij} \ = \ -\partial_i\partial_j(\Phi+\Psi)+\left(2\mathcal{H}^2+\mathcal{H}'+\Psi''-\triangle\Psi+\right.\\
\left.2(\mathcal{H}'+2\mathcal{H}^2)(\Psi-\Phi)+\mathcal{H}(5\Psi'-\Phi')\right)\delta_{ij} \ , \\
R_{i0} \ = \ 2\partial_i(\mathcal{H}\Phi-\Psi') \ .
\end{gathered}
\eeq
The stress--energy tensor of inflaton is
\beq
T_{\mu\nu} \ = \ \partial_\mu\phi\partial_\nu\phi-g_{\mu\nu}(\frac{1}{2}\partial_\alpha\phi\partial^\alpha\phi+V(\phi)) \ .
\eeq
Assuming localized field perturbations on uniform background:
\beq
\phi \ = \ \bar{\phi}(\tau)+\delta\phi(\tau,r) \ ,
\eeq
we get $T_{ij}=0$ for $i\neq j$. As a result, the $ij$-components of Einstein equations with $i\neq j$ are simply
\beq
\partial_i\partial_j(\Phi+\Psi) \ = \ 0 \ ,
\eeq
which is equivalent to $\Psi=-\Phi$ due to the boundary condition $\Phi|_{r\rightarrow\infty}=\Psi|_{r\rightarrow\infty}=0$.

The Ricci scalar is therefore
\beq
R \ = \ \frac{1}{a^2}\left(6(\mathcal{H}^2+\mathcal{H}')+2\triangle\Phi-6\Phi''-24\mathcal{H}\Phi'-12(\mathcal{H}'+\mathcal{H}^2)\Phi\right) \ .
\eeq
The background Einstein equations are
\beq
\begin{gathered}
\frac{3}{2}\mathcal{H}^2=4\pi G\left(\frac{1}{2}{\phi'}^2+a^2V(\bar{\phi})\right) \ ,\\
-\frac{1}{2}\mathcal{H}^2-\mathcal{H}'\ = \ 4\pi G\left(\frac{1}{2}\bar{\phi}'^2-a^2V(\bar{\phi})\right) \ ,
\end{gathered}
\eeq
and equations for the perturbations are given by
\beq
\begin{gathered}
\triangle\Phi-3\mathcal{H}\Phi'=4\pi G\left({\bar{\phi}'}\delta\phi'+a^2\delta\phi V'(\bar{\phi})+2\Phi a^2V(\bar{\phi})\right) \ ,\\
\Phi''+3H\Phi'+2\Phi(\mathcal{H}^2+2\mathcal{H}') \ = \ \\
4\pi G\left(\delta\phi'\bar{\phi}'-2\Phi(\bar{\phi}')^2-a^2V'(\bar{\phi})\delta\phi+2\Phi a^2V\right) \ ,\\
\partial_i\left(\mathcal{H}\Phi+\Phi'\right) \ = \ 4\pi G\bar{\phi}'\partial_i\delta\phi \ .
\end{gathered}
\eeq
The Klein--Gordon equation is
\beq
\bar{\phi}''+2\mathcal{H}\bar{\phi}'+a^2V'=0
\eeq
for background, and
\beq
\delta\phi''+2\mathcal{H}\delta\phi'-\triangle\delta\phi-4\Phi'\bar{\phi}'+2a^2\Phi V'+a^2V''\delta\phi \ = \ 0 \ ,
\eeq
for the perturbations.

Then, we can take the second Einstein equation for perturbations and substitute the definitions of $V(\phi)$ and $\Phi$ from the background Einstein equations and the definition of $V'(\phi)$ from the Klein-Gordon equation to obtain:
\beq
\triangle\Phi-\left(\mathcal{H}-\frac{\mathcal{H}'}{\mathcal{H}}\right)\Phi' \ = \ 4\pi G\left(\bar{\phi}'\delta\phi'-\delta\phi(\bar{\phi}''-\frac{\mathcal{H}'}{\mathcal{H}}\bar{\phi}')\right) \ .
\eeq
Substituting $\mathcal{H}-\frac{\mathcal{H}'}{\mathcal{H}}$ for $\frac{4\pi G}{\mathcal{H}}\bar{\phi}'^2$, based on background Einstein equations, we find
\beq
\triangle\Phi \ = \ 4\pi G\frac{\bar{\phi}'^2}{\mathcal{H}}\left(\Phi+\frac{\mathcal{H}}{\bar{\phi}'}\delta\phi\right)'
\eeq
Finally, taking the third Einstein equation and acting on it with the Laplace equation, we obtain
\beq
\begin{gathered}
\left(\Phi+\frac{\mathcal{H}}{\bar{\phi}'}\delta\phi\right)''+2\left(\mathcal{H}-\frac{\mathcal{H}'}{\mathcal{H}}+\frac{\bar{\phi}''}{\bar{\phi}'}\right)*\\
\left(\Phi+\frac{\mathcal{H}}{\bar{\phi}'}\delta\phi\right)'-\triangle\left(\Phi+\frac{\mathcal{H}}{\bar{\phi}'}\delta\phi\right) \ = \ 0 \ ,
\end{gathered}
\eeq
or, alternatively,
\beq
u''-\triangle u-\frac{z''}{z}u \ = \ 0 \ ,
\eeq
where
\beq
u \ = \ a\left(\delta\phi+\frac{\bar{\phi}'}{\mathcal{H}}\Phi\right)
\eeq
is the so--called Mukhanov--Sasaki variable, and
\beq
z \ = \ \frac{a}{\mathcal{H}}\bar{\phi}' \ = \ \frac{a}{H}\dot{\bar{\phi}} \ .
\eeq
This equation corresponds to the effective action
\beq
\mathcal{S} \ = \ \int d\tau d^3x \ \frac{1}{2}\left[u'^2-(\vec{\nabla}u)^2+\frac{z''}{z}u^2\right] \ ,
\eeq
which is basically a scalar field with a mass that changes in time.

In perfect de Sitter space, $a=-(H\tau)^{-1}$, and $\dot{\bar{\phi}}=0$; however, in a quasi--de--Sitter space with a slowly rolling field, we can see from expressions~\eqref{Slowroll} and~\eqref{Epsiloneta} that $z\propto a\sqrt{\epsilon}$, and therefore
\beq\label{z}
\frac{z''}{z} \ \approx \ \frac{2}{\tau^2} \ .
\eeq
In Fourier space, the equation has the form
\beq
u_k'' \ + \ \left(k^2-\frac{2}{\tau^2}\right)u_k \ = \ 0 \ .
\eeq
For perturbations beyond the Hubble scale ($k\gg aH$), we can neglect the last term and canonically quantize the field. To do this, we define the conjugate momentum:
\beq
\pi \ = \ \frac{\delta\mathcal{L}}{\delta u'} \ = \ u' \ ,
\eeq
and then perform the Fourier series expansion:
\beq
\hat{u}(\tau,\vec{r}) \ = \ \int \ \frac{d^3k}{(2\pi)^{3/2}}\left[u_k(\tau)\hat{a}_ke^{i\vec{k}\vec{r}} \ + \ u^*_k(\tau)\hat{a}^\dagger_ke^{-i\vec{k}\vec{r}}\right] \ .
\eeq
However, as the perturbations stretch out beyond the horizon, the last term starts to dominate over $k^2$, and the field condenses, acquiring a VEV corresponding to the classical solution of the equation.

The solution is given by
\beq
\alpha\left(1-\frac{i}{k\tau}\right)e^{-ik\tau} \ + \ \beta\left(1+\frac{i}{k\tau}\right)e^{ik\tau} \ .
\eeq
In Minkowski space, the positive--energy mode is
\beq\label{Mink}
\frac{1}{\sqrt{2k}}e^{-ik\tau} \ ,
\eeq
and therefore, to obtain correct asymptotics, we have to set $\alpha=\frac{1}{\sqrt{2k}}, \beta=0$.

More generically, if we take into account the higher--order corrections from slow--roll parameters, the coefficient in~\eqref{z} would be different from 2, and the equation can be written in the form
\beq
u_k'' \ + \ \left(k^2-\frac{\nu^2-1/4}{\tau^2}\right)u_k \ = \ 0 \ 
\eeq
without loss of generality (the standard de Sitter case corresponds to $\nu=\frac{3}{2}$). The solution to this equation is
\beq\label{Usol}
u \ = \ \sqrt{-\tau}\left(\alpha H^{(1)}_\nu(-k\tau)+\beta H^{(2)}_\nu(-k\tau)\right) \ ,
\eeq
where $H^{(1)}_\nu$ and $H^{(2)}_\nu$ are Hankel functions of the first and second kind (for power--law inflation, these solutions are exact~\cite{Lucchin:1984yf}). Using their asymptotic forms for $-k\tau\gg1$:
\beq
\begin{gathered}
H^{(1)}_\nu(-k\tau) \ \sim \ \sqrt{\frac{2}{-k\tau\pi}}\exp\left(-ik\tau-i\frac{\pi}{2}\nu-i\frac{\pi}{4}\right) \ ,\\
H^{(2)}_\nu(-k\tau) \ \sim \ \sqrt{\frac{2}{-k\tau\pi}}\exp\left(ik\tau+i\frac{\pi}{2}\nu+i\frac{\pi}{4}\right) \ ,
\end{gathered}
\eeq
and matching the expression~\eqref{Usol} to~\eqref{Mink}, we obtain the condition
\beq
\alpha=\frac{\sqrt{\pi}}{2}\exp\left(i\frac{\pi}{2}\nu+i\frac{\pi}{4}\right) \ , \ \beta \ = \ 0 \ .
\eeq
Then, we use the asymptotic form of $H^{1}_\nu$ at $-k\tau\ll1$:
\beq
H^{(1)}_\nu(-k\tau) \ \sim \ -i\left(\frac{2}{-k\tau}\right)^\nu\frac{\Gamma(\nu)}{\pi} 
\eeq
to obtain the result
\beq
u \ = \ 2^{\nu-2}k^{-\nu}\left(-\tau\right)^{\frac{1}{2}-\nu}\frac{\Gamma(\nu)}{\Gamma(\frac{3}{2})}\exp\left(i\frac{\pi}{2}\nu-i\frac{\pi}{4}-ik\tau\right) \ .
\eeq
Knowing the value of $u$, we can easily compute the so-called comoving curvature perturbation $\mathcal{R}$:
\beq
\mathcal{R} \ = \ \Phi+\frac{\mathcal{H}}{\bar{\phi}'}\delta\phi \ = \ \frac{u}{z} \ ,
\eeq
and calculate its power spectrum:
\beq
P_{\mathcal{R}} \ = \ \frac{k^3}{2\pi^2}|\mathcal{R}|^2 \ = \ \frac{H^4}{4\pi^2\dot{\phi}^2}\left(\frac{\Gamma(\nu)}{\Gamma(\frac{3}{2})}\right)^2\left(\frac{k}{2Ha}\right)^{3-2\nu} \ ,
\eeq
which may also be written in terms of the slow--roll parameter:
\beq
P_{\mathcal{R}} \ = \ \frac{H^2G}{\pi\epsilon}\left(\frac{\Gamma(\nu)}{\Gamma(\frac{3}{2})}\right)^2\left(\frac{k}{2Ha}\right)^{3-2\nu} \ .
\eeq
We can define the so--called spectral index $n_s$:
\beq
n_s \ = \ 4-2\nu \ .
\eeq
The case of $n_s=1$ (equivalent to $\nu=\frac{3}{2}$) corresponds to the idealized de Sitter model, and small deviations from 1 can be measured to test competing inflation models. According to the recent Planck data, the value of $n_s$ is $0.965\pm0.004$~\cite{PLANCK}, which is significantly different from 1.

Finally, we have to repeat the same procedure for the tensor perturbations. For this purpose, we can use the expression~\eqref{EffGR} at first (quadratic) order, and decompose it into two polarizations, which basically gives us an action for two massless scalar fields:
\beq
\begin{gathered}
S_{eff} \ = \ -\frac{1}{64\pi G}\int d\tau d^3 x \ a^2(\tau) \ \left[(\partial_\alpha h_{+})^2+(\partial_\alpha h_{x})^2\right] \ .
\end{gathered}
\eeq
For each of these fields, we can redefine the variables:
\beq
v \ = \ \frac{a}{4\sqrt{2\pi G}}h_{+,x}
\eeq
to obtain two copies of an action, which is almost exactly Mukhanov--Sasaki:
\beq
\mathcal{S} \ = \ \int d\tau d^3x \ \frac{1}{2}\left[u'^2-(\vec{\nabla}u)^2+\frac{a''}{a}u^2\right] \ .
\eeq
Now, $\frac{a''}{a}=\frac{\mu^2-1/4}{\tau^2}$, where, once again, $\mu$ depends on the slow--roll parameters, and in the idealized case we have $\mu=\frac{3}{2}$. Retracing the same steps as for the scalar perturbations, one finds the spectrum of the metric perturbations
\beq
P_{T} \ = \ \frac{16H^2G}{\pi}\left(\frac{\Gamma(\mu)}{\Gamma(\frac{3}{2})}\right)^2\left(\frac{k}{2Ha}\right)^{3-2\mu} \ ,
\eeq
with the additional factor of 2 due to the presence of two fields. We can also introduce the tensorial spectral index:
\beq
n_T \ = \ 3-2\mu \ . 
\eeq
In the de Sitter limit, the tensor--to--scalar ratio is simply
\beq
r \ = \ \frac{P_T}{P_\mathcal{R}} \ = \ 16\epsilon \ .
\eeq
The observations of the Planck collaboration constrain the value of $r$ to be smaller than $0.06$~\cite{PLANCK}.

In scenarios motivated by String Theory, the inflaton is a combination of the dilaton field and a ubiquitous modulus of the Calabi-Yau manifold, known as the breathing mode. In the aforementioned scenario of brane SUSY breaking, and in fact in all three tachyon-free ten-dimensional strings with broken supersymmetry~\cite{so16so16,0primeB,bsb}, the dilaton potential would have a term of the form~\eqref{DilPot}, with a ``critical'' value of the exponent (in the two types of orientifolds) and a hyper-critical one (in the $SO(16)\times SO(16)$ string) that make an early climbing phase inevitable, at least within the low-energy theory. On the other hand, for $\gamma \leq \frac{3}{2}$ the field could exhibit both descending and climbing behaviors, which makes a climbing phase a telltale sign of SUSY breaking in String Theory.

In this fashion, String Theory and SUSY breaking can provide a possible clue for the onset of inflation, although not for the actual inflationary phase. The scalar, after the early climbing, would collect some energy driving it to descend the potential, and if this were corrected by terms capable of bringing it to slow-roll climbing the subsequent steps would lead naturally to the onset of inflation. For example, the simplest resulting scenarios would approach the Lucchin-Matarrese attractor~\cite{LM}, thus leading to power-like inflation. A simple, albeit not realistic, analytic scenario to this end is provided by the $\left(\gamma,\frac{1}{\gamma}\right)$ model in~\cite{fss}.

Interestingly, an early climbing phase could have left some tangible signs in the sky. With a sufficiently short inflationary period (say 50-60 e-folds or so), an early fast-roll period would induce a natural damping of scalar (and tensor) perturbations for low values of $\ell$~\cite{climbing}, so that signs of this early phase might be accessible in the sky. The damping effect appears consistent with Planck data~\cite{PLANCK}, where a sizable lack of power has long been noted in the quadrupole, and the effect improves if one masks the region around the galactic plane a bit further than is usually done, and especially so for even multipoles. In the best possible scenario, this behavior could signal some incomplete cleaning of the available data in that region, and the new generation of experiments may say more on all this.

\chapter{Dark Matter}\label{C2}
\section{Observational evidence for dark matter}\label{C2S1}
The discovery of dark matter (DM) dates to the 1930s, when the Swiss-American astronomer Fritz Zwicky estimated the velocity dispersion of the Coma Cluster and found the theoretical values to be much smaller than what was actually observed; in order to match the observations, the gravitating mass of the cluster would have to be considerably larger than the luminous mass~\cite{Zwicky:1933gu,Zwicky:1937zza}. The work of Zwicky was followed by a paper of Sinclair Smith, who estimated the total mass of the Virgo cluster from the velocities of the galaxies rotating around it, assuming that their rotation be circular and using the formula for the centripetal acceleration from classical mechanics~\cite{Smith:1936mlg}:
\beq
a \ = \ \frac{v^2}{r} \ . \label{e1}
\eeq
On the other hand, the gravitational acceleration created by the cluster at large distances from it is given by
\beq
a \ = \ \frac{GM_{c}}{r^2} \ , \label{e2}
\eeq
so equating~\ref{e1} and~\ref{e2}, we obtain
\beq
M_c \ = \ \frac{rv^2}{G} \ , \label{e3}
\eeq
or, equivalently,
\beq
v \ = \ \sqrt{\frac{GM_c}{r}} \ . \label{RC}
\eeq
Alternatively, this result may be derived from the virial theorem for the Coulomb potential. The cluster mass estimated according to~\eqref{e3} turned out to be about two orders of magnitude larger than the average nebula mass.

\begin{figure}
    \centering
	\includegraphics[scale=0.45]{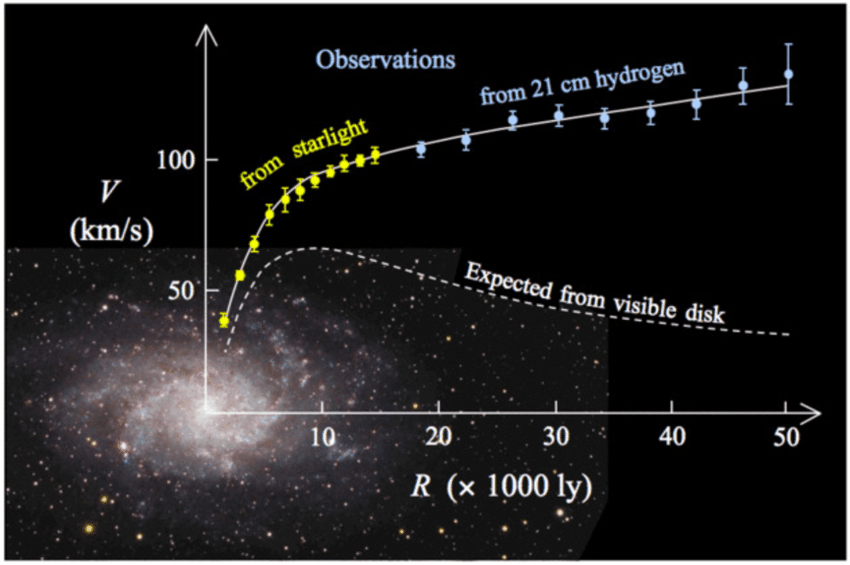}
	\caption{The rotation curve of the galaxy M33: dashed - rotation curve inferred from the galaxy's baryonic profile, solid - observed rotation curve}
\label{M33}
\end{figure}

\begin{figure}
    \centering
	\includegraphics[scale=0.55]{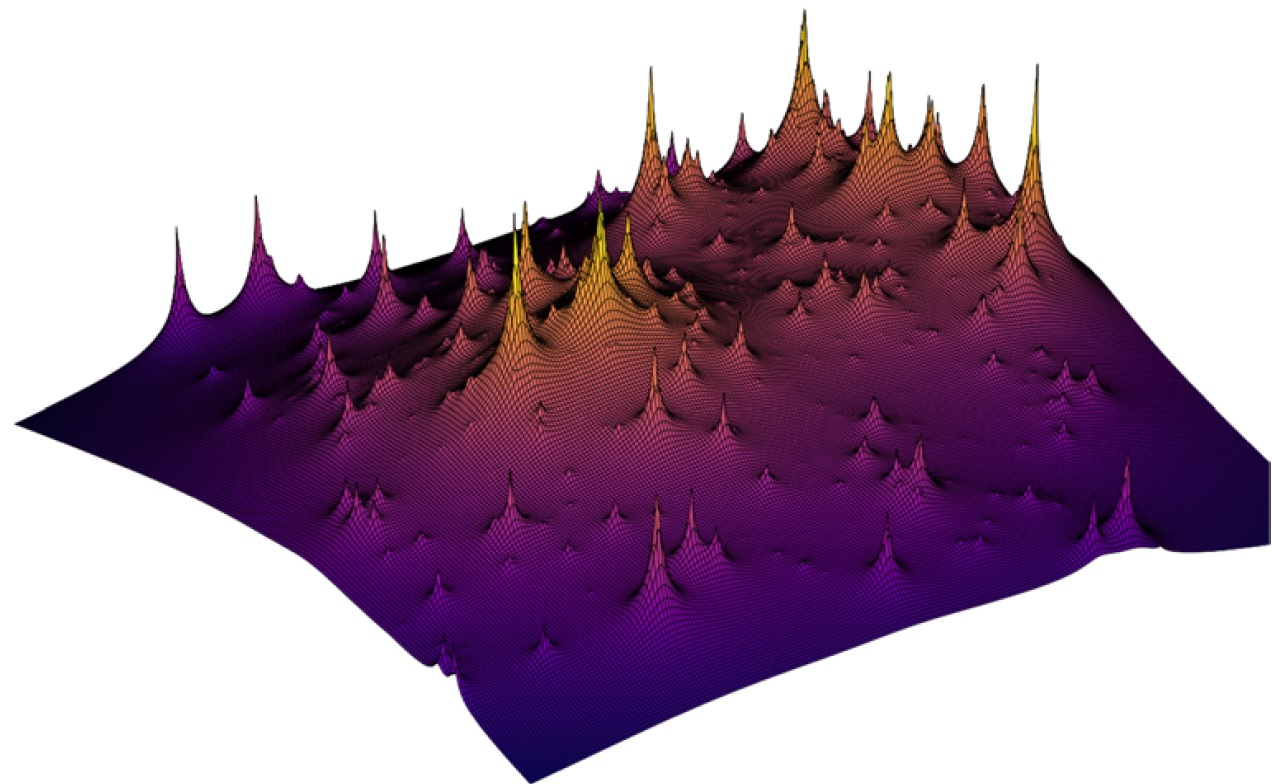}
	\caption{Lensing--derived matter distribution map in the Abell 2744 cluster (sharp peaks correspond to baryonic matter, the flat background corresponds to dark matter)}
\label{LensingMap}
\end{figure}

However, the first systematic evidence of the existence of the non-luminous matter was provided only in the 1970s by Vera Rubin, Kenneth Ford, and Norbert Thonnard, who studied the rotation curves of the galaxies, i.e. the rotation velocities of galactic objects versus the distance of these objects from the center of the galaxy~\cite{Rubin:1978kmz,Rubin:1980zd}. At radii that are large enough to enclose almost all baryonic matter in the galaxy, the rotation curves should be given by~\ref{RC}; however, instead of decreasing, they reach an almost constant value (they `flatten``). While they are not exactly flat, they decrease much slower than predicted by~\ref{RC}, and in certain cases like the M33 galaxy, they can even increase (fig.~\ref{M33}). This would imply either a dark halo consisting of a yet unknown type of matter surrounding the baryonic disk or a modification of the laws of gravity at large distances and/or small accelerations.
Although Rubin herself favored the second solution, over the next few decades it became clear that dark matter is necessary to explain a multitude of observational phenomena. First of all, dark matter is essential for primordial structure formation: in the early Universe before recombination, baryonic matter interacts with radiation, which inhibits the growth of baryonic perturbations, and therefore the ``skeleton`` of the Universe has to consist of DM. Second, the matter content of the Universe can be deduced from the acoustic peaks in the CMB spectrum; the ratio of the second peak's amplitude to the first defines the percentage of baryonic matter (i.e. matter which has both self-gravitation and pressure), while the third peak determines the fraction of dark matter (i.e. matter which has self-gravitation but no pressure); according to Planck data, the former constitutes about 5\% of the critical density of the Universe, and the latter is around 27\%~\cite{PLANCK}. Third, gravitational lensing indicates the presence of dark matter halos around baryonic matter, and even allows to create ``maps`` of dark matter distributions~\cite{Natarajan:2017sbo,Tyson:1998vp} (fig.~\ref{LensingMap}). Besides, when galaxy clusters collide, like the Bullet Cluster and the Abell 520, dark matter is clearly separated from the baryons: clumps of baryonic hot gas from the two clusters interact with each other, which slows their relative motion, while the DM halos pass through each other without interaction~\cite{Markevitch:2003at,Mahdavi:2007yp}. Finally, there is hot gas in clusters which would have evaporated without the gravitational wells created by dark matter~\cite{Buote:2003tw}, and the so-called Kaiser effect. The latter refers to the distortion of the redshifts of the galaxies due to their peculiar motion: using this distortion, it is possible to calculate the rotation velocities and thus the depth of the gravitational wells; calculations show that the total matter density is about 30\% of the critical density, in accordance with the $\Lambda$CDM model and with CMB~\cite{Peacock:2001gs}.

If dark matter were a particle, it would be logical to assume that it was in thermal equilibrium with the Standard Model particles since reheating, and decoupled at some point after that. Based on this assumption, we can divide it into three types: hot, warm and cold, depending on its mass and temperature. If its temperature is smaller than its mass during the decoupling (i.e. it is non--relativistic), it is known as cold dark matter (CDM). If the temperature is larger than the mass during both the decoupling and the radiation-matter transition (i.e. it remains relativistic during the radiation-dominated epoch), it is hot dark matter (HDM). Finally, there is the intermediate case, known as warm dark matter (WDM), when the decoupling temperature is larger than the mass, but the matter-radiation transition temperature is smaller than the mass.

\begin{figure}
    \centering
	\includegraphics[scale=0.9]{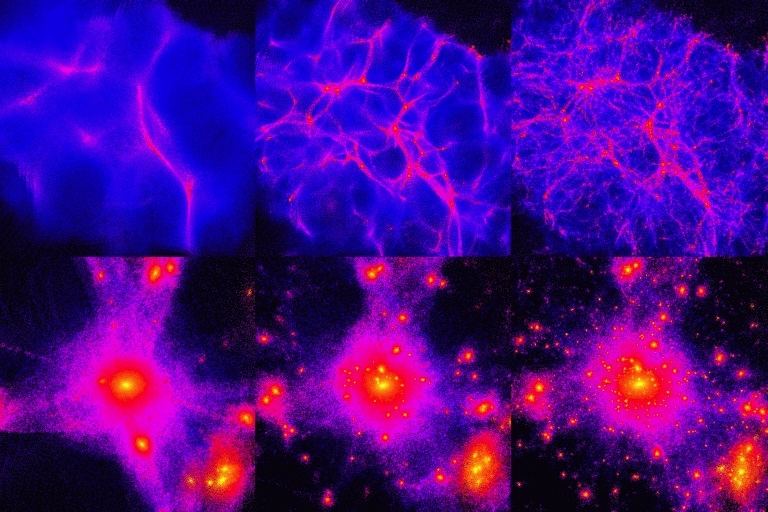}
	\caption{From left to right: structure formation with hot, warm, cold dark matter}	
\end{figure}	
\section{Discrepancies within the cold dark matter model}\label{C2S2}

The cold dark matter (CDM) model describes physics at intergalactic scales (i.e. large-scale structure) with high precision; however, it has a few discrepancies at smaller (galactic) scales. These discrepancies may have purely astrophysical explanations, but they have also been interpreted as possible indications of some unknown properties of dark matter.

The first one of these discrepancies is the problem of ``missing satellites``: CDM-based computer simulations predict that large galaxies like the Milky Way should have far more satellite dwarf galaxies than the astronomers actually observe~\cite{Klypin:1999uc}. It may be possible to resolve this problem by stating that we simply do not observe the extra satellite galaxies because they lack visible stars, but this would lead to another issue, known as ``too big to fail``. The essence of ``too big to fail`` is that some of the observed satellite galaxies are too massive not to have visible stars. 

Simulations suggest that the problem could be resolved by baryonic feedback: if one includes baryonic matter, gravitational and tidal forces from the baryonic disk of the main galaxy would destroy small satellite galaxies~\cite{Kim:2017iwr,Garrison-Kimmel:2017zes,Brooks:2012ah}.

The second problem with CDM is the core-cusp issue: in simulated galaxies, the dark matter density profiles are close to the so-called Navarro-Frank-White profile (NFW)~\cite{Navarro:1996gj}:
\beq
\rho_{NFW} \ = \ \frac{\rho_0}{\frac{r}{r_s}(1+\frac{r}{r_s})^2} \ ,
\label{NFW}
\eeq 
with the density becoming infinite at the center (the ``cusp``), while fits of observed rotation curves favor profiles with a ``core``, i.e. an almost constant density profile at the center. These profiles are better approximated by the Burkert anzatz~\cite{Burkert:1995yz}:
\beq
\rho_B \ = \ \frac{\rho_0}{(1+\frac{r}{r_0})(1+\frac{r^2}{r_0^2})} \ . \label{IT}
\eeq
\begin{figure}
    \centering
	\includegraphics[scale=0.4]{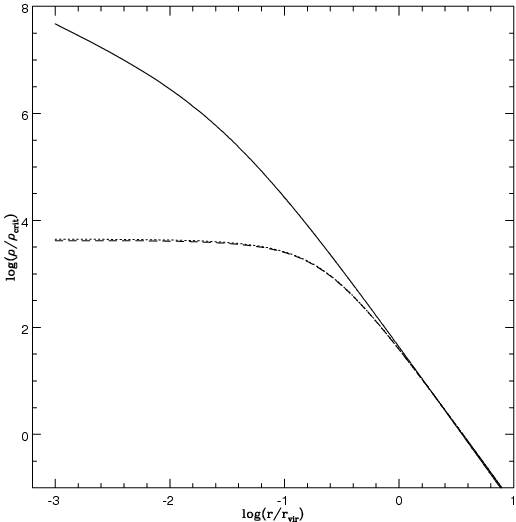}
	\includegraphics[scale=0.7]{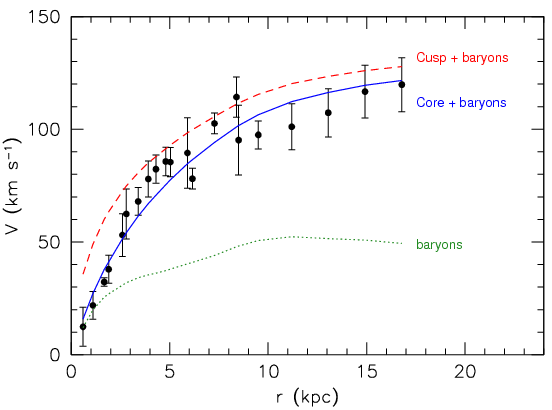}
	\caption{Left panel: cored Burkert density profile (dashed line), and cusped NFW density profile (solid line). Right panel: a typical galaxy rotation curve, where the black dots are data points, the green dotted curve illustrates the baryonic disk, the blue curve is a data fit with Burkert and baryons, while the red curve is a data fit with NFW and baryons.}
\end{figure}
As is the case with ``missing satellites``, astrophysical observations seemingly favor a solution based on baryonic feedback. According to those observations, dwarf galaxies with intensive star formation processes in the center appear to have shallow cores (corresponding to the thermalization of dark matter), while those with a lower star formation rate have more ``cuspy`` profiles. This can be explained by the fact that intensive star formation leads to an outflow of baryonic matter from the center; as a result, the gravitational wells become shallower, resulting in dark matter depletion~\cite{Read:2018fxs,Pontzen:2011ty}.

Another phenomenon unexplained within the CDM model is the Tully-Fisher relation, according to which the integrated luminosity of a galaxy, proportional to its baryonic mass, is also proportional to the fourth power of its asymptotic rotational velocity (the velocity at the far end of the rotation curve)~\cite{Tully:1977fu}:
\beq
M_b \ \propto \ L \ \propto \ v_a^4 \ .
\label{eqn1}
\eeq
\begin{figure}
	\begin{center}
		\includegraphics{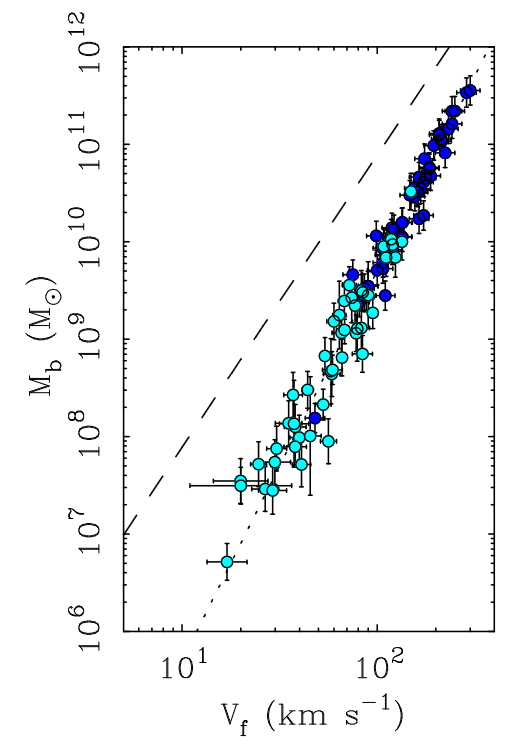}
	\end{center}
	\caption{The Tully-Fisher relation: circles - data (dark-blue - star-dominated galaxies, light-blue - gas-dominated galaxies), dotted line - $v^4$ fit (empirical), dashed line - $v^3$ fit (theoretical)}
\end{figure}
Using the standard formula for the centripetal acceleration, we can see that the total acceleration acting on a test particle is given by
\beq
a_{tot}(r) \ = \ \frac{v_a^2}{r} \ . \label{a}
\eeq
At the same time, the acceleration created by the baryons far away from the baryonic disk is
\beq
a_B(r) \ = \ \frac{GM_B}{r^2} \ .
\eeq
Thus the expression~\eqref{eqn1} can be rewritten in the following form, known as the radial acceleration relation (RAR)~\cite{McGaugh:2016leg}:
\beq
a_{tot}(r) \ = \ \sqrt{a_0a_B(r)} \ ,
\label{RAR}
\eeq
where $a_0$ is a constant of the order of the Hubble parameter:
\beq
a_0\ \approx \ \frac{1}{2\pi}cH_0 \ = \ \frac{c^2}{2\pi}\sqrt{\frac{\Lambda}{3}} \ .
\eeq
The standard collapse theory would predict a cubic power, and moreover, the asymptotic velocity depends almost exclusively on the physics of the dark halo, so that the correlation with baryonic matter would be expected to be weak over there, which should result in a much larger dispersion. Cosmological simulations suggest that the Tully-Fisher relation may also emerge from baryonic, in particularly supernova, feedback~\cite{Steinmetz:1998gr,DeRossi:2010ey}; besides, studies of rotation curves of dwarf disk spirals (DDS) and low surface brightness galaxies (LSBs) indicate that the Tully-Fisher relation is a limiting case of a more general phenomenological formula, the so-called GGBX relationship~\cite{DiPaolo:2018mae}.

The reason for the correspondence between $a_0$ and $H_0$ remains unclear, however, since the former is defined by galaxy-scale dynamics, and the latter emerges at much larger (cosmological) scales.

There is also the possibility to address all three of these issues postulating new dark matter properties: namely, for the ``missing satellites`` one would need a mechanism to suppress the high wavenumbers in the primordial perturbation power spectrum, for the core-cusp one would need a small-scale repulsive force (for instance, DM self-interaction or Heisenberg uncertainty pressure which can become significant for very light particles), and for the Tully-Fisher and radial acceleration relation, it would be necessary to either introduce a new interaction between baryonic matter and dark matter or to replace dark matter with a modification of Newtonian laws. A number of these approaches are reviewed in chapter 2; however, given the aforementioned baryonic feedback effects, it remains unclear whether these models would be preferable over CDM, or whether they would be tightly constrained.

\begin{figure}
    \centering
	\includegraphics[scale=0.8]{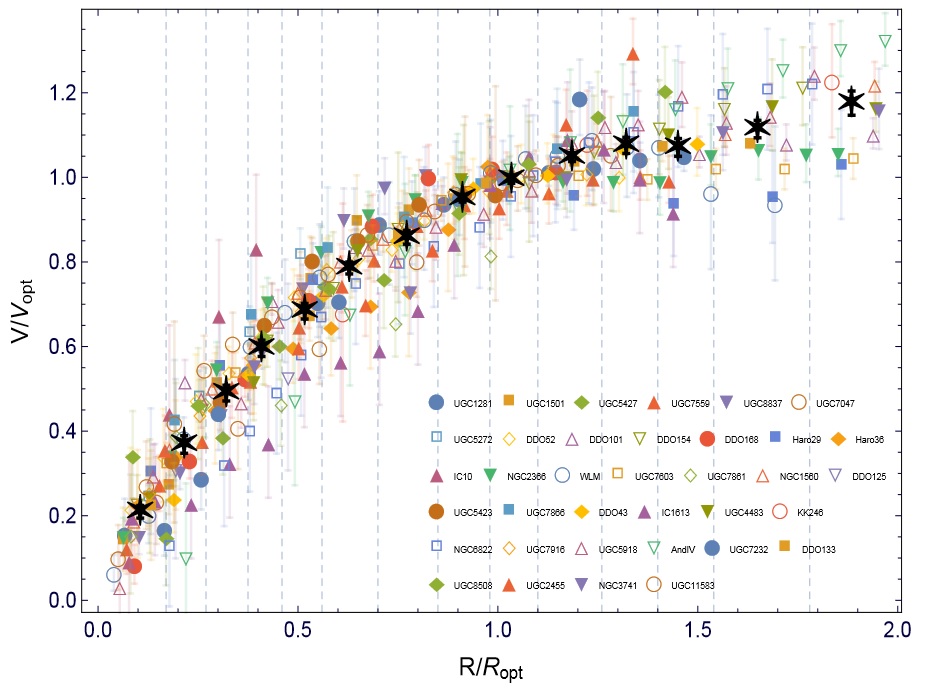}
	\caption{The universal rotation curve of 36 dwarf disk galaxies in normalized units (the black stars correspond to the URC)}
	\label{URC}
\end{figure}

Finally, there is the so-called universal rotation curve (URC). It has been demonstrated that a number of rotation curves of galaxies within the same class (spirals, dwarf disc galaxies) look almost the same in normalized units~\cite{Karukes:2016eiz,Persic:1995ru,Salucci:2007tm} (fig.~\ref{URC}): the radius is normalized to the optical radius of the galaxy, and the velocity is normalized to its value at $R_{opt}$:
\beq
V_{opt} \ = \ V(R_{opt}) \ .
\eeq
Since the optical radius is a parameter characterizing only the distribution of luminous matter, the URCs, just like the Tully-Fisher, imply a tight correlation between baryonic and dark matter distributions, and have also been interpreted as a possible sign of the additional non-gravitational interaction between them~\cite{Salucci:2017cet}. 

\section{Models of Dark Matter}\label{C2S3}

\subsection{Weakly interacting massive particles (WIMPs)}\label{C2S3S1}
If the interactions between dark--matter particles and the Standard Model particles were not solely due to gravity, all particles should have been in thermal equilibrium in the Early Universe before decoupling.
Dark matter manifests itself today in non--relativistic regimes, so that its number density scales with the cube of the temperature.

One is thus led to express its value at the decoupling temperature either as a function of its current number density
\beq
n(T_d) \ = \ \frac{\rho_c\Omega_\chi}{m_\chi}\left(\frac{T_d}{T}\right)^3 \label{ntd1} \ ,
\eeq
or via the Maxwell-Boltzmann distribution, which applies to bosons and fermions in the non--relativistic regime, where $E \simeq m_\chi + \frac{p^2}{2 m_\chi}$:
\beq
\begin{gathered}
n(T_d) \ = \ g_\chi \left[\frac{m_\chi T_d}{2\pi}\right]^{3/2}e^{-\frac{m_\chi}{T_d}} \ . \label{ntd2}
\end{gathered}
\eeq
Here $g_\chi$ is the typical number of degrees of freedom per dark matter particle, and combining eqs.~\eqref{ntd1} and~\eqref{ntd2} gives
\be
\left(\frac{\rho_c \Omega_\chi}{m_\chi T^3}\right)\ =\ g_\chi \ \left(\frac{x}{2\pi}\right)^{3/2}\ e^{-x} \ ,
\eeq
with
\beq
x \ = \ \frac{m_\chi}{T_d} \ .
\eeq
Taking into account the value $\Omega_\chi\sim0.27$ from PLANCK observations,
\beq
x^{3/2}e^{-x}\sim\frac{10^{-8} GeV}{m_\chi} \ ,
\eeq
so that $x\sim25-30$ for $m_\chi$ in the range 100 GeV--10 TeV.

Decoupling occurred when the interaction rate
\beq
\Gamma \ =\ <\sigma v>n(T_d)
\eeq
became comparable to the Hubble rate
\beq
H \ = \ \frac{T_d^2}{\bar{M}_P} \ ,
\eeq
so that
\beq
<\sigma v> \ = \ \frac{T^3x}{\bar{M}_P\rho_c\Omega_\chi}\sim3*10^{-26} cm^3/s \ ,
\eeq
and for typical velocities $v\sim0.1 c$ one ends up with the Weak--interaction cross section
\beq
\sigma\sim\frac{\alpha^2}{m^2_w} \ .
\eeq
This coincidence, known as the ``WIMP miracle``, has motivated numerous experiments aimed at the direct detection of WIMP scattering on targets.

The WIMPs should have an additional conserved quantum number to prevent their decays into lighter Standard Model particles. In supersymmetric models, this is the so-called R-parity, discussed in Section~\ref{SUSY}. Its value is +1 for ordinary particles and -1 for the superpartners, which therefore cannot decay into each other, and the lightest supersymmetric particle (LSP) would be stable. Within the minimal supersymmetric Standard Model (MSSM), the most natural candidate for the LSP is the lightest neutralino, a superposition of b--ino, w--ino, and two higgs--inos, the superpartners of the $B$-boson, the $W^0$-boson, and the Higgs boson,
\beq
\chi \ = \ \alpha\tilde{B}+\beta\tilde{W}^0+\gamma\tilde{H}_1^0+\delta\tilde{H}_2^0 \ .
\eeq
Other candidates include s-neutrinos (in extensions beyond the MSSM) and gravitinos (in supergravity). Note that the masses of SUSY candidates are generally of order
\beq
\frac{\Lambda_{SSB}^2}{\bar{M}_P} \ ,
\eeq
where $\Lambda_{SSB}$ is the scale of supersymmetry breaking.
Likewise, in models with compactified extra dimensions DM naturally arises from internal excitations, and the conserved quantum number is the Kaluza-Klein parity (reflection symmetry in a compactified dimension) which is +1 for SM particles and -1 for KK modes.
\subsubsection{Detection of WIMPs}
Three possible ways of detecting WIMPs have been widely explored. The first method is model independent: that is, to look for WIMP-nucleus coherent scattering, with an expected signal that should display an annual modulation due to Earth's motion around the Sun through the galactic plane~\cite{Goodman:1984dc}. The other methods are model dependent, and involve collider searches looking for signatures of specific supersymmetric particles, or indirect searches looking for some possible excess of cosmic particles originating from two--WIMP annihilations into SM particles. Notice that only the last two options are viable for gravitinos, which can interact with nuclei only via strongly suppressed higher-order diagrams.
\begin{figure}
\centering
\includegraphics[scale=0.58]{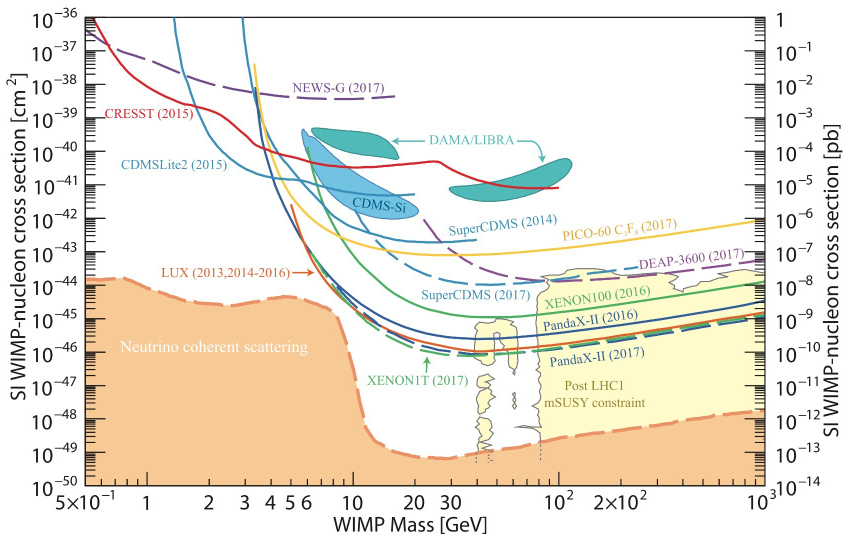}
\caption{Experimental constraints on WIMP parameters originating from null results, together with the location of the DAMA/LIBRA signals.}
\label{WIMP}
\end{figure}
With a notable exception, the experiments looking for the scattering of WIMPs on nuclei have provided no evidence for it so far, and this negative result has been turned into stringent constraints on the WIMPs' mass and cross section (fig.~\ref{WIMP}).
The notable exception is due to the Gran Sasso DAMA/LIBRA collaboration, which has reported for about two decades by now a signal with an annual modulation and the proper peak times, for which no other viable explanation appears possible~\cite{Bernabei:2008yi}. The superconductor-based CDMS-Si experiment has also reported some events deviating from the background level~\cite{Ahmed:2009zw,Agnese:2013rvf,Agnese:2014aze}, but their significance remains unclear, since its more sensitive version SuperCDMS gave only negative results. Positive results reported by two other experiments, CoGeNT and CRESST-II, have been explained by previously underestimated backgrounds~\cite{Davis:2015vla}. The fact that the DAMA experiment was the only one to use a NaI(Tl) target, while the other experiments employed either superconductors or liquid gases, leaves open a couple of ``loopholes``, related either to exotic composite ``dark atoms`` that form decaying bound states with sodium nuclei but not with those of Xe or He~\cite{Khlopov:2010ik}, or to errors in low--energy quenching factor estimates for noble gases. The recent negative results of the COSINE-100 experiment, which comes close to the DAMA set-up~\cite{Adhikari:2018ljm}, have raised the hope of closing the loophole, but the experiment does not have the proper sensitivity. As an alternative, it was also suggested that the DAMA signal may reflect seasonal variations of the helium in the underground environment penetrating the photomultipliers rather than DM, an option potentially supported by the fact that DM-Ice, another NaI(Tl)-based experiment located at the South Pole, detected no seasonal variations so far~\cite{Ferenc:2019esv}.
\begin{figure}
\centering
	\includegraphics[scale=0.3]{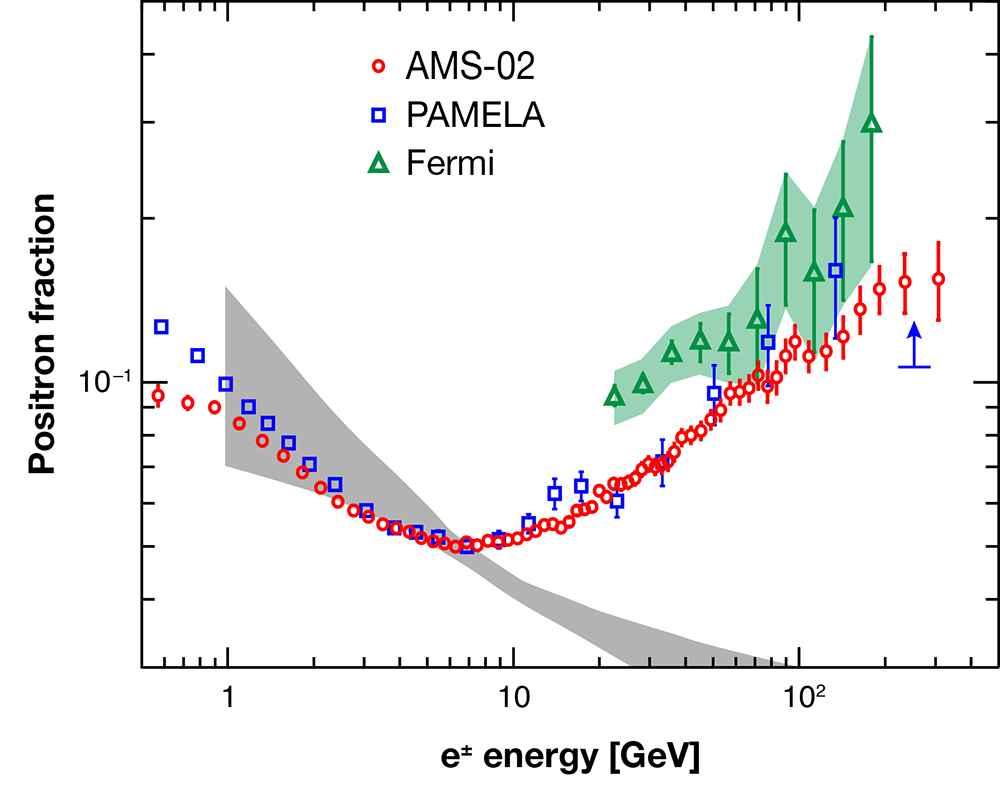}
	\caption{Cosmic positron excess measured by PAMELA, Fermi, and AMS.}
\label{Positron}
\end{figure}
Collider searches have also found no evidence of supersymmetric particles so far around the electroweak scale or beyond, up to a few TeVs. However, particles of this kind could also manifest themselves indirectly, through two-WIMP annihilation into Standard Model particles. On the other hand, the cosmic positron excess observed by the PAMELA, Fermi and AMS collaborations (fig.~\ref{Positron}) has been interpreted as possible evidence in favour of the DAMA scenario: the ``dark atom`` story, if correct, would also produce a signature of this sort~\cite{Khlopov:2010ik}.

\subsection{Primordial black holes (PBHs)}\label{C2S3S2}
Primordial black holes (PBHs) are expected to have formed in the post-inflationary Universe via the gravitational collapse of perturbations~\cite{Zeldovich:1967lct,Hawking:1971ei}. Dark matter, if this were its composition, could have only gravitational interactions, consistently with the difficulties met with its direct detection, and accretion on primordial black holes would also provide a rationale for the presence of supermassive black holes at the centres of galaxies~\cite{Silk:2018rry}.  PBHs of different masses, from microscopic to astrophysical scales, are detectable, in principle, by various means. These include surely gravitational lensing, but also Hawking radiation for small enough ones. Moreover, PBHs could have left some imprints in CMB anisotropies, in reionization observations (Square Kilometre Array), and could play a role in gravitational wave emission from mergers (fig. 1.3)~\cite{Carr:2018rid}.
\begin{figure}
	\includegraphics[scale=0.6]{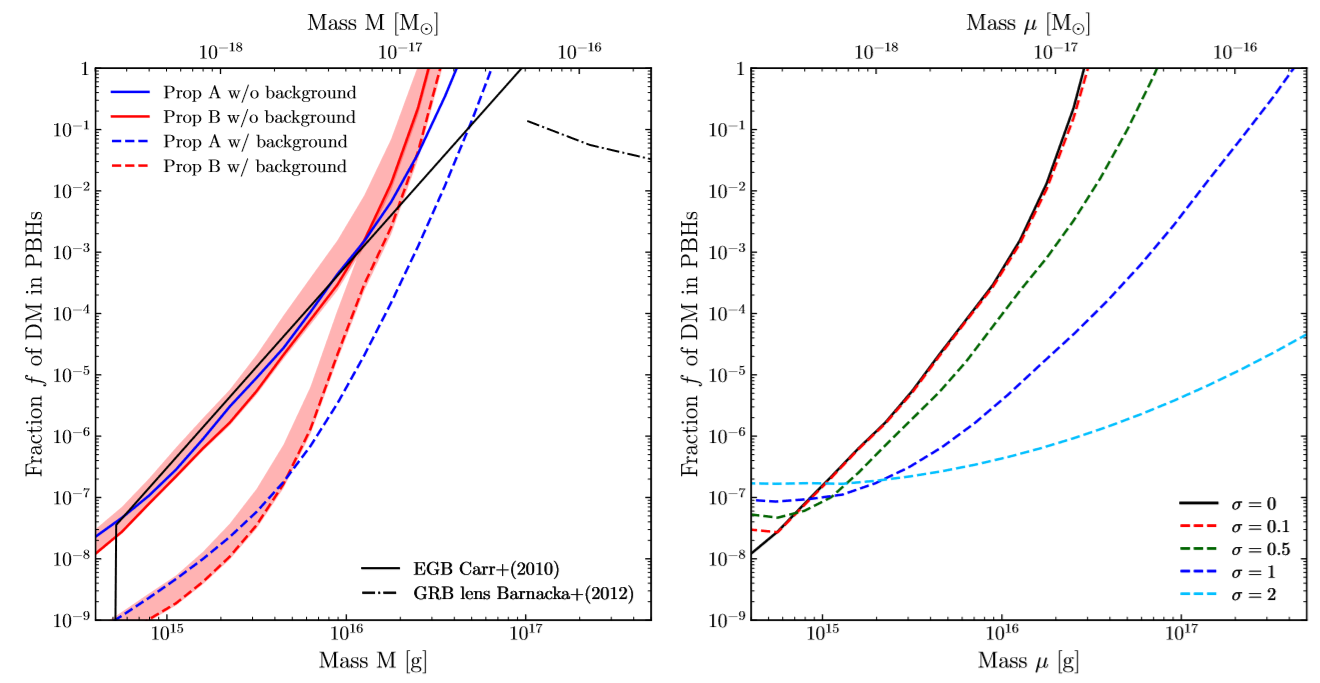}
	\caption{Constraints on the fraction of microscopic black holes in dark matter. Left panel: fixed DM mass with two possible models of $e^\pm$ propagation (A and B), considered with and without astrophysical background; right panel: log-normal BH mass distribution (predicted by inflationary models).}
\label{PBH}
\end{figure}
Astrophysical--scale black holes with masses $\ge0.01M_\odot$ have been constrained by gravitational lensing observations to constitute no more than 40\% of dark matter~\cite{Zumalacarregui:2017qqd}, although this result has been called into question due to supernova systematics~\cite{Garcia-Bellido:2017imq}. Moreover, it was argued that in models with a considerable fraction of stellar--mass PBHs, the remaining dark matter cannot consist of WIMPs, because these black holes would tend to form WIMP cores around them, producing gamma ray signals from annihilations that are not observed. And, conversely, in models with WIMPs the fraction of dark matter associated to large PBHs should be negligible~\cite{Adamek:2019gns}. Smaller, particle-scale BHs are constrained, in principle, by Voyager-1 observations of electron or positron Hawking radiation (fig.~\ref{PBH}). In general, considerations of this type lead to the conclusion that BHs with masses $\le10^{16} g$ can make up not more than 0.1\% of DM)~\cite{Boudaud:2018hqb}. However, these constraints may be relaxed in models with extra dimensions, in which gravity behaves differently at small scales, so that the evaporation rate is smaller. For instance, estimates based on Hawking's semi--classical analysis indicate that PBHs with masses $\le10^{-19} M_\odot$ should have completely evaporated by now, but they could have survived in models with extra dimensions~\cite{Keeton:2006di}. Moreover, non--evaporating tiny black holes could also provide clues on the behavior of quantum gravity, as their behavior would deviate sizeably from the semiclassical predictions~\cite{Helfer:2003va}.
\begin{figure}
\centering
\includegraphics[scale=0.7]{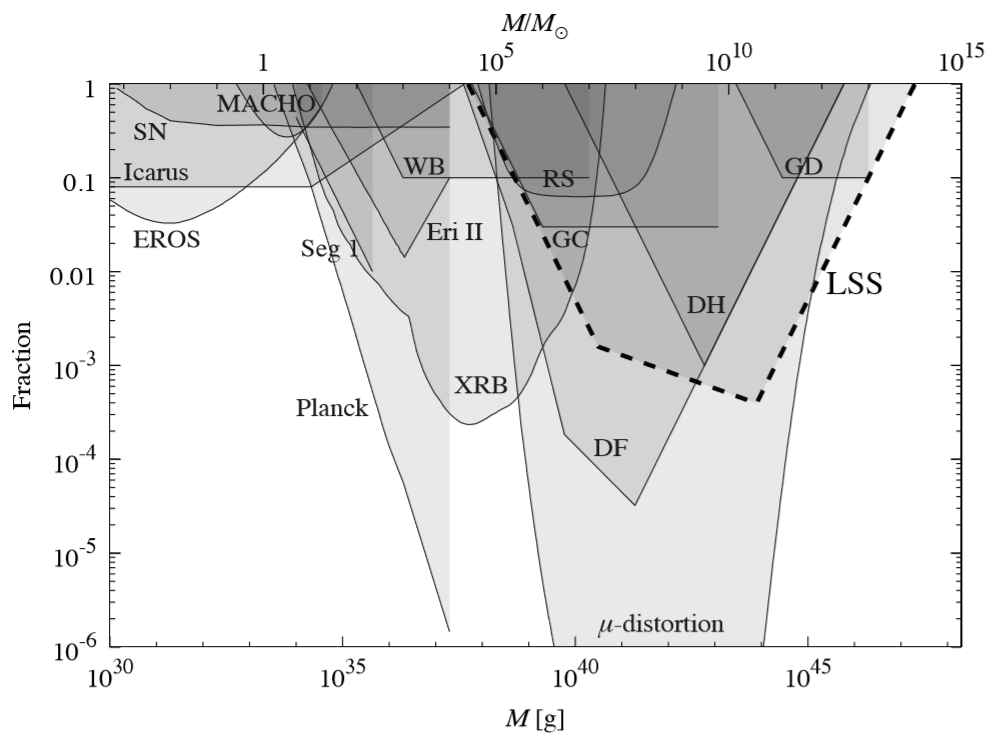}
\caption{Constraint on the PBH fraction in dark matter from various sources: microlensing of stars in the LMC (MACHO/EROS) and in the giant arcs of cluster lenses (Icarus); microlensing of supernovae (SN); millilensing of radio sources (RS); disruption of wide binaries (WB), globular clusters (GC) and star clusters in Eridanus (Eri II); disk—heating (DH) and disruption of the dwarf galaxy Segue 1 (Seg 1); dynamical friction drag of halo objects (DF) and galaxy disruption in clusters (GD); accretion constraints from CMB anisotropy (PLANCK) and X--ray binaries (XRB). Also shown is the range of masses of PBHs formation from Gaussian primordial fluctuations that are excluded by the $\mu$--distortion constraints: the primordial fluctuations, from which the PBHs are expected to form, would be damped by photons (the Silk damping), which in turn would distort the Bose-Einstein photon energy spectrum, giving them nonzero chemical potential~\cite{Nakama:2017xvq}. The large--scale structure is shown as a broken bold line.}
\label{PBH2}
\end{figure}
Generally, the constraints leave three possible mass ranges for PBHs~\cite{Carr:2019yxo} (figs.~\ref{PBH2} and~\ref{PBH3}):
\begin{enumerate}
  \item the intermediate mass range A ($10-10^3 M_\odot$);
  \item the sublunar mass range B ($10^{20}-10^{24}$ g);
  \item the asteroidal mass range C ($10^{16}-10^{17}$ g),
\end{enumerate}
and with quantum gravitational effects the range C can in principle go down to the Planck scale.
It has been suggested that the black holes detected by LIGO may in fact be primordial, due to their low spin~\cite{Garcia-Bellido:2017imq}.
\begin{figure}
\centering
\includegraphics[scale=0.8]{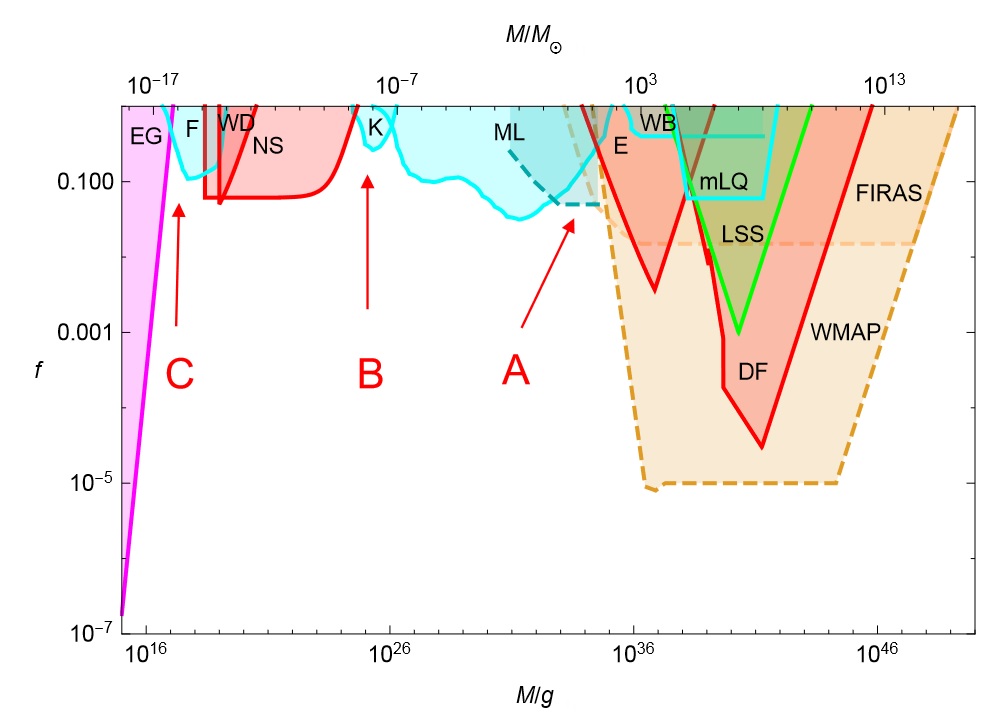}
\caption{Constraints on the black hole fraction from various observations (EG - extragalactic gamma rays, F- femtolensing of gamma-ray bursts, WD - white dwarf explosions, NS - neutron star capture, K - Kepler microlensing of stars, ML - MACHO/EROS/OGLE microlensing of stars and quasar microlensing, E - star cluster survival in Eridanus II, WB - wide-binary disruption, DF - dynamical friction on halo objects, mLQ - millilensing of quasars, LSS - large-scale structure generation through Poisson effects, WMAP, FIRAS - accretion effects). They leave only three possible mass ranges for PBHs: the intermediate mass range A ($10-10^3 M_\odot$), the sublunar mass range B ($10^{20}-10^{24}$ g), and the asteroid mass range C ($10^{16}-10^{17}$ g); the EG constraint may be invalidated by quantum gravitational effects.}
\label{PBH3}
\end{figure}
\subsection{Warm dark matter and sterile neutrinos}\label{C2S3S3}
Warm dark matter (WDM) is a form of DM that is relativistic at decoupling and becomes non-relativistic only around the radiation--matter transition. Simulations show that WDM produces cored profiles~\cite{Lovell:2011rd}, while its power spectrum is suppressed at large wavenumbers (small scales).

WDM candidates include gravitinos, an option that we have already discussed in the WIMP section, and sterile neutrinos motivated by the seesaw mechanism. Sterile neutrinos are detectable by modifications of neutrino oscillation processes, and the neutrino experiments LSND and MiniBooNE have reported some anomalies which may hint to the existence of a fourth neutrino species~\cite{Aguilar-Arevalo:2018gpe,Aguilar:2001ty}.
\subsection{Fuzzy dark matter/axion-like particles (ALPs)}\label{C2S3S4}
The so-called ``fuzzy dark matter`` provides another possible solution for the core-cusp and ``missing satellites`` issues. It is an ultralight ($m\sim10^{-21}-10^{-22}$ eV) axion--like scalar boson that gives rise to a galaxy--scale Bose--Einstein condensate. It behaves, to a large extent, as a classical scalar field described by the nonlinear Schrödinger--Poisson equation system~\cite{Hui:2016ltb}:
\beq
\begin{gathered}
i\partial_t\psi \ = \ -\frac{1}{2m}\triangle\psi+mU\psi \ , \\
\triangle U \ = \ 4\pi m|\psi|^2 \ .
\end{gathered}
\eeq
In the Madelung representation
\beq
\begin{gathered}
\rho \ = \ m|\psi|^2 \ ,\\
\vec{v} \ = \ \frac{1}{m}\vec{\nabla}\theta \ ,
\end{gathered}
\eeq
where $\theta$ is the phase of the wavefunction,
\beq
\psi \ = \ |\psi|e^{i\theta} \ ,
\eeq
the Schrödinger equation is equivalent to the Euler equations:
\beq
\begin{gathered}
\partial_t\rho+\vec{\nabla}(\rho\vec{v}) \ = \ 0 \ ,\\
\partial_t\vec{v}+(\vec{v}\vec{\nabla})\vec{v} \ = \ -\vec{\nabla}(U+Q) \ , \label{FP}
\end{gathered}
\eeq
with the additional ``quantum potential``
\beq
Q \ = \ -\frac{1}{2m^2}\frac{\triangle\sqrt{\rho}}{\sqrt{\rho}} \ ,
\eeq
describing a repulsive quantum pressure that reflects the Heisenberg uncertainty and prevents the formation of cusps. The ground state of the Schrödinger-Poisson system can be regarded as a soliton, and has a fixed minimal radius. The ``missing satellites`` problem is explained by the suppression of the power spectrum at large wave numbers or small scales, and via tunnelling effects, absent in the classical case, which result in the gradual evaporation of FDM solitonic systems. It was also shown that this model can reproduce the Tully-Fisher relation and the radial acceleration relation~\cite{Lee:2019ums}. Namely, assuming the system is in equilibrium:
\beq
\partial_t\vec{v} \ = \ 0 \ ,
\eeq
and given that at large radii the rotation velocity is close to a constant (the rotation curves ``flatten``):
\beq
\nabla_i\vec{v}\approx0 \ ,
\eeq
we obtain from \eqref{FP} that the gravitational force from the baryonic matter and the dark halo should be counterbalanced by the quantum pressure:
\beq
g_B(r)+g_D(r) \ = \ g_Q \ ,
\eeq
with $g_Q$ given by
\beq
\frac{1}{2m^2}\nabla\left(\frac{\triangle\sqrt{\rho}}{\sqrt{\rho}}\right)\approx\frac{1}{2m^2\xi^3} \ , \label{FK}
\eeq
where $\xi$ is the characteristic size of the galaxy. Near the center of the galaxy $g_B(r)$ is large (much larger than $g_Q$), and $g_D(r)$ is almost zero; at larger radii $g_B$ rapidly decreases and $g_D$ increases until the total gravitational acceleration becomes constant and equal to $g_Q$; the transition point $r_*$ is given by the condition:
\beq
g_B(r_*) \ = \ g_D(r_*) \ = \ \frac{g_Q}{2} \ .
\eeq
Numerical results indicate that $r_*$ is about the half-light radius, i.e.
\beq
M_b(r_*) \ = \ \frac{M_{b}}{2} \ .
\eeq
This means that
\beq
g_Q \ = \ \frac{GM_B}{r_*^2} \ ,
\eeq
and combining this with \eqref{FK}, we obtain:
\beq
r_* \ = \ \sqrt{2m^2\xi^3GM_B} \ .
\eeq
The asymptotic rotation velocity is therefore given by:
\beq
v_a\approx\left(\frac{GM_B}{2m^2\xi^3}\right)^{1/4} \ ,
\eeq
which is equivalent to the Tully-Fisher relation
\beq
M_B \ = \ Cv_a^4 \ ,
\eeq
with
\beq
C \ = \ \frac{2m^2\xi^3}{G} \ .
\eeq
Since $\xi$ varies for different galaxies, the Tully-Fisher relation is only approximate, which is consistent with observations.

Other phenomenological effects include the relaxation of FDM halos from higher excited states into a soliton state, which tend to inhibit black hole mergers, disk thickening and pseudobulges, dark--bright soliton interference patterns in dark halo collisions, and decelerated dynamical friction possibly resolving the Fornax globular cluster puzzle. The latter refers to the fact that the globular clusters within the Fornax dwarf spheroidal are expected to have spiraled to the center of the galaxy due to dynamical friction, which is not actually observed.

Simulations confirm that FDM produces large-scale structure identical to CDM, with cores rather than cusps around galaxy centres~\cite{Schive:2014dra}. They also show the condensation and formation of Bose stars in initially homogeneous distributions of FDM particles. These stars may explode, producing relativistic axions and radiophotons; the latter may explain the so-called fast radio bursts (FRBs), the ARCADE2 excess radio spectrum, and the EDGES 21 cm anomaly~\cite{Levkov:2018kau}. The latter has to do with the absorption depth being proportional to the ratio of the photon temperature and the spin temperature of the hydrogen:
\beq
I\propto\frac{T_\gamma}{T_S} \ .
\eeq
An influx of photons would increase $T_\gamma$, resulting in larger absorption magnitude~\cite{Fraser:2018acy}.

However, the model is pretty tightly constrained by results from the Lyman-$\alpha$ forest. This approach is based on the fact that clouds of neutral hydrogen absorb photons with the wavelength corresponding to the Lyman-$\alpha$ spectral line of hydrogen, and these absorption lines are redshifted as radiation propagates through space. Therefore, by measuring the photon spectrum from a distant source, it may be possible to determine the positions of the hydrogen clouds from the redshifts of absorption lines and thereby calculate the large-scale structure power spectrum. Comparisons of these results with fuzzy dark matter power spectrum simulations gives a lower bound for the boson mass of around $2*10^{-21}$ eV and leaves little room to solve the small--scale problems~\cite{Irsic:2017yje,Kobayashi:2017jcf}. Moreover, the aforementioned dwarf galaxy observations by Read et al. put also some constraints on ``warm`` and ``fuzzy`` dark matter, since not all observed galaxies have shallow cores~\cite{Read:2018fxs}. In addition, the existence of coherent states on such a large scale raises problematic queries in connection with quantum measurement theory, since the usual hierarchy of scales between a quantum system and the measuring device is reversed. Notice, however, that this problem also exists for the inflationary theory of structure formation~\cite{Helfer:2018ylo}. 
\subsubsection{Axions}\label{C2S3S4S1}
Probably the best--motivated type of fuzzy dark matter candidate is an axion. Many axions emerge in String Theory from compactifications of gauge $p$-forms, either from the dualization of form components with indices tangent to the Minkowski spacetime (model-independent axion) or from the zero modes of components with indices tangent to the internal space (model-dependent axions). The axions are characterized by an axion decay constant $F$, in the range
\beq
F \ = \ 10^9 - 10^{12} GeV
\eeq
for cosmological reasons. Indeed, if $F$ were smaller, too many axions would be produced in astrophysical systems, which would cause, for instance, red giants to cool too rapidly. On the other hand, if it were larger, there would be too large a fraction of axion dark matter. Generally, in (closed) String Theory $F$ is of order $10^{16} GeV$, and the problem is to reduce its value~\cite{Svrcek:2006yi}. An overview of different approaches is given in the Table~\ref{tab:Axions}. Notice that, in certain cases, the relevant decay constant is $\frac{F}{k}$ or $\frac{F}{k'}$, where k is the level of Standard Model embedding in heterotic string symmetry group, and
\beq
k'\ =\ \frac{1}{16\pi^2}\int_Y \left[ tr_1(F\wedge F)-\frac{1}{2} (R\wedge R)\right] \ .
\eeq
Axions couple to the electromagnetic field, and for this reason the QCD axion may be detectable with resonant microwave cavities. Moreover, axion--baryon scattering or axion--photon conversion have been proposed as explanations for the 21 cm anomaly: gravitational scattering of baryons with an axionic Bose-Einstein condensate would decrease the spin temperature~\cite{Sikivie:2018tml}, while the resonant conversion of axions into photons in intergalactic magnetic fields would increase the radiation temperature~\cite{Moroi:2018vci}. However, the axion--photon coupling is tightly constrained by the ABRACADABRA axion search experiment~\cite{Ouellet:2018beu} and by cosmic observations~\cite{Ivanov:2018byi}.
\begin{table}
\footnotesize
\begin{tabularx}{\textwidth}{|X|X|X|} \hline
		Theory type & Constraint on axion coupling (model-independent) & Constraint on axion coupling (model-dependent) \\ \hline
		Weakly coupled heterotic string & $\frac{F}{k}\gtrsim\frac{\alpha_cM_P}{2\pi\sqrt{2}}\approx1.1*10^{16}$ GeV\newline\newline ($\alpha_c\sim\frac{1}{25}$) & General isotropic Calabi-Yau: \newline \newline $\frac{F}{k}\gtrsim\frac{\alpha^{1/3}_cM_P}{2\pi\sqrt{2}k^{4/3}}\approx10^{17}$ GeV \newline \newline 4-manifold fibered over Riemann 2-surface:\newline\newline$\frac{F}{k'}\gtrsim\frac{k\alpha_cM_P}{k'2\pi\sqrt{2}}\approx10^{15}$  GeV\newline\newline($\alpha_c\sim{1}{25}$, $k=1$, $k'=12$)\\ \hline
		Heterotic M-theory & $\frac{F}{k}\gtrsim\frac{\alpha_cM_P}{2\pi\sqrt{2}\epsilon}\approx\newline \newline\frac{1}{\epsilon}*1.1*10^{16} $ GeV\newline\newline($\epsilon\approx1$ for small intervals and $\approx0.5$ for large ones) & Fibering over Riemann 2-surface:\newline\newline $\frac{F}{k'}\gtrsim\frac{k\alpha_cM_P}{2\pi\sqrt{2}}\approx\newline \newline 1.1*10^{16}$ GeV\newline\newline(k=1) \\ \hline
		Heterotic M-theory with warped compactification (standard) & $\frac{F}{k}\lesssim\frac{k\alpha_cM_P}{2\pi\sqrt{q}}\approx1.1*10^{16}$ GeV & $\frac{F}{k}\gtrsim\frac{q\alpha_cM_P}{3\pi}\sim10^{16}$ GeV\newline\newline ($q\sim1$) \\ \hline
		Heterotic M-theory with warped compactification (non-standard) & $F\gtrsim M_{11}\frac{\sqrt{3}(k\alpha_c)^{1/6}}{2\sqrt{\pi q}}$ \newline\newline (phenomenologically acceptable values for $3.5*10^9$ GeV$\le M_{11}\le3.5*10^{12}$ GeV, $2.3*10^4*l_{11}\le\rho\le4.2*10^6*l_{11}$) & $F=M_P\frac{4}{3\pi q(k\alpha_c)^{1/3}}(\frac{l_{11}}{\pi\rho})^2$\newline\newline (phenomenologically acceptable values for $1.1*10^{12}$ GeV$\le M_{11}\le1.1*10^{14}$ GeV, $1.7*10^3*l_{11}\le\pi\rho\le5.4*10^4*l_{11}$) \\ \hline
		Anomalous U(1) in compactified heterotic string & Two-axion model:\newline\newline$F_a=\frac{k\alpha_cM_P}{2\pi\sqrt{2}}\sim\newline \newline1.1*10^{16}$ GeV\newline\newline $F_b= M_P\sqrt{\frac{k\alpha_cTr(B)}{24\pi q}}\sim F_a$ &  \\ \hline
		Intersecting branes (vanishing cycle) &  & $F=M_P(\frac{l_s}{R})^q\sqrt{\frac{xg_s^2}{8\pi^2}}$,\newline \newline $M_P^2=\frac{4\pi R^6}{g_s^2l_s^8}$\newline\newline$(M_s\gg M_{GUT}$, $R\gg l_s)$ \\ \hline
		Intersecting branes (anisotropic Calabi-Yau) &  & $F=\frac{\sqrt{x}\alpha_cM_P}{2\pi\sqrt{2}}\sim10^{16}$ GeV \\ \hline
		Type I string & $F=\frac{M_P}{\sqrt{2}}*\frac{\alpha_{GUT}}{2\pi}=\newline \newline1.1*10^{16}$ GeV & \\ \hline
		\end{tabularx}
		\end{table}
		\begin{table}
		\footnotesize
		\begin{tabularx}{\textwidth}{|X|X|X|} \hline
		Theory type & Constraint on axion coupling (model-independent) & Constraint on axion coupling (model-dependent) \\ \hline
		M-theory on a manifold of $G_2$ holonomy & &$F^2=\frac{xR}{2\pi l^3_{11}}$, $M^2_P=\frac{4\pi R^7}{l_{11}^9}$\newline\newline ($\frac{3}{F}\le l_{11}\le\frac{10}{F} \ , \ 50*l_{11}\le R\le 500*l_{11}$) \newline\newline For a vanishing cycle:\newline\newline$F^2=\frac{R}{l^3_{11}}\frac{16x}{3^{5/2}\pi N}$\newline\newline(for $x\sim1$ and N=5: $\frac{2}{F}\le l_{11}\le\frac{6}{F} \ , \  55*l_{11}\le R\le 550*l_{11}$)\newline\newline For an anisotropic X factorized into isotropic F and B:\newline\newline$M_P^2=\frac{4\pi V_FV_B}{l_{11}^9} \ , \newline\newline F^2=\frac{xV_X}{2\pi l^3_{11}R_F^{2b}R_B^{6-2b}}\sim(\alpha_CM_P)^2$ \\ \hline
		D3-branes & C0 RR-form:\newline\newline $F=\frac{\alpha_cM_P}{2\pi\sqrt{2}}\sim10^{16}$ GeV & \\ \hline
		Intersecting D-branes with small cycles (IIB) &  & $F=\sqrt{\frac{x}{6}}\frac{R}{l_s^2}$, $M^2_P=\frac{4\pi R^6}{g_s^2l_s^8}$\newline \newline ($1.5*10^5$ GeV$\le\M_s\sqrt{g_s}\le5*10^9$ GeV$, \ 1.6*10^4*l_s\ge\frac{R}{\sqrt{g_s}}\ge5*10^2*l_s$) \\ \hline
		Intersecting D-branes with small cycles (IIA) &  & $F^2=\frac{3x}{4\pi^2l_s^2}\ln(\frac{R}{r_0}),     \newline \newline M^2_P=\frac{4\pi R^6}{g_s^2l_s^8}$\newline\newline ($r_0\sim l_s$ is the cycle radius)\newline \newline
		$1.4*10^9$ GeV$\le M_s\le1.8*10^{12}$ GeV, $800*l_s\ge\frac{R}{\sqrt{g_s}}\ge73*l_s$ \\ \hline
	\end{tabularx}
	\caption{Constraints on axion decay constant in various string-motivated models.} \label{tab:Axions}
	\end{table}

\subsection{Dark sector and mirror dark matter}\label{C2S3S5}
The diversity of dark matter properties, inferred from observations, may imply that dark matter is an ensemble of particles rather than one specific type of particle. Some of them may belong to the categories discussed before, like WIMPs or axion-like particles, and could be detected in the same way, by nuclear scattering, EM cavities, or ``light shine through wall`` experiments; however, in some models like the $E8\otimes E8$ heterotic String Theory, this sector may be decoupled from SM particles and interact with them only via gravity. In that case, the ``hidden sector`` may be explored through portals like the Higgs portal and the photon-dark photon mixing, and interactions with baryons through the latter term may allow to reproduce the Tully-Fisher relation~\cite{Chashchina:2016wle}.

Another telltale signature of a ``dark sector`` is the self-interaction of dark matter, which would create a sort of ``friction``, affecting the collisions of two dark matter halos; such a friction was not observed in collisions of numerous galaxy clusters, which constrains the ``dark sector`` models by putting an upper limit on the self-interaction cross section~\cite{Harvey:2015hha,Massey:2017cwf}.
In models produced by String Theory and M-theory, there are often several hidden sectors which are connected to the visible sector by one or more portals; it was estimated that in such a model, a ``visible-sector`` particle (like a WIMP) would most likely decay into a hidden-sector one (which makes phenomenology more difficult, and may explain the negative results of WIMP search experiments)~\cite{Acharya:2016fge}.\\
It is also probable that the dark sector particles can decay into SM particles; this possibility is discussed in the next section.
\subsection{Dynamical dark matter}\label{C2S3S6}
Decaying, or dynamical, dark matter is not a separate model; in fact, almost any dark matter model can be supplied with an interaction term  which would make the particles decay. There exist some tensions between the high-redshift (CMB) and low-redshift measurements of the cosmological parameters, which are alleviated if the amount of dark matter is smaller by 2-5 percent at lower redshifts~\cite{Chudaykin:2016yfk}; however, if an abundant fraction of dark matter decays into SM particles, it would be an issue for CMB and BBN nucleosynthesis. Therefore dynamical dark matter is generally expected to comprise a ``dark sector`` in which the decay widths into SM particles are balanced against abundances~\cite{Dienes:2011ja}. One possibility for DDM is axion KK modes~\cite{Dienes:2011sa}.\\
Components of DDM may be detectable by usual WIMP and axion search experiments (though they would provide a different, more complex signal in WIMP detectors), as well as by specified surface collider detectors aimed at long-lived particles~\cite{Curtin:2018ees}. Decays of dark matter have also been proposed as explanations for positron excess and 21 cm anomaly.

Conversely, some data analysis of galaxy rotation curves at higher redshifts was interpreted to imply that the galaxies contained less dark matter at $0.6<z<2.6$, and their dynamics were mostly baryon-driven~\cite{Genzel:2017jgd}; however, more recent studies contradict this notion, suggesting that the dark matter density was more or less the same during this period~\cite{Tiley:2018}.\\
Alternatively, the discrepancy between Hubble constant measurements can be explained by either statistical errors or faster expansion of the early Universe; the latter may be due to additional light non-interacting degrees of freedom (an additional ``dark component``)~\cite{Aylor:2018drw} or time-changing density of dark energy (some recent observations of quasars suggest that it may be growing with time)~\cite{Risaliti:2018reu}.

\chapter{A critique of modified gravity alternatives to dark matter}\label{C3}
\section{A brief introduction to MOND}\label{C3S1}
\begin{figure}
\centering
\begin{adjustbox}{width={\textwidth},totalheight={\textheight},keepaspectratio}
	\includegraphics[scale=0.3]{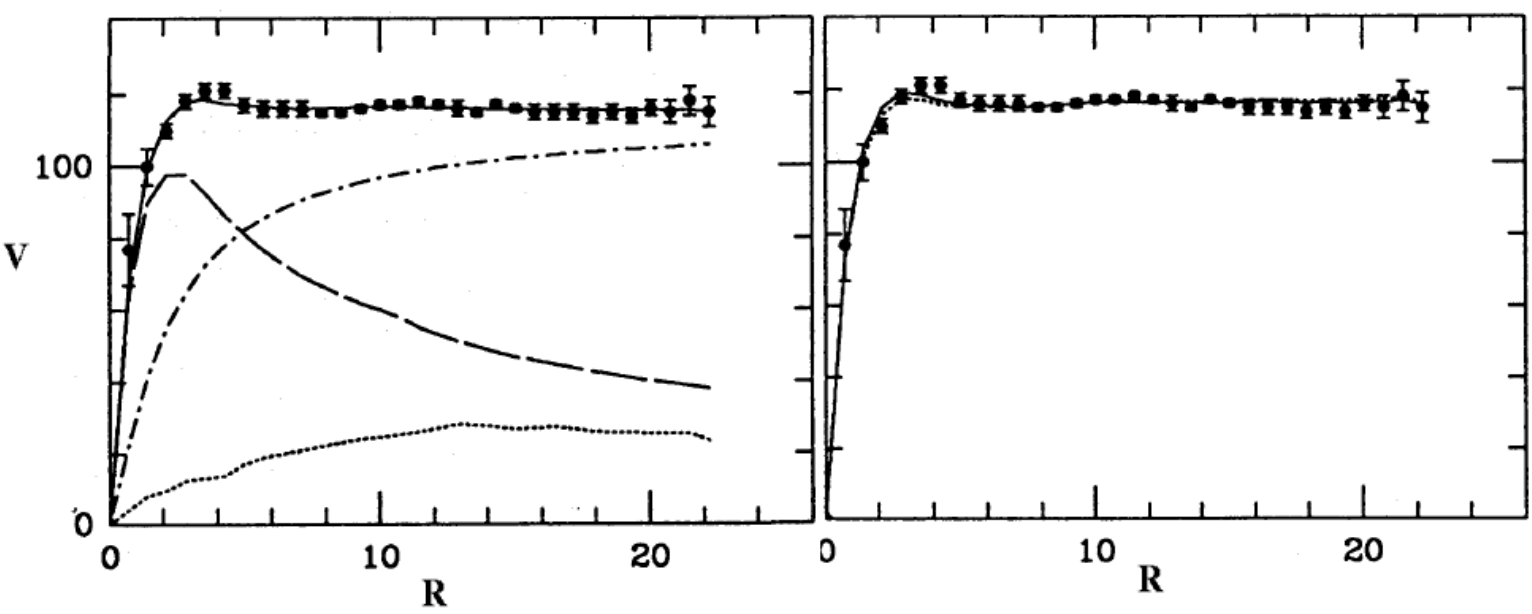}
\end{adjustbox}
	\caption{Rotation curve fit for a real galaxy NGC 6503; left three-component Newtonian gravity fit (stellar+gas+DM), right: one-component MOND fit}
\label{NGC6503}
\end{figure}
An alternative to dark matter, which directly incorporates the radial acceleration relation, is the so-called modified Newtonian dynamics (MOND), proposed by M. Milgrom in 1981~\cite{Milgrom:1983ca}. Within the MOND paradigm, the gravitational force is created only by baryonic matter, but gravity itself is modified in such a way that the physical gravitational acceleration is related to the Newtonian value by\\
\beq
a \ = \ a_N\nu\left(\frac{a_N}{a_0}\right) \ ,
\eeq
where $\nu$ is a function with the asymptotics:
\beq
\nu(x) \sim 1 \quad (x\gg 1) \ ,\qquad
\nu(x) \sim \frac{1}{\sqrt{x}} \quad (x\ll 1) \ ,
\eeq
and $a_0$ is a fixed acceleration scale of order $\sqrt{\Lambda}\propto H_0$. 

\begin{figure}
\begin{center}
\includegraphics[scale=0.3]{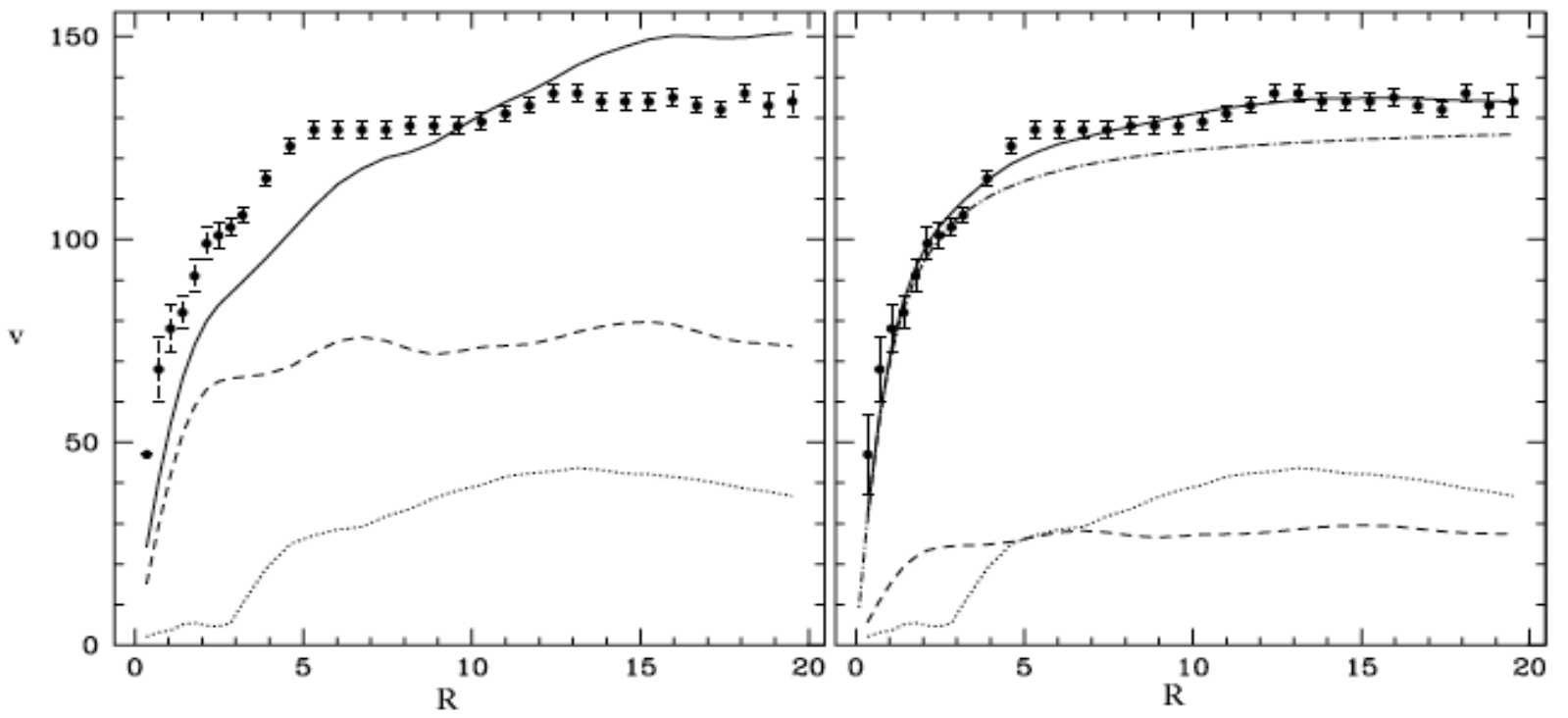}
\end{center}
\caption{``Fake`` galaxy fit (rotation curve of NGC2403+photometry of UGC 128); left: two-component MOND fit (stellar+gas); right: three-component Newtonian fit (stellar+gas+DM)}
\label{Fake}
\end{figure}

The current observational status of MOND and RAR is controversial. One common choice of $\nu(x)$, given by
\beq
\nu(x) \ = \ \frac{1}{1-e^{-\sqrt{x}}} \ ,
\eeq
has provided good fit to a number of galaxies from the SPARC sample~\cite{McGaugh:2016leg,Lelli:2016zqa}. Other choices of $\nu$ have also been used to fit certain galaxies with high precision (fig.~\ref{NGC6503})~\cite{Scarpa:2006cm}. Interestingly, since MOND has fewer parameters than dark matter models, it fails to fit ``fake`` galaxies, i.e. those which use the rotation curve of one galaxy and baryonic mass distribution of the other (fig.~\ref{Fake}).

However, there are certain counterexamples which are better explained by the dark matter model, for instance, the dwarf ``twins`` Carina and Draco which require different dark matter profiles for almost the same radial light profile\cite{Read:2018fxs,Randriamampandry:2014eoa}; in addition, a study of fifteen dwarf and spiral galaxies shows a correlation between $a_0$ and the extrapolated disk surface brightness, which contradicts the notion that $a_0$ should be a constant~\cite{Pointecouteau:2005mr}. A couple of galaxies, known as ultra-diffuse galaxies, appear to strongly deviate from the Tully-Fisher relation and have a very small M/L ratio, i.e. almost no dark matter~\cite{Danieli:2019zyi}. The Bullet Cluster is frequently seen as another counterexample against modified gravity~\cite{Clowe:2006eq,Markevitch:2003at}, although its high collision velocities are also challenging to explain within $\Lambda$CDM~\cite{Lee:2010hja,Thompson:2014zra,Bouillot:2014hda}, and MOND was conjectured to be a better framework for reproducing them~\cite{Angus:2007qj}. Another pressing issue is to explain a number of phenomena like CMB and primordial structure formation that find a natural explanation within the dark matter paradigm. Moreover, most rotation curves are not \emph{exactly} flat at large distances, and have either a slightly decreasing or a slightly increasing slope, implying that MOND can only be true as an approximation of some underlying phenomena~\cite{Persic:1995ru,Salucci:2007tm}.

One can derive a Lagrangian formulation of MOND noting that the ``deep MOND`` dynamics ought to be scale invariant, since eq.~\eqref{eqn1} does not change under the transformations
\beq
\vec{r} \ \rightarrow \ \lambda\vec{r} \ , \ t \ \rightarrow \ \lambda t \ . \label{redefinitions}
\eeq
Alternatively, the MOND dynamics depends only on the quantity $GMa_0$, whose dimension is $[L/T]^4$, and not on $G$, $M$, and $a_0$ separately~\cite{Milgrom:2008cs}.
A typical action principle for Newtonian mechanics is
\beq
\mathcal{S}_N \ = \ -\frac{1}{8\pi G}\int dt\,d^3x \ (\vec{\nabla}\Phi)^2-\int dt\,d^3x \ \rho\Phi+\int dt\,d^3x \ \frac{\rho v^2}{2} \ ,
\eeq
and in order to arrive at scale-invariant equations, all terms should have identical scaling dimensions. Under eq.~\eqref{redefinitions} the third term scales as $\lambda$ ($\rho\propto\lambda^{-3}$), and $\Phi$ should be scale-invariant to grant the same scaling dimension to the second term. However, this choice would make the first term scale as $\lambda^2$, and therefore, to recover MOND, one should change either the third term or the first one. The first approach, known as modified inertia (MI), still lacks a complete theoretical description, because the modified acceleration is a functional of the whole particle trajectory. This means that the corresponding equations of motion would be \emph{non local} in time, and the Lagrangian would not contain a finite number of derivatives of $\vec{r}$~\cite{Milgrom:1992hr,Milgrom:1998sy,Milgrom:2005mc}.
However, within the second approach, known as modified gravity, one has the option of altering the first term in order to make it scale like $\lambda$, according to
\beq\label{MG}
\mathcal{S}_{MG} \ = \ -\,\frac{1}{12\pi Ga_0}\int dt\,d^3x \ \left((\vec{\nabla}\Psi)^2\right)^{3/2}-\int dt\,d^3x \ \rho\Psi+\int dt\,d^3x \ \frac{\rho v^2}{2} \ .
\eeq
This model is known as AQUAL (AQUAdratic Lagrangian), and gives the modified Poisson equation
\beq\label{MP}
\vec{\nabla}(|\vec{\nabla}\Psi|\vec{\nabla}{\Psi}) \ = \ 4\pi Ga_0\rho \ ,
\eeq
which, for a point-like mass
\beq
\rho(r) \ = \ M\delta(\vec{r}) \ ,
\eeq
yields indeed the spherically symmetric solution
\beq\label{MOND_Pot}
\Psi(r) \ = \ -\sqrt{a_0GM}\ln(r)
\eeq
and the radial acceleration~\eqref{RAR}. Since we already know that a field like $\Phi$ can emerge in General Relativity as a perturbation of the metric tensor $g_{\mu\nu}$, it is natural to try and understand the relativistic origin of $\Psi$.
\section{Relativistic completions of MOND}\label{C3S2}
\subsection{f(R) gravity}\label{C3S2S1}
The most straightforward approach would be to directly modify Einstein's gravity along the lines of $f(R)$ gravity to incorporate $\Phi$ and $\Psi$ as different limits of the same function. Outside the source, the equations of motion for f(R) gravity are
\beq
\begin{gathered}
f'(R)R_{\mu\nu} \ - \ \frac{1}{2}f(R)g_{\mu\nu} \ + \ (g_{\mu\nu}g^{\alpha\beta} \ - \ \delta_\mu^\alpha\delta_\nu^\beta)*\\
\left(f``'(R)\partial_\alpha R\partial_\beta R \ + \ f``(R)\partial_\alpha\partial_\beta R-f``(R)\Gamma^\gamma_{\alpha\beta}\partial_\gamma R\right) \ = \ 0 \ .
\end{gathered}
\eeq
For a static and spherically symmetric system, they can be written in the form
\beq\label{f}
\begin{gathered}
f'R_{00}-\frac{1}{2}fg_{00}+g_{00}g^{rr}\left(f``'(R'(r))^2 \ + \ f``R``(r)\right) \ - \\ 
f``R'g_{00}(g^{rr}\Gamma^r_{rr} \ + \ g^{\theta\theta}\Gamma^r_{\theta\theta} \ + \ g^{\phi\phi}\Gamma^r_{\phi\phi}) \ = \ 0 \ , \\
f'R_{rr} \ - \ \frac{1}{2}fg_{rr} \ - \ f``R'g_{rr}(g^{00}\Gamma^r_{00} \ + \ g^{\theta\theta}\Gamma^r_{\theta\theta} \ + \ g^{\phi\phi}\Gamma^r_{\phi\phi}) \ = \ 0 \ , \\
f'\left(g_{\theta\theta}R_{00} \ - \ g_{00}R_{\theta\theta}\right) \ + \ f``R'\left(g_{\theta\theta}\Gamma^r_{00}-g_{00}\Gamma^r_{\theta\theta}\right) \ = \ 0 \ .
\end{gathered}
\eeq
The first equation is the $tt$-component, the second is the $rr$-component, and the third is the combination of $tt$- and $\theta\theta$-components. It is also useful to write the trace equation,
\beq\label{eqtrace}
f'R \ - \ 2f \ + \ 3\Box f' \ = \ 0 \ ,
\eeq
although it is not independent of the three above.

Assuming time-independence and spherical symmetry, one can use Schwarzschild coordinates, letting
\beq
ds^2 \ = \ -(1+2\Psi)dt^2 \ + \ (1+2\Theta)dr^2 \ + \ r^2(d\theta^2 \ + \ \sin^2\theta d\phi^2) \ . 
\eeq
To first order in $\Psi$ and $\Theta$, the Ricci tensor and its trace are then
\beq
\begin{gathered}
R_{00} \ = \ \Psi`` \ + \ \frac{2}{r}\Psi' \ , \ R_{rr} \ = \ -\Psi`` \ + \ \frac{2}{r}\Theta' \ , \\
R_{\theta\theta} \ = \ 2\Theta \ + \ r(\Theta' \ - \ \Psi') \ , \ R_{\phi\phi} \ = \ \sin^2\theta R_{\theta\theta} \ , \\
R \ = \ -2\Psi`` \ + \ \frac{4}{r}(\Theta' \ - \ \Psi') \ + \ \frac{4}{r^2}\Theta \ ,
\end{gathered}
\eeq
while the relevant Christoffel symbols are
\beq
\begin{gathered}
\Gamma^r_{00} \ = \ \Psi' \ , \ \Gamma^r_{rr} \ = \ \Theta' \ , \ \Gamma^r_{\theta\theta} \ = \ -r \ , \ \Gamma^r_{\phi\phi} \ = \ -r\sin^2\theta \ .
\end{gathered}
\eeq
Now, let us restrict our analysis to functions that are analytic in the vicinity of $R=0$:
\beq
f(R) \ = \ \sum^\infty_{n=1}c_nR^n
\eeq
(n=0 is not considered, since this term can be absorbed into the cosmological constant). If one considers the case $c_1\neq0$, the leading-order term would be given by $R$, and, upon substitution into the equations~\eqref{f}, one obtains just the classical Schwarzschild solution. However, for $c_1=0$ the leading-order term is $R^\lambda$, with $\lambda>1$. Therefore the last equation in \eqref{f} demands that, to leading order in $\Psi$ and $\Theta$, $R'=0$. Now, if one wanted that $\Psi$ be the MOND potential, the preceding equations determine a corresponding form for $\Theta$,
\beq
\Psi(r) \ = \ C \,\ln(r) \ ,
\eeq
\beq
\begin{gathered}
\Theta \ = \ \frac{C}{2} \left( 1 \ + \ \frac{k_1}{r} \ + \ k_2r^2 \right) \ , \qquad  R \ = \ 12\,k_2\,C \ .
\end{gathered}
\label{lambdaless}
\eeq
For $\lambda >1$, the last term, proportional to $k_2$, would not be present, and the special case $k_1=k_2=0$ of this result reproduces the one obtained in~\cite{Mendoza:2015una}. Since the MOND solution belongs to the large family defined by the condition $R=0$, it exists for \emph{all} $\lambda>1$. At any rate,  $C$ is our expansion parameter in perturbation theory, and therefore one should demand that
\beq
\left|C \right| \ \ll \ 1 \ .
\eeq
On the other hand, $C$ enters the MOND term $\sqrt{GMa_0}$, so that the preceding condition translates into the upper bound
\beq
M \ \ll \ M_P\frac{L_H}{l_P} \ \sim \ 10^{23} M_\odot \ .
\eeq
This is well above typical galaxy masses, which are of the order $10^{10}-10^{12} M_\odot$, so that the MOND approximation appears justified, in this context, on galaxy scales.

However, it is less evident how to ``glue`` the MOND solution to the Schwarzschild one. We have in mind two possible ways to do it: the first is to select a function $f(R)$ that converges to $R$ at small radii and to $R^\lambda$ at large radii, and the second is to use the same $f(R)\propto R^\lambda$ at all distances, while resorting to the \emph{solution} $\Psi$ that combines the MOND potential and Schwarzschild potentials. 
The first approach appears problematic, since $R=0$ in both the Schwarzschild and MOND regimes. Hence, even if one chooses $k_2\neq0$ in order to have a nonzero Ricci scalar in the MOND regime, it seems unclear why at larger distances $\Psi$ and $\Theta$ would ``choose`` to converge from the Schwarzschild solution with $R=0$ to the MOND solution with $R\neq0$, even leaving aside the need for the particular MOND--value of $C$, $\sqrt{GMa_0}$, and we do not see how to ascertain whether any of the solutions possess attractor properties. 
The second approach is more viable, because the equation $R'=0$ is linear in $\Psi$ and $\Theta$ at first order, so that any combination of  the Schwarzschild potential and MOND potentials,
\beq
\begin{gathered}
\Psi \ = \ C_1 \ln(\frac{r}{r_0}) \ + \ \frac{C_2}{r} \ ,\\
\Theta \ = \ \frac{C_1}{2}(1 \ + \ \frac{k_1}{r} \ + \ k_2r^2) \ - \ \frac{C_2}{r} \ .
\end{gathered}
\eeq
is also a solution, where $r_0$ is an arbitrary length scale. However, how can one link the constant to the singular behavior as $r\to 0$? Namely, supplementing eq.~\eqref{eqtrace} with a source $\rho=M\delta(\vec{r})$ and integrating it yields, to leading order,
\beq\label{v2}
(r^2R^{\lambda-1})\Big|_{r=0} \ = \ \frac{2}{3(\lambda-1)}GM \ .
\eeq
But the MOND potential is sub--dominant with respect to the Schwarzschild term, and therefore there is apparently no way to obtain the condition $C_2=GM$. Our conclusion is therefore the following: \emph{assuming weak field limit and analyticity of $f(R)$ near $R=0$, there is no way to obtain a unique solution for a point mass that would be, at the same time, distinct from the Schwarzschild one}. This conclusion also applies to more generic non--analytic power--law functions:
\beq
f(R) \ = \ R^\lambda \ ,
\eeq
with non--integer $\lambda$. One example is the choice  $\lambda=\frac{3}{2}$, suggested by an order-of-magnitude approach and by scaling invariance~\cite{Bernal:2011qz}. 

Besides, it has been demonstrated in~\cite{Soussa:2003sc} that under a number of reasonable assumptions (such as the stability of gravitational theory), $f(R)$ gravity cannot account for gravitational lensing. This can be easily understood from the fact that $f(R)$ theories are dual to scalar-tensor theories of the form~\cite{Sotiriou:2008rp}
\beq
\mathcal{S} \ = \ - \int \ d^4x \ \sqrt{-\tilde{g}} \ \left(\frac{1}{2\kappa}\tilde{R}\ + \ \frac{1}{2}\tilde{g}^{\mu\nu}\partial_\mu\phi\partial_\nu\phi \ + \ U(\phi) \right) \ + \ \int \ d^4x \ \sqrt{-g} \ \mathcal{L}(g_{\mu\nu}) \ ,
\eeq
where
\beq
\tilde{g}_{\mu\nu} \ = \ e^{\sqrt{\frac{2}{3}\kappa}\phi}g_{\mu\nu} \ , \ U(\phi) \ = \ \frac{Rf' \ - \ f}{2\kappa f'^2} \ , \ e^{\sqrt{\frac{2}{3}\kappa}\phi} \ = \ f'(R) \ ,
\eeq
and $\kappa=8\pi G$. In particular, for $f=KR^\lambda$, the potential is
\beq
U(\phi) \ = \ \frac{\lambda-1}{2\kappa}K^{\frac{1}{1-\lambda}}\lambda^{\frac{\lambda}{1-\lambda}}\exp(\frac{2-\lambda}{\lambda-1}\sqrt{\frac{2}{3}\kappa}\phi) \ .
\eeq
On the scalar-tensor side, one can also try to define a MOND-like solution, taking $\phi$ to be the MOND field $\Psi$. Outside the source, the equation of motion for $\phi$ is
\beq
\frac{1}{r^2}\partial_r(r^2\partial_r\phi) \ = \ \frac{2-\lambda}{\sqrt{6\kappa}}K^{\frac{1}{1-\lambda}}\lambda^{\frac{\lambda}{1-\lambda}}\exp(\frac{2-\lambda}{\lambda-1}\sqrt{\frac{2}{3}\kappa}\phi) \ ,
\eeq
and therefore, letting
\beq
\phi \ = \ C\ln(r) \ ,
\eeq
leads to
\beq
\frac{C}{r^2} \ \propto \ r^{\frac{2-\lambda}{\lambda-1}\sqrt{\frac{2}{3}\kappa}C} .
\eeq
For consistency, one should require 
\beq
C \ = \ \frac{\lambda-1}{2-\lambda}\sqrt{\frac{3}{2\kappa}} \ ,
\eeq
and in order to reproduce the MOND prefactor $C=\sqrt{GMa_0}$ one should make $\lambda$ a function of the mass M. This feature was also observed in~\cite{Saffari:2007xc}, and while it allows bypassing the no-go theorem on gravitational lensing, it makes the theory somewhat baroque and ill-defined, for general matter Lagrangians.
Alternatively, for $f(R)=R+KR^\lambda$, the potential is
\beq
U(\phi) \ = \ \frac{\lambda-1}{2\kappa}K^{\frac{1}{1-\lambda}}\lambda^{\frac{\lambda}{1-\lambda}}e^{-2\sqrt{\frac{2}{3}\kappa}\phi}(e^{\sqrt{\frac{2}{3}\kappa}\phi}-1)^{\frac{\lambda}{\lambda-1}} \ ,
\eeq
but the structure of the equation is the same as in the previous case, and suffers from the same problem.
\subsection{Covariant emergent gravity (CEG)}\label{C3S2S2}
An alternative approach would be to consider $\Psi$ a separate field distinct from $\Phi$ (the aforementioned scalar dual of $f(R)$ gravity is one particular example of such a theory). Since $\Phi$ and $\Psi$ act on baryonic matter independently from each other, the total acceleration would be given by
\beq
a \ = \ a_N+\sqrt{a_Na_0}
\eeq
at leading order in $\Phi$ and $\Psi$, corresponding to the interpolation function
\beq
\nu(x) \ = \ 1+\frac{1}{\sqrt{x}} \ .
\eeq
This interpolation function has been tested in~\cite{Hossenfelder:2018vfs} for a number of galaxies from the SPARC data sample~\cite{Lelli:2016zqa}, and shown to provide good agreement with the data~\footnote{However, it cannot be accurate at large $x$, or, equivalently, at small radii (otherwise the corrections to Newton's law would be too large), which means the MOND contribution should somehow ``turn off`` at large gradients of $\Psi$~\cite{Bruneton:2007si}.}.

In this case, one would need to ensure that this new field interacts with light and reproduces the gravitational lensing observations, and that its energy is bounded from below. Absence of superluminal modes would also be a desirable feature of the theory, albeit, as was shown in~\cite{Babichev:2007dw,Bruneton:2006gf}, their presence does not automatically lead to acausality. This requirement can be stated as follows. If one perturbs the equations of motion around a stable field configuration, letting
\begin{equation}
\Psi\ =\ \Psi_0(\vec{r})\ +\ \delta\Psi(t,\vec{r}) \ ,
\end{equation}
the eikonal approximation
\begin{equation}
\delta\Psi\ = \ A\,e^{i\phi} \ ,    
\end{equation}
where $A$ is a slowly varying amplitude and $\phi$ is a quickly oscillating phase, would yield for the wave vector the dispersion relation
\begin{equation}
\omega\ =\ \partial_0\,\phi \ , \qquad  k_i\ =\ \partial_i\,\phi \ ,
\end{equation}
and the group velocity, defined as
\begin{equation}
v_g \ = \ \frac{\partial\omega}{\partial|\vec{k}|} \ ,
\end{equation}
should not be larger than 1.

The energy, which needs to be bounded from below, can be defined either in terms of the canonical stress-energy tensor:\\
\beq
\left(T_{00}\right)_C \ = \ \left(-\frac{\delta\mathcal{L}_\Psi}{\delta\partial^\mu\Psi^a}\partial_\nu\Psi^a+g_{\mu\nu}\mathcal{L}_\Psi\right)\Big|_{00} \ ,
\eeq
where $\mathcal{L}_\Psi$ is the Lagrangian of $\Psi$, and $a$ are its internal indices, or the gravitational one:\\
\beq
\left(T_{00}\right)_G \ = \ -\frac{2}{\sqrt{-g}}\frac{\delta}{\delta g^{\mu\nu}}(\sqrt{-g}\mathcal{L}_\Psi)\Big|_{00} \ .
\eeq
In most instances these two definitions coincide, but as we shall see, this is not always the case. 

The simplest option to reproduce the equation~\eqref{MP} would be to construct a Lagrangian that yields~\eqref{MG} in the static non-relativistic limit. The fractional power of the kinetic term ($\frac{3}{2}$) may either be present in the covariant formulation of the theory, or emerge in this specific limit.

In the first case, one can consider
\beq \label{Lagr}
\mathcal{L}_\Psi \ = \ \chi^{3/2} \ ,
\eeq
where $\chi=\partial_\mu\Psi^a\partial^\mu\Psi^a$. This class of theories is known as \emph{k-essence theories}, and though fractional kinetic terms may appear exotic from a field theory perspective, their emergence is not uncommon in physics. The best known example is the action of a relativistic particle:
\beq
\mathcal{S}=-m \int \ \ dt \ \ \sqrt{1-\vec{v}^2(t)} \ ,
\eeq%
Other examples include the Dirac-Born-Infeld action~\cite{Born:1934gh}, which plays a role in String Theory~\cite{Fradkin:1985qd,Tseytlin:1999dj}, close to the critical field values, the phononic action of the unitary Fermi gas (UFG)~\cite{Son:2005rv}, and the Hamiltonian of fractional quantum mechanics~\cite{Laskin:2002zz}, although the last two are non-relativistic examples. Insofar as these kinetic terms contain only contributions of the form $(\partial\psi)^2$, they do not spoil the Cauchy problem, because the equations of motion do not include derivatives of order higher than two. Schematically, the EOMs of a Lagrangian proportional to $\chi^\lambda$ read
\beq
\frac{\delta}{\delta\psi}(\partial\psi)^{2\lambda} \ = \ -2\lambda(2\lambda-1)(\partial\psi)^{2\lambda-2}\partial^2\psi \ .
\eeq
It should be noted that a Lagrangian of the form $\chi^\lambda$, with non-integer $\lambda$, is generally defined only for $\chi>0$. The region of negative $\chi$ is not accessible, as is the case for the superluminal regime of a relativistic particle, and its boundary signals the breakdown of the effective field theory. However, it is also possible, in principle, to try and circumvent this limitation considering kinetic terms of the form $\chi^\alpha|\chi|^\beta$, where $\alpha+\beta=\lambda>0$, $\alpha$ is an integer and $\beta$ is positive. The last requirement does not lead to any loss of generality, since $|\chi|^2=\chi^2$. For such a theory, the canonical stress-energy tensor is given by:
\beq
(T_{00})_C=-\left(\lambda\frac{\delta\chi}{\delta\partial^0\Psi^a}\partial_0\Psi^a+\chi\right)\chi^{\alpha-1}|\chi|^\beta \ \ ,
\eeq
taking into account that terms proportional to $\sgn'(\chi)=\delta(\chi)$ vanish for positive values of $\lambda$, while we are excluding, to begin with, negative values of $\lambda$, which would result in a theory that is unbounded at $\chi=0$. Proceeding along these lines, one can conclude that the choice $\mathcal{L}\propto|\chi|^{\lambda-1}\chi$ can encompass both positive and negative values of $\chi$ (our particular choice, $\lambda=\frac{3}{2}$, would yield $\chi\sqrt{|\chi|}$). The same logic can be applied to the gravitational stress-energy tensor:
\beq
(T_{00})_G=-\left(2\lambda\frac{\delta\chi}{\delta g^{\mu\nu}}\Big|_{00}+\chi\right)\chi^{\alpha-1}|\chi|^\beta \ .
\eeq
In addition to the kinetic term, we would also need a coupling term. Given that $\rho$ is actually the 00-component of baryonic matter's stress-energy tensor, this coupling term can have the form
\beq\label{Int}
\mathcal{L}_{int}\ = \ h^{\mu\nu}(\Psi)\,T_{\mu\nu} \ .
\eeq
Given that $T_{\mu\nu}$ is proportional to the first-order functional derivative of the matter Lagrangian over $g_{\mu\nu}$, the presence of this interaction term is equivalent, at first order, to coupling baryonic matter to the effective metric $\tilde{g}_{\mu\nu}$:
\beq
\mathcal{S}_m=\int \ d^4x \ \sqrt{-\tilde{g}} \ \mathcal{L}_m(\tilde{g}_{\mu\nu}) \ ,
\eeq
given by
\beq\label{Metr}
\tilde{g}_{\mu\nu}=g_{\mu\nu}-2h_{\mu\nu} \ .
\eeq

Our next step would be to determine how does $\Psi$ transform under the Lorentz group. Again, the easiest option would be to make it a scalar; this model, known as RAQUAL (relativistic AQUAL), was proposed in~\cite{Bekenstein:1984tv}. A generic scalar field Lagrangian of the form
\beq
\mathcal{L} \ = \ -(\partial_\mu\Psi\partial^\mu\Psi)^\lambda
\eeq
yields the stress-energy tensor
\beq
T_{00}=(\partial^\mu\Psi\partial_\mu\Psi)^{\lambda-1}\left[(2\lambda-1)(\dot{\Psi})^2+(\vec{\nabla}\Psi)^2\right] \ 
\eeq
(the canonical and gravitational definitions are equivalent in this case). The expression within square brackets is positive definite for all $\lambda\ge\frac{1}{2}$, but the prefactor can become negative for even $\lambda$ when
\beq
\dot{\Psi}^2 \ > \ |\vec{\nabla}\Psi|^2 \ .
\eeq
We already stated that one should exclude this region when considering non-integer $\lambda$, so that $T_{00}$ is positive definite for all positive and non-even (\emph{i.e.} either odd or non-integer) $\lambda$, including the special value $\frac{3}{2}$ that plays a role in the models of interest.

Alternatively, for a Lagrangian of the form
\beq
\mathcal{L}=-|\partial^\mu\Psi\partial_\mu\Psi|^{\lambda-1}(\partial^\mu\Psi\partial_\mu\Psi)
\eeq
that incorporates both positive and negative values of $(\partial^\mu\Psi\partial_\mu\Psi)$, the energy density would be given by
\beq
T_{00}=|\partial^\mu\Psi\partial_\mu\Psi|^{\lambda-1}\left[(2\lambda-1)(\dot{\Psi})^2+(\vec{\nabla}\Psi)^2\right] \ .
\eeq
In this theory the energy density is bounded from below for all $\lambda\ge\frac{1}{2}$ (even values of $\lambda$ are not a problem, since the prefactor cannot be negative).

However, as was shown in~\cite{Bekenstein:2004ne}, RAQUAL contains superluminal modes, and more importantly, a scalar can only couple to the \emph{trace} of the stress-energy tensor: 
\beq
\mathcal{L}= -\frac{1}{12\pi Ga_0}(\partial^\mu\Psi\partial_\mu\Psi)^{3/2}-\Psi T^\alpha_\alpha \ ,
\eeq
corresponding to a conformal factor of the metric:
\beq
\tilde{g}_{\mu\nu} \ = \ e^{-2\Psi}g_{\mu\nu} \ .
\eeq
As a result, it cannot account for gravitational lensing.

A more complicated option would be to introduce $\Psi$ as the time component of a vector field $u_\mu$. This theory is known as ``covariant emergent gravity`` (CEG)~\cite{Hossenfelder:2017eoh} due to its conjectured connection to emergent gravity, a framework inspired by the AdS/CFT correspondence~\cite{ADSCFT}. Within the emergent gravity paradigm, dark energy is ascribed to an elastic medium, understood as an artifact of quantum gravity, while the ``MOND force``, represented by the vector field, is regarded as the medium's response to baryonic matter~\cite{Verlinde:2016toy}. In a corpuscular approach to quantum gravity, the ``dark medium`` was also interpreted as a Bose-Einstein condensate of gravitons~\cite{Tuveri:2019zor,Giusti:2019wdx}.

The CEG Lagrangian is yielded by~\eqref{Lagr}, with $\chi$ given by
\beq
\chi \ = \ \alpha(\nabla_\alpha u^\alpha)^2+\beta(\nabla_\mu u_\nu)(\nabla^\mu u^\nu)+\gamma(\nabla_\mu u_\nu)(\nabla^\nu u^\mu) \ . \label{N}
\eeq
For any choice of parameters other than the one in standard Maxwell theory ($\alpha=\beta+\gamma=0$), the kinetic term would contain Christoffel symbols. Therefore, even on flat spacetime, the gravitational stress-energy tensor would be different from the canonical one due to additional terms proportional to variations of these symbols.

More precisely, for a generic vector field theory with a Lagrangian $\propto\chi^\lambda$ the canonical stress-energy tensor would be
\beq
(T_{\mu\nu})_C=-2\lambda\left(\alpha(\partial_\alpha u^\alpha)(\partial_\nu u_\mu)+\beta(\partial_\mu u^\alpha)(\partial_\nu u_\alpha)+\gamma(\partial^\alpha u_\mu)(\partial_\nu u_\alpha)\right)\chi^{\lambda-1}+g_{\mu\nu}\chi^\lambda \ .
\eeq
This tensor is not symmetric, but, generalizing the result for the Maxwell case, one can symmetrize it by the addition of a term of the form
\beq
-2\lambda\left(\alpha(\partial_\alpha u^\alpha)(\partial_\mu u_\nu)+\beta(\partial_\alpha u_\mu)(\partial^\alpha u_\nu)+\gamma(\partial^\alpha u_\nu)(\partial_\mu u_\alpha)\right)\chi^{\lambda-1} \ ,
\eeq
which is a total divergence when the field equations are satisfied. One is thus led to consider
\beq
\begin{gathered}
\frac{(T_{00})_C}{\chi^{\lambda-1}} \ = \ (4\lambda-1)(\alpha+\beta+\gamma)(\partial_0u_0)^2-(\alpha+\beta+\gamma)\sum_i(\partial_iu_i)^2-\\
2(2\lambda-1)\alpha\partial_0u_0\sum_i\partial_iu_i-2\alpha\sum_{i\neq j}(\partial_iu_i)(\partial_ju_j)-\beta\sum_{i\neq j}(\partial_iu_j)^2-\\
2\gamma\sum_{i\neq j}(\partial_iu_j)(\partial_ju_i)-(2\lambda-1)\beta\sum_i(\partial_0u_i)^2-\\
(2\lambda-1)\beta\sum_i(\partial_iu_0)^2-2(2\lambda-1)\gamma\sum_i(\partial_iu_0)(\partial_0u_i) \ .
\end{gathered}
\eeq
This quadratic form can be presented as a 16 x 16 matrix. To guarantee the stability of the theory, one would need to require all of the matrix's eigenvalues to have the same sign (if they are all negative, the kinetic term should be taken with the opposite sign to ensure positivity). The matrix is comprised of blocks: one 4 x 4 matrix for the products of $\partial_0 u_0$ and $\partial_iu_i$ ($i=1,2,3$), three identical 2 x 2 blocks for products of $\partial_0u_i$ and $\partial_iu_0$, and three identical 2 x 2 blocks for products of $\partial_iu_j$ and $\partial_ju_i$. The latter type of blocks have the form:
\beq
-
\begin{pmatrix}
	\beta & \gamma \\
	\gamma & \beta
\end{pmatrix} \ ,
\eeq
with the eigenvalues
\beq
\begin{gathered}
\Lambda_{1,2} \ = \ -\beta\pm\gamma \ .
\end{gathered}
\eeq
The other type of 2 x 2 blocks is
\beq
(1-2\lambda)
\begin{pmatrix}
	\beta & \gamma \\
	\gamma & \beta
\end{pmatrix} \ ,
\eeq
with the eigenvalues
\beq
\begin{gathered}
\Lambda_{3,4} \ = \ -(2\lambda-1)(\beta\pm\gamma) \ .
\end{gathered}
\eeq
Finally, the 4 x 4 matrix is given by
\beq
\begin{pmatrix}
	(4\lambda-1)(\alpha+\beta+\gamma) & (1-2\lambda)\alpha & (1-2\lambda)\alpha & (1-2\lambda)\alpha \\
    (1-2\lambda)\alpha & -(\alpha+\beta+\gamma) & -\alpha & -\alpha \\
    (1-2\lambda)\alpha & -\alpha & -(\alpha+\beta+\gamma) & -\alpha \\
    (1-2\lambda)\alpha & -\alpha & \alpha & -(\alpha+\beta+\gamma) \\
\end{pmatrix} \ .
\eeq
Two of its eigenvalues are
\beq
\Lambda_{5,6} \ = \ -\beta-\gamma \ , \label{l56}
\eeq
and the remaining two are determined by the quadratic equation
\beq
\Lambda^2+B\Lambda+C=0 \ ,
\eeq
with
\beq
\begin{gathered}
B \ = \ 3\alpha+\delta-(4\lambda-1)(\alpha+\delta) \ , \\
C \ = \ -(4\lambda-1)(\delta+2\alpha)^2-12(\lambda-1)(\lambda-\frac{1}{3})\alpha^2 \ ,
\end{gathered}
\eeq
with $\delta=\beta+\gamma$. For the two eigenvalues to have the same sign, $C$ should not be negative, but the only way to ensure this for $\lambda>1$ is to set both $\alpha$ and $\delta$ to zero, opting for a Maxwell-type theory.

Now, we turn to the gravitational stress-energy tensor, which appears more relevant in this context, since the canonical one is generally defined for field theories on flat spacetime, and within the MOND framework gravitational effects are clearly non-negligible. The gravitational tensor is given by
\beq
\frac{(T_{\mu\nu})_G}{2\chi^{\lambda-1}} \ = \ -\lambda\frac{\delta\chi}{\delta g^{\mu\nu}}+\frac{1}{2}g_{\mu\nu}\chi \ ,
\eeq
with
\beq
\begin{gathered}
\frac{\delta\chi}{\delta g^{\mu\nu}} \ = \ \alpha g_{\mu\nu}(\partial_\alpha u^\alpha)^2+\frac{1}{2}(2\alpha+\beta+\gamma)(\partial_\alpha u^\alpha)(\partial_\mu u_\nu+\partial_\nu u_\mu)+\beta(\partial_\mu u_\alpha)(\partial_\nu u^\alpha)-\\
\gamma(\partial_\alpha u_\mu)(\partial^\alpha u_\nu)+\frac{\gamma-\beta}{2}\left((\partial_\alpha u_\mu)(\partial_\nu u^\alpha)+(\partial_\alpha u_\nu)(\partial_\mu u^\alpha)\right)+O(\partial^2u).
\end{gathered}
\eeq
To compute the variation of $\chi$, we also took into account contributions involving the Christoffel symbols $\Gamma\propto\partial g$, since after integrating by parts they yield terms that survive even in flat spacetime backgrounds. There are also terms proportional to $u(\partial^2 u)$, which vanish, interestingly, for the MOND solution with $\lambda=\frac{3}{2}$. In detail, for a configuration of the form 
\beq
u_0=\phi(\vec{r}) \ , \ u_i=0 \ ,    
\eeq
with $\phi\propto r^\kappa$, these terms are proportional to
\beq
(\beta+\gamma)\partial_i\left((\partial_i\phi)((\partial_j\phi)^2)^{\lambda-1}\right) \ \propto \ \partial_i\left(\frac{x_i}{r}r^{(\kappa-1)(2\lambda-1)}\right) \ ,
\eeq
which vanishes for
\beq
\lambda \ = \ \frac{1}{2}-\frac{1}{\kappa-1} \ .
\eeq
For a logarithmic potential ($\kappa=0$), this occurs precisely if $\lambda=\frac{3}{2}$.

However, for more generic solutions, there is a stability issue, which manifests itself even with linear field configurations of the type 
\beq
u_\mu \ = \ A_\mu+B_{\mu\nu}x^\nu \ , \label{w}
\eeq
for which the terms $\propto\partial^2u$ also vanish.

It should be noted that these special configurations solve the equations of motion, which involve \textit{second-order} derivatives of $u$~\footnote{~\cite{Hossenfelder:2017eoh} and~\cite{Dai:2017guq} propose either a mass term or a quartic self-interaction term for $u_\mu$, but their contribution to the EOMs can be made negligible if we consider very small values of $A_\mu$ and a region of spacetime sufficiently close to the origin, so that the $B_{\mu\nu}x^\nu$ are small even for large enough values of $B_{\mu\nu}$.}. For the configurations in eq.~\eqref{w}, one thus obtains the gravitational energy--momentum tensor
\beq
\begin{gathered}
\frac{(T_{00})_G}{2\chi^{\lambda-1}} = \left((3\lambda-\frac{1}{2})\alpha+(\lambda-\frac{1}{2})(\beta+\gamma)\right)(\partial_0u_0)^2 \ +\\
((1-4\lambda)\alpha-\lambda(\beta+\gamma))(\partial_0u_0)(\sum_i\partial_iu_i) \ + \\
\left((\lambda-\frac{1}{2})\alpha-\frac{1}{2}(\beta+\gamma)\right)\sum_i(\partial_iu_i)^2+(2\lambda-1)\alpha\sum_{i\neq j}(\partial_iu_i)(\partial_ju_j) \ -\\
\frac{\beta}{2}\sum_{i\neq j}(\partial_iu_j)^2-\gamma\sum_{i\neq j}(\partial_iu_j)(\partial_ju_i)+\left(\lambda\gamma+\frac{\beta}{2}\right)\sum_i(\partial_iu_0)^2 \ + \\
\left(\frac{1}{2}-\lambda\right)\beta\sum_i(\partial_0u_i)^2+\left(\gamma(1-\lambda)+\lambda\beta\right)\sum_i(\partial_0u_i)(\partial_iu_0) \ .
\end{gathered}
\eeq
For this quadratic form, the corresponding matrix block for products of $\partial_iu_j$ and $\partial_ju_i$ would be:
\beq
-\frac{1}{2}
\begin{pmatrix}
	\beta & \gamma \\
	\gamma & \beta
\end{pmatrix}
\eeq
with eigenvalues
\beq
\Lambda_{3,4} \ = \ -\frac{1}{2}(\beta\pm\gamma) \ .
\eeq
One can introduce the new variables
\beq
\begin{gathered}
\delta \ = \ \frac{1}{2}(\beta+\gamma) \ , \ \xi \ = \ \frac{1}{2}(\beta-\gamma) \ ,
\end{gathered}
\eeq
and the energy positivity condition requires that $\delta$ and $\xi$ have the same sign.

The other $2 \times 2$ block corresponds to the products of $\partial_0u_i$ and $\partial_iu_0$ (for each i), is
\beq
\begin{pmatrix}
	\lambda\gamma+\frac{\beta}{2} & \frac{\gamma}{2}(1-\lambda)+\lambda\frac{\beta}{2}\\
	\frac{\gamma}{2}(1-\lambda)+\lambda\frac{\beta}{2} & (1-2\lambda)\frac{\beta}{2}
\end{pmatrix} \ ,
\eeq
and its two eigenvalues have the same sign if
\beq
-\lambda^2\delta^2\ -\ 2\left(\lambda\,-\,\frac{1}{2}\right)\delta\,\xi\ \ge \ 0 \ .
\eeq
Since the first term in $b$ is negative definite for $\delta\neq0$, the second should be positive to satisfy the condition. But for $\lambda\ge\frac{1}{2}$, it can only be positive if $\delta\xi<0$, which cannot be true, since $\delta$ and $\xi$ should have the same sign, as we have seen. Therefore the only option is $\delta=0$, \emph{i.e.} $\gamma=-\beta$.

Finally, for products of $\partial_0u_0$ and $\partial_iu_i$ there is the 4 x 4 block
\beq
\begin{pmatrix}
	(3\lambda-\frac{1}{2})\alpha+2(\lambda-\frac{1}{2})\delta & -(2\lambda-\frac{1}{2})\alpha-\lambda\delta & -(2\lambda-\frac{1}{2})\alpha-\lambda\delta & -(2\lambda-\frac{1}{2})\alpha-\lambda\delta \\
	-(2\lambda-\frac{1}{2})\alpha-\lambda\delta & (\lambda-\frac{1}{2})\alpha-\delta & (\lambda-\frac{1}{2})\alpha & (\lambda-\frac{1}{2})\alpha \\
	-(2\lambda-\frac{1}{2})\alpha-\lambda\delta & (\lambda-\frac{1}{2})\alpha & (\lambda-\frac{1}{2})\alpha-\delta & (\lambda-\frac{1}{2})\alpha \\
	-(2\lambda-\frac{1}{2})\alpha-\lambda\delta & (\lambda-\frac{1}{2})\alpha & (\lambda-\frac{1}{2})\alpha & (\lambda-\frac{1}{2})\alpha-\delta
\end{pmatrix} \ . \label{e}
\eeq
One eigenvalue of this matrix,
\beq
\Lambda \ = \ -\ \delta \ ,
\eeq
is doubly degenerate, and if one now sets $\delta$ to zero, as required by the preceding discussion, the remaining two eigenvalues are determined by the quadratic equation
\beq
\Lambda^2 \ - \ 2(3\lambda-1)\alpha\,\Lambda\  - \ 3\lambda^2\alpha^2 \ = \ 0 \ .
\eeq
Following the same logic as in the previous case, one should require that the last term be non-negative for the two eigenvalues to have the same sign, but this is not possible unless $\alpha=0$. In conclusion, one is left is Maxwell's choice of parameters $\alpha=0, \beta=-\gamma$. All other options, including the choice made in CEG ($\alpha=\frac{4}{3}, \beta=\gamma=-\frac{1}{2}$)~\cite{Hossenfelder:2017eoh}, result in $(T_{00})_G$ unbounded from below within the class of configurations that we have explored.

The aforecited argument has several loopholes: first of all, since the relevant quantity is not $(T_{00})_G$ but $\frac{(T_{00})_G}{\chi^{\lambda-1}}$, this analysis suffices only for odd integer values of $\lambda$ (including the Maxwell case $\lambda=1$). If, instead, $\lambda$ is either even or non-integer (the latter is relevant for CEG), one should check whether $\frac{(T_{00})_G}{\chi^{\lambda-1}}$ can become negative in the region where $\chi$ is positive. Otherwise, for even $\lambda$, the sign change would be compensated by the sign change of $\chi^{\lambda-1}$, and $(T_{00})_G$ would remain positive, while for non-integer $\lambda$, regions with negative $\chi$ are simply removed from the configuration space. However, it is possible to construct field configurations that satisfy the following conditions:
\beq
\label{v}
\begin{gathered}
\frac{\delta\chi}{\delta g^{\mu\nu}}\Big|_{00}\ge 0 \ , \\
\chi\ge 0 \ .
\end{gathered}
\eeq
For example, for negative $\beta$ and $\gamma$ (as is the case in CEG), a static configuration with
\beq
u_0 \ = \ A+B_{\mu}x^\mu \ ,\\
u_i \ = \ 0
\eeq
would lead to a negative $(T_{00})_G$,
\beq
(T_{00})_G \ \propto \ \left(\beta+2\lambda\gamma\right)(\sum_i(B_i)^2)\chi^{\lambda-1} \ ,
\eeq
while
\beq
\chi \ = \ -\beta\sum_i(B_i)^2
\eeq
would be positive. The coefficients $B_i$ can be arbitrarily large, and this means the theory is not bounded from below.

Following~\cite{Hossenfelder:2017eoh}, one can take this argument one step further, supplementing it with the condition that the gravitational stress-energy tensor be covariantly conserved on the simplest non-flat background, de Sitter space:
\beq
ds^2=-dt^2+e^{2Ht}d\vec{r}^2 \ .
\eeq
For a field configuration of the form
\beq
u_0=C \ , \ u_i=0 \ ,
\eeq
the components of the gravitational stress-energy tensor are now
\beq
\begin{gathered}
(T_{00})_G\propto9\alpha(2\lambda-1)-3(\beta+\gamma) \ ,\\
(T_{ii})_G\propto\left(9\alpha+3(\beta+\gamma)(1-4\lambda)\right)e^{2Ht} \ ,
\end{gathered}
\eeq
and lead to the conservation condition
\beq
\beta\,+\,\gamma \ = \ -\,\frac{3(\lambda-1)}{2\lambda-1}\,\alpha \ .
\eeq
However, for the Maxwell choice of parameters, one can see that  
\beq
\begin{gathered}
\chi \ = \ \gamma\left(\sum_i(\partial_0u_i-\partial_iu_0)^2-\sum_{i,j}(\partial_iu_j-\partial_ju_i)^2\right) \ ,\\
(T_{00})_G \ \propto \ \frac{\gamma}{2}\sum_{i,j}(\partial_iu_j-\partial_ju_i)^2+\gamma(\lambda-\frac{1}{2})(\partial_iu_0-\partial_0u_i)^2 \ ,
\end{gathered}
\eeq
so that the positivity condition is only satisfied for $\lambda\ge\frac{1}{2}$. Following a similar logic, one can also see that theories with even $\lambda$ ($\lambda=2n$) are problematic, even for Maxwell's choice of parameters. Since $\frac{(T_{00})_G}{\chi^{\lambda-1}}$ is positive definite, negative values of $\chi$, \emph{i.e.} such that
\beq
\begin{gathered}
(\partial_iu_j-\partial_ju_i)^2 \ > \ (\partial_0u_i-\partial_iu_0)^2
\end{gathered}
\eeq
would result in negative values of $(T_{00})_G$.
Therefore, our conclusion is that the only consistent vector field theories rest on Maxwell's choice of parameters ($\alpha=0, \beta=-\gamma$), and on odd or non-integer values of $\lambda$, with $\lambda\ge\frac{1}{2}$.
Moreover, for the Lagrangian based on $|\chi|^{\lambda-1}\chi$, which can also be defined for negative values of $\chi$, $\lambda$ can be even-valued, but again it cannot be smaller than $\frac{1}{2}$.

Continuing to focus on the gravitational energy--momentum tensor, one may wonder whether one could circumvent the problem introducing Lagrange multipliers, at the cost of altering the dynamical content of the vector model, or replacing the condition that $(T_{00})_G$ be positive with the weaker condition that some linear combination of $(T_{00})_G$ and $(T_{0i})_G$ be positive~\cite{Babichev:2018uiw,Esposito-Farese:2019vlh}~\footnote{I am grateful to prof. G.~Esposito--Farese for a useful correspondence on this issue.}. However, for the MOND case, neither of these approaches can eliminate contributions leading to negative energy density, such as the terms $\propto(\partial_0u_i)^2$. For example, taking a configuration with $u_0=C$ and $u_i=B_it$, the $T_{0i}$ components would vanish. 

Turning to the Lagrange multipliers, if one wanted to constrain only first derivatives, a scrutiny of the available option shows that the quantities that could be used, which include $(\partial_\alpha u_\mu)(\partial^\alpha u_\nu)$, also contribute to the MOND term, and there appears to be no way to remove the negative contributions without affecting it.

We can now address {the issue of superluminal modes}. The equations of motion for a Lagrangian of the form
\begin{equation}
\mathcal{L} \sim \ (D_{\mu\nu}D^{\mu\nu})^\lambda \ , \label{maxwell_lambda}
\end{equation}
where $D_{\mu\nu}$ is Maxwell's field strength for $u_\mu$, are 
\begin{equation}\label{EQS}
\partial_\mu\Big[D^{\mu\nu}(D_{\alpha\beta}D^{\alpha\beta})^{\lambda-1}\Big]\ =\ J^\nu \ ,
\end{equation}
and in absence of sources they reduce to
\begin{equation}
(D_{\mu\nu}D^{\mu\nu})^{\lambda-1}\left[\partial^\mu D_{\mu\nu}+2(\lambda-1)\frac{D_{\mu\nu}D^{\gamma\delta}}{(D_{\alpha\beta}D^{\alpha\beta})}\partial^\mu D_{\gamma\delta}\right] \ = \ 0 \ .
\end{equation}
A perturbation $\delta u$ of a static background configuration of the form
\begin{equation}
u_0=\Psi(\vec{r}) \ , \ \qquad u_i=0
\end{equation}
in the eikonal approximation and in the Lorenz gauge
\begin{equation}
k_\mu \delta u^\mu=0
\end{equation}
yields
\begin{equation}
\begin{gathered}
(\vec{k}^2-\omega^2)\delta u_0+2(\lambda-1)(\vec{n}\vec{k})\left((\vec{n}\vec{k})\delta u_0-\omega(\vec{n}\delta\vec{u})\right)=0 \ ,\\
(\vec{k}^2-\omega^2)(\vec{n}\delta \vec{u})+2(\lambda-1)\omega\left((\vec{n}\vec{k})A_0-\omega(\vec{n}\vec{A})\right)=0 \ , \label{vector_caus}
\end{gathered}
\end{equation}
where
\begin{equation}
n_i=\frac{\partial_i\Psi}{\sqrt{(\partial_j\Psi)^2}}
\end{equation}
is a spacelike vector. Combining eqs.~\eqref{vector_caus} gives a quadratic equation for $\omega^2$, with solutions
\begin{equation}
\omega_1^2=\vec{k}^2 \ , \qquad \omega_2^2=\frac{\vec{k}^2+2(\lambda-1)(\vec{n}\vec{k})^2}{2\lambda-1} \ .
\end{equation}
This means that the group velocity can vary between 1 and $\frac{1}{\sqrt{2\lambda-1}}$, and therefore theories with $\lambda\ge1$ have no superluminal modes.

In principle, it is possible to construct a MOND solution with a Lagrangian of the form~\eqref{maxwell_lambda} by introducing an interaction term of the form
\beq\label{INT}
\mathcal{L}_{int}\ = \ \xi\frac{u_\mu u_\nu}{u}\,T^{\mu\nu} \ ,
\eeq
as in~\cite{Hossenfelder:2017eoh}, corresponding to the effective metric 
\beq
\tilde{g}_{\mu\nu} \ = \ g_{\mu\nu} - 2\xi\frac{u_\mu u_\nu}{u}
\eeq
and considering a static field configuration
\beq
u_0 \ = \ \Psi(\vec{r}) \ , \ u_i \ = \ 0 \ .
\eeq
However, this choice breaks the gauge invariance of $u_\mu$, and is thus inconsistent with the equations following from the types of Lagrangians of eq.~\eqref{maxwell_lambda}, which grant a lower bound on the energy density, as we have seen. The effective current should be conserved, and this condition is generally not satisfied for 
\beq
J^\mu\ = \ \frac{\delta\mathcal{L}_{int}}{\delta u_{\mu}}\ \propto \ 2\, \frac{u_\nu}{u}T^{\mu\nu}\ +\ \frac{u^\mu u_\alpha u_\beta}{u^3}T^{\alpha\beta} \ .
\eeq
Gauge invariance is guaranteed, however, if the coupling term only involves $D_{\mu\nu}$, Maxwell's field strength for the $u_\mu$ field. Let us therefore consider the class of couplings
\beq
\mathcal{L}_{int}\ = \ F\left(-D_{\gamma\delta}D^{\gamma\delta}\right)D_\mu^\alpha D_{\nu\alpha}T^{\mu\nu} \ , \label{sources1}
\eeq
where the function $F$, which we allow out of despair to recover MOND, will be specified shortly.
For a static point mass, one would look for a spherically symmetric field configuration
\beq
\label{Anz}
u_0=\Psi(r) \ , \ u_i=0
\eeq
and the emergence of a MOND--like potential would demand, for consistency, that
\beq
\label{LogPot}
F(\Psi'^2)\ \Psi'^2\propto\ln(r) \ ,
\eeq
since the source accompanying $T_{\mu\nu}$ in this case ought to play the role of a scalar potential. Away from a point source the Lagrangian of eq.~\eqref{maxwell_lambda} would yield the field equation
\beq
\partial_r(r^2(\Psi')^{2\lambda-1})=0 \ \Rightarrow \ \Psi'=Cr^{\frac{2}{1-2\lambda}} \ ,
\eeq
and the scaling symmetry demands that $\lambda=\frac{3}{2}$.
The sought logarithmic potential \eqref{LogPot} and gauge invariance would conspire into a non--local dressing for the source coupling of the type
\beq
F(x)\ =\ \frac{\ln(x^2)}{x^2} \ \Rightarrow\  \mathcal{L}_{int}\propto T^{\mu\nu}\frac{D_\mu^\rho D_{\nu\rho}}{D_{\alpha\beta}D^{\alpha\beta}}\ln\left(-D_{\gamma\delta}D^{\gamma\delta}\right) \ ,
\eeq
which appears indeed rather baroque, a substantial overkill.

In principle, one could concoct an even more elaborate theory with a larger number of fields, such as TEVES (tensor-vector-scalar gravity), an extension of RAQUAL that contains two fields: a scalar conformal factor of the metric and a vector field which accounts for the gravitational lensing~\cite{Bekenstein:2004ne}. However, according to the LIGO--Virgo gravitational wave observations, gravity and light should propagate on the same geodesics, which rules out theories like CEG and TEVES that have matter and gravity couple to conformally unrelated metrics~\cite{Boran:2017rdn}.
\subsection{Superfluid dark matter (SfDM)}\label{C3S2S3}
Finally, we should consider the possibility that the fractional power of the kinetic term emerges specifically in the non--relativistic limit. One possible way to implement this is via symmetry breaking, as proposed in~\cite{Berezhiani:2015bqa}. One should start with a Lagrangian of the form\\ 
\beq
\mathcal{L} \ = \ -\frac{1}{2}\left(|D_\mu\Phi|^2+m^2\Phi^2\right)-\frac{\Lambda^4}{6(\Lambda_c^2+|\Phi|^2)^6}\left(|D_\mu\Phi|^2+m^2\Phi^2\right)^3 \ 
\eeq
with a $U(1)$ symmetry. This Lagrangian satisfies the bounded-from-below condition:
\beq
\label{a2}
\begin{gathered}
T_{00} \ = \ \frac{1}{2}\left(|D_0\Phi|^2+|D_i\Phi|^2+m^2|\Phi|^2\right) \ + \ \frac{\Lambda^4}{6(\Lambda_c^2+|\Phi|^2)^6}*\\
(|D_\mu\Phi|^2+m^2|\Phi|^2)^2\left(|D_i\Phi|^2+5|D_0\Phi|^2+m^2|\Phi|^2\right)\ge 0 \ .
\end{gathered}
\eeq
Once the symmetry is spontaneously broken and $\Phi$ acquires a nonzero non--relativistic VEV:
\beq\label{VEV}
\Phi \ = \ \rho e^{i(\theta+mt)} \ ,
\eeq
one can integrate out $\rho$ to obtain
\beq
\mathcal{L}_\theta \ = \ \frac{2\Lambda(2m)^{3/2}}{3}X\sqrt{|X|}
\eeq
with
\beq
X=\dot{\theta}-mV-\frac{(\nabla\theta)^2}{2m} \ ,
\eeq
where $V$ is the external gravitational potential. Then, adding a coupling to baryonic matter\\
\beq \label{Sfint}
\mathcal{L}_{int} \ = \ -\alpha\frac{\Lambda}{M_P}\theta\rho_b \ ,
\eeq
one would obtain MONDian behaviour. This model is known as superfluid dark matter (SfDM); at finite temperatures, SfDM would effectively be a mixture of superfluid and CDM-like normal fluid that can account for gravitational lensing~\cite{Hossenfelder:2018iym}. Besides, due to its ability to behave as an ordinary scalar field with a mass of order $\sim$eV, it could, in principle, account for phenomena like CMB and structure formation that are unexplained in most versions of MOND. Nonetheless, it is not clear how to obtain the interaction term~\eqref{Sfint} from the relativistic form of the theory. One could implement $\theta$ as a conformal factor of the metric, as in RAQUAL, but it would be necessary then to express it as a function of $\Phi$:
\beq
\theta \ = \ \frac{i}{2}\ln\left(\frac{\Phi^*}{\Phi}\right) \ - \ mt \ ,
\eeq
and this function is non-covariant and ill-defined at $\Phi=0$. This problem is generic in superfluid/symmetry breaking models since Goldstone bosons usually emerge as phases of complex fields~\cite{Mistele:2019byy}. Alternatively, one could couple $\theta$ to baryonic density instead of matter density, in which case our Noether current would be the baryonic current, rather than the stress-energy tensor. This would make SfDM a field theory that \emph{imitates} the effects of modified gravity rather than an actual modified gravity theory; however, its complete theoretical description is still lacking~\cite{Alexander:2018fjp}.

One more potential difficulty with this approach has to do with the phase transition mechanism. Since \eqref{a2} is equal to zero at $\Phi=0$ regardless of the presence of the gravitational potential, and is larger than zero in the broken phase, according to \eqref{VEV}, the broken phase ends up having a higher energy density than the unbroken one.
\section{Conclusion}
To sum up the results of this section, we have demonstrated the impossibility to construct a relativistic version of MOND (or, for that matter, \emph{any} infrared modification of gravity) either through a direct modification of general relativity along the lines of $f(R)$ theories or through a disformal transformation of metric. Therefore, as long as the modified gravity Lagrangian is required to satisfy certain assumptions (locality, Lorentz invariance, and fixed number of dimensions $D=4$), there are only two possible ways to reproduce MOND:
\begin{itemize}
    \item A model along the lines of RAQUAL that couples matter to a metric with a conformal factor \emph{plus} a CDM-like mass component to account for gravitational lensing. The scalar field and the mass component may be either separate entities or two different manifestations of the same phenomenon (SfDM is one example of the latter approach, though its relativistic description is incomplete so far).
    \item ``Fake modified gravity``, i.e. a theories of fields that don't directly couple to matter, but interact with it \emph{through} gravity, producing a MOND-like potential. Such fields can be non--minimally coupled to gravity, as is the case for bimetric MOND~\cite{Milgrom:2009gv}. Einstein aether theory also belongs to this category, though it does break the Lorentz symmetry~\cite{Zlosnik:2006zu}.
\end{itemize}

Finally, there is the possibility that the radial acceleration relation emerges through dynamical effects rather than a new physical law. In most cases it has to do with a specific type of dark matter that has some particular property (such as dipolar~\cite{Blanchet:2009zu}, dissipative~\cite{Chashchina:2016wle}, or fuzzy dark matter~\cite{Lee:2019ums}). However, in the next chapter we propose the existence of a purely geometric effect that turns up simply from the spatial distribution of DM, regardless of its specific nature. Should this effect actually exist, it would also imply that standard CDM model is preferable over modified gravity or self-interacting dark matter (SiDM).
\chapter{Elongated mass distributions and their imprint on rotation curves}\label{C4}
\section{Gravitational effects of filaments and prolate dark halos}\label{C4S1}
Our proposal is rooted in the simple physical observation that the gravitational potential of an infinitely thin wire, given by
\beq
V\left(r\right) \ = \ G\,\mu \int_{-\,\ell_0}^{\ell_0} \frac{dz}{\sqrt{r^2+z^2}} \ = \ G\,\mu\,\ln\left(\frac{\sqrt{\ell_0^2+r^2}+\ell_0}{\sqrt{\ell_0^2+r^2}-\ell_0}\right) \ , \label{line_potential}
\eeq
where $\mu$ is a mass per unit length and $\ell_0$ is the half--length of the wire,
exhibits an interesting transition between a $\log$--like behavior for $r\ll \ell_0$ and the standard $\frac{1}{r}$ behavior for $r\gg \ell_0$. In the former region the elongated mass distribution flattens the field lines in the radial plane, mimicking a two--dimensional log--like behavior, before the standard monopole term finally dominates at larger distances. Therefore an object moving on a stable circular orbit around such a wire would have the velocity
\beq
\label{Wire_vel}
v_{DM}^2(r) \ = \ -\,r\,\frac{\partial V\left(r\right)}{\partial r} \ = \ \frac{2\,G\,\mu\, \ell_0}{\sqrt{\ell_0^2+r^2}} \ .
\eeq
Cylindrical and elliptical mass distributions also yield quasi-logarithmic potentials. Therefore, if the dark matter halos of galaxies have a prolate shape, with the elongation along the Z-axis, they would yield approximately flat rotation curves at distances larger than the minor axis but smaller than the major one. There are several independent indications from kinematic data~\cite{Hayashi:2014nra,Bowden:2016bwq,Hattori:2019lgu} and gravitational lensing~\cite{Hoekstra:2003pn,Oguri:2004in} that a number of galaxies and galaxy clusters, including the Milky Way and Andromeda (M31), may host bulged dark halos. Prolate halo shapes were also observed in simulations of collisionless CDM~\cite{Dubinski:1991bm}, and were conjectured to result from either halo merging~\cite{Allgood:2005eu} or hierarchical structure formation, resulting in the collapse of matter along filaments rather than sheets~\cite{Bett:2006zy}. On the contrary, DM self-interactions tend to favor rounder halo shapes~\cite{Yoshida:2000uw}, and the MOND framework can only mimic spherical or slightly oblate DM distributions~\cite{Read:2005if}. As a result, ascertaining the actual presence of elongated halos would also provide additional evidence to discriminate among different scenarios (it's not clear however whether this would also impinge on hybrid scenarios such as superfluid dark matter~\cite{Berezhiani:2015bqa}).

The potential interest in these considerations lies in the fact that they rest solely on standard laws of gravity, and in particular on its Newtonian limit, without the need for any infrared modifications.

For the purposes of our analysis, we consider the Navarro-Frenk-White (NFW)~\cite{Navarro:1996gj} and Burkert~\cite{Burkert:1995yz} profiles that have been introduced in~\ref{C2S2}, and that are among the most popular spherically symmetric smooth distributions used for dark matter in galaxies. Making the simple replacement
\beq
r \ = \ \sqrt{x^2+y^2+z^2} \ \to \ \sqrt{x^2+y^2+q\,z^2} \ , \label{deformed_r}
\eeq
one can deform these spherical distributions into prolate ones for $q< 1$, or oblate ones for $q>1 $. 
The resulting contributions to the rotation velocity are determined by
\beq
-\frac{\partial V}{\partial r} \ = \ 2G\int_0^\infty \ dz \ \int^\infty_0 \ dr'r'\rho \left(\sqrt{r'^2+q^2z^2}\right) \ \int_0^{2\pi} \  \frac{d\phi \left(r-r'\cos\phi\right)}{(r'^2+r^2+z^2-2rr'\cos\phi)^{3/2}} \ ,
\eeq
and performing the angular integral one can cast them in the form
\beq
\begin{gathered}\label{VDM}
v^2_{DM}(r) \ = \ -r\,\frac{\partial V}{\partial r} \ = \ 2G\int_0^\infty dz \ \int^\infty_0 dr'\,\frac{r'\rho \left(\sqrt{r'^2+q^2z^2}\right)}{\sqrt{(r-r')^2+z^2}}*\\
\left[F\left(\pi \ \Big| \ -\frac{4rr'}{(r-r')^2+z^2}\right) \ - \ \frac{r'^2+z^2-r^2}{(r+r')^2+z^2}\,E\left(\pi\Big|-\frac{4rr'}{(r-r')^2+z^2}\right)\right] \ ,
\end{gathered}
\eeq
where 
\beq
F\left(\phi \ | \ x\right) \ = \ \int_0^\phi \ \frac{d\theta}{\sqrt{1-x^2\sin^2\theta}} \ ,\qquad 
E\left(\phi \ | \ x \right) \ = \ \int_0^\phi \ d\theta \ \sqrt{1-x^2\sin^2\theta} 
\eeq
are incomplete elliptic integrals of the first and second kind.
\begin{figure}[ht]
\centering
	\includegraphics[width=90mm]{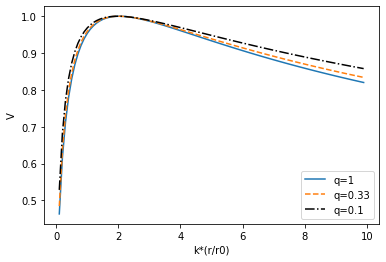}
	\caption{Dark--matter contributions to the rotation velocity for the NFW profile produced by a spherical halo (blue solid line), a prolate halo with a major-to-minor axis ratio 3 (orange dashed line), and a prolate halo with a major-to-minor axis ratio 10 (black dash-dotted line). The velocity is always normalized to unity at the peak, and the radius is normalized to $\frac{r_0}{k}$, with the coefficient $k$ always chosen so that the peak lies at $k\big(\frac{r}{r_0}\big)=2$. The actual value of $r_0$ depends on the galaxy.}
	\label{fig:NFW}
\end{figure}
Eq.~\eqref{VDM} is a complicated expression, which becomes however far simpler in the standard spherical limit ($q\rightarrow1$), and also in the limit of infinite elongation away from the galactic plane ($q\rightarrow0$). 
In the former case, the DM rotation velocity is determined by
\beq
v^2_{DM}(r) \ = \ \frac{4\pi G}{r}\int_0^r \ dr' \ r'^2 \rho(r') \ ,
\eeq
which results in
\beq\label{Vsph}
\begin{gathered}
v^2_{NFW}(r) \ = \ \frac{4\pi G\rho_0r_0^3}{r}\left[\ln(1+\frac{r}{r_0})-\frac{r}{r+r_0}\right] \ ,\\ 
v^2_B(r) \ = \ \frac{\pi G\rho_0r_0^3}{r}\left[\ln\left(1+\frac{r^2}{r_0^2}\right)+2\log\left(1+\frac{r}{r_0}\right)-2\arctan\left(\frac{r}{r_0}\right)\right]
\end{gathered}
\eeq
for the NFW and Burkert profiles.
\begin{figure}[ht]
\centering
	\includegraphics[width=90mm]{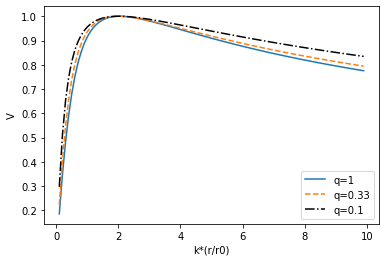}
	\caption{Dark--matter contributions to the rotation velocity for the Burkert profile produced by a spherical halo (blue solid line), a prolate halo with a major-to-minor axis ratio 3 (orange dashed line), and a prolate halo with a major-to-minor axis ratio 10 (black dash-dotted line). The velocity is always normalized to unity at the peak, and the radius is normalized to $\frac{r_0}{k}$, with the coefficient $k$ always chosen so that the peak lies at $k\big(\frac{r}{r_0}\big)=2$. The actual value of $r_0$ depends on the galaxy.}
	\label{fig:Burkert}
\end{figure}
On the other hand, in the limit of infinite elongation the rotation velocity reflects the two--dimensional version of Gauss's theorem, so that
\beq
v^2_{DM}(r) \ = \ 4\pi G \int_0^r \ dr' \ r'\rho(r') \ ,
\eeq
where $r=\sqrt{x^2+y^2}$. In detail, for the deformed NFW and Burkert profiles
\beq
\begin{gathered}
v^2_{NFW}(r) \ = \ 4\pi G\rho_0r^2_0\, \frac{r}{r+r_0} \ ,\\
v^2_B(r) \ = \ \pi G\rho_0r^2_0\left[\ln\left(1+\frac{r^2}{r_0^2}\right)-2\log\left(1+\frac{r}{r_0}\right)+2\arctan\left(\frac{r}{r_0}\right)\right] \ .
\end{gathered}
\eeq
From these expressions one can see that, if the density profile $\rho$ has a characteristic radial scale $r_0$ beyond which it tends smoothly to zero, eventually $v^2_{NFW, B}\propto r^{-1}\ln{r}$ in the spherical limit but $v^2_{NFW,B}\propto {\mathrm{const}}$ in the limit of infinite elongation. Even in the presence of smooth mass distributions, the effect we are after can therefore grant the emergence of flat rotation curves for distances scales $r_0 < r < q^{-1}r_0$, if $q<1$.
On the other hand, inside dark matter distributions one would observe growing rotation curves in both limits, with $v_{NFW}^2\propto r$ and $v_B^2\propto r^2$ for NFW and Burkert profiles. For generic bulged profiles the integrals must be computed numerically, and figs.~\ref{fig:NFW} and \ref{fig:Burkert} compare the results thus obtained with the spherical limits for the two cases of NFW and Burkert profiles.

Our conclusion if therefore that, \emph{for both NFW and Burkert profiles the dark--matter contributions to rotation velocities exhibit steep rises followed by shallow declines, and prolate halos yield steeper rises and shallower declines than the standard spherical ones}. This behavior complies to the picture that we have advocated at the beginning of the chapter.
\section{SPARC data analysis and interpretation of the results}\label{C4S2}
Before starting the quantitative analysis, it is useful to introduce a different parametrization of velocities and density profiles, which rests on the definition of the virial mass of the halo:
\begin{equation}
M \ = \ \int d^3r \ \rho(\vec{r}) \ = \ \frac{4\pi}{q}\int_0^{r_c} dr \ r^2\rho(r) \ .
\end{equation}
Here we have defined $r_c$, in an admittedly arbitrary but reasonable manner, demanding that the average DM density within the ellipsoid be 200 times larger than the critical density of the Universe~\cite{White:2000jv}:
\begin{equation}\label{Mcr}
M \ = \ 200 \ * \ \frac{4}{3}\pi\frac{\rho_cr_c^3}{q} \ = \ 100\frac{H^2}{qG}r_c^3 \ ,
\end{equation}
where we have taken the Hubble parameter $H$ to be $0.072$ km/s/kpc. For the NFW and Burkert profiles, the halo masses are
\begin{equation}
\begin{gathered}
M_{NFW} \ = \ \frac{4\pi\rho_0r_0^3}{q} \left[\ln(1+\frac{r_c}{r_0})-\frac{r_c}{r_c+r_0}\right] \ , \\
M_B \ = \ \frac{\pi\rho_0r_0^3}{q}\left[\ln(1+\frac{r_c^2}{r_0^2})+2\ln(1+\frac{r_c}{r_0})-2\arctan(\frac{r_c}{r_0})\right] \ .
\end{gathered}
\end{equation}
Equating these expressions to~\eqref{Mcr} gives
\begin{equation}\label{NFW2}
\rho_0 \ = \ \frac{25H^2}{\pi G}\frac{C^3}{\ln(1+C)-\frac{C}{1+C}}
\end{equation}
for the NFW case, and
\begin{equation}\label{B2}
\rho_0 \ = \ \frac{100H^2}{\pi G}\frac{C^3}{\ln(1+C^2)+2\ln(1+C)-2\arctan(C)}
\end{equation}
for the Burkert case, where the parameter
\begin{equation}
C \ = \ \frac{r_c}{r_0}
\end{equation}
is known as concentration. For $q=1$, one can easily plug this expression into~\eqref{Vsph} to obtain that the rotation velocity at $r_c$ is given by
\begin{equation}
v_c \ = \ 10Hr_0C \ 
\end{equation}
for \emph{any} spherical DM profile. We can now rewrite~\eqref{Vsph} in terms of the new parameters, $C$ and $v_c$:
\begin{equation}
\label{NFWB}
\begin{gathered}
v^2_{NFW}(r) \ = \ v_c^2 \frac{v_c}{10Hr}\frac{\ln(1+10HCv^{-1}_cr)-10HCv^{-1}_cr(1+10HCv^{-1}_cr)^{-1}}{\ln(1+C)-C(1+C)^{-1}} \ , \\
\\
v^2_B(r) \ = \ v_c^2 \frac{v_c}{10Hr}\frac{\ln(1+(10HCv^{-1}_cr)^2)+2\ln(1+10HCv^{-1}_cr)-2\arctan(10HCv^{-1}_cr)}{\ln(1+C^2)+2\ln(1+C)-2\arctan(C)} \ .
\end{gathered}
\end{equation}
Likewise, for very small values of $q$ close to zero
\begin{equation}
v^2_{NFW}(r) \ \approx \ v_c^2\left(\frac{1+C}{C}\right)\left(\frac{r}{r+r_0}\right) \ ,
\end{equation}
with
\begin{equation}
r_0 \ = \ \frac{v_c}{10HC^2}\sqrt{(1+C)\ln(1+C)-C} \ ,
\end{equation}
and
\begin{equation}
v_B^2(r) \ \approx \ v_c^2\left(\frac{\ln\left(1+\frac{r^2}{r_0^2}\right)-2\ln\left(1+\frac{r}{r_0}\right)+2\arctan\left(\frac{r}{r_0}\right)}{\ln\left(1+C^2\right)-2\ln\left(1+C\right)+2\arctan\left(C\right)}\right) \ ,
\end{equation}
with
\begin{equation}
r_0 \ = \ \frac{v_c}{10H}\left(\frac{\ln(1+C^2)+2\ln(1+C)-2\arctan(C)}{\ln(1+C^2)-2\ln(1+C)+2\arctan(C)}\right)^{1/2} \ .
\end{equation}
For generic values of $q$, the expression for rotation velocity would be more complicated,
\begin{equation}
\label{vdef}
v^2(r,v_c,C,q) \ = \ v_c^2\frac{F\left(\sqrt{2G\rho_0F(C,q)}v_c^{-1}r,q\right)}{F(C,q)} \ ,
\end{equation}
where $F$ is the function
\begin{equation}
\begin{gathered}
F(x,q) \ = \ q^{-1} \int_0^\infty d\xi \ \xi^2\bar{\rho} (\xi) \ \int^{\pi/2}_0 \frac{d\theta\, \cos{\theta}}{\sqrt{(x-\xi\cos\theta)^2+(\xi/q)^2\sin^2\theta}}*\\
\left[F\left(\pi \ \Big| u\right) \ - \ \frac{\xi^2\cos^2\theta+(\xi/q)^2\sin^2\theta-x^2}{(x+\xi\cos\theta)^2+(\xi/q)^2\sin^2\theta}\,E\left(\pi\Big|u\right)\right]    \ ,
\end{gathered}
\end{equation}
with
\begin{equation}
u \ = \ -\frac{4x\xi\cos\theta}{(x-\xi\cos\theta)^2+(\xi/q)^2\sin^2\theta} \ , \qquad \bar{\rho}(x) \ = \ \frac{\rho(xr_0)}{\rho_0} \ ,
\end{equation}
and $\rho_0$ is either~\eqref{NFW2} or~\eqref{B2}, depending on the choice of profile.

In order to select a suitable sample of galaxies from the SPARC database, we have resorted to the following criteria:
\begin{itemize}
  \item The galaxy should have the quality flag $Q=1${, which means high quality HI (the 21-cm line from neutral atomic hydrogen) or hybrid HI/H$\alpha$ (the strongest emission line of ionized hydrogen) data}~\cite{Lelli:2016zqa};
  \item The galaxy's inclination should be equal to or larger than 30${}^o$~\cite{Lelli:2016zqa};
  \item The number of data points should be equal to or larger than 10. This requirement is admittedly arbitrary, but if the length of the curve is too small, identifying the ``flattening`` effect we are after and discriminating among competing models becomes very difficult. A similar choice was made in an earlier work~\cite{Bondarenko:2020mpf};
\end{itemize}
These criteria leave 84 galaxies out of 175, on which the deformations we intend to explore are expected to lead to more sizable effects, if they are relevant at all to them.

We then performed a five--parameter fit with the function
\begin{equation}
\small
\label{Vstring}
v(r; Y_D,Y_B,v_c,C,\mu)=\sqrt{Y_Dv^2_D(r)+Y_Bv_B^2(r)+v_G(r)|v_G(r)|+v^2_{DM}(r,v_c,C)+2G\mu} \ ,
\end{equation} 
using the Python \textit{SciPy.Optimize.Curve\textunderscore Fit} package. The five parameters are the mass-to-light ratios $Y_D$ and $Y_B$, bounded from below at 0.1, the two parameters $v_c$ and $C$ of the NFW and Burkert profiles of eq.~\eqref{NFWB}, and finally the string tension parameter $\mu$. The disk, bulge, and gas contributions $v_{D,B,G}$ are taken from the SPARC database; notice that the absolute value is needed for the gas contribution, because it can become negative at small radii if the gas distribution is significantly depressed in the innermost regions, so that the gravitational pull from outwards is stronger than from inwards~\cite{Lelli:2016zqa}. The galaxy UGC 01281 has particularly large negative values of $v_G$ at small radii, rendering the fit unstable; for this reason, we excluded it from the sample, confining our attention to a total of 83 objects. Given that each object was fitted by two profiles (NFW and Burkert), this means 166 models in total, which is indeed a large number.

Then, taking the fit results as a starting point, we performed an MCMC analysis on them resorting to the Python package \textit{emcee}~\cite{ForemanMackey:2012ig}. As in~\cite{Li:2020iib}, we imposed flat priors on $V_c$ and $C$, with $10<V_c<500$ and $0<C<1000$. We also included two additional parameters, namely the inclination of the galaxy $i$ and the distance to the galaxy $D$, and imposed Gaussian priors on them, with their mean values and standard deviations taken from the SPARC database. The former affects the observed velocities and their errors as
\begin{equation}\label{Inc}
v_{obs}' \ = \ \frac{sin(i)}{sin(i')}\,v_{obs} \ , \ \delta v_{obs}' \ = \ \frac{sin(i)}{sin(i')}\,\delta v_{obs} \ ,   
\end{equation}
while the latter impacts the disk, gas, and bulge components:
\begin{equation}
v'_{d,b,g} \ = \ \sqrt{\frac{D'}{D}}\, v_{d,b,g} \ .   
\end{equation}
We chose our posterior probability to be proportional to $e^{-L}$, with
\begin{equation}
\begin{gathered}
L \ = \ \frac{1}{2}\chi^2 \ + \ \frac{1}{2}\left(\frac{\log_{10}(Y_D)-log_{10}(0.5)}{0.1}\right)^2 \ + \\\
\frac{1}{2}\left(\frac{\log_{10}(Y_B)-log_{10}(0.7)}{0.1}\right)^2 \ + \ \frac{1}{2}\left(\frac{D-D_0}{\delta D}\right)^2 \ + \ \frac{1}{2}\left(\frac{i-i_0}{\delta i}\right)^2 \ ,
\end{gathered}
\end{equation}
with $\chi$-square given by
\begin{equation}
\chi^2 \ = \ \sum_i\left(\frac{v(x_i)-v_{obs}(x_i)}{\delta v_{obs}(x_i)}\right)^2 \ ,\\
\end{equation}
and
\begin{equation}
\begin{gathered}
v(r; Y_D,Y_B,v_c,C,i,D)=\frac{sin(i)}{sin(i_0)}*\\
\sqrt{(D/D_0)\left(Y_Dv^2_D(r)+Y_Bv_B^2(r)+v_G(r)|v_G(r)|\right)+v^2_{DM}(r,v_c,C)} \ 
\end{gathered}
\end{equation}
(the prefactor $\frac{sin(i)}{sin(i_0)}$ is due to~\eqref{Inc}; we multiplied both the numerator and the denominator in $\chi^2$ by $\frac{sin(i_0)}{sin(i)}$).

The first term in $L$ is the likelihood function, and all the rest are priors (we omit the normalization factor, since it is irrelevant for our purpose). Namely, as in~\cite{Li:2020iib}, we imposed Gaussian priors on $D$ and $i$, with the mean values and errors given by the SPARC database, and lognormal priors on $Y_D$ and $Y_B$, with mean values of $0.5$ and $0.7$ and error values of $0.1$. The third term, which has to do with bulge, is omitted for bulgeless galaxies.
\begin{figure}[ht]
\centering
\begin{adjustbox}{width={\textwidth},totalheight={\textheight},keepaspectratio}
\begin{tabular}{cc}
		\includegraphics[width=70mm]{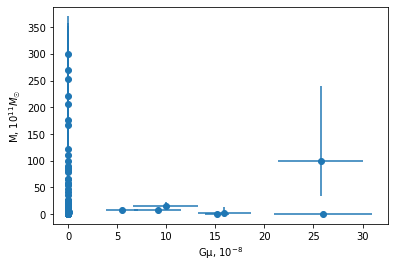} &
		\includegraphics[width=70mm]{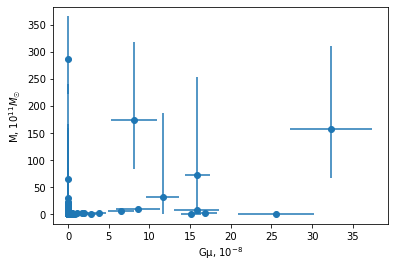} \\
	\end{tabular}
\end{adjustbox}
	\caption{Left panel: string tension versus dark halo mass for the NFW profiles; right panel: string tension versus dark halo mass for the Burkert profiles. In the NFW model, there is no obvious correlation between the presence of a string and larger halo masses, while in the Burkert model, one can observe some degree of correlation. The asymmetric error bars on the masses are due to the fact that the masses are proportional to $v_c^3$, and $(v_c-\delta v_c)^3$ and $(v_c+\delta v_c)^3$ are not equidistant from $v_c^3$.}
		\label{MassString}
\end{figure}
\begin{figure}[ht]
\centering
\begin{adjustbox}{width={\textwidth},totalheight={\textheight},keepaspectratio}
\begin{tabular}{cc}
		\includegraphics[width=75mm]{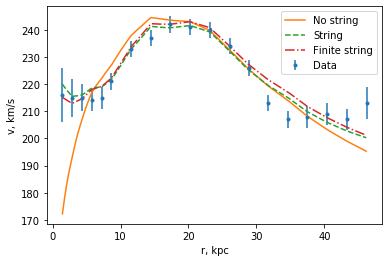} &
		\includegraphics[width=75mm]{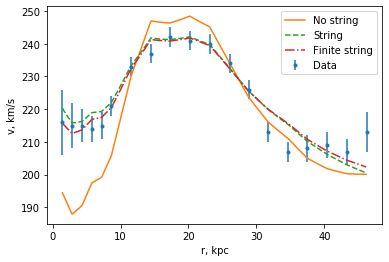} \\
	\end{tabular}
\end{adjustbox}
	\caption{Rotation curves for the galaxy NGC 5371: data points (blue dots with error bars), fit with a DM halo (solid orange line), a DM halo plus an infinite string-like object at the origin (dashed green line), and a DM halo plus a finite string-like object of length 200 kpc (dash-dotted red line). Left panel: NFW profile; right panel: Burkert profile.}
		\label{NGC5371}
\end{figure}
\begin{figure}[ht]
\centering
\begin{adjustbox}{width={\textwidth},totalheight={\textheight},keepaspectratio}
\begin{tabular}{cc}
		\includegraphics[width=75mm]{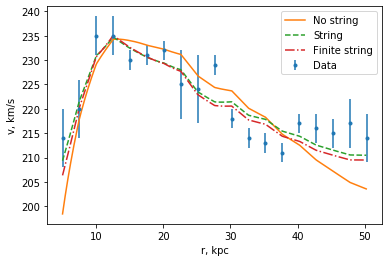} &
		\includegraphics[width=75mm]{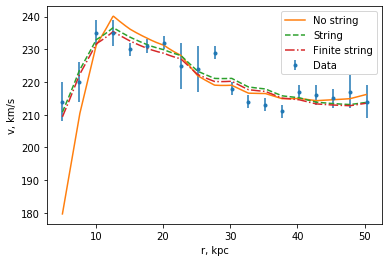} \\
	\end{tabular}
\end{adjustbox}
	\caption{Rotation curves for the galaxy NGC 5907: data points (blue dots with error bars), fit with a DM halo (solid orange line), a DM halo plus an infinite string-like object at the origin (dashed green line), and a DM halo plus a finite string-like object of length 200 kpc (dash-dotted red line). Left panel: NFW profile; right panel: Burkert profile.}
		\label{NGC5907}
\end{figure}

Following~\cite{Li:2020iib}, we initialized the MCMC chains with 200 random walkers and ran 500 burn-in iterations, before resetting the sampler and running another 2000 iterations. For fits that have a minimum at $G\mu=0$, where $G \mu$ is the tension term in eq.~\eqref{Vstring}, we ran only a 6-parameter MCMC ($v_c, C, Y_D, Y_B, D$, and $i$), while for those that have a minimum at $G\mu>0$ (69 in total), we also performed a 7-parameter MCMC with $G\mu$ as the seventh parameter. Then, using the obtained values of $L$, we computed the Bayesian and Akaike information criteria (BIC and AIC) to compare the evidence for competing models. BIC is given by
\begin{equation}
BIC \ = \ k\ln(n)+2L \ , 
\end{equation}
where $k$ is the number of parameters (6 and 7 for bulged and bulgeless galaxies, respectively, when the filament is present, and 5 and 6, respectively, when the filament is absent), and $n$ is the number of data points. Likewise, AIC is defined as 
\begin{equation}
AIC \ = \ 2k+2L \ .
\end{equation}
For cases when non--zero values of $G\mu$ yielded improvement of BIC and/or AIC, we also considered a more physically realistic situation when the filament at the center of the galaxy has finite length of 200 kpc, which means that the factor of $2G\mu$ in~\eqref{Vstring} was replaced with $2G\mu\left(1+(r/100\right)^2)^{-1/2}$, as per~\ref{Wire_vel}. Based on the result, we selected 23-25 galaxies that have \textit{some} evidence for a ``wire`` at the center (values of fit parameters are given in Tables \ref{tab:table1},~\ref{tab:table2},~\ref{tab:table3}), and split them into three groups: those with strong evidence, when considerable fit improvement is shown for both NFW and Burkert profiles, implying a model--independent feature (9 galaxies for BIC, 11 for AIC; Tables ~\ref{tab:table4} and~\ref{tab:table8}); those with moderate evidence, where the addition of the filament improves fit quality for the better--fitting profile (6 galaxies for BIC, 7 for AIC; Tables~\ref{tab:table5} and~\ref{tab:table9}); those with scant evidence, where the filament improves the fitting quality for the worse--fitting profile (8 galaxies for BIC, 7 for AIC; Tables~\ref{tab:table6} and~\ref{tab:table10}). The values of BIC and AIC for the galaxies that show~\textit{no} improvement upon the addition of the filament are given in Tables~\ref{tab:table9} and~\ref{tab:table11}.
\begin{figure}[ht]
\centering
	\includegraphics[width=120mm]{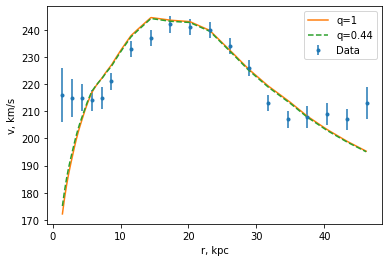}
	\caption{The rotation curve of NGC 5371, fitted by a spherical dark halo (orange solid curve) and by a deformed halo (green dashed curve). The result is given for the NFW profile.}
	\label{NGC5371_q_NFW}
\end{figure}
\begin{figure}[ht]
\centering
	\includegraphics[width=120mm]{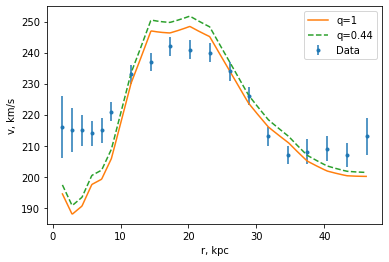}
	\caption{The rotation curve of NGC 5371, fitted by a spherical dark halo (orange solid curve) and by a deformed halo (green dashed curve). The result is given for the Burkert profile.}
	\label{NGC5371_q_B}
\end{figure}

There are numerous physical motivations to consider the presence of elongated, string--like mass distributions at the center of galaxies. First and foremost, black holes can produce relativistic jets comprised of gas~\cite{elongated}, and given the gravitational nature of the effect, dark--matter jets can in principle exist as well. One of our primary examples, the galaxy NGC 5371, is classified as a LINER, which may imply the presence of an active galactic nucleus~\cite{Rush:1993qz}. Alternatively, such an object could be a tidal stream: for instance, another key example NGC 5907 is known to host an extended stellar tidal stream structure~\cite{MartinezDelgado:2008cx,Dokkum}. One more prominent example, NGC 2841, has a polar ring orthogonal to the galaxy plane, which is probably a result of interaction with another galaxy~\cite{Kaneda:2007uw,Afanasiev:1998xi}. Finally, there are exotic candidates like cosmic strings: a mechanism for the migration of cosmic strings to the center of galaxies was proposed in~\cite{Chudnovsky:1986hc,Vilenkin:2018zol}, and interestingly, the values of $G\mu$ obtained in our fits are below the upper constraint on cosmic string tension obtained from Planck observations, which is around $7.8*10^{-7}$~\cite{PLANCK}. Cosmic strings have also been proposed as the cause for the distortions in pulsar signals observed by the NANOGrav collaboration; if true, this result may also lend indirect support to the conjecture of their presence in galaxies~\cite{Blasi:2020mfx,Ellis:2020ena,Arzoumanian:2020vkk}. A filament of this type, observed at the center of Milky Way, has also been conjectured to be either a black hole jet or a cosmic string~\cite{morris}. The Milky Way hosts several more such objects, which have been interpreted as jets produced by shock waves within the regions of intensive star formation~\cite{YusefZadeh:2003qx}.

One could also hypothesize that these ``strings`` originate from intergalactic filaments of the large--scale structure. In this case, there would be a correlation between larger masses of the dark halos and the presence of a ``string`` at the center, with the largest galaxies located at the nodes of the filaments. However, based on our data sample, we cannot reach a clear conclusion on whether or not such correlation exists: there appear to be some indications to this effect in the Burkert model, while they do not emerge in the NFW model (Fig.~\ref{MassString}).

Two most notable examples of galaxies belonging to the first category are NGC 5371 and NGC 5907, for both of which the fit improves considerably due to the presence of the filament (Figs.~\ref{NGC5371} and~\ref{NGC5907}). Though BIC is more stringent than AIC, the conclusions for both criteria are similar in most cases (the few counterexamples, such as NGC 3521 and UGC 03205, yield only marginal improvement in AIC that could be attributed to statistical uncertainty). Likewise, the addition of a cutoff does not have any significant impact on either BIC or AIC for most galaxies.

\begin{figure}[ht]
\centering
	\includegraphics[width=120mm]{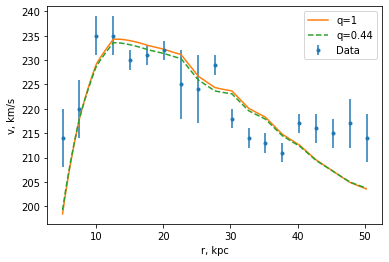}
	\caption{The rotation curve of NGC 5907, fitted by a spherical dark halo (orange solid curve) and by a deformed halo (green dashed curve). The result is given for the NFW profile (for the Burkert profile, a deformed halo is disfavored in comparison to the spherical one).}
	\label{NGC5907_q}
\end{figure}

Taking the aforementioned galaxies NGC 5371 and NGC 5907 as benchmarks, we also fitted them with the alternative model that involves prolate dark matter halos. In this case, the rotation velocity would be
\begin{equation}
\small
\label{Vhalo}
v(r; Y_D,Y_B,v_c,C,\mu)=\sqrt{Y_Dv^2_D(r)+Y_Bv_B^2(r)+v_G(r)|v_G(r)|+v^2_{DM}\left(r,v_c,C,q\right)} \ ,
\end{equation}
with $v^2_{DM}$ given by~\eqref{vdef}. We opted for a flat prior on $q$ in the range from 1/3 to 1, i.e. between the limiting cases of a spherical halo and halo with a major-to-minor axis ratio of 3. The MCMC analysis results, along with the values of BIC and AIC, are given in Table~\ref{tab:table12}. As can be seen from these values and from the examples of the same two galaxies NGC 5371 and NGC 5907 (figs.~\ref{NGC5371_q_NFW},~\ref{NGC5371_q_B}, and~\ref{NGC5907_q}), the deformation of the halo gives only a small correction to the fit, unlike the filament at the center.

Finally, for the same two galaxies, we have explored the possibility that the halo deformation and the filament be simultaneously present. However, this model is not supported by the data analysis: for the NGC 5371 (both NFW and Burkert) and NGC 5907 NFW, the best fit obtains for $q=1$ once one incorporates both parameters $q$ and $G\mu$. For NGC 5907 with Burkert profile, there is a minimum at $q<1$, but the BIC and AIC results in this case are worse than for a filament without the deformation (the values of the fitting parameters, BIC and AIC for this case are given in Table~\ref{tab:table13}).

One should note however that we have worked with simple deformations of two specific, if very popular, density profiles, and confined our analysis to only two galaxies. The prefatory conclusion, based on both theoretical considerations and fit results for these two examples, is that the improvement in the fit quality due to halo deformation is marginal at best. This implies that the rotation curve analysis is not conclusive, and may probably be more useful as an \textit{exclusionary} method to filter out the galaxies that are unlikely to host non--spherical dark halos. In principle, the galaxies that exhibit~\textit{some} preference for prolate shapes, especially when the indication obtains for both NFW and Burkert profiles, can be tested independently by other observations, and in particular by the detection of kinematic stellar groups~\cite{RojasNino:2011xh,Rojas-Nino:2015qna}.
\newpage
\begin{table}
\footnotesize
\begin{tabularx}{\textwidth}{|| X || X | X | X | X | X | X |} \hline
    Galaxy & $V_{200}$, km/s & $C_{200}$ & $Y_d$ & $Y_b$ & D, Mpc & i, $^o$ \\ \hline\hline
    D631-7, NFW & 372.9$\pm$88 & 0.6$\pm$0.45 & 0.33$\pm$0.07 & 0 & 7.62$\pm$0.18 & 53.65$\pm$3.04 \\ \hline
    DDO064, NFW & 224.07$\pm$ 176.15 & 1.69$\pm$2.22 & 0.53$\pm$0.14 & 0 & 8.02$\pm$2 & 60.69$\pm$4.94 \\ \hline
    DDO064, B & 39$\pm$41.51 & 22.43$\pm$3.9 & 0.55$\pm$0.15 & 0 & 7.34$\pm$2.09 & 60.33$\pm$5 \\ \hline
    DDO161, NFW & 99.65$\pm$21.74 & 2.23$\pm$0.82 & 0.36$\pm$0.08 & 0 & 7.94$\pm$1.5 & 69.91$\pm$8.96 \\ \hline
    DDO161, B & 55.19$\pm$3.34 & 10.26$\pm$1.11 & 0.52$\pm$0.12 & 0 & 7.78$\pm$1.58 & 69.67$\pm$9.5 \\ \hline
    ESO079-G014, NFW & 304.46$\pm$ 116.15 & 2.62$\pm$1.34 & 0.63$\pm$0.12 & 0 & 32.61$\pm$5.78 & 79.13$\pm$4.77 \\ \hline
    ESO079-G014, B & 116.59$\pm$4.45 & 20.73$\pm$1.22 & 0.48$\pm$0.11 & 0 & 23.72$\pm$6.24 & 78.27$\pm$4.89 \\ \hline
    ESO116-G012, NFW & 124.76$\pm$ 27.32 & 6.91$\pm$1.52 & 0.49$\pm$0.11 & 0 & 13.47$\pm$3.51 & 73.86$\pm$3 \\ \hline
    ESO116-G012, B & 68.69$\pm$2.67 & 24.68$\pm$2.4 & 0.55$\pm$0.12 & 0 & 13.87$\pm$3.25 & 74.02$\pm$3 \\ \hline
    ESO563-G021, NFW & 454.45$\pm$ 32.71 & 2.38$\pm$0.44 & 0.79$\pm$0.09 & 0 & 70.47$\pm$7.8 & 83.31$\pm$2.81 \\ \hline
    ESO563-G021, B & 191.92$\pm$2.64 & 33.77$\pm$0.77 & 0.32$\pm$0.08 & 0 & 31.78$\pm$9.85 & 82.38$\pm$3.06 \\ \hline
    F568-3, NFW & 314.87$\pm$ 118.35 & 0.71$\pm$0.75 & 0.56$\pm$0.13 & 0 & 84.86$\pm$8.01 & 47.8$\pm$8.42 \\ \hline
    F568-3, B & 100.35$\pm$ 33.65 & 20.91$\pm$7.49 & 0.51$\pm$0.13 & 0 & 81.28$\pm$8.04 & 29.88$\pm$9.28 \\ \hline
    F568-V1, NFW & 92.77$\pm$32.18 & 15.93$\pm$5.55 & 0.54$\pm$0.14 & 0 & 80.71$\pm$7.94 & 36.96$\pm$10.28 \\ \hline
    F568-V1, B & 72.71$\pm$21.53 & 34.15$\pm$9.62 & 0.53$\pm$0.14 & 0 & 80.73$\pm$8.11 & 36.89$\pm$10.23 \\ \hline
    F571-8, NFW & 179.63$\pm$27.7 & 6.74$\pm$0.96 & 0.46$\pm$0.13 & 0 & 3.02$\pm$4.29 & 83.7$\pm$3.87 \\ \hline
    F571-8, B & 104$\pm$4.53 & 22.28$\pm$1.09 & 0.37$\pm$0.1 & 0 & 24.48$\pm$7.89 & 83.41$\pm$4.07 \\ \hline
    F574-1, NFW & 93.93$\pm$15.31 & 7.75$\pm$1.51 & 0.56$\pm$0.15 & 0 & 96.59$\pm$9.58 & 63.5$\pm$10.11 \\ \hline
    F574-1, B & 61.96$\pm$6.69 & 24.3$\pm$2.86 & 0.52$\pm$0.13 & 0 & 96.72$\pm$9.32 & 63.16$\pm$10.23 \\ \hline
    F583-1, NFW & 111.52$\pm$30 & 4.57$\pm$0.95 & 0.54$\pm$0.14 & 0 & 95.71$\pm$9.63 & 60.95$\pm$10.71 \\ \hline
    F583-1, B & 61.86$\pm$7.72 & 18.51$\pm$2 & 0.52$\pm$0.13 & 0 & 95.48$\pm$9.66 & 61.18$\pm$10.51 \\ \hline
    F583-4, NFW & 53.1$\pm$238.29 & 12.53$\pm$10.32 & 0.67$\pm$0.21 & 0 & 61.62$\pm$10.88 & 62.11$\pm$9.49 \\ \hline
    F583-4, B & 55.5$\pm$243.17 & 11.85$\pm$11.18 & 0.68$\pm$0.21 & 0 & 61.65$\pm$10.98 & 61.87$\pm$9.67 \\ \hline
    IC4202, NFW & 153.1$\pm$5.15 & 18.43$\pm$0.59 & 0.85$\pm$0.16 & 0.16$\pm$0.03 & 51.72$\pm$9.4 & 89.33$\pm$0.47 \\ \hline
    IC4202, B & 140.67$\pm$1.06 & 43.87$\pm$0.5 & 0.52$\pm$0.14 & 0.67$\pm$0.19 & 0.67$\pm$1.2 & 89.34$\pm$0.47 \\ \hline
    NGC 0024, NFW & 69.17$\pm$3.7 & 19.01$\pm$2.77 & 0.72$\pm$0.2 & 0 & 7.38$\pm$0.36 & 64.7$\pm$2.98 \\ \hline
    NGC 0024, B & 56.21$\pm$1.62 & 50.94$\pm$2.93 & 0.48$\pm$0.12 & 0 & 7.25$\pm$0.36 & 93.44$\pm$3.02 \\ \hline
    NGC 0100, NFW & 295.85$\pm$ 133.84 & 1.64$\pm$1.21 & 0.49$\pm$0.11 & 0 & 13.66$\pm$3.63 & 88.78$\pm$0.76 \\ \hline
    NGC 0100, B & 63.66$\pm$6.93 & 18.13$\pm$2.35 & 0.52$\pm$0.12 & 0 & 13.52$\pm$3.84 & 88.8$\pm$0.75 \\ \hline
    NGC 0801, NFW & 290.68$\pm$ 115.25 & 1.01$\pm$0.83 & 0.62$\pm$0.06 & 0 & 82.66$\pm$7.28 & 80.05$\pm$0.98 \\ \hline
    NGC 0801, B & 169.34$\pm$ 18.81 & 5.9$\pm$0.69 & 0.64$\pm$0.06 & 0 & 83.32$\pm$7.36 & 80.02$\pm$1 \\ \hline
    \end{tabularx}
    \end{table}
    \newpage
    \begin{table}
    \footnotesize
    \begin{tabularx}{\textwidth}{|| X || X | X | X | X | X | X |} \hline
    Galaxy & $V_{200}$, km/s & $C_{200}$ & $Y_d$ & $Y_b$ & D, Mpc & i, $^o$ \\ \hline\hline
    NGC 0891, NFW & 127.98$\pm$4.72 & 21.72$\pm$2.7 & 0.21$\pm$0.03 & 0.53$\pm$0.07 & 9.18$\pm$0.5 & 89.32$\pm$0.48 \\ \hline
    NGC 0891, B & 112.02$\pm$2.08 & 34.5$\pm$2.75 & 0.23$\pm$0.03 & 0.63$\pm$0.07 & 9.35$\pm$0.49 & 89.33$\pm$0.47 \\ \hline
    NGC 1003, NFW & 114.94$\pm$6.6 & 3.9$\pm$0.49 & 0.65$\pm$0.11 & 0 & 12.03$\pm$2.18 & 66.88$\pm$4.97 \\ \hline
    NGC 1003, B & 92.2$\pm$3.32 & 9.13$\pm$0.5 & 0.74$\pm$0.12 & 0 & 14.89$\pm$2.25 & 68.04$\pm$4.88 \\ \hline
    NGC 1090, NFW & 116.54$\pm$5.59 & 9.8$\pm$1.98 & 0.52$\pm$0.11 & 0 & 30.77$\pm$6.98 & 63.43$\pm$2.95 \\ \hline
    NGC 1090, B & 99$\pm$2.8 & 29.58$\pm$2.56 & 0.37$\pm$0.1 & 0 & 23.7$\pm$8.54 & 63.24$\pm$3.01 \\ \hline
    NGC 2403, NFW & 96.59$\pm$3.54 & 17.73$\pm$0.91 & 0.4$\pm$0.05 & 0 & 2.78$\pm$0.16 & 53.59$\pm$3.04 \\ \hline
    NGC 2403, B & 85.76$\pm$2.51 & 20.3$\pm$0.58 & 0.9$\pm$0.7 & 0 & 2.98$\pm$0.15 & 59.02$\pm$3.02 \\ \hline
    NGC 2841, NFW & 228.6$\pm$12.42 & 6.16$\pm$0.85 & 0.97$\pm$0.1 & 0.86$\pm$0.09 & 15.4$\pm$1.19 & 79.87$\pm$6.06 \\ \hline
    NGC 2841, B & 196.23$\pm$5.92 & 11.76$\pm$0.7 & 1.11$\pm$0.11 & 0.87$\pm$0.09 & 15.73$\pm$1.18 & 80.61$\pm$5.63 \\ \hline
    NGC 2903, NFW & 108.42$\pm$2.53 & 30.45$\pm$2.68 & 0.43$\pm$0.11 & 0 & 3.49$\pm$0.98 & 65.3$\pm$3.08 \\ \hline
    NGC 2903, B & 101.12$\pm$2.4 & 33.34$\pm$2.63 & 0.46$\pm$0.09 & 0 & 5.17$\pm$1.11 & 65.51$\pm$3.01 \\ \hline
    NGC 2955, NFW & 188.02$\pm$ 21.95 & 12.83$\pm$4.02 & 0.32$\pm$0.06 & 0.91$\pm$0.16 & 92.85$\pm$9.46 & 47.34$\pm$5.88 \\ \hline
    NGC 2955, B & 159.43$\pm$ 13.18 & 19.86$\pm$2.59 & 0.34$\pm$0.07 & 0.98$\pm$0.16 & 94.23$\pm$9.23 & 50.74$\pm$5.47 \\ \hline
    NGC 2998, NFW & 136.78$\pm$6.7 & 14.67$\pm$3.86 & 0.42$\pm$0.12 & 0 & 65.11$\pm$10.3 & 57.81$\pm$2 \\ \hline
    NGC 2998, B & 141.72$\pm$9.22 & 9.1$\pm$1.24 & 0.73$\pm$0.1 & 0 & 78.03$\pm$8.64 & 58.6$\pm$1.94 \\ \hline
    NGC 3109, NFW & 428.44$\pm$ 50.86 & 0.48$\pm$0.23 & 0.43$\pm$0.1 & 0 & 1.38$\pm$0.07 & 74.03$\pm$4.45 \\ \hline
    NGC 3109, B & 56.49$\pm$3.18 & 16.65$\pm$0.75 & 0.53$\pm$0.14 & 0 & 1.33$\pm$0.07 & 69.67$\pm$4.97 \\ \hline
    NGC 3198, NFW & 110.52$\pm$2.44 & 10.46$\pm$0.8 & 0.46$\pm$0.06 & 0 & 13.58$\pm$1.32 & 72.72$\pm$3.09 \\ \hline
    NGC 3198, B & 99.94$\pm$2.03 & 15.04$\pm$1.12 & 0.66$\pm$0.07 & 0 & 14.61$\pm$1.28 & 73.56$\pm$2.89 \\ \hline
    NGC 3521, NFW & 325.63$\pm$ 115.27 & 3.57$\pm$2.14 & 0.51$\pm$0.1 & 0 & 8.13$\pm$1.56 & 75.03$\pm$4.98 \\ \hline
    NGC 3521, B & 176.92$\pm$ 85.74 & 14.3$\pm$2.4 & 0.53$\pm$0.11 & 0 & 8.55$\pm$1.59 & 75.06$\pm$4.95 \\ \hline
    NGC 3741, NFW & 79.66$\pm$30.24 & 3.63$\pm$0.96 & 0.52$\pm$0.12 & 0 & 3.21$\pm$0.17 & 69.91$\pm$3.98 \\ \hline
    NGC 3741, B & 36.36$\pm$1.77 & 17.43$\pm$0.93 & 0.8$\pm$0.18 & 0 & 3.29$\pm$0.17 & 71.68$\pm$3.75 \\ \hline
    NGC 3893, NFW & 116.7$\pm$15.31 & 16.64$\pm$5.13 & 0.43$\pm$0.08 & 0 & 16.55$\pm$2.36 & 48.41$\pm$1.98 \\ \hline
    NGC 3893, B & 94.83$\pm$6.67 & 31.94$\pm$6.83 & 0.46$\pm$0.08 & 0 & 17$\pm$2.31 & 48.56$\pm$2.01 \\ \hline
    NGC 3917, NFW & 273.87$\pm$ 144.01 & 1.42$\pm$1.31 & 0.86$\pm$0.11 & 0 & 20.88$\pm$2.16 & 79.17$\pm$2 \\ \hline
    NGC 3917, B & 81.24$\pm$2.92 & 23.89$\pm$1.39 & 0.47$\pm$0.1 & 0 & 17.01$\pm$2.43 & 78.89$\pm$1.98 \\ \hline
    NGC 3972, NFW & 235.87$\pm$ 148.52 & 4.1$\pm$2.41 & 0.5$\pm$0.11 & 0 & 17.7$\pm$2.48 & 76.98$\pm$1 \\ \hline
    NGC 3972, B & 74.76$\pm$5.46 & 26.74$\pm$2.64 & 0.48$\pm$0.11 & 0 & 17.43$\pm$2.47 & 77$\pm$1.01 \\ \hline
    \end{tabularx}
    \end{table}
    \newpage
    \begin{table}
    \footnotesize
    \begin{tabularx}{\textwidth}{|| X || X | X | X | X | X | X |} \hline
    Galaxy & $V_{200}$, km/s & $C_{200}$ & $Y_d$ & $Y_b$ & D, Mpc & i, $^o$ \\ \hline\hline
    NGC 4088, NFW & 272.96$\pm$ 144.36 & 2.06$\pm$2.01 & 0.42$\pm$0.07 & 0 & 16.18$\pm$2.24 & 68.71$\pm$1.98 \\ \hline
    NGC 4088, B & 271.66$\pm$ 149.94 & 7.6$\pm$1.67 & 0.46$\pm$0.07 & 0 & 17.23$\pm$2.14 & 68.84$\pm$1.98 \\ \hline
    NGC 4100, NFW & 100.81$\pm$3.93 & 19.33$\pm$3.23 & 0.52$\pm$0.08 & 0 & 17.82$\pm$2.3 & 73.01$\pm$2 \\ \hline
    NGC 4100, B & 91.03$\pm$2.19 & 37.21$\pm$3.41 & 0.45$\pm$0.08 & 0 & 16.83$\pm$2.33 & 72.81$\pm$1.95 \\ \hline
    NGC 4157, NFW & 260.98$\pm$ 143.5 & 2.6$\pm$2.05 & 0.5$\pm$0.07 & 0.67$\pm$0.16 & 17.06$\pm$2.18 & 81.68$\pm$3.08 \\ \hline
    NGC 4157, B & 152.63$\pm$ 49.93 & 10.16$\pm$2.32 & 0.54$\pm$0.08 & 0.68$\pm$0.16 & 17.78$\pm$2.17 & 81.98$\pm$3 \\ \hline
    NGC 4183, NFW & 75.09$\pm$3.89 & 11.97$\pm$1.79 & 0.55$\pm$0.13 & 0 & 17.97$\pm$2.36 & 81.97$\pm$1.98 \\ \hline
    NGC 4183, B & 63.71$\pm$2.07 & 25.76$\pm$2.13 & 0.5$\pm$0.12 & 0 & 17.56$\pm$2.44 & 81.91$\pm$2 \\ \hline
    NGC 4217, NFW & 163.81$\pm$37.2 & 8.68$\pm$2.73 & 0.88$\pm$0.23 & 0.26$\pm$0.05 & 13.8$\pm$2.39 & 85.92$\pm$1.92 \\ \hline
    NGC 4217, B & 105.18$\pm$3.56 & 31.8$\pm$3.5 & 0.57$\pm$0.15 & 0.34$\pm$0.09 & 9.29$\pm$2.66 & 85.85$\pm$1.93 \\ \hline
    NGC 4559, NFW & 94.78$\pm$7.03 & 9.5$\pm$2.23 & 0.46$\pm$0.11 & 0 & 6.1$\pm$2.26 & 66.97$\pm$1.01 \\ \hline
    NGC 4559, B & 80.05$\pm$5.06 & 14.62$\pm$2.44 & 0.52$\pm$0.11 & 0 & 9.16$\pm$1.9 & 67$\pm$1 \\ \hline
    NGC 5005, NFW & 243.04$\pm$ 157.17 & 6.76$\pm$6.28 & 0.51$\pm$0.1 & 0.56$\pm$0.09 & 15.98$\pm$1.37 & 67.6$\pm$2 \\ \hline
    NGC 5005, B & 174.6$\pm$ 192.93 & 18.18$\pm$13.19 & 0.55$\pm$0.09 & 0.58$\pm$0.09 & 16.38$\pm$1.33 & 67.77$\pm$1.95 \\ \hline
    NGC 5033, NFW & 131.01$\pm$1.74 & 22.12$\pm$1.87 & 0.4$\pm$0.09 & 0.54$\pm$0.13 & 11.41$\pm$2.63 & 65.95$\pm$1 \\ \hline
    NGC 5033, B & 122.76$\pm$1.48 & 26.38$\pm$1.67 & 0.46$\pm$0.1 & 0.52$\pm$0.11 & 15.14$\pm$2.57 & 65.97$\pm$1 \\ \hline
    NGC 5055, NFW & 120.5$\pm$6.1 & 13.19$\pm$1.12 & 0.28$\pm$0.03 & 0 & 9.96$\pm$0.29 & 59.03$\pm$5 \\ \hline
    NGC 5055, B & 115$\pm$5.54 & 14.98$\pm$0.89 & 0.39$\pm$0.05 & 0 & 9.96$\pm$0.3 & 59.35$\pm$4.84 \\ \hline
    NGC 5371, NFW & 121.35$\pm$3.81 & 27.32$\pm$1.26 & 0.37$\pm$0.09 & 0 & 27.16$\pm$6.57 & 52.48$\pm$2.04 \\ \hline
    NGC 5371, B & 430.04$\pm$ 50.88 & 3.48$\pm$0.26 & 0.84$\pm$0.15 & 0 & 30.77$\pm$4.98 & 52.82$\pm$2 \\ \hline
    NGC 5585, NFW & 98.78$\pm$8.41 & 6.7$\pm$0.86 & 0.41$\pm$0.11 & 0 & 3.89$\pm$1.57 & 50.51$\pm$2.03 \\ \hline
    NGC 5585, B & 64.28$\pm$2.55 & 16.95$\pm$0.74 & 0.54$\pm$0.11 & 0 & 7.93$\pm$1.43 & 51.09$\pm$1.92 \\ \hline
    NGC 5907, NFW & 133.61$\pm$1.33 & 23.7$\pm$1.93 & 0.28$\pm$0.05 & 0 & 16.61$\pm$0.9 & 87.53$\pm$1.53 \\ \hline
    NGC 5907, B & 194.36$\pm$21 & 5.95$\pm$0.45 & 0.96$\pm$0.05 & 0 &17.78$\pm$0.86 & 87.63$\pm$1.47 \\ \hline 
    NGC 5985, NFW & 176.2$\pm$4.26 & 29.59$\pm$1.04 & 0.45$\pm$0.12 & 0.71$\pm$0.18 & 11.18$\pm$8.41 & 59.61$\pm$2.01 \\ \hline
    NGC 5985, B & 155.17$\pm$3.69 & 48.06$\pm$1.81 & 0.31$\pm$0.09 & 0.87$\pm$0.23 & 24.66$\pm$10.42 & 59.61$\pm$2.07 \\ \hline
    NGC 6195, NFW & 335.59$\pm$ 98.84 & 2.27$\pm$1.34 & 0.39$\pm$0.06 & 0.77$\pm$0.09 & 123.7$\pm$11.78 & 60.51$\pm$4.72 \\ \hline
    NGC 6195, B & 182.02$\pm$ 35.68 & 9.92$\pm$2.06 & 0.42$\pm$0.08 & 0.79$\pm$0.1 & 126.13$\pm$11.68 & 61.12$\pm$4.72 \\ \hline
    NGC 6503, NFW & 80.77$\pm$1.55 & 14.44$\pm$0.97 & 0.4$\pm$0.04 & 0 & 6.15$\pm$0.3 & 73.67$\pm$2.04 \\ \hline
    NGC 6503, B & 74.01$\pm$1.01 & 20.6$\pm$0.77 & 0.57$\pm$0.04 & 0 & 6.25$\pm$0.3 & 73.91$\pm$2.01 \\ \hline
    \end{tabularx}
    \end{table}
    \newpage
    \begin{table}
    \footnotesize
    \begin{tabularx}{\textwidth}{|| X || X | X | X | X | X | X |} \hline
    Galaxy & $V_{200}$, km/s & $C_{200}$ & $Y_d$ & $Y_b$ & D, Mpc & i, $^o$ \\ \hline\hline
    NGC 6674, NFW & 409.98$\pm$62.3 & 0.56$\pm$0.37 & 0.82$\pm$0.12 & 0.88$\pm$0.22 & 65.14$\pm$7.78 & 60.61$\pm$4.5 \\ \hline
    NGC 6674, B & 258.74$\pm$ 35.11 & 4.25$\pm$0.39 & 0.89$\pm$0.14 & 0.76$\pm$0.18 & 68.39$\pm$8.27 & 61.24$\pm$5.14 \\ \hline
    NGC 6946, NFW & 117.59$\pm$ 10.15 & 10.21$\pm$2.09 & 0.5$\pm$0.08 & 0.56$\pm$0.09 & 5.05$\pm$0.93 & 37.63$\pm$1.93 \\ \hline
    NGC 6946, B & 90.83$\pm$6.65 & 19.7$\pm$5.06 & 0.57$\pm$0.09 & 0.65$\pm$0.11 & 5.06$\pm$0.88 & 37.52$\pm$1.96 \\ \hline
    NGC 7331, NFW & 199.66$\pm$ 13.59 & 6.7$\pm$0.92 & 0.42$\pm$0.05 & 0.62$\pm$0.15 & 12.7$\pm$1.34 & 74.59$\pm$2 \\ \hline
    NGC 7331, B & 159.65$\pm$4.55 & 14.34$\pm$0.72 & 0.48$\pm$0.06 & 0.61$\pm$0.14 & 12.83$\pm$1.31 & 74.55$\pm$2.04 \\ \hline
    NGC 7793, NFW & 78.52$\pm$51.24 & 6.75$\pm$4.21 & 0.52$\pm$0.1 & 0 & 3.65$\pm$0.18 & 54.18$\pm$7.34 \\ \hline
    NGC 7793, B & 47.57$\pm$6.78 & 19.34$\pm$4.55 & 0.58$\pm$0.1 & 0 & 3.65$\pm$0.18 & 55.38$\pm$6.89 \\ \hline
    NGC 7814, NFW & 116.22$\pm$3.57 & 29.14$\pm$2.82 & 0.54$\pm$0.15 & 0.63$\pm$0.05 & 14.33$\pm$0.66 & 89.31$\pm$0.48 \\ \hline
    NGC 7814, B & 136.08$\pm$9.18 & 16.81$\pm$2.87 & 0.54$\pm$0.14 & 0.56$\pm$0.05 & 14.25$\pm$0.66 & 89.35$\pm$0.46 \\ \hline
    UGC 00128, NFW & 109.63$\pm$10.5 & 8.88$\pm$0.81 & 0.48$\pm$0.11 & 0 & 61.64$\pm$9 & 52.28$\pm$7.28 \\ \hline
    UGC 00128, B & 94.64$\pm$8.41 & 16.14$\pm$1.18 & 0.35$\pm$0.08 & 0 & 64.49$\pm$8.76 & 55.41$\pm$7.44 \\ \hline
    UGC 00731, NFW & 58.64$\pm$6.3 & 9.43$\pm$1.23 & 0.54$\pm$0.14 & 0 & 11.52$\pm$3.36 & 56.69$\pm$3 \\ \hline
    UGC 00731, B & 40.8$\pm$2.59 & 25.7$\pm$1.61 & 0.53$\pm$0.14 & 0 & 15.78$\pm$3.17 & 57.61$\pm$2.93 \\ \hline
    UGC 02487, NFW & 204.06$\pm$ 17.34 & 11.82$\pm$5.55 & 0.78$\pm$0.2 & 0.59$\pm$0.13 & 74.03$\pm$9.61 & 38.96$\pm$3.6 \\ \hline
    UGC 02487, B & 182.71$\pm$ 12.33 & 16.39$\pm$5.32 & 0.67$\pm$0.17 & 0.76$\pm$0.16 & 77.45$\pm$9.39 & 41.01$\pm$3.72 \\ \hline
    UGC 02885, NFW & 258.15$\pm$ 36.39 & 4.74$\pm$1.54 & 0.5$\pm$0.1 & 0.97$\pm$0.12 & 83.69$\pm$7.28 & 64.76$\pm$3.88 \\ \hline
    UGC 02885, B & 209.72$\pm$ 14.71 & 10.45$\pm$2 & 0.58$\pm$0.11 & 0.98$\pm$0.12 & 86.21$\pm$7 & 65.65$\pm$3.74 \\ \hline
    UGC 03205, NFW & 193.53$\pm$26.1 & 4.28$\pm$1.2 & 0.62$\pm$0.08 & 0.92$\pm$0.12 & 60.81$\pm$7 & 68.47$\pm$3.74 \\ \hline
    UGC 03205, B & 157.72$\pm$8.93 & 9.56$\pm$0.66 & 0.68$\pm$0.08 & 0.89$\pm$0.12 & 65.07$\pm$7.12 & 68.92$\pm$3.76 \\ \hline
    UGC 03546, NFW & 127.79$\pm$8.31 & 16.06$\pm$2.93 & 0.51$\pm$0.1 & 0.59$\pm$0.12 & 22.87$\pm$4.7 & 52.48$\pm$4.89 \\ \hline
    UGC 03546, B & 112.97$\pm$5.62 & 23.18$\pm$2.55 & 0.56$\pm$0.12 & 0.6$\pm$0.12 & 24.94$\pm$4.66 & 53.14$\pm$4.51 \\ \hline
    UGC 04278, NFW & 380.04$\pm$ 83.06 & 1.42$\pm$0.64 & 0.41$\pm$0.09 & 0 & 9.26$\pm$2.48 & 87.94$\pm$1.43 \\ \hline
    UGC 04278, B & 102.13$\pm$ 78.33 & 14.9$\pm$1.63 & 0.7$\pm$0.17 & 0 & 11.3$\pm$2 & 87.98$\pm$1.4 \\ \hline 
    UGC 05005, NFW & 196.03$\pm$ 139.86 & 1.64$\pm$2.02 & 0.52$\pm$0.13 & 0 & 54.08$\pm$10.51 & 37.83$\pm$9.35 \\ \hline
    UGC 05005, B & 86.79$\pm$24.66 & 10.53$\pm$4.43 & 0.53$\pm$0.14 & 0 & 55.16$\pm$10.64 & 39.49$\pm$10.33 \\ \hline
    \end{tabularx}
    \end{table}
    \newpage
    \begin{table}
    \footnotesize
    \begin{tabularx}{\textwidth}{|| X || X | X | X | X | X | X |} \hline
    Galaxy & $V_{200}$, km/s & $C_{200}$ & $Y_d$ & $Y_b$ & D, Mpc & i, $^o$ \\ \hline\hline
    UGC 05716, NFW & 57.47$\pm$8.46 & 9.07$\pm$1.05 & 0.48$\pm$0.11 & 0 & 22.87$\pm$4.92 & 55.13$\pm$8.82 \\ \hline
    UGC 05716, B & 40.15$\pm$3.2 & 15.36$\pm$1.39 & 0.76$\pm$0.17 & 0 & 30.9$\pm$4.16 & 67.65$\pm$7.3 \\ \hline
    UGC 05721, NFW & 48.6$\pm$3.06 & 28.2$\pm$3.47 & 0.48$\pm$0.11 & 0 & 5.59$\pm$1.79 & 60.12$\pm$5.02 \\ \hline
    UGC 05721, B & 40.1$\pm$2 & 53.09$\pm$5.42 & 0.54$\pm$0.13 & 0 & 6.72$\pm$1.62 & 60.98$\pm$4.88 \\ \hline
    UGC 05750, NFW & 154.5$\pm$ 150.93 & 1.32$\pm$1.55 & 0.57$\pm$0.15 & 0 & 62.01$\pm$11.62 & 65.19$\pm$9.64 \\ \hline
    UGC 05750, B & 63.09$\pm$13.52 & 11.57$\pm$1.93 & 0.53$\pm$0.14 & 0 & 58.98$\pm$11.5 & 62.9$\pm$10.1 \\ \hline
    UGC 06446, NFW & 58.36$\pm$5.74 & 14.51$\pm$2.25 & 0.54$\pm$0.14 & 0 & 12.06$\pm$3.52 & 50.74$\pm$2.98 \\ \hline
    UGC 06446, B & 44.8$\pm$2.96 & 31.84$\pm$3 & 0.55$\pm$0.15 & 0 & 14.15$\pm$3.43 & 51.34$\pm$3 \\ \hline
    UGC 06614, NFW & 263.47$\pm$ 62.82 & 4.51$\pm$2.58 & 0.51$\pm$0.13 & 0.73$\pm$0.18 & 87.88$\pm$8.86 & 28.83$\pm$4.59 \\ \hline
    UGC 06614, B & 179.27$\pm$ 22.54 & 11.43$\pm$3.18 & 0.52$\pm$0.13 & 0.74$\pm$0.18 & 88.38$\pm$8.66 & 31.88$\pm$4.73 \\ \hline
    UGC 06786, NFW & 148.47$\pm$4.06 & 19.79$\pm$1.7 & 0.4$\pm$0.08 & 0.81$\pm$0.16 & 21.09$\pm$4.56 & 63.26$\pm$3 \\ \hline
    UGC 06786, B & 140.21$\pm$3.77 & 20.24$\pm$1.37 & 0.67$\pm$0.1 & 0.74$\pm$0.11 & 35.41$\pm$4.75 & 
    64.49$\pm$2.86 \\ \hline
    UGC 06917, NFW & 104.4$\pm$23.82 & 7.55$\pm$1.86 & 0.51$\pm$0.12 & 0 & 17.57$\pm$2.48 & 55.83$\pm$2 \\ \hline
    UGC 06917, B & 64.2$\pm$3.49 & 24.1$\pm$2.18 & 0.52$\pm$0.13 & 0 & 17.97$\pm$2.43 & 55.89$\pm$1.98 \\ \hline
    UGC 06930, NFW & 80.84$\pm$18.16 & 10.89$\pm$3.56 & 0.53$\pm$0.14 & 0 & 18.01$\pm$2.45 & 31.13$\pm$5.07 \\ \hline
    UGC 06930, B & 64.11$\pm$11.65 & 24.12$\pm$5.33 & 0.53$\pm$0.13 & 0 & 18.08$\pm$2.47 & 31.21$\pm$5 \\ \hline
    UGC 06983, NFW & 79.09$\pm$6.44 & 12.7$\pm$2.06 & 0.52$\pm$0.13 & 0 & 17.76$\pm$2.47 & 48.96$\pm$1 \\ \hline
    UGC 06983, B & 63.51$\pm$2.6 & 27.05$\pm$2.45 & 0.54$\pm$0.14 & 0 & 18.33$\pm$2.41 & 49$\pm$1 \\ \hline 
    UGC 07125, NFW & 55.55$\pm$5.34 & 5.15$\pm$0.92 & 0.52$\pm$0.13 & 0 & 14.28$\pm$4.82 & 88.02$\pm$1.38 \\ \hline
    UGC 07125, B & 43.05$\pm$2.67 & 12.41$\pm$1.32 & 0.53$\pm$0.13 & 0 & 17.42$\pm$ 4.13 & 87.94$\pm$1.45 \\ \hline
    UGC 07151, NFW & 68.2$\pm$38.11 & 6.95$\pm$2.96 & 0.8$\pm$0.17 & 0 & 7$\pm$0.34 & 87.98$\pm$1.41 \\ \hline
    UGC 07151, B & 37.02$\pm$1.73 & 31.77$\pm$2.83 & 0.56$\pm$0.14 & 0 & 6.87$\pm$0.34 & 87.98$\pm$1.42 \\ \hline
    UGC 07323, NFW & 300.09$\pm$ 132.71 & 1.51$\pm$1.16 & 0.48$\pm$0.11 & 0 & 8.43$\pm$2.4 & 47.13$\pm$3 \\ \hline
    UGC 07323, B & 145.34$\pm$ 227.66 & 11.35$\pm$4.5 & 0.61$\pm$0.12 & 0 & 9.85$\pm$1.94 & 47.67$\pm$2.94 \\ \hline
    \end{tabularx}
    \end{table}
    \newpage
    \begin{table}
    \footnotesize
    \begin{tabularx}{\textwidth}{|| X || X | X | X | X | X | X |} \hline
    Galaxy & $V_{200}$, km/s & $C_{200}$ & $Y_d$ & $Y_b$ & D, Mpc & i, $^o$ \\ \hline\hline
    UGC 07399, NFW & 75.98$\pm$6.86 & 17.77$\pm$2.2 & 0.51$\pm$0.13 & 0 & 8.26$\pm$2.49 & 54.69$\pm$3.05 \\ \hline
    UGC 07399, B & 51.32$\pm$2.34 & 45.01$\pm$3.43 & 0.57$\pm$0.15 & 0 & 9.61$\pm$2.41 & 55.21$\pm$2.98 \\ \hline
    UGC 07524, NFW & 83.53$\pm$10.96 & 5.63$\pm$0.73 & 0.49$\pm$0.12 & 0 & 4.71$\pm$0.24 & 45.12$\pm$2.98 \\ \hline
    UGC 07603, NFW & 72.51$\pm$30.11 & 8.57$\pm$2.63 & 0.51$\pm$0.12 & 0 & 5.11$\pm$1.31 & 77.95$\pm$3 \\ \hline
    UGC 07603, B & 37.06$\pm$2.02 & 33.48$\pm$3.14 & 0.51$\pm$0.12 & 0 & 4.6$\pm$1.32 & 77.93$\pm$3.04 \\ \hline
    UGC 08286, NFW & 59.06$\pm$2.37 & 14.78$\pm$1.36 & 0.69$\pm$0.19 & 0 & 6.54$\pm$0.21 & 88.04$\pm$1.39 \\ \hline
    UGC 08286, B & 46$\pm$0.75 & 36.14$\pm$1.4 & 0.45$\pm$0.1 & 0 & 6.49$\pm$0.21 & 88.02$\pm$1.39 \\ \hline
    UGC 08490, NFW & 50.41$\pm$2.45 & 21.58$\pm$2.59 & 0.52$\pm$0.13 & 0 & 4.65$\pm$0.52 & 49.79$\pm$2.98 \\ \hline
    UGC 08490, B & 43.36$\pm$1.86 & 33.53$\pm$5.87 & 0.81$\pm$0.2 & 0 & 5.04$\pm$0.51 & 51.5$\pm$2.95 \\ \hline
    UGC 08550, NFW & 43.59$\pm$4.32 & 12.34$\pm$2.09 & 0.58$\pm$0.15 & 0 & 7.38$\pm$1.78 & 87.96$\pm$1.43 \\ \hline
    UGC 08550, B & 31.6$\pm$1.35 & 31.43$\pm$3.3 & 0.56$\pm$0.16 & 0 & 8.06$\pm$1.85 & 88.01$\pm$1.4 \\ \hline
    UGC 09133, NFW & 174.93$\pm$9.95 & 5.94$\pm$0.45 & 0.7$\pm$0.1 & 0.49$\pm$0.07 & 61.36$\pm$7.79 & 54.39$\pm$4.89 \\ \hline
    UGC 09133, B & 162.21$\pm$8.57 & 8.42$\pm$0.49 & 0.77$\pm$0.12 & 0.42$\pm$0.07 & 67.29$\pm$8.12 & 55.98$\pm$4.35 \\ \hline
    UGC 11455, NFW & 400.63$\pm$ 66.56 & 2.13$\pm$0.75 & 0.52$\pm$0.07 & 0 & 79.59$\pm$9.94 & 89.33$\pm$0.47 \\ \hline
    UGC 11455, B & 161.53$\pm$2.13 & 28.8$\pm$1.34 & 0.32$\pm$0.07 & 0 & 51.07$\pm$11.26 & 89.32$\pm$0.48 \\ \hline
    UGC 11820, NFW & 92.59$\pm$16.58 & 4.98$\pm$1.16 & 0.61$\pm$0.18 & 0 & 17.12$\pm$4.78 & 40.45$\pm$8.3 \\ \hline
    UGC 11820, B & 66.52$\pm$27.16 & 6.07$\pm$2.17 & 1.38$\pm$0.18 & 0 & 24.35$\pm$4.36 & 56.01$\pm$7.46 \\ \hline
    UGC 11914, NFW & 438.47$\pm$ 43.18 & 6.47$\pm$1.23 & 0.36$\pm$0.06 & 1$\pm$0.17 & 16.12$\pm$3.07 & 31.8$\pm$2.92 \\ \hline
    UGC 11914, B & 447.85$\pm$ 37.31 & 18.57$\pm$1.47 & 0.49$\pm$0.08 & 0.87$\pm$0.14 & 17.93$\pm$3.35 & 32.88$\pm$3.18 \\ \hline
    UGC 12632, NFW & 57.49$\pm$6.68 & 9.08$\pm$1.39 & 0.54$\pm$0.14 & 0 & 9.58$\pm$2.87 & 45.6$\pm$3 \\ \hline
    UGC 12632, B & 44.27$\pm$3.43 & 21.69$\pm$1.73 & 0.53$\pm$0.14 & 0 & 10.3$\pm$2.76 & 45.92$\pm$3.01 \\ \hline
        \end{tabularx}
        \end{table}
\newpage
\begin{table}
\footnotesize
\begin{tabularx}{\textwidth}{|| X || X | X | X | X | X | X |} \hline
    Galaxy & $V_{200}$, km/s & $C_{200}$ & $Y_d$ & $Y_b$ & D, Mpc & i, $^o$ \\ \hline\hline
    UGC 12732, NFW & 76.41$\pm$ 14.86 & 7.26$\pm$1.35 & 0.54$\pm$0.14 & 0 & 14.53$\pm$3.73 & 39.49$\pm$5.73 \\ \hline
    UGC 12732, B & 43.89$\pm$ 5.57 & 16.55$\pm$ 2.47 & 0.58$\pm$0.17 & 0 & 19.97$\pm$ 3.11 & 47.69$\pm$ 4.76 \\ \hline
\end{tabularx}
        \caption{The values of fitting parameters with errors for the selected 83 galaxies from the SPARC sample, given a spherical dark matter profile \textit{without} a filament at the center, obtained from the MCMC analysis.}
        \label{tab:table1}
\end{table}
\clearpage
\begin{table}
\footnotesize
\begin{tabularx}{\textwidth}{|| X || X | X | X | X | X | X | X |} \hline
    Galaxy & $V_{200}$, km/s & $C_{200}$ & $Y_d$ & $Y_b$ & D, Mpc & i, $^o$ & $G\mu, 10^{-8}$  \\ \hline\hline
   DDO064, B & 46.64$\pm$ 135.33 & 20.49$\pm$4.41 & 0.55$\pm$0.14 & 0 & 7.38$\pm$2.08 & 60.36$\pm$5.06 & 0.02$\pm$0.01 \\ \hline
   DDO161, B & 56.11$\pm$4.38 & 10.94$\pm$1.35 & 0.48$\pm$0.11 & 0 & 6.78$\pm$1.74 & 65.48$\pm$ 10.06 & 0.06$\pm$0.05 \\ \hline
   ESO079-G014, B & 118.04$\pm$ 4.46 & 20.57$\pm$1.22 & 0.47$\pm$0.11 & 0 & 22.33$\pm$6.37 & 78.1$\pm$4.92 & 0.18$\pm$0.21 \\ \hline
   ESO116-G012, B & 68.92$\pm$2.28 & 25.51$\pm$2.1 & 0.5$\pm$0.12 & 0 & 11.76$\pm$3.44 & 73.78$\pm$3.02 & 0.14$\pm$0.08 \\ \hline
   F583-4, B & 105.87$\pm$ 252.16 & 9.48$\pm$7.1 & 0.63$\pm$0.17 & 0 & 60.29$\pm$ 10.41 & 59.32$\pm$9.4 & 0.13$\pm$0.1 \\ \hline
   NGC 1090, B & 99.23$\pm$ 2.85 & 29.9$\pm$2.45 & 0.38$\pm$0.1 & 0 & 19.68$\pm$8.95 & 63.1$\pm$3.02 & 0.29$\pm$0.29 \\ \hline
   NGC 2403, B & 83.49$\pm$ 2.62 & 21.53$\pm$ 0.67 & 0.8$\pm$0.07 & 0 & 2.93$\pm$0.15 & 57.86$\pm$2.96 & 0.44$\pm$0.07 \\ \hline
   NGC 2841, NFW & 318.86$\pm$ 102.29 & 0.48$\pm$0.67 & 0.85$\pm$0.11 & 0.46$\pm$0.07 & 13.86$\pm$1.34 & 72.97$\pm$8.51 & 25.76$\pm$4.33 \\ \hline
   NGC 2841, B & 366.79$\pm$ 92.74 & 2.46$\pm$0.47 & 0.79$\pm$0.12 & 0.39$\pm$0.06 & 13.33$\pm$ 5.05 & 70.1$\pm$9.22 & 32.35$\pm$5.05 \\ \hline
   NGC 2903, B & 92.81$\pm$2.02 & 79.03$\pm$2.33 & 0.49$\pm$0.13 & 0 & 0.1$\pm$0.18 & 65.58$\pm$ 2.97 & 1.92$\pm$0.51 \\ \hline
   NGC 2955, NFW & 168.35$\pm$ 21.3 & 13.71$\pm$6.18 & 0.41$\pm$0.08 & 0.7$\pm$0.13 & 92.12$\pm$9.67 & 43.76$\pm$5.98 & 9.95$\pm$3.29 \\ \hline
   NGC 2955, B & 141.83$\pm$ 15.19 & 19.05$\pm$4.13 & 0.41$\pm$0.09 & 0.75$\pm$0.15 & 93.37$\pm$9.58 & 47.88$\pm$5.74 & 8.6$\pm$2.65 \\ \hline
   NGC 2998, NFW & 129.46$\pm$ 11.49 & 7.67$\pm$3.25 & 0.47$\pm$0.1 & 0 & 66.34$\pm$9.86 & 57.74$\pm$1.97 & 5.46$\pm$1.63 \\ \hline
   NGC 2998, B & 115.14$\pm$ 9.19 & 10.4$\pm$2.46 & 0.53$\pm$0.1 & 0 & 70.17$\pm$9.49 & 57.88$\pm$2 & 6.5$\pm$1.6 \\ \hline
   NGC 3109, B & 58.3$\pm$3.77 & 16.24$\pm$0.8 & 0.53$\pm$0.13 & 0 & 1.33$\pm$0.07 & 69.56$\pm$5 & 0.01$\pm$0.01 \\ \hline
   NGC 3198, B & 94.51$\pm$2.38 & 16.76$\pm$1.72 & 0.52$\pm$0.08 & 0 & 14.11$\pm$1.29 & 73.06$\pm$3.02 & 1.13$\pm$0.32 \\ \hline
   NGC 3521, NFW & 125.88$\pm$ 67.84 & 10.04$\pm$ 11.06 & 0.46$\pm$0.11 & 0 & 5.72$\pm$1.66 & 74.1$\pm$4.98 & 6.87$\pm$2.34 \\ \hline
   NGC 3521, B & 132.39$\pm$ 57.78 & 15.33$\pm$4.16 & 0.49$\pm$0.11 & 0 & 7.04$\pm$1.68 & 74.22$\pm$5.29 & 5.19$\pm$2.89 \\ \hline
   NGC 3972, B & 77.13$\pm$7.06 & 25.05$\pm$2.9 & 0.47$\pm$0.11 & 0 & 17.2$\pm$2.48 & 76.97$\pm$1.01 & 0.22$\pm$0.25 \\ \hline
   NGC 4088, B & 260.57$\pm$ 154.94 & 7.35$\pm$1.79 & 0.43$\pm$0.07 & 0 & 16.86$\pm$2.25 & 68.74$\pm$1.96 & 1.06$\pm$1.12 \\ \hline
   NGC 4183, B & 61.66$\pm$2.67 & 20.96$\pm$3.14 & 0.5$\pm$0.12 & 0 & 17.43$\pm$2.47 & 81.91$\pm$2.01 & 0.87$\pm$0.44 \\ \hline
   NGC 4559, B & 77.24$\pm$4.52 & 15.5$\pm$2.7 & 0.48$\pm$0.11 & 0 & 7.67$\pm$2.09 & 66.95$\pm$1 & 0.66$\pm$0.49 \\ \hline
   NGC 5005, NFW & 181.72$\pm$ 193.81 & 17.12$\pm$ 10.72 & 0.52$\pm$0.1 & 0.54$\pm$0.09 & 15.88$\pm$ 1.37 & 67.54$\pm$1.93 & 2.65$\pm$3.16 \\ \hline
   NGC 5005, B & 241.66$\pm$ 158.14 & 6.18$\pm$5.88 & 0.49$\pm$0.09 & 0.52$\pm$0.09 & 15.64$\pm$ 1.37 & 67.38$\pm$1.95 & 2.35$\pm$3 \\ \hline
   NGC 5055, NFW & 115.92$\pm$ 6.26 & 14.61$\pm$1.59 & 0.25$\pm$0.03 & 0 & 9.95$\pm$0.3 & 57.5$\pm$4.72 & 1.46$\pm$0.79 \\ \hline
   \end{tabularx}
   \end{table}
\newpage
\begin{table}
\footnotesize
\begin{tabularx}{\textwidth}{|| X || X | X | X | X | X | X | X |} \hline
    Galaxy & $V_{200}$, km/s & $C_{200}$ & $Y_d$ & $Y_b$ & D, Mpc & i, $^o$ & $G\mu, 10^{-8}$ \\ \hline\hline
   NGC 5055, B & 104.44$\pm$ 7.02 & 16.46$\pm$1.54 & 0.34$\pm$0.05 & 0 & 9.9$\pm$0.31 & 55.53$\pm$5 & 3.86$\pm$1.66 \\ \hline
   NGC 5371, NFW & 15.63$\pm$5.95 & 417.54$\pm$ 386.61 & 0.42$\pm$0.08 & 0 & 31.54$\pm$5.97 & 52.52$\pm$2 & 15.16$\pm$1.24 \\ \hline
   NGC 5371, B & 15.43$\pm$5.78 & 406.9$\pm$ 390.98 & 0.42$\pm$0.08 & 0 & 31.8$\pm$6 & 52.54$\pm$1.96 & 15.16$\pm$1.24 \\ \hline 
   NGC 5585, NFW & 98.54$\pm$7.07 & 7.06$\pm$0.75 & 0.44$\pm$0.12 & 0 & 2$\pm$1.6 & 50.49$\pm$2 & 0.06$\pm$0.03 \\ \hline
   NGC 5585, B & 62.3$\pm$2.09 & 19.47$\pm$0.93 & 0.47$\pm$0.11 & 0 & 5.57$\pm$1.42 & 50.56$\pm$1.93 & 0.17$\pm$0.04 \\ \hline
   NGC 5907, NFW & 78.98$\pm$ 84.52 & 1.17$\pm$58.94 & 0.46$\pm$0.09 & 0 & 16.95$\pm$0.89 & 87.48$\pm$1.56 & 15.91$\pm$2.69 \\ \hline
   NGC 5907, B & 283.42$\pm$ 146.09 & 1.63$\pm$0.65 & 0.5$\pm$0.06 & 0 & 17.06$\pm$0.88 & 87.63$\pm$1.49 & 15.89$\pm$1.53 \\ \hline
   NGC 5985, B & 137.75$\pm$ 5.21 & 37.59$\pm$2.68 & 0.44$\pm$0.12 & 0.71$\pm$0.18 & 10.24$\pm$8.79 & 59.57$\pm$2 & 15.79$\pm$2.72 \\ \hline
   NGC 6195, NFW & 311.97$\pm$ 106.97 & 1.96$\pm$1.38 & 0.38$\pm$0.06 & 0.67$\pm$0.11 & 118.61$\pm$ 12.31 & 58.22$\pm$4.95 & 6.32$\pm$5.04 \\ \hline
   NGC 6503, NFW & 76.07$\pm$3.17 & 13.63$\pm$1.15 & 0.36$\pm$0.04 & 0 & 6.11$\pm$0.3 & 73.51$\pm$2.05 & 0.98$\pm$0.73 \\ \hline 
   NGC 6503, B & 59.9$\pm$3.89 & 17.89$\pm$1.34 & 0.35$\pm$0.06 & 0 & 6.12$\pm$0.31 & 73.56$\pm$2.03 & 2.82$\pm$0.64 \\ \hline
   NGC 6674, B & 379.05$\pm$ 83.56 & 2.94$\pm$0.49 & 0.8$\pm$0.14 & 0.78$\pm$0.19 & 63.34$\pm$8.2 & 58.71$\pm$5.27 & 8.11$\pm$2.86 \\ \hline
   NGC 6946, B & 85.9$\pm$4.19 & 23.21$\pm$3.61 & 0.54$\pm$0.1 & 0.61$\pm$0.12 & 4.35$\pm$0.93 & 37.09$\pm$1.93 & 1.77$\pm$0.75 \\ \hline
   NGC 7331, NFW & 266.34$\pm$ 150.1 & 1.95$\pm$3.23 & 0.34$\pm$0.07 & 0.56$\pm$0.13 & 12.04$\pm$1.46 & 74.36$\pm$2.05 & 11.73$\pm$3.9 \\ \hline
   NGC 7331, B & 153.93$\pm$ 147.69 & 6.66$\pm$4.25 & 0.3$\pm$0.09 & 0.56$\pm$0.13 & 11.47$\pm$1.56 & 74.32$\pm$2.05 & 17.25$\pm$3.84 \\ \hline
   NGC 7814, NFW & 129.38$\pm$ 12.47 & 13.68$\pm$3.75 & 0.53$\pm$0.14 & 0.5$\pm$0.06 & 14.15$\pm$0.66 & 89.3$\pm$0.5 & 4.44$\pm$4.76 \\ \hline
   NGC 7814, B & 108.73$\pm$ 178.49 & 17.85$\pm$8.26 & 0.55$\pm$0.15 & 0.45$\pm$0.12 & 14.11$\pm$0.67 & 89.33$\pm$0.47 & 10.89$\pm$6.71 \\ \hline
   UGC 00128, B & 88.28$\pm$9.36 & 11.21$\pm$1.42 & 0.48$\pm$0.11 & 0 & 60.24$\pm$9.44 & 50.29$\pm$7.84 & 3.76$\pm$0.89 \\ \hline
   UGC 00731, B & 43.01$\pm$3.38 & 20.83$\pm$2.58 & 0.53$\pm$0.14 & 0 & 13.39$\pm$3.4 & 56.75$\pm$3 & 0.27$\pm$0.13 \\ \hline
   UGC 03205, NFW & 210.81$\pm$ 66.83 & 2.78$\pm$1.44 & 0.63$\pm$0.09 & 0.71$\pm$0.13 & 54.66$\pm$8.02 & 67.19$\pm$3.8 & 6.21$\pm$2.76 \\ \hline
   UGC 03205, B & 216.1$\pm$ 171.82 & 4.49$\pm$1.42 & 00.65$\pm$0.1 & 0.48$\pm$0.11 & 51.14$\pm$8.4 & 66.67$\pm$4.15 & 11.63$\pm$2.04 \\ \hline
   UGC 04278, B & 120.15$\pm$ 120.7 & 14.43$\pm$1.46 & 0.63$\pm$0.16 & 0 & 9.93$\pm$2.16 & 88.01$\pm$1.4 & 0.09$\pm$0.07 \\ \hline 
   UGC 05005, B & 104.66$\pm$ 47.21 & 12.41$\pm$7.29 & 0.53$\pm$0.14 & 0 & 54.25$\pm$ 10.76 & 31.32$\pm$12.5 & 0.26$\pm$0.58 \\ \hline
   UGC 05716, NFW & 59.55$\pm$9.01 & 8.61$\pm$1.13 & 0.47$\pm$0.11 & 0 & 22.58$\pm$4.97 & 52.94$\pm$9.09 & 0.1$\pm$0.13 \\ \hline
   UGC 05716, B & 46.27$\pm$6.14 & 15.01$\pm$1.74 & 0.52$\pm$0.13 & 0 & 22.86$\pm$5.13 & 53.4$\pm$8.9 & 0.6$\pm$0.27 \\ \hline
   \end{tabularx}
   \end{table}
\newpage
\begin{table}
\footnotesize
\begin{tabularx}{\textwidth}{|| X || X | X | X | X | X | X | X |} \hline
    Galaxy & $V_{200}$, km/s & $C_{200}$ & $Y_d$ & $Y_b$ & D, Mpc & i, $^o$ & $G\mu, 10^{-8}$ \\ \hline\hline
    UGC 05721, B & 40.09$\pm$2.1 & 53.2$\pm$5.29 & 0.51$\pm$0.12 & 0 & 6.24$\pm$1.66 & 60.09$\pm$5.06 & 0.09$\pm$0.1 \\ \hline
   UGC 06614, B & 213.26$\pm$ 57.86 & 10.62$\pm$3.49 & 0.52$\pm$0.13 & 0.71$\pm$0.18 & 88.13$\pm$8.73 & 24.08$\pm$6.04 & 13.45$\pm$ 20.04 \\ \hline
   UGC 06786, NFW & 127.8$\pm$8.02 & 18.66$\pm$2.13 & 0.47$\pm$0.11 & 0.65$\pm$0.16 & 10.72$\pm$5.66 & 63.04$\pm$3.06 & 9.12$\pm$2.43 \\ \hline
   UGC 06786, B & 94.47$\pm$4.54 & 27.57$\pm$2.25 & 0.51$\pm$0.13 & 0.64$\pm$0.18 & 3.73$\pm$4.49 & 63.3$\pm$3.01 & 16.79$\pm$1.53 \\ \hline
   UGC 06917, B & 65.16$\pm$4.23 & 22.18$\pm$2.61 & 0.49$\pm$0.12 & 0 & 17.61$\pm$2.47 & 55.71$\pm$2 & 0.3$\pm$0.35 \\ \hline
    UGC 06930, B & 67.46$\pm$ 15.17 & 20.9$\pm$6.4 & 0.52$\pm$0.14 & 0 & 17.97$\pm$2.45 & 28.94$\pm$5.1 & 1.01$\pm$1.18 \\ \hline
   UGC 07323, B & 156.25$\pm$ 213.52 & 12.4$\pm$3.36 & 0.55$\pm$0.13 & 0 & 8.67$\pm$2.18 & 46.73$\pm$3.05 & 0.15$\pm$0.12 \\ \hline
   UGC 07399, B & 52.37$\pm$2.62 & 41.02$\pm$3.88 & 0.53$\pm$0.14 & 0 & 8.64$\pm$2.49 & 54.7$\pm$2.96 & 0.38$\pm$0.25 \\ \hline
   UGC 07524, B & 51.56$\pm$3.5 & 18.83$\pm$1.23 & 0.51$\pm$0.13 & 0 & 4.74$\pm$0.24 & 45.16$\pm$2.95 & 0.06$\pm$0.05 \\ \hline
   UGC 08490, B & 41.69$\pm$1.93 & 33.93$\pm$4.07 & 0.54$\pm$0.14 & 0 & 4.64$\pm$0.53 & 49.47$\pm$3.03 & 0.55$\pm$0.19 \\ \hline
   UGC 08550, B & 31.97$\pm$1.69 & 27.04$\pm$3.81 & 0.56$\pm$0.15 & 0 & 7.68$\pm$1.82 & 87.98$\pm$1.41 & 0.14$\pm$0.09 \\ \hline
   UGC 09133, NFW & 61.16$\pm$ 16.26 & 10.15$\pm$8.36 & 0.75$\pm$0.13 & 0.34$\pm$0.06 & 43.82$\pm$9.03 & 47.43$\pm$5.45 & 26.03$\pm$5.03 \\ \hline
   UGC 09133, B & 70.13$\pm$ 9.48 & 17.37$\pm$ 5.14 & 0.67$\pm$0.13 & 0.4$\pm$0.07 & 44.35$\pm$8.35 & 47.31$\pm$5.32 & 25.53$\pm$4.69 \\ \hline
   UGC 11820, NFW & 102.14$\pm$ 18.31 & 4.69$\pm$1.1 & 0.57$\pm$0.15 & 0 & 17.28$\pm$4.72 & 37.91$\pm$7.17 & 0.14$\pm$0.12 \\ \hline
   UGC 11820, B & 74.55$\pm$ 13.89 & 14.34$\pm$2.84 & 0.59$\pm$0.17 & 0 & 15$\pm$5.42 & 34.9$\pm$9.12 & 0.48$\pm$0.29 \\ \hline
   UGC 12632, B & 44.52$\pm$3.71 & 19.65$\pm$2.2 & 0.53$\pm$0.14 & 0 & 9.9$\pm$2.76 & 45.44$\pm$3.04 & 0.19$\pm$0.18 \\ \hline
   UGC 12732, B & 52.93$\pm$9.48 & 15.06$\pm$2.35 & 0.54$\pm$0.14 & 0 & 16.84$\pm$3.62 & 41.98$\pm$5.67 & 0.36$\pm$0.24 \\ \hline
   \end{tabularx}
    \caption{The values of fitting parameters with errors for a spherical dark matter profile with an infinite filament at the center, obtained from the MCMC analysis (the values are given only for the galaxies and profiles that have the fit minimum at $G\mu\neq0$).}
    \label{tab:table2}
\end{table}
\newpage
\begin{table}
\footnotesize
\begin{tabularx}{\textwidth}{|| X || X | X | X | X | X | X | X |} \hline
    Galaxy & $V_{200}$, km/s & $C_{200}$ & $Y_d$ & $Y_b$ & D, Mpc & i, $^o$ & $G\mu, 10^{-8}$  \\ \hline\hline
   ESO116-G012 & 68.9$\pm$2.28 & 25.53$\pm$2.12 & 0.5$\pm$0.12 & 0 & 11.77$\pm$3.47 & 73.65$\pm$3 & 0.14$\pm$0.08 \\ \hline
   NGC 2403, B & 83.83$\pm$2.69 & 21.59$\pm$0.67 & 0.81$\pm$0.07 & 0 & 2.93$\pm$0.16 & 57.54$\pm$3.03 & 0.44$\pm$0.07 \\ \hline
   NGC 2841, NFW & 389.84$\pm$ 73.89 & 0.69$\pm$0.51 & 0.86$\pm$0.12 & 0.49$\pm$0.07 & 13.98$\pm$1.3 & 74.29$\pm$8.31 & 23.46$\pm$3.98 \\ \hline
   NGC 2841, B & 400.86$\pm$ 68.93 & 3.24$\pm$0.4 & 0.79$\pm$0.11 & 0.39$\pm$0.06 & 13.37$\pm$1.29 & 71.69$\pm$9.35 & 31.83$\pm$4.67 \\ \hline
   NGC 2903, B & 93.06$\pm$1.98 & 79.08$\pm$2.33 & 0.49$\pm$0.13 & 0 & 0.1$\pm$0.16 & 65.61$\pm$3.01 & 1.8$\pm$0.53 \\ \hline
   NGC 2955, NFW & 171.13$\pm$ 23.17 & 13$\pm$5.59 & 0.41$\pm$0.09 & 0.7$\pm$0.13 & 92.17$\pm$9.94 & 44.02$\pm$5.74 & 9.89$\pm$3.15 \\ \hline
   NGC 2955, B & 143.66$\pm$ 14.47 & 19.2$\pm$4 & 0.42$\pm$0.09 & 0.76$\pm$0.15 & 93.33$\pm$9.31 & 47.5$\pm$5.67 & 8.72$\pm$2.68 \\ \hline
   NGC 2998, NFW & 132.92$\pm$ 12.6 & 7.12$\pm$3.14 & 0.47$\pm$0.09 & 0 & 66.67$\pm$9.83 & 57.71$\pm$1.97 & 5.59$\pm$1.58 \\ \hline
   NGC 2998, B & 118.79$\pm$ 8.89 & 10.09$\pm$2.15 & 0.53$\pm$0.1 & 0 & 70.01$\pm$9.26 & 57.92$\pm$1.98 & 6.49$\pm$1.59 \\ \hline
   NGC 3198, B & 95.34$\pm$2.38 & 16.36$\pm$1.69 & 0.54$\pm$0.08 & 0 & 14.17$\pm$1.3 & 73.07$\pm$3 & 1.06$\pm$0.33 \\ \hline
   NGC 3521, NFW & 140.55$\pm$ 101.3 & 8.36$\pm$10.16 & 0.46$\pm$0.11 & 0 & 6.01$\pm$1.73 & 73.86$\pm$5.1 & 6.55$\pm$2.44 \\ \hline
   NGC 3521, B & 132.12$\pm$ 65.33 & 15.32$\pm$4.38 & 0.48$\pm$0.11 & 0 & 7.09$\pm$1.64 & 74.38$\pm$5.14 & 5.31$\pm$2.92 \\ \hline
   NGC 4183, B & 61.7$\pm$2.67 & 21$\pm$3.16 & 0.5$\pm$0.12 & 0 & 17.48$\pm$2.44 & 81.87$\pm$2 & 0.87$\pm$0.44 \\ \hline
   NGC 5371, NFW & 59.74$\pm$ 27.39 & 2.75$\pm$8.89 & 0.43$\pm$0.1 & 0 & 29.48$\pm$6.48 & 52.37$\pm$2 & 15.2$\pm$1.63 \\ \hline
   NGC 5371, B & 280.03$\pm$ 150.68 & 1.43$\pm$0.52 & 0.46$\pm$0.1 & 0 & 29.76$\pm$5.74 & 52.56$\pm$2.03 & 14.73$\pm$1.55 \\ \hline
   NGC 5585, NFW & 98.62$\pm$7.02 & 7.04$\pm$0.74 & 0.43$\pm$0.12 & 0 & 2.04$\pm$1.62 & 50.46$\pm$2.02 & 0.06$\pm$0.03\\ \hline
   NGC 5585, B & 62.34$\pm$2.09 & 19.43$\pm$0.94 & 0.47$\pm$0.11 & 0 & 5.58$\pm$1.4 & 50.57$\pm$2.01 & 0.17$\pm$0.04 \\ \hline
   NGC 5907, NFW & 179.91$\pm$ 81.59 & 0.39$\pm$0.68 & 0.51$\pm$0.07 & 0 & 16.94$\pm$0.87 & 87.58$\pm$1.51 & 14.24$\pm$2.07 \\ \hline
   NGC 5907, B & 318.63$\pm$ 123.19 & 2.28$\pm$0.48 & 0.5$\pm$0.07 & 0 & 17.07$\pm$0.88 & 87.61$\pm$1.48 & 15.54$\pm$1.78 \\ \hline
   NGC 5985, B & 138.89$\pm$ 5.14 & 37.05$\pm$2.82 & 0.45$\pm$0.12 & 0.71$\pm$0.18 & 9.32$\pm$8.4 & 59.6$\pm$2.03 & 16.06$\pm$2.9 \\ \hline
   NGC 6503, B & 60.67$\pm$3.8 & 17.71$\pm$1.36 & 0.36$\pm$0.06 & 0 & 6.13$\pm$0.3 & 73.59$\pm$2.02 & 2.77$\pm$0.65 \\ \hline
   NGC 6674, B & 388.56$\pm$ 76.74 & 3.12$\pm$0.42 & 0.79$\pm$0.13 & 0.77$\pm$0.19 & 63.47$\pm$7.94 & 58.73$\pm$5.19 & 8.12$\pm$2.93 \\ \hline
   NGC 6946, B & 85.83$\pm$4.2 & 23.04$\pm$3.54 & 0.54$\pm$0.09 & 0.61$\pm$0.1 & 4.35$\pm$0.86 & 37.15$\pm$1.95 & 1.77$\pm$0.74 \\ \hline
   NGC 7331, NFW & 273.14$\pm$ 143.04 & 2.1$\pm$2.88 & 0.35$\pm$0.07 & 0.57$\pm$0.13 & 12.04$\pm$1.48 & 74.37$\pm$2.02 & 11.15$\pm$3.69 \\ \hline
   NGC 7331, B & 163.5$\pm$ 168.52 & 6.51$\pm$4.19 & 0.31$\pm$0.08 & 0.55$\pm$0.12 & 11.44$\pm$1.5 & 74.21$\pm$2.07 & 17.56$\pm$3.43 \\ \hline
   \end{tabularx}
   \end{table}
\newpage
\begin{table}
\footnotesize
\begin{tabularx}{\textwidth}{|| X || X | X | X | X | X | X | X |} \hline
    Galaxy & $V_{200}$, km/s & $C_{200}$ & $Y_d$ & $Y_b$ & D, Mpc & i, $^o$ & $G\mu, 10^{-8}$  \\ \hline\hline
   UGC 00128, B & 93.35$\pm$ 10.32 & 11.14$\pm$ 1.45 & 0.48$\pm$0.12 & 0 & 59.19$\pm$ 9.67 & 47.59$\pm$7.68 & 4.28$\pm$1.11 \\ \hline
   UGC 00731, B & 42.9$\pm$3.4 & 20.85$\pm$ 2.62 & 0.53$\pm$0.14 & 0 & 13.47$\pm$3.4 & 56.84$\pm$2.96 & 0.26$\pm$0.13 \\ \hline
   UGC 03205, NFW & 218.93$\pm$ 79.55 & 2.24$\pm$1.58 & 0.63$\pm$0.08 & 0.71$\pm$0.13 & 54.66$\pm$8.02 & 67.19$\pm$3.91 & 5.89$\pm$2.75 \\ \hline
   UGC 03205, B & 228.69$\pm$ 166.37 & 4.67$\pm$1.37 & 0.66$\pm$0.1 & 0.48$\pm$0.11 & 50.95$\pm$8 & 66.57$\pm$ 3.79 & 11.68$\pm$ 2.01 \\ \hline
   UGC 05716, B & 46.72$\pm$ 6.69 & 15.08$\pm$ 1.92 & 0.52$\pm$0.13 & 0 & 22.81$\pm$5.03 & 52.85$\pm$9.5 & 0.62$\pm$0.29 \\ \hline
   UGC 06786, NFW & 128.78$\pm$ 7.66 & 18.15$\pm$ 2.03 & 0.47$\pm$0.11 & 0.65$\pm$0.16 & 10.67$\pm$5.36 & 62.92$\pm$3 & 9.41$\pm$2.49 \\ \hline
   UGC 06786, B & 95.67$\pm$ 5.43 & 26.15$\pm$2.41 & 0.52$\pm$0.13 & 0.63$\pm$0.18 & 3.98$\pm$5.81 & 63.29$\pm$3.11 & 17.1$\pm$1.56 \\ \hline
   UGC 08490, B & 41.66$\pm$ 1.95 & 33.88$\pm$ 3.98 & 0.54$\pm$0.14 & 0 & 4.65$\pm$0.53 & 49.54$\pm$3.08 & 0.55$\pm$0.19 \\ \hline
   UGC 09133, NFW & 173.51$\pm$ 25.76 & 0.53$\pm$0.52 & 0.76$\pm$0.11 & 0.35$\pm$0.05 & 47.09$\pm$ 8.46 & 48.9$\pm$3.42 & 23.49$\pm$3.31 \\ \hline
   UGC 09133, B & 142.54$\pm$ 13.9 & 2.36$\pm$0.47 & 0.78$\pm$0.14 & 0.32$\pm$0.05 & 44.52$\pm$ 8.15 & 48.62$\pm$4.22 & 26.18$\pm$3.81 \\ \hline
   UGC 11820, NFW & 102.36$\pm$ 19.08 & 4.73$\pm$1.12 & 0.57$\pm$0.15 & 0 & 17.3$\pm$4.83 & 37.77$\pm$7.31 & 0.14$\pm$0.12 \\ \hline
   UGC 11820, B & 75.59$\pm$ 16.17 & 14.54$\pm$3.22 & 0.59$\pm$0.18 & 0 & 15.22$\pm$5.39 & 34.33$\pm$8.92 & 0.5$\pm$0.33 \\ \hline
   UGC 12732, B & 53.36$\pm$9.58 & 15.02$\pm$2.31 & 0.54$\pm$0.15 & 0 & 16.92$\pm$3.54 & 41.66$\pm$5.73 & 0.37$\pm$0.25 \\ \hline
   \end{tabularx}
    \caption{The values of fitting parameters with errors for a spherical dark matter profile with an finite filament of length 200 kpc at the center, obtained from the MCMC analysis (the values are given only for the galaxies and profiles that have the fit minimum at $G\mu\neq0$ \textit{and} demonstrate improvement in BIC and AIC values upon the addition of a filament).}
    \label{tab:table3}
\end{table}
\newpage
\begin{table}
\begin{tabularx}{\textwidth}{|| X || X | X | X || X | X | X |} \hline
    Galaxy & NFW, w/o string & NFW, string & NFW, finite string & B, w/o string & B, string & B, finite string  \\ \hline\hline
    NGC 2841 & 105.02 & 91.36 & 98.41 & 121.81 & 93.17 & 93.88 \\ \hline
    NGC 2955 & 98.16 & 73.1 & 72.83 & 91.85 & 68.87 & 68.63 \\ \hline
    NGC 2998 & 42.58 & 23.86 & 23.75 & 44.65 & 23.7 & 23.35 \\ \hline
    NGC 5371 & 88.94 & 40.78 & 42.72 & 161.36 & 40.63 & 38.9 \\ \hline
    NGC 5585 & 145.45 & 143.37 & 143.35 & 110.84 & 88.45 & 88.43 \\ \hline
    NGC 5907 & 95.8 & 63.1 & 65.16 & 107.7 & 59.96 & 59.49 \\ \hline
    UGC 06786 & 52.85 & 51.16 & 51.73 & 113.81 & 58.8 & 61.33 \\ \hline
    UGC 09133 & 477.21 & 439.18 & 440.31 & 531.1 & 428.28 & 434.16 \\ \hline
    UGC 11820 & 23.6 & 22.18 & 22.7 & 48.59 & 34.7 & 34.49 \\ \hline
    \end{tabularx}
    \caption{The values of the Bayesian criterion for three cases: spherical DM halo, spherical DM halo with an infinite filament at the origin, and spherical DM halo with a finite filament of length 200 kpc at the origin. The values are given for 9 galaxies that demonstrate BIC improvement upon the addition of the filament for \textit{both} NFW and Burkert profiles (we refer to this as \textit{strong} evidence for the filament): the left portion of the table concerns NFW halos, while the right portion contains corresponding results for the Burkert halos.}
    \label{tab:table4}
\end{table}
\newpage
\begin{table}
\begin{tabularx}{\textwidth}{|| X || X | X | X || X | X | X |} \hline
    Galaxy & NFW, w/o string & NFW, string & NFW, finite string & B, w/o string & B, string & B, finite string  \\ \hline\hline
    ESO116-G012 & 44.01 & --- & --- & 23.54 & 22.88 & 22.85 \\ \hline
    NGC 2903 & 204.52 & --- & --- & 227.75 & 192.35 & 194.25 \\ \hline
    NGC 3198 & 74.66 & --- & --- & 76.75 & 68.93 & 70.32 \\ \hline
    NGC 6674 & 45.12 & --- & --- & 40.38 & 40.29 & 40.86 \\ \hline
    NGC 6946 & 114.4 & --- & --- & 111.69 & 103.48 & 103.09 \\ \hline
    UGC 05716 & 32.39 & 34.03 & --- & 40.36 & 27.73 & 28.42 \\ \hline
    \end{tabularx}
    \caption{The values of the Bayesian criterion for three cases: spherical DM halo, spherical DM halo with an infinite filament at the origin, and spherical DM halo with a finite filament of length 200 kpc at the origin. The values are given for 6 galaxies that demonstrate BIC improvement for either NFW or Burkert profile, compared to the standard case for the \textit{better}-fitting profile (we refer to this as \textit{moderate} evidence for the filament): the left portion of the table concerns NFW halos, while the right portion contains corresponding results for the Burkert halos. The cases when the initial fit minimum is at $G\mu=0$ are marked with dashes.}
    \label{tab:table5}
\end{table}
\newpage
\begin{table}
\begin{tabularx}{\textwidth}{|| X || X | X | X || X | X | X |} \hline
    Galaxy & NFW, w/o string & NFW, string & NFW, finite string & B, w/o string & B, string & B, finite string  \\ \hline\hline
    NGC 2403 & 648.36 & --- & --- & 822.35 & 756.14 & 757.05  \\ \hline
    NGC 4183 & 19.34 & --- & --- & 23.72 & 23.17 & 23.24 \\ \hline
    NGC 5985 & 81.61 & --- & --- & 109.89 & 82.55 & 85.52 \\ \hline
    NGC 6503 & 58.48 & 61.26 & --- & 83.28 & 70.67 & 71.77 \\ \hline
    UGC 00128 & 76.63 & 86 & --- & 176.42 & 86 & 89.43 \\ \hline
    UGC 00731 & 15.81 & --- & --- & 17.18 & 16.47 & 16.45 \\ \hline
    UGC 03205 & 176.43 & 176.59 & --- & 190.57 & 184.33 & 185.54 \\ \hline
    UGC 08490 & 20.3 & --- & --- & 33 & 24.63 & 24.65 \\ \hline
    UGC 12732 & 17.56 & --- & --- & 24.16 & 22.52 & 22.57 \\ \hline
    \end{tabularx}
    \caption{The values of the Bayesian criterion for three cases: spherical DM halo, spherical DM halo with an infinite filament at the origin, and spherical DM halo with a finite filament of length 200 kpc at the origin. The values are given for 9 galaxies that demonstrate BIC improvement for either NFW or Burkert profile, compared to the standard case for the \textit{worse}--fitting profile (we refer to this as \textit{weak} evidence for the filament): the left portion of the table concerns NFW halos, while the right portion contains corresponding results for the Burkert halos. The cases when the initial fit minimum is at $G\mu=0$ are marked with dashes.}
    \label{tab:table6}
\end{table}
\newpage
\begin{table}
\begin{tabularx}{\textwidth}{|| X || X | X | X || X | X | X |} \hline
    Galaxy & NFW, w/o string & NFW, string & B, w/o string & B, string  \\ \hline\hline
    DDO 064 & 19.67 & --- & 17.96 & 20.39 \\ \hline
    DDO 161 & 53.64 & --- & 27.23 & 29.37 \\ \hline
    ESO079-G014 & 61.49 & --- & 26.2 & 29.04 \\ \hline
    F583-4 & 22.18 & --- & 21.65 & 22.81 \\ \hline
    NGC 1090 & 68.71 & --- & 47.26 & 50.05 \\ \hline
    NGC 3109 & 210.82 & ---  & 20.31 & 23.01  \\ \hline
    NGC 3521 & 27.85 & 28 & 26.28 & 26.85 \\ \hline
    NGC 3972 & 20.82 & --- & 17.56 & 20.53 \\ \hline
    NGC 4088 & 17.76 & --- & 17.12 & 20.5 \\ \hline
    NGC 4559 & 25.22 & --- & 24.46 & 27.52 \\ \hline
    NGC 5005 & 19.92 & 23.84 & 19.87 & 24.26 \\ \hline
    NGC 5055 & 84.92 & 90.16 & 67.64 & 70 \\ \hline
    NGC 6195 & 55.6 & 59.13 & 51.23 & --- \\ \hline
    NGC 7331 & 48.35 & 62.96 & 62.05 & 69.51 \\ \hline
    NGC 7814 & 28.57 & 30.46 & 25.53 & 70.88 \\ \hline
    UGC 04278 & 45.26 & --- & 30.23 & 32.43 \\ \hline
    UGC 05005 & 15.19 & --- & 12.89 & 16.89 \\ \hline
    UGC 05721 & 35.72 & ---  & 23.86 & 26.8  \\ \hline
    UGC 06614 & 18 & --- & 16.35 & 21.42 \\ \hline
    UGC 06917 & 19.3 & --- & 13.83 & 17 \\ \hline
    UGC 06930 & 14.82 & --- & 13.11 & 16.08 \\ \hline
    UGC 07323 & 17 & --- & 15.56 & 16.62 \\ \hline
    UGC 07399 & 19.71 & --- & 18.42 & 18.85 \\ \hline
    UGC 07524 & 40.54 & --- & 23.85 & 26.47 \\ \hline
    UGC 08550 & 19.2 & --- & 20.74 & 21.15 \\ \hline
    UGC 12632 & 18.06 & --- & 14.48 & 17.38 \\ \hline
    \end{tabularx}
    \caption{The values of the Bayesian criterion for two cases: spherical DM halo and spherical DM halo with an infinite filament at the origin. The values are given for 26 galaxies that demonstrate \textit{no} BIC improvement for either NFW or Burkert profile (i.e. \textit{no} evidence for the filament): the left portion of the table concerns NFW halos, while the right portion contains corresponding results for the Burkert halos. The cases when the initial fit minimum is at $G\mu=0$ are marked with dashes.} 
    \label{tab:table7}
\end{table}
\newpage
\renewcommand\thetable{8}
\begin{table}
\begin{tabularx}{\textwidth}{|| X || X | X | X || X | X | X |} \hline
    Galaxy & NFW, w/o string & NFW, string & NFW, finite string & B, w/o string & B, string & B, finite string  \\ \hline\hline
    NGC 2841 & 93.54 & 77.98 & 85.03 & 110.34 & 79.79 & 80.79 \\ \hline
    NGC 2955 & 91.1 & 64.86 & 64.58 & 84.78 & 60.62 & 60.39 \\ \hline
    NGC 2998 & 39.76 & 20.47 & 20.36 & 41.82 & 23.7 & 19.96 \\ \hline
    NGC 3521 & 19.28 & 17.7 & 17.66 & 17.71 & 16.57 & 16.59 \\ \hline
    NGC 5371 & 84.22 & 35.11 & 37.06 & 156.64 & 34.97 & 33.23 \\ \hline
    NGC 5585 & 139.56 & 136.3 & 136.28 & 104.95 & 81.38 & 81.36 \\ \hline
    NGC 5907 & 91.08 & 57.43 & 59.49 & 102.98 & 54.29 & 53.82 \\ \hline
    UGC 03205 & 165.2 & 163.49 & 164.42 & 179.34 & 171.23 & 172.44 \\ \hline
    UGC 06786 & 42.01 & 38.51 & 39.08 & 102.97 & 46.16 & 48.69 \\ \hline
    UGC 09133 & 463.9 & 423.65 & 424.77 & 517.78 & 412.74 & 418.62 \\ \hline
    UGC 11820 & 22.1 & 20.37 & 20.88 & 47.08 & 32.88 & 32.67 \\ \hline
    \end{tabularx}
    \caption{The values of the Akaike criterion for three cases: spherical DM halo, spherical DM halo with an infinite filament at the origin, and spherical DM halo with a finite filament of length 200 kpc at the origin. The values are given for 11 galaxies that demonstrate AIC improvement upon the addition of the filament for \textit{both} NFW and Burkert profiles (we refer to this as \textit{strong} evidence for the filament): the left portion of the table concerns NFW halos, while the right portion contains corresponding results for the Burkert halos.}
    \label{tab:table8}
\end{table}
\newpage
\begin{table}
\begin{tabularx}{\textwidth}{|| X || X | X | X || X | X | X |} \hline
    Galaxy & NFW, w/o string & NFW, string & NFW, finite string & B, w/o string & B, string & B, finite string  \\ \hline\hline
    ESO116-G012 & 40.47 & --- & --- & 20 & 18.64 & 18.6 \\ \hline
    NGC 2903 & 196.89 & --- & --- & 220.12 & 183.19 & 185.09 \\ \hline
    NGC 3198 & 65.85 & --- & --- & 67.95 & 58.37 & 59.76 \\ \hline
    NGC 5985 & 72.63 & --- & --- & 100.91 & 72.08 & 75.04 \\ \hline
    NGC 6674 & 40.87 & --- & --- & 36.13 & 35.33 & 35.91 \\ \hline
    NGC 6946 & 102.04 & --- & --- & 99.33 & 89.05 & 88.67 \\ \hline
    UGC 05716 & 29.96 & 31.12 & --- & 37.93 & 24.83 & 25.51 \\ \hline
    \end{tabularx}
    \caption{The values of the Akaike criterion for three cases: spherical DM halo, spherical DM halo with an infinite filament at the origin, and spherical DM halo with a finite filament of length 200 kpc at the origin. The values are given for 7 galaxies that demonstrate AIC improvement for either NFW or Burkert profile, compared to the standard case for the \textit{better}-fitting profile (we refer to this as \textit{moderate} evidence for the filament): the left portion of the table concerns NFW halos, while the right portion contains corresponding results for the Burkert halos. The cases when the initial fit minimum is at $G\mu=0$ are marked with dashes.}
    \label{tab:table9}
\end{table}
\newpage
\begin{table}
\begin{tabularx}{\textwidth}{|| X || X | X | X || X | X | X |} \hline
    Galaxy & NFW, w/o string & NFW, string & NFW, finite string & B, w/o string & B, string & B, finite string  \\ \hline\hline
    NGC 2403 & 636.91 & --- & --- & 810.89 & 742.39 & 743.3 \\ \hline
    NGC 4183 & 13.66 & --- & --- & 18.04 & 16.36 & 16.42 \\ \hline
    NGC 6503 & 51.31 & 52.65 & --- & 76.11 & 62.06 & 63.16 \\ \hline
    UGC 00128 & 71.18 & --- & --- & 170.96 & 79.45 & 82.88 \\ \hline
    UGC 00731 & 13.38 & --- & --- & 14.76 & 13.56 & 13.54 \\ \hline
    UGC 08490 & 13.3 & --- & --- & 26 & 16.22 & 16.24 \\ \hline
    UGC 12732 & 13.7 & --- & --- & 20.3 & 17.89 & 17.93 \\ \hline
    \end{tabularx}
    \caption{The values of the Akaike criterion for three cases: spherical DM halo, spherical DM halo with an infinite filament at the origin, and spherical DM halo with a finite filament of length 200 kpc at the origin. The values are given for 7 galaxies that demonstrate AIC improvement for either NFW or Burkert profile, compared to the standard case for the \textit{worse}--fitting profile (we refer to this as \textit{weak} evidence for the filament): the left portion of the table concerns NFW halos, while the right portion contains corresponding results for the Burkert halos. The cases when the initial fit minimum is at $G\mu=0$ are marked with dashes.}
    \label{tab:table10}
\end{table}
\newpage
\begin{table}
\begin{tabularx}{\textwidth}{|| X || X | X | X || X | X | X |} \hline
    Galaxy & NFW, w/o string & NFW, string & B, w/o string & B, string \\ \hline\hline
    DDO064 & 16.47 & --- & 14.76 & 16.55 \\ \hline
    DDO 161 & 46.47 & --- & 20.06 & 20.77 \\ \hline
    ESO079-G014 & 57.95 & --- & 22.66 & 24.79 \\ \hline
    F583-4 & 19.76 & --- & 19.23 & 19.9 \\ \hline
    NGC 1090 & 62.82 & --- & 41.37 & 43 \\ \hline
    NGC 3109 & 204.72 & --- & 14.21 & 15.7 \\ \hline
    NGC 3972 & 19.31 & --- & 16.04 & 18.72 \\ \hline
    NGC 4088 & 15.33 & --- & 14.7 & 17.59 \\ \hline
    NGC 4559 & 17.89 & --- & 17.13 & 18.72 \\ \hline
    NGC 5005 & 14.58 & 17.61 & 14.53 & 18.03 \\ \hline
    NGC 5055 & 78.26 & 82.17 & 60.98 & 62 \\ \hline
    NGC 6195 & 48.78 & 51.18 & 44.42 & --- \\ \hline
    NGC 7331 & 38.84 & 51.88 & 52.55 & 58.42 \\ \hline
    NGC 7814 & 23.23 & 24.22 & 20.19 & 64.64 \\ \hline
    UGC 04278 & 39.16 & --- & 24.13 & 25.12 \\ \hline
    UGC 05005 & 13.2 & --- & 10.9 & 14.48 \\ \hline
    UGC 05721 & 30.05 & --- & 18.18 & 20 \\ \hline
    UGC 06446 & 13.35 & --- & 16.53 & 16.53 \\ \hline
    UGC 06614 & 14.61 & --- & 12.96 & 17.46 \\ \hline
    UGC 06917 & 17.31 & --- & 11.84 & 14.61 \\ \hline
    UGC 06930 & 13.31 & --- & 11.6 & 14.26 \\ \hline
    UGC 07323 & 15.49 & --- & 14.05 & 14.8 \\ \hline
    UGC 07399 & 18.19 & --- & 16.9 & 17.03 \\ \hline
    UGC 07524 & 33.37 & --- & 16.68 & 17.86 \\ \hline
    UGC 08550 & 17.21 & --- & 18.75 & 18.77 \\ \hline
    UGC 12632 & 14.52 & --- & 10.94 & 13.13 \\ \hline
    \end{tabularx}
    \caption{The values of the Akaike criterion for two cases: spherical DM halo and spherical DM halo with an infinite filament at the origin. The values are given for 26 galaxies that demonstrate \textit{no} AIC improvement for either NFW or Burkert profile (i.e. \textit{no} evidence for the filament): the left portion of the table concerns NFW halos, while the right portion contains corresponding results for the Burkert halos. The cases when the initial fit minimum is at $G\mu=0$ are marked with dashes.}
    \label{tab:table11}
\end{table}
\newpage
\begin{table}
\begin{tabularx}{\textwidth}{|| X || X | X | X | X | X | X | X | X | X | X | X | X | X | X |} \hline
 \footnotesize
    Galaxy & $V_{200}$, km/s & $C_{200}$ & $Y_d$ & $Y_b$ & D, Mpc & i, $^o$ & q & BIC & AIC \\ \hline\hline
   NGC 5371, NFW & 131.73 $\pm$5.21 & 26.9$\pm$ 1.67 & 0.38$\pm$ 0.09 & 0 & 27.4$\pm$ 6.76 & 52.46$\pm$ 2.03 & 0.41$\pm$ 0.18 & 86.81 & 81.14 \\ \hline
   NGC 5371, B & 440.08 $\pm$43.49 & 3.25$\pm$ 0.22 & 0.86$\pm$ 0.14 & 0 & 30.96$\pm$ 5.02 & 52.76$\pm$ 1.97 & 0.7$\pm$ 0.21 & 158.5 & 152.83 \\ \hline
   NGC 5907, NFW & 144.76 $\pm$2.94 & 23.21$\pm$ 2.05 & 0.28$\pm$ 0.05 & 0 & 16.6$\pm$ 0.89 & 87.59$\pm$ 1.5 & 0.44$\pm$ 0.21 & 94.36 & 88.69 \\ \hline
   NGC 5907, B & 202.33 $\pm$ 23.55 & 5.6$\pm$ 0.49 & 0.96$\pm$ 0.05 & 0 & 17.82$\pm$ 0.86 & 87.62$\pm$ 1.47 & 0.66$\pm$ 0.23 & 110.77 & 105.1 \\ \hline
   \end{tabularx}
    \caption{The values of fitting parameters with errors, BIC and AIC for the 2 galaxies, obtained from the MCMC analysis, for a deformed dark matter profile.}
\label{tab:table12}
\end{table}
\clearpage
\begin{table}
\begin{tabularx}{\textwidth}{|| X | X | X | X | X | X | X | X | X | X | X | X |} \hline
    $V_{200}$, km/s & $C_{200}$ & $Y_d$ & $Y_b$ & D, Mpc & i, $^o$ & $G\mu$, $10^{-8}$ & q & BIC & AIC \\ \hline\hline
    307.71 $\pm$ 141.33 & 1.46$\pm$ 0.64 & 0.5$\pm$ 0.06 & 0 & 17.05$\pm$ 0.86 & 87.5$\pm$ 1.52 & 15.84$\pm$ 1.48 & 0.68$\pm$ 0.22 & 62.59 & 55.98 \\ \hline
    \end{tabularx}
    \caption{The values of fitting parameters with errors, BIC and AIC for NGC 5907 (Burkert profile), obtained from the MCMC analysis, for a deformed dark matter profile plus a filament.}
\label{tab:table13}
\end{table}

\begin{chapter}{Conclusions}\label{C5}
We can now conclude with a brief summary of the original results obtained in this Thesis. 

In Chapter~\ref{C3}, we have derived some significant constraints on the relativistic completions of MOND, which weaken somewhat the case for modified gravity and point towards alternative explanations for the radial acceleration relation, such as the interplay between dark and luminous matter. Namely, we have shown that the energy, in the model known as covariant emergent gravity (CEG), is unbounded from below, that $f(R)$ modifications of gravity cannot yield the transition between the Newtonian regime and the MOND regime, and we have stressed that superfluid dark matter has an unclear mechanism of phase transition and coupling to baryons. Combining these results with an earlier work~\cite{Boran:2017rdn}, we have derived a generic ``no--go`` statement that applies rather generally to infrared modifications of gravity. In our opinion, these results put significant constraints on relativistic extensions of MOND--based frameworks, lending credence to alternative models that can explain the flattening of rotation curves and RAR via dynamical effects. 

In Chapter~\ref{C4}, we have hypothesized the existence of one such phenomenon, namely the presence of bulges or string--like filaments at the centers of galaxies that would produce quasi--logarithmic gravitational potentials, resulting in near--constant rotation velocities at large distances (the elongation of dark halos themselves yields a similar, albeit smaller effect). To test our proposal, we have performed a detailed analysis of about 83 galaxy rotation curves from the SPARC catalogue, collecting some interesting hints for the presence of elongated string--like objects in about 10--25 of them, with the improvement in fit quality (measured as the decrease of $\chi^2$) of about 40--70\% in best cases. From an astrophysical point of view, these objects could be interpreted as black--hole relativistic jets comprising luminous or dark matter; they could also be connected to features like tidal streams, polar rings, and larger--scale intergalactic filaments. A more exotic candidate would be a cosmic string: it has been proposed in~\cite{Chudnovsky:1986hc,Vilenkin:2018zol} that primordial cosmic strings can migrate to the center of galaxies, and the data from pulsar observations by NANOGrav was also interpreted in the light of the cosmic string hypothesis~\cite{Blasi:2020mfx,Ellis:2020ena,Arzoumanian:2020vkk}. On the other hand, contrary to our original expectations, our current results indicate that the shape of the dark halo, and thus the presence of possible bulges, is not clearly discernible from galaxy rotation curves, yielding an improvement of only about 6--7\%. Further work will be needed to elucidate these issues, and we expect that the more refined galaxy surveys from the Euclid mission will help to shed more light on them~\cite{Laureijs:2011gra}.

Given the scarcity of phenomenological clues on the actual nature of dark matter, it is essential to try and interpret properly the few available ones. These include the radial acceleration relation and the flattening of rotation curves at large distances, and our motivation was to try and see to what extent, without invoking low--energy modifications of gravity, one can attain a physical picture of these phenomena. We feel that, to some extent at least, we have been able to move farther in this direction.
\end{chapter}


\begin{thebibliography}{9}
\bibitem{SM}
For reviews see:
J. F. Donoghue, E. Golowich, B.R. Holstein, “Dynamics of the Standard Model,“ Cambridge University Press (1994);
W. Cottingham, D. Greenwood, “An Introduction to the Standard Model of Particle Physics,“ Cambridge University Press (2007);
M. Schwartz, “Quantum Field Theory and the Standard Model,“ Cambridge University Press (2014);
D. Goldberg, “The Standard Model in a Nutshell,“ Princeton University Press (2017).

\bibitem{GR}
D.~Hilbert,
Gott. Nachr. \textbf{27} (1915), 395-407;
A.~Einstein,
Annalen Phys. \textbf{49} (1916) no.7, 769-822.
For reviews see:
S. Weinberg, “Gravitation and Cosmology: Principles and Applications of the General Theory of Relativity,“ John Wiley \& Sons, Inc. (1972);
C. Misner, K. Thorne, J. Wheeler, “Gravitation,“ W. H. Freeman, Princeton University Press (1973);
R. Wald, “General Relativity,“ The University of Chicago Press (1984);
S. Carroll, “Spacetime and Geometry: An Introduction to General Relativity,” San Francisco: Addison-Wesley (2004).

\bibitem{QFT}
M.~Born and P.~Jordan,
Z. Phys. \textbf{34} (1925) no.1, 858-888;
M.~Born, W.~Heisenberg and P.~Jordan,
Z. Phys. \textbf{35} (1926) no.8-9, 557-615;
P.~A.~M.~Dirac,
Proc. Roy. Soc. Lond. A \textbf{114} (1927), 243.
For reviews see:
M. Peskin, D. Schroeder, “An Introduction to Quantum Field Theory,” Westview Press (1995);
S. Weinberg, “The Quantum Theory of Fields,” Cambridge University Press (1995);
M. Srednicki, “Quantum Field Theory,” Cambridge University Press (2007);
D. Tong, “Lectures on Quantum Field Theory”.

\bibitem{Cosmology}
A.~Friedmann,
Z. Phys. \textbf{10} (1922), 377-386;
Z. Phys. \textbf{21} (1924), 326-332.
For reviews see:
V. Mukhanov, “Physical foundations of cosmology,” Cambridge, UK: Univ. Pr. (2005);
S. Weinberg, “Cosmology,” Oxford, UK: Oxford Univ. Pr. (2008);
D. S. Gorbunov and V. A. Rubakov, “Introduction to the theory of the early Universe: Hot Big Bang Theory,”; 
A. Dolgov and C. Bambi, “Introduction to particle cosmology,” Springer-Verlag Berlin Heidelberg (2016).

\bibitem{Inflation}
A.~H.~Guth,
Phys. Rev. D \textbf{23} (1981), 347-356;
A.~D.~Linde,
Phys. Lett. B \textbf{108} (1982), 389-393.
For reviews see:
N. Bartolo, E. Komatsu, S. Matarrese and A. Riotto,
Phys. Rept. 402 (2004) 103
[arXiv:astro-ph/0406398];
V. Mukhanov, “Physical foundations of cosmology,” Cambridge, UK: Univ. Pr. (2005);
S. Weinberg, “Cosmology,” Oxford, UK: Oxford Univ. Pr. (2008);
D. H. Lyth and A. R. Liddle, “The primordial density perturbation: Cosmology, inflation and the origin of structure,” Cambridge, UK: Cambridge Univ. Pr. (2009);
D. S. Gorbunov and V. A. Rubakov, “Introduction to the theory of the early Universe: Cosmological perturbations and inflationary theory,”; 
J. Martin, C. Ringeval and V. Vennin,
[arXiv:1303.3787 [astro-ph.CO]].

\bibitem{Chib_Mukh}
V.~F.~Mukhanov and G.~V.~Chibisov,
JETP Lett. \textbf{33} (1981), 532-535.

      \bibitem{PauliLuders}
      J.~S.~Schwinger,
      Phys. Rev. \textbf{82} (1951), 914-927;
      G.~Lüders,
      Kong. Dan. Vid. Sel. Mat. Fys. Med. \textbf{28N5} (1954) no.5, 1-17;
      G.~Luders and B.~Zumino,
Phys. Rev. \textbf{106} (1957), 385-386.
      For a review, see:
      O.~W.~Greenberg,
      Found. Phys. \textbf{36} (2006), 1535-1553
      [arXiv:hep-ph/0309309].

     \bibitem{gs}
     M.~H.~Goroff and A.~Sagnotti,
Phys. Lett. B \textbf{160} (1985), 81-86;
     Nucl. Phys. B \textbf{266} (1986), 709-736.
     
     \bibitem{Zwicky:1933gu}
	F.~Zwicky,
	Helv.\ Phys.\ Acta {\bf 6} (1933) 110
	[Gen.\ Rel.\ Grav.\  {\bf 41} (2009) 207].
	
	\bibitem{Zwicky:1937zza}
	F.~Zwicky,
	Astrophys.\ J.\  {\bf 86} (1937) 217.
	
	\bibitem{Smith:1936mlg}
	S.~Smith,
	Astrophys.\ J.\  {\bf 83} (1936) 23.
	
	\bibitem{Rubin:1978kmz}
	V.~C.~Rubin, W.~K.~Ford, Jr. and N.~Thonnard,
	Astrophys.\ J.\  {\bf 225} (1978) L107.
	
	\bibitem{Rubin:1980zd}
	V.~C.~Rubin, N.~Thonnard and W.~K.~Ford, Jr.,
	Astrophys.\ J.\  {\bf 238} (1980) 471.
     
     \bibitem{Natarajan:2017sbo}
	P.~Natarajan {\it et al.},
	Mon.\ Not.\ Roy.\ Astron.\ Soc.\  {\bf 468} (2017) no.2,  1962
	[arXiv:1702.04348 [astro-ph.GA]].
	
	\bibitem{Tyson:1998vp}
	J.~A.~Tyson, G.~P.~Kochanski and I.~P.~Dell'Antonio,
	Astrophys.\ J.\  {\bf 498} (1998) L107
	[arXiv:astro-ph/9801193].
	
	\bibitem{Markevitch:2003at}
	M.~Markevitch {\it et al.},
	Astrophys.\ J.\  {\bf 606} (2004) 819
	[arXiv:astro-ph/0309303].
	
	\bibitem{Mahdavi:2007yp}
	A.~Mahdavi, H.~y.~Hoekstra, A.~y.~Babul, D.~y.~Balam and P.~Capak,
	Astrophys.\ J.\  {\bf 668} (2007) 806
	[arXiv:0706.3048 [astro-ph]].
	
	\bibitem{Buote:2003tw}
	D.~A.~Buote and A.~D.~Lewis,
	Astrophys.\ J.\  {\bf 604} (2004) 116
	[arXiv:astro-ph/0312109].
     
     \bibitem{Peacock:2001gs}
	J.~A.~Peacock {\it et al.},
	Nature {\bf 410} (2001) 169
	[arXiv:astro-ph/0103143].
     
     \bibitem{PLANCK}
P.~A.~R.~Ade {\it et al.} [Planck Collaboration],
  Astron.\ Astrophys.\  {\bf 594} (2016) A13
  [arXiv:1502.01589 [astro-ph.CO]];
  Astron.\ Astrophys.\  {\bf 594} (2016) A16
  [arXiv:1506.07135 [astro-ph.CO]];
  Astron.\ Astrophys.\  {\bf 571} (2014) A25
  [arXiv:1303.5085 [astro-ph.CO]];
  N.~Aghanim {\it et al.} [Planck Collaboration],
  Astron.\ Astrophys.\  {\bf 594} (2016) A11
  [arXiv:1507.02704 [astro-ph.CO]];
  Astron. Astrophys. \textbf{641} (2020) A6
  [arXiv:1807.06209 [astro-ph.CO]].
     
    \bibitem{Minkowski:1977sc}
      P.~Minkowski,
      Phys. Lett. B \textbf{67} (1977), 421-428.
      
        \bibitem{Peccei:1977hh}
      R.~D.~Peccei and H.~R.~Quinn,
      Phys. Rev. Lett. \textbf{38} (1977), 1440-1443.
      
      \bibitem{Kim:1979if}
      J.~E.~Kim,
      Phys. Rev. Lett. \textbf{43} (1979), 103.
      
      \bibitem{Shifman:1979if}
      M.~A.~Shifman, A.~I.~Vainshtein and V.~I.~Zakharov,
      Nucl. Phys. B \textbf{166} (1980), 493-506.
      
          \bibitem{Gervais:1971ji}
    J.~L.~Gervais and B.~Sakita,
    Nucl. Phys. B \textbf{34} (1971), 632-639.
    
    \bibitem{Volkov:1972jx}
D.~V.~Volkov and V.~P.~Akulov,
JETP Lett. \textbf{16} (1972), 438-440.
    
\bibitem{Golfand:1971iw}
Y.~A.~Golfand and E.~P.~Likhtman,
JETP Lett. \textbf{13} (1971), 323-326.

\bibitem{Zeldovich:1967lct}
Y.~B.~Zel'dovich and I.~D.~Novikov,
Soviet Astron. AJ (Engl. Transl. ), \textbf{10} (1967), 602.

\bibitem{Hawking:1971ei}
S.~Hawking,
Mon. Not. Roy. Astron. Soc. \textbf{152} (1971), 75.

	\bibitem{Milgrom:1983ca}
	M.~Milgrom,
	Astrophys.\ J.\  {\bf 270} (1983) 365.
    For a review see:
    B.~Famaey and S.~McGaugh,
    Living Rev.\ Rel.\  {\bf 15} (2012) 10
    [arXiv:1112.3960 [astro-ph.CO]].
    
\bibitem{ADSCFT}    
J.~M.~Maldacena,
Adv. Theor. Math. Phys. \textbf{2} (1998), 231-252
[arXiv:hep-th/9711200];
S.~S.~Gubser, I.~R.~Klebanov and A.~M.~Polyakov,
Phys. Lett. B \textbf{428} (1998), 105-114
[arXiv:hep-th/9802109];
E.~Witten,
Adv. Theor. Math. Phys. \textbf{2} (1998), 253-291
[arXiv:hep-th/9802150].
For reviews see:
O.~Aharony, S.~S.~Gubser, J.~M.~Maldacena, H.~Ooguri and Y.~Oz,
Phys. Rept. \textbf{323} (2000), 183-386
[arXiv:hep-th/9905111];
M. Ammon, J. Erdmenger, “Gauge/Gravity Duality: Foundations and Applications,“ Cambridge University Press (2015);
H. Nastase, “Introduction to the AdS/CFT Correspondence,“ Cambridge University Press (2015).
 
     \bibitem{Verlinde:2016toy}
	E.~P.~Verlinde,
	SciPost Phys.\  {\bf 2} (2017) no.3,  016
	[arXiv:1611.02269 [hep-th]].
		
    \bibitem{Berezhiani:2015bqa}
	L.~Berezhiani and J.~Khoury,
	Phys.\ Rev.\ D {\bf 92} (2015) 103510
	[arXiv:1507.01019 [astro-ph.CO]].
	
	\bibitem{Lelli:2016zqa}
    F.~Lelli, S.~S.~McGaugh and J.~M.~Schombert,
    Astron. J. \textbf{152} (2016) 157
    [arXiv:1606.09251 [astro-ph.GA]], and references therein.

\bibitem{SUGRA}
D.~Z.~Freedman, P.~van Nieuwenhuizen and S.~Ferrara,
  Phys.\ Rev.\ {\bf D 13} (1976) 3214;
S.~Deser and B.~Zumino,
  Phys.\ Lett.\ {\bf B 62} (1976) 335.
For a recent review see:
 D.~Z.~Freedman and A.~Van Proeyen,
  Cambridge, UK: Cambridge Univ. Pr. (2012) 607 p.
  For an elementary review, with a brief discussion of some key historical steps, see:
S.~Ferrara and A.~Sagnotti,
Riv. Nuovo Cim. \textbf{40} (2017) no.6, 279-295
[arXiv:1702.00743 [hep-th]].

\bibitem{Scherk:1974ca}
J.~Scherk and J.~H.~Schwarz,
Nucl. Phys. B \textbf{81} (1974), 118-144.

\bibitem{Yoneya:1974jg}
T.~Yoneya,
Prog. Theor. Phys. \textbf{51} (1974), 1907-1920.

\bibitem{stringtheory}
For reviews see: M.~B.~Green, J.~H.~Schwarz and E.~Witten, ``SuperString Theory'', 2 vols., Cambridge, UK: Cambridge Univ. Press (1987); J.~Polchinski, ``String Theory'', 2 vols. Cambridge, UK: Cambridge Univ. Press (1998);  C.~V.~Johnson, ``D-branes,'' USA: Cambridge Univ. Press (2003) 548 p; B.~Zwiebach, ``A first course in String Theory''
Cambridge, UK: Cambridge Univ. Press (2004); K.~Becker, M.~Becker and J.~H.~Schwarz,
``String Theory and M-theory: A modern introduction'' Cambridge, UK: Cambridge Univ.
Press (2007); E.~Kiritsis, ``String Theory in a nutshell'', Princeton, NJ: Princeton Univ. Press (2007);
P.~West, ``Introduction to strings and branes,'' Cambridge: Cambridge Univ. Press (2012).

\bibitem{Witten:1995ex}
E.~Witten,
Nucl. Phys. B \textbf{443} (1995), 85-126
[arXiv:hep-th/9503124].

      \bibitem{climbing}
      E.~Dudas, N.~Kitazawa and A.~Sagnotti,
 Phys.\ Lett.\ {\bf B 694} (2010) 80
[arXiv:1009.0874 [hep-th]];
      E.~Dudas, N.~Kitazawa, S.~P.~Patil and A.~Sagnotti,
      JCAP \textbf{05} (2012), 012
      [arXiv:1202.6630 [hep-th]];
     N.~Kitazawa and A.~Sagnotti,
     JCAP \textbf{04} (2014), 017
     [arXiv:1402.1418 [hep-th]];
     A.~Gruppuso and A.~Sagnotti,
  Int.\ J.\ Mod.\ Phys.\ D {\bf 24} (2015) no.12,  1544008
  [arXiv:1506.08093 [astro-ph.CO]];
  A.~Gruppuso, N.~Kitazawa, N.~Mandolesi, P.~Natoli and A.~Sagnotti,
  Phys.\ Dark Univ.\  {\bf 11} (2016) 68
  [arXiv:1508.00411 [astro-ph.CO]];
A.~Gruppuso, N.~Kitazawa, M.~Lattanzi, N.~Mandolesi, P.~Natoli and A.~Sagnotti,
Phys. Dark Univ. \textbf{20} (2018), 49-64
[arXiv:1712.03288 [astro-ph.CO]].
	
	\bibitem{Klypin:1999uc}
	A.~A.~Klypin, A.~V.~Kravtsov, O.~Valenzuela and F.~Prada,
	Astrophys.\ J.\  {\bf 522} (1999) 82
	[arXiv:astro-ph/9901240].
	
	\bibitem{McGaugh:2016leg}
	S.~McGaugh, F.~Lelli and J.~Schombert,
	Phys.\ Rev.\ Lett.\  {\bf 117} (2016) no.20,  201101
	[arXiv:1609.05917 [astro-ph.GA]].
	
	\bibitem{Karukes:2016eiz}
	E.~V.~Karukes and P.~Salucci,
	Mon.\ Not.\ Roy.\ Astron.\ Soc.\  {\bf 465} (2017) no.4,  4703
	[arXiv:1609.06903 [astro-ph.GA]].
	
	\bibitem{Persic:1995ru}
    M.~Persic, P.~Salucci and F.~Stel,
    Mon.\ Not.\ Roy.\ Astron.\ Soc.\  {\bf 281} (1996) 27
    [arXiv:astro-ph/9506004].
    
    \bibitem{Salucci:2007tm}
    P.~Salucci, A.~Lapi, C.~Tonini, G.~Gentile, I.~Yegorova and U.~Klein,
    Mon.\ Not.\ Roy.\ Astron.\ Soc.\  {\bf 378} (2007) 41
    [arXiv:astro-ph/0703115].
	
	\bibitem{Goodman:1984dc}
	M.~W.~Goodman and E.~Witten,
	Phys.\ Rev.\ D {\bf 31} (1985) 3059.
	
	\bibitem{Hui:2016ltb}
	L.~Hui, J.~P.~Ostriker, S.~Tremaine and E.~Witten,
	Phys.\ Rev.\ D {\bf 95} (2017) no.4,  043541
	[arXiv:1610.08297 [astro-ph.CO]].
	
	\bibitem{Svrcek:2006yi}
	P.~Svrcek and E.~Witten,
	JHEP {\bf 0606} (2006) 051
	[arXiv:hep-th/0605206].
	
	\bibitem{Acharya:2016fge}
	B.~S.~Acharya, S.~A.~R.~Ellis, G.~L.~Kane, B.~D.~Nelson and M.~J.~Perry,
	Phys.\ Rev.\ Lett.\  {\bf 117} (2016) 181802
	[arXiv:1604.05320 [hep-ph]].
	
	\bibitem{Dienes:2011ja}
	K.~R.~Dienes and B.~Thomas,
	Phys.\ Rev.\ D {\bf 85} (2012) 083523
	[arXiv:1106.4546 [hep-ph]].
	
	\bibitem{Dienes:2011sa}
	K.~R.~Dienes and B.~Thomas,
	Phys.\ Rev.\ D {\bf 85} (2012) 083524
	[arXiv:1107.0721 [hep-ph]].
	
    \bibitem{Hossenfelder:2017eoh}
	S.~Hossenfelder,
	Phys.\ Rev.\ D {\bf 95} (2017) no.12,  124018
	[arXiv:1703.01415 [gr-qc]].
	    
    \bibitem{Boran:2017rdn}
    S.~Boran, S.~Desai, E.~O.~Kahya and R.~P.~Woodard,
    Phys.\ Rev.\ D {\bf 97} (2018) no.4,  041501
    [arXiv:1710.06168 [astro-ph.HE]].

\bibitem{Fukuda:1998mi}
Y.~Fukuda \textit{et al.} [Super-Kamiokande],
Phys. Rev. Lett. \textbf{81} (1998), 1562-1567
[arXiv:hep-ex/9807003].

      \bibitem{Baker:2006ts}
      C.~A.~Baker, D.~D.~Doyle, P.~Geltenbort, K.~Green, M.~G.~D.~van der Grinten, P.~G.~Harris, P.~Iaydjiev, S.~N.~Ivanov, D.~J.~R.~May and J.~M.~Pendlebury, \textit{et al.}
      Phys. Rev. Lett. \textbf{97} (2006), 131801
      [arXiv:hep-ex/0602020].

      \bibitem{Ellis:1973yv}
      H.~G.~Ellis,
      J. Math. Phys. \textbf{14} (1973), 104-118.

      \bibitem{Bronnikov:1973fh}
      K.~A.~Bronnikov,
      Acta Phys. Polon. B \textbf{4} (1973), 251-266.

      \bibitem{Morris:1988cz}
      M.~S.~Morris and K.~S.~Thorne,
      Am. J. Phys. \textbf{56} (1988), 395-412.

     \bibitem{tv}
     G.~'t Hooft and M.~J.~G.~Veltman,
     Ann. Inst. H. Poincare Phys. Theor. A \textbf{20} (1974), 69-94.

\bibitem{so16so16}
L.~J.~Dixon and J.~A.~Harvey,
  Nucl.\ Phys.\ B {\bf 274} (1986) 93;
L.~Alvarez-Gaume, P.~H.~Ginsparg, G.~W.~Moore and C.~Vafa,
  Phys.\ Lett.\ B {\bf 171} (1986) 155.
  
\bibitem{0primeB}
A.~Sagnotti,
[arXiv:hep-th/9509080 [hep-th]],
Nucl. Phys. B Proc. Suppl. \textbf{56} (1997), 332-343
[arXiv:hep-th/9702093 [hep-th]].
\bibitem{LM}
F.~Lucchin and S.~Matarrese,
Phys. Rev. D \textbf{32} (1985), 1316
doi:10.1103/PhysRevD.32.1316.

\bibitem{fss}
P.~Fr\'e, A.~Sagnotti and A.~S.~Sorin,
Nucl. Phys. B \textbf{877} (2013), 1028-1106
[arXiv:1307.1910 [hep-th]].

	\bibitem{Ostrogradsky:1850fid}
    M.~Ostrogradsky,
    Mem.\ Acad.\ St.\ Petersbourg {\bf 6} (1850) no.4,  385.
	
	\bibitem{Woodard:2006nt}
    R.~P.~Woodard,
    Lect.\ Notes Phys.\  {\bf 720} (2007) 403
    [arXiv:astro-ph/0601672].
	
	\bibitem{Stelle:1977ry}
    K.~S.~Stelle,
    Gen.\ Rel.\ Grav.\  {\bf 9} (1978) 353.
    
    \bibitem{Sotiriou:2008rp}
    T.~P.~Sotiriou and V.~Faraoni,
    Rev.\ Mod.\ Phys.\  {\bf 82} (2010) 451
    [arXiv:0805.1726 [gr-qc]].
    
    \bibitem{Coleman:1967ad}
S.~R.~Coleman and J.~Mandula,
Phys. Rev. \textbf{159} (1967), 1251-1256.

\bibitem{Fayet:1977yc}
P.~Fayet,
Phys. Lett. B \textbf{69} (1977), 489.

\bibitem{VolkovSoroka}
D.~V.~Volkov and V.~A.~Soroka,
JETP Lett. \textbf{18} (1973), 312-314.

\bibitem{Chamseddine:1982jx}
A.~H.~Chamseddine, R.~L.~Arnowitt and P.~Nath,
Phys. Rev. Lett. \textbf{49} (1982), 970.

\bibitem{Rarita:1941mf}
W.~Rarita and J.~Schwinger,
Phys. Rev. \textbf{60} (1941), 61.

\bibitem{Cremmer:1978ds}
E.~Cremmer and B.~Julia,
Phys. Lett. B \textbf{80} (1978), 48,
Nucl. Phys. B \textbf{159} (1979), 141-212

\bibitem{Cremmer:1978km}
E.~Cremmer, B.~Julia and J.~Scherk,
Phys. Lett. B \textbf{76} (1978), 409-412.

\bibitem{Bern:2011qn}
For a review see: 
Z.~Bern, J.~J.~Carrasco, L.~J.~Dixon, H.~Johansson and R.~Roiban,
Fortsch. Phys. \textbf{59} (2011), 561-578
[arXiv:1103.1848 [hep-th]].

\bibitem{Banks:2012dp}
T.~Banks,
[arXiv:1205.5768 [hep-th]].

\bibitem{Bern:2018jmv}
Z.~Bern, J.~J.~Carrasco, W.~M.~Chen, A.~Edison, H.~Johansson, J.~Parra-Martinez, R.~Roiban and M.~Zeng,
Phys. Rev. D \textbf{98} (2018) no.8, 086021
[arXiv:1804.09311 [hep-th]].

\bibitem{Veneziano:1968yb}
G.~Veneziano,
Nuovo Cim. A \textbf{57} (1968), 190-197.

\bibitem{Polchinski:1995mt}
J.~Polchinski,
Phys. Rev. Lett. \textbf{75} (1995), 4724-4727
[arXiv:hep-th/9510017].

\bibitem{Gliozzi:1976qd}
F.~Gliozzi, J.~Scherk and D.~I.~Olive,
Nucl. Phys. B \textbf{122} (1977), 253-290.

\bibitem{Dai:1989ua}
J.~Dai, R.~G.~Leigh and J.~Polchinski,
Mod. Phys. Lett. A \textbf{4} (1989), 2073-2083.

\bibitem{Dine:1989vu}
M.~Dine, P.~Y.~Huet and N.~Seiberg,
Nucl. Phys. B \textbf{322} (1989), 301-316.

\bibitem{Gross:1984dd}
D.~J.~Gross, J.~A.~Harvey, E.~J.~Martinec and R.~Rohm,
Phys. Rev. Lett. \textbf{54} (1985), 502-505.

\bibitem{Sagnotti:1987tw}
A.~Sagnotti, in Cargese '87, ``Non-Perturbative Quantum Field
Theory'', eds. G. Mack et al (Pergamon Press, 1988), p. 521,
[arXiv:hep-th/0208020];
G.~Pradisi and A.~Sagnotti,
Phys.\ Lett.\ {\bf B 216} (1989) 59;
P.~Horava,
Nucl.\ Phys.\ {\bf B 327} (1989) 461;
Phys.\ Lett.\ {\bf B 231} (1989) 251;
M.~Bianchi and A.~Sagnotti,
Phys.\ Lett.\ {\bf B 247} (1990) 517;
M.~Bianchi and A.~Sagnotti,
Nucl.\ Phys.\ {\bf B 361} (1991) 519;
M.~Bianchi, G.~Pradisi and A.~Sagnotti,
Nucl.\ Phys.\ {\bf B 376} (1992) 365;
A.~Sagnotti,
 Phys.\ Lett.\  {\bf B 294} (1992) 196
 [arXiv:hep-th/9210127].
For reviews see: E.~Dudas,
Class.\ Quant.\ Grav.\  {\bf 17} (2000) R41 [arXiv:hep-ph/0006190];
C.~Angelantonj and A.~Sagnotti,
Phys.\ Rept.\  {\bf 371} (2002) 1 [Erratum-ibid.\  {\bf 376} (2003)
339] [arXiv:hep-th/0204089].

\bibitem{Sen:1994fa}
A.~Sen,
Int. J. Mod. Phys. A \textbf{9} (1994), 3707-3750
[arXiv:hep-th/9402002].

\bibitem{Townsend:1995kk}
P.~K.~Townsend,
Phys. Lett. B \textbf{350} (1995), 184-187
[arXiv:hep-th/9501068].

\bibitem{Horava:1996ma}
P.~Horava and E.~Witten,
Nucl. Phys. B \textbf{475} (1996), 94-114
[arXiv:hep-th/9603142].

\bibitem{Horava:1995qa}
P.~Horava and E.~Witten,
Nucl. Phys. B \textbf{460} (1996), 506-524
[arXiv:hep-th/9510209].

\bibitem{bsb}
S.~Sugimoto,
Prog.\ Theor.\ Phys.\  {\bf 102} (1999) 685 [arXiv:hep-th/9905159];
I.~Antoniadis, E.~Dudas and A.~Sagnotti,
Phys.\ Lett.\ {\bf B 464} (1999) 38 [arXiv:hep-th/9908023];
C.~Angelantonj,
Nucl.\ Phys.\ {\bf B 566} (2000) 126 [arXiv:hep-th/9908064];
G.~Aldazabal and A.~M.~Uranga,
JHEP {\bf 9910} (1999) 024 [arXiv:hep-th/9908072];
C.~Angelantonj, I.~Antoniadis, G.~D'Appollonio, E.~Dudas and
A.~Sagnotti,
Nucl.\ Phys.\ {\bf B 572} (2000) 36 [arXiv:hep-th/9911081];
E.~Dudas and J.~Mourad,
  Phys.\ Lett.\ B {\bf 514} (2001) 173
  [arXiv:hep-th/0012071];
  G.~Pradisi and F.~Riccioni,
  Nucl.\ Phys.\ B {\bf 615} (2001) 33
  [arXiv:hep-th/0107090];
  N.~Kitazawa,
  JHEP {\bf 1804} (2018) 081
  [arXiv:1802.03088 [hep-th]].
  For a recent review see:
J.~Mourad and A.~Sagnotti,
[arXiv:1711.11494 [hep-th]].

\bibitem{Starobinsky:1979ty}
A.~A.~Starobinsky,
JETP Lett. \textbf{30} (1979), 682-685.

\bibitem{Lucchin:1984yf}
F.~Lucchin and S.~Matarrese,
Phys. Rev. D \textbf{32} (1985), 1316.
	
	\bibitem{Kim:2017iwr}
	S.~Y.~Kim, A.~H.~G.~Peter and J.~R.~Hargis,
	Phys.\ Rev.\ Lett.\  {\bf 121} (2018) no.21,  211302
	[arXiv:1711.06267 [astro-ph.CO]].
	
	\bibitem{Garrison-Kimmel:2017zes}
	S.~Garrison-Kimmel {\it et al.},
	Mon.\ Not.\ Roy.\ Astron.\ Soc.\  {\bf 471} (2017) no.2,  1709
	[arXiv:1701.03792 [astro-ph.GA]].
	
	\bibitem{Brooks:2012ah}
	A.~M.~Brooks, M.~Kuhlen, A.~Zolotov and D.~Hooper,
	Astrophys.\ J.\  {\bf 765} (2013) 22
	[arXiv:1209.5394 [astro-ph.CO]].
	
	\bibitem{Navarro:1996gj}
    J.~F.~Navarro, C.~S.~Frenk and S.~D.~M.~White,
    Astrophys. J. \textbf{490} (1997) 493
    [arXiv:astro-ph/9611107].
	
	\bibitem{Burkert:1995yz}
    A.~Burkert,
    IAU Symp. \textbf{171} (1996), 175
    [arXiv:astro-ph/9504041].
	
	\bibitem{Read:2018fxs}
	J.~I.~Read, M.~G.~Walker and P.~Steger,
	[arXiv:1808.06634 [astro-ph.GA]].
	
	\bibitem{Pontzen:2011ty}
	A.~Pontzen and F.~Governato,
	Mon.\ Not.\ Roy.\ Astron.\ Soc.\  {\bf 421} (2012) 3464
	[arXiv:1106.0499 [astro-ph.CO]].
	
	\bibitem{Tully:1977fu}
	R.~B.~Tully and J.~R.~Fisher,
	Astron.\ Astrophys.\  {\bf 54} (1977) 661.
	
	\bibitem{Steinmetz:1998gr}
	M.~Steinmetz and J.~Navarro,
	Astrophys.\ J.\  {\bf 513} (1999) 555
	[arXiv:astro-ph/9808076].
	
	\bibitem{DeRossi:2010ey}
	M.~E.~De Rossi, P.~B.~Tissera and S.~E.~Pedrosa,
	Astron.\ Astrophys.\  {\bf 519} (2010) A89
	[arXiv:1005.4960 [astro-ph.CO]].
	
	\bibitem{DiPaolo:2018mae}
	C.~Di Paolo, P.~Salucci and J.~P.~Fontaine,
	Astrophys.\ J.\  {\bf 873} (2019) no.2,  106
	[arXiv:1810.08472 [astro-ph.GA]].
	
	\bibitem{Salucci:2017cet}
	P.~Salucci and N.~Turini,
	[arXiv:1707.01059 [astro-ph.CO]].
	
	\bibitem{Bernabei:2008yi}
	R.~Bernabei {\it et al.} [DAMA Collaboration],
	Eur.\ Phys.\ J.\ C {\bf 56} (2008) 333
	[arXiv:0804.2741 [astro-ph]].
	
	\bibitem{Ahmed:2009zw}
	Z.~Ahmed {\it et al.} [CDMS-II Collaboration],
	Science {\bf 327} (2010) 1619
	[arXiv:0912.3592 [astro-ph.CO]].
	
	\bibitem{Agnese:2013rvf}
	R.~Agnese {\it et al.} [CDMS Collaboration],
	Phys.\ Rev.\ Lett.\  {\bf 111} (2013) no.25,  251301
	[arXiv:1304.4279 [hep-ex]].
	
	\bibitem{Agnese:2014aze}
	R.~Agnese {\it et al.} [SuperCDMS Collaboration],
	Phys.\ Rev.\ Lett.\  {\bf 112} (2014) no.24,  241302
	[arXiv:1402.7137 [hep-ex]].
	
	\bibitem{Davis:2015vla}
	J.~H.~Davis,
	Int.\ J.\ Mod.\ Phys.\ A {\bf 30} (2015) no.15,  1530038
	[arXiv:1506.03924 [hep-ph]].
	
    \bibitem{Khlopov:2010ik}
    M.~Y.~Khlopov, A.~G.~Mayorov and E.~Y.~Soldatov,
    Prespace.\ J.\  {\bf 1} (2010) 1403
    [arXiv:1012.0934 [astro-ph.CO]].

	\bibitem{Adhikari:2018ljm}
	G.~Adhikari {\it et al.},
	Nature {\bf 564} (2018) no.7734,  83.
	
	\bibitem{Ferenc:2019esv}
	D.~Ferenc, D.~Ferenc Šegedin, I.~Ferenc Šegedin and M.~Šegedin Ferenc,
	[arXiv:1901.02139 [physics.ins-det]].
	
	\bibitem{Silk:2018rry}
	J.~Silk,
	Found.\ Phys.\  {\bf 48} (2018) no.10,  1305.

    \bibitem{Carr:2018rid}
    B.~Carr and J.~Silk,
    Mon.\ Not.\ Roy.\ Astron.\ Soc.\  {\bf 478} (2018) no.3,  3756
    [arXiv:1801.00672 [astro-ph.CO]].
	
	\bibitem{Zumalacarregui:2017qqd}
	M.~Zumalacarregui and U.~Seljak,
	Phys.\ Rev.\ Lett.\  {\bf 121} (2018) no.14,  141101
	[arXiv:1712.02240 [astro-ph.CO]].
	
	\bibitem{Garcia-Bellido:2017imq}
	J.~Garcia-Bellido, S.~Clesse and P.~Fleury,
	Phys.\ Dark Univ.\  {\bf 20} (2018) 95
	[arXiv:1712.06574 [astro-ph.CO]].
	
	\bibitem{Adamek:2019gns}
	J.~Adamek, C.~T.~Byrnes, M.~Gosenca and S.~Hotchkiss,
	[arXiv:1901.08528 [astro-ph.CO]].
	
	\bibitem{Boudaud:2018hqb}
	M.~Boudaud and M.~Cirelli,
	[arXiv:1807.03075 [astro-ph.HE]].
	
	\bibitem{Keeton:2006di}
	C.~R.~Keeton and A.~O.~Petters,
	Phys.\ Rev.\ D {\bf 73} (2006) 104032
	[arXiv:gr-qc/0603061].
	
	\bibitem{Helfer:2003va}
	A.~D.~Helfer,
	Rept.\ Prog.\ Phys.\  {\bf 66} (2003) 943
	[arXiv:gr-qc/0304042].
	
	\bibitem{Carr:2019yxo}
	B.~Carr,
	[arXiv:1901.07803 [astro-ph.CO]].
	
	\bibitem{Nakama:2017xvq}
	T.~Nakama, B.~Carr and J.~Silk,
	Phys.\ Rev.\ D {\bf 97} (2018) no.4,  043525
	[arXiv:1710.06945 [astro-ph.CO]].
	
	\bibitem{Lovell:2011rd}
	M.~R.~Lovell {\it et al.},
	Mon.\ Not.\ Roy.\ Astron.\ Soc.\  {\bf 420} (2012) 2318
	[arXiv:1104.2929 [astro-ph.CO]].
	
	\bibitem{Aguilar-Arevalo:2018gpe}
	A.~A.~Aguilar-Arevalo {\it et al.} [MiniBooNE Collaboration],
	Phys.\ Rev.\ Lett.\  {\bf 121} (2018) no.22,  221801
	[arXiv:1805.12028 [hep-ex]].
	
	\bibitem{Aguilar:2001ty}
	A.~Aguilar-Arevalo {\it et al.} [LSND Collaboration],
	Phys.\ Rev.\ D {\bf 64} (2001) 112007
	[arXiv:hep-ex/0104049].
	
	\bibitem{Lee:2019ums}
	J.~W.~Lee, H.~C.~Kim and J.~Lee,
	[arXiv:1901.00305 [astro-ph.GA]].
	
	\bibitem{Schive:2014dra}
	H.~Y.~Schive, T.~Chiueh and T.~Broadhurst,
	Nature Phys.\  {\bf 10} (2014) 496
	[arXiv:1406.6586 [astro-ph.GA]].
	
	\bibitem{Levkov:2018kau}
	D.~G.~Levkov, A.~G.~Panin and I.~I.~Tkachev,
	Phys.\ Rev.\ Lett.\  {\bf 121} (2018) no.15,  151301
	[arXiv:1804.05857 [astro-ph.CO]].
	
	\bibitem{Fraser:2018acy}
	S.~Fraser {\it et al.},
	Phys.\ Lett.\ B {\bf 785} (2018) 159
	[arXiv:1803.03245 [hep-ph]].
	
	\bibitem{Irsic:2017yje}
	V.~Iršič, M.~Viel, M.~G.~Haehnelt, J.~S.~Bolton and G.~D.~Becker,
	Phys.\ Rev.\ Lett.\  {\bf 119} (2017) no.3,  031302
	[arXiv:1703.04683 [astro-ph.CO]].
	
	\bibitem{Kobayashi:2017jcf}
	T.~Kobayashi, R.~Murgia, A.~De Simone, V.~Iršič and M.~Viel,
	Phys.\ Rev.\ D {\bf 96} (2017) no.12,  123514
	[arXiv:1708.00015 [astro-ph.CO]].
	
	\bibitem{Helfer:2018ylo}
	A.~D.~Helfer,
	Phys.\ Rev.\ D {\bf 98} (2018) no.6,  065015
	[arXiv:1809.04946 [gr-qc]].
	
	\bibitem{Sikivie:2018tml}
	P.~Sikivie,
	Phys.\ Dark Univ.\  {\bf 24} (2019) 100289
	[arXiv:1805.05577 [astro-ph.CO]].
	
	\bibitem{Moroi:2018vci}
	T.~Moroi, K.~Nakayama and Y.~Tang,
	Phys.\ Lett.\ B {\bf 783} (2018) 301
	[arXiv:1804.10378 [hep-ph]].
	
	\bibitem{Ouellet:2018beu}
	J.~L.~Ouellet {\it et al.},
	[arXiv:1810.12257 [hep-ex]].
	
	\bibitem{Ivanov:2018byi}
	M.~M.~Ivanov, Y.~Y.~Kovalev, M.~L.~Lister, A.~G.~Panin, A.~B.~Pushkarev, T.~Savolainen and S.~V.~Troitsky,
	JCAP {\bf 1902} (2019) no.02,  059
	[arXiv:1811.10997 [astro-ph.CO]].
	
	\bibitem{Chashchina:2016wle}
    O.~Chashchina, R.~Foot and Z.~Silagadze,
    Phys. Rev. D \textbf{95} (2017) no.2, 023009
    [arXiv:1611.02422 [astro-ph.GA]].
	
	\bibitem{Harvey:2015hha}
	D.~Harvey, R.~Massey, T.~Kitching, A.~Taylor and E.~Tittley,
	Science {\bf 347} (2015) 1462
	[arXiv:1503.07675 [astro-ph.CO]].
	
	\bibitem{Massey:2017cwf}
	R.~Massey {\it et al.},
	Mon.\ Not.\ Roy.\ Astron.\ Soc.\  {\bf 477} (2018) no.1,  669
	[arXiv:1708.04245 [astro-ph.CO]].
	
	\bibitem{Chudaykin:2016yfk}
	A.~Chudaykin, D.~Gorbunov and I.~Tkachev,
	Phys.\ Rev.\ D {\bf 94} (2016) 023528
	[arXiv:1602.08121 [astro-ph.CO]].
	
	\bibitem{Curtin:2018ees}
	D.~Curtin, K.~R.~Dienes and B.~Thomas,
	Phys.\ Rev.\ D {\bf 98} (2018) no.11,  115005
	[arXiv:1809.11021 [hep-ph]].
	
	\bibitem{Genzel:2017jgd}
	R.~Genzel {\it et al.},
	Nature {\bf 543} (2017) 397
	[arXiv:1703.04310 [astro-ph.GA]].
	
	\bibitem{Tiley:2018}
	A. L.~Tiley {\it et al.},
	[arXiv:1811.05982 [astro-ph.GA]].
	
	\bibitem{Aylor:2018drw}
	K.~Aylor, M.~Joy, L.~Knox, M.~Millea, S.~Raghunathan and W.~L.~K.~Wu,
	[arXiv:1811.00537 [astro-ph.CO]].
	
	\bibitem{Risaliti:2018reu}
	G.~Risaliti and E.~Lusso,
	Nat.\ Astron.\
	[arXiv:1811.02590 [astro-ph.CO]].
	
	\bibitem{Scarpa:2006cm}
    R.~Scarpa,
    AIP Conf. Proc. \textbf{822} (2006) no.1, 253-265
    [arXiv:astro-ph/0601478].
	
	\bibitem{Randriamampandry:2014eoa}
	T.~Randriamampandry and C.~Carignan,
	Mon.\ Not.\ Roy.\ Astron.\ Soc.\  {\bf 439} (2014) no.2,  2132
	[arXiv:1401.5619 [astro-ph.GA]].
	
	\bibitem{Pointecouteau:2005mr}
	E.~Pointecouteau and J.~Silk,
	Mon.\ Not.\ Roy.\ Astron.\ Soc.\  {\bf 364} (2005) 654
	[arXiv:astro-ph/0505017].
	
	\bibitem{Danieli:2019zyi}
	S.~Danieli, P.~van Dokkum, C.~Conroy, R.~Abraham and A.~J.~Romanowsky,
	[arXiv:1901.03711 [astro-ph.GA]].
	
	\bibitem{Clowe:2006eq}
    D.~Clowe, M.~Bradac, A.~H.~Gonzalez, M.~Markevitch, S.~W.~Randall, C.~Jones and D.~Zaritsky,
    Astrophys.\ J.\  {\bf 648} (2006) L109
    [arXiv:astro-ph/0608407].
	
	\bibitem{Lee:2010hja}
    J.~Lee and E.~Komatsu,
    Astrophys.\ J.\  {\bf 718} (2010) 60
    [arXiv:1003.0939 [astro-ph.CO]].
    
    \bibitem{Thompson:2014zra}
    R.~Thompson, R.~Davé and K.~Nagamine,
    Mon.\ Not.\ Roy.\ Astron.\ Soc.\  {\bf 452} (2015) no.3,  3030
    [arXiv:1410.7438 [astro-ph.CO]].
    
    \bibitem{Bouillot:2014hda}
    V.~R.~Bouillot, J.~M.~Alimi, P.~S.~Corasaniti and Y.~Rasera,
    Mon.\ Not.\ Roy.\ Astron.\ Soc.\  {\bf 450} (2015) 145
    [arXiv:1405.6679 [astro-ph.CO]].
    
    \bibitem{Angus:2007qj}
    G.~W.~Angus and S.~S.~McGaugh,
    Mon.\ Not.\ Roy.\ Astron.\ Soc.\  {\bf 383} (2008) 417
    [arXiv:0704.0381 [astro-ph]].
    
    \bibitem{Milgrom:2008cs}
	M.~Milgrom,
	Astrophys.\ J.\  {\bf 698} (2009) 1630
	[arXiv:0810.4065 [astro-ph]].
	
	\bibitem{Milgrom:1992hr}
    M.~Milgrom,
    Annals Phys.\  {\bf 229} (1994) 384
    [arXiv:astro-ph/9303012].
	
	\bibitem{Milgrom:1998sy}
    M.~Milgrom,
    Phys.\ Lett.\ A {\bf 253} (1999) 273
    [arXiv:astro-ph/9805346].
	
	\bibitem{Milgrom:2005mc}
    M.~Milgrom,
    EAS Publ.\ Ser.\  {\bf 20} (2006) 217
    [arXiv:astro-ph/0510117].
	
	\bibitem{Mendoza:2015una}
    S.~Mendoza,
    J.\ Phys.\ Conf.\ Ser.\  {\bf 600} (2015) no.1,  012045.
    
    \bibitem{Bernal:2011qz}
	T.~Bernal, S.~Capozziello, J.~C.~Hidalgo and S.~Mendoza,
	Eur.\ Phys.\ J.\ C {\bf 71} (2011) 1794
	[arXiv:1108.5588 [astro-ph.CO]].
    
    \bibitem{Soussa:2003sc}
    M.~E.~Soussa and R.~P.~Woodard,
    Phys.\ Lett.\ B {\bf 578} (2004) 253
    [arXiv:astro-ph/0307358].
	
	\bibitem{Saffari:2007xc}
    Y.~Sobouti,
    Astron.\ Astrophys.\  {\bf 464} (2007) 921
    Erratum: [Astron.\ Astrophys.\  {\bf 472} (2007) 833]
    [arXiv:0704.3345 [astro-ph], astro-ph/0603302].
    
    \bibitem{Hossenfelder:2018vfs}
    S.~Hossenfelder and T.~Mistele,
    Int.\ J.\ Mod.\ Phys.\ D {\bf 27} (2018) no.14,  1847010
    [arXiv:1803.08683 [gr-qc]].
    
    \bibitem{Bruneton:2007si}
    J.~P.~Bruneton and G.~Esposito-Farese,
    Phys. Rev. D \textbf{76} (2007), 124012
    [erratum: Phys. Rev. D \textbf{76} (2007), 129902]
    [arXiv:0705.4043 [gr-qc]].
	
    \bibitem{Babichev:2007dw}
    E.~Babichev, V.~Mukhanov and A.~Vikman,
    JHEP \textbf{02} (2008), 101
    [arXiv:0708.0561 [hep-th]].
    
    \bibitem{Bruneton:2006gf}
    J.~P.~Bruneton,
    Phys.\ Rev.\ D {\bf 75} (2007) 085013
    [arXiv:gr-qc/0607055].
	
	\bibitem{Born:1934gh}
    M.~Born and L.~Infeld,
    Proc.\ Roy.\ Soc.\ Lond.\ A {\bf 144} (1934) no.852,  425.
    
    \bibitem{Fradkin:1985qd}
    E.~S.~Fradkin and A.~A.~Tseytlin,
    Phys.\ Lett.\  {\bf 163B} (1985) 123.
	
    \bibitem{Tseytlin:1999dj}
    A.~A.~Tseytlin,
    In *Shifman, M.A. (ed.): The many faces of the superworld* 417-452
    [arXiv:hep-th/9908105].
	
    \bibitem{Son:2005rv}
    D.~T.~Son and M.~Wingate,
    Annals Phys.\  {\bf 321} (2006) 197
    [arXiv:cond-mat/0509786].
    
    \bibitem{Laskin:2002zz}
    N.~Laskin,
    Phys.\ Rev.\ E {\bf 66} (2002) 056108
    [arXiv:quant-ph/0206098].
    
    \bibitem{Bekenstein:1984tv}
    J.~Bekenstein and M.~Milgrom,
    Astrophys. J. \textbf{286} (1984), 7-14.
    
    \bibitem{Bekenstein:2004ne}
    J.~D.~Bekenstein,
    Phys.\ Rev.\ D {\bf 70} (2004) 083509
    Erratum: [Phys.\ Rev.\ D {\bf 71} (2005) 069901]
    [arXiv:astro-ph/0403694].
    
    \bibitem{Tuveri:2019zor}
	M.~Tuveri and M.~Cadoni,
	Phys.\ Rev.\ D {\bf 100} (2019) no.2,  024029
	[arXiv:1904.11835 [gr-qc]].
	
	\bibitem{Giusti:2019wdx}
	A.~Giusti,
	Int.\ J.\ Geom.\ Meth.\ Mod.\ Phys.\  {\bf 16} (2019) no.03,  1930001.
    
    \bibitem{Dai:2017guq}
    D.~C.~Dai and D.~Stojkovic,
    Phys.\ Rev.\ D {\bf 96} (2017) no.10,  108501
    [arXiv:1706.07854 [gr-qc]].
    
    \bibitem{Babichev:2018uiw}
    E.~Babichev, C.~Charmousis, G.~Esposito-Farèse and A.~Lehébel,
    Phys. Rev. D \textbf{98} (2018) no.10, 104050
    [arXiv:1803.11444 [gr-qc]].
    
    \bibitem{Esposito-Farese:2019vlh}
    G.~Esposito-Farese,
    [arXiv:1905.04586 [gr-qc]].
	
	\bibitem{Hossenfelder:2018iym}
	S.~Hossenfelder and T.~Mistele,
	JCAP {\bf 1902} (2019) 001
	[arXiv:1809.00840 [astro-ph.GA]].
	
	\bibitem{Mistele:2019byy}
    T.~Mistele,
    JCAP {\bf 1911} (2019) 039
    [arXiv:1909.05710 [astro-ph.GA]].
	
	\bibitem{Alexander:2018fjp}
    S.~Alexander, E.~McDonough and D.~N.~Spergel,
    JCAP {\bf 1805} (2018) 003
    [arXiv:1801.07255 [hep-th]].
    
    \bibitem{Milgrom:2009gv}
    M.~Milgrom,
    Phys.\ Rev.\ D {\bf 80} (2009) 123536
    [arXiv:0912.0790 [gr-qc]].
    
    \bibitem{Zlosnik:2006zu}
    T.~G.~Zlosnik, P.~G.~Ferreira and G.~D.~Starkman,
    Phys. Rev. D \textbf{75} (2007), 044017
    [arXiv:astro-ph/0607411].
    
    \bibitem{Blanchet:2009zu}
    L.~Blanchet and A.~Le Tiec,
    Phys. Rev. D \textbf{80} (2009), 023524
    [arXiv:0901.3114 [astro-ph.CO]].
	
	\bibitem{Hayashi:2014nra}
    K.~Hayashi and M.~Chiba,
    Astrophys.\ J.\  \textbf{789} (2014) 62
    [arXiv:1405.4606 [astro-ph.GA]].
	
	\bibitem{Bowden:2016bwq}
    A.~Bowden, N.~Evans and A.~Williams,
    Mon.\ Not.\ Roy.\ Astron.\ Soc.\  \textbf{460} (2016) 329
    [arXiv:1604.06885 [astro-ph.GA]].
    
    \bibitem{Hattori:2019lgu}
    K.~Hattori and M.~Valluri,
    [arXiv:1909.03321 [astro-ph.GA]].
    
    \bibitem{Hoekstra:2003pn}
    H.~Hoekstra, H.~K.~C.~Yee and M.~D.~Gladders,
    Astrophys. J. \textbf{606} (2004) 67.
    [arXiv:astro-ph/0306515].
    
    \bibitem{Oguri:2004in}
    M.~Oguri,
    [arXiv:astro-ph/0408573].

    \bibitem{Dubinski:1991bm}
    J.~Dubinski and R.~G.~Carlberg,
    Astrophys. J. \textbf{378} (1991) 496.
    
    \bibitem{Allgood:2005eu}
    B.~Allgood, R.~A.~Flores, J.~R.~Primack, A.~V.~Kravtsov, R.~H.~Wechsler, A.~Faltenbacher and J.~S.~Bullock,
    Mon. Not. Roy. Astron. Soc. \textbf{367} (2006) 1781
    [arXiv:astro-ph/0508497].
    
    \bibitem{Bett:2006zy}
    P.~Bett, V.~Eke, C.~S.~Frenk, A.~Jenkins, J.~Helly and J.~Navarro,
    Mon. Not. Roy. Astron. Soc. \textbf{376} (2007) 215
    [arXiv:astro-ph/0608607].
    
  \bibitem{Yoshida:2000uw}
    N.~Yoshida, V.~Springel, S.~D.~M.~White and G.~Tormen,
    Astrophys. J. Lett. \textbf{544} (2000) L87
    [arXiv:astro-ph/0006134].
    
    \bibitem{Read:2005if}
    J.~I.~Read and B.~Moore,
    Mon.\ Not.\ Roy.\ Astron.\ Soc.\  {\bf 361} (2005) 971
    [arXiv:astro-ph/0501273].

    \bibitem{White:2000jv}
    M.~J.~White,
    Astron. Astrophys. \textbf{367} (2001) 27
    [arXiv:astro-ph/0011495].
    
    \bibitem{Bondarenko:2020mpf}
    K.~Bondarenko, A.~Sokolenko, A.~Boyarsky, A.~Robertson, D.~Harvey and Y.~Revaz,
    JCAP \textbf{01} (2021), 043
    [arXiv:2006.06623 [astro-ph.CO]].
    
    \bibitem{ForemanMackey:2012ig}
    D.~Foreman-Mackey, D.~W.~Hogg, D.~Lang and J.~Goodman,
    Publ. Astron. Soc. Pac. \textbf{125} (2013), 306-312
    [arXiv:1202.3665 [astro-ph.IM]].
    
    \bibitem{Li:2020iib}
    P.~Li, F.~Lelli, S.~McGaugh and J.~Schombert,
    Astrophys. J. Suppl. \textbf{247} (2020) no.1, 31
    [arXiv:2001.10538 [astro-ph.GA]].
    
    \bibitem{elongated}
    J.~Craig Wheeler, ``Cosmic Catastrophes: Exploding Stars, Black Holes, and Mapping the Universe'' (Cambridge Univ. Press, Cambridge, 2007).
    
    \bibitem{Rush:1993qz}
    B.~Rush, M.~A.~Malkan and L.~Spinoglio,
    Astrophys. J. Suppl. \textbf{89} (1993), 1
    [arXiv:astro-ph/9306013].
        
    \bibitem{MartinezDelgado:2008cx}
    D.~Martinez-Delgado, J.~Penarrubia, R.~J.~Gabany, I.~Trujillo, S.~R.~Majewski and M.~Pohlen,
    Astrophys. J. \textbf{689} (2008) 184
    [arXiv:0805.1137 [astro-ph]].
    
    \bibitem{Dokkum}
    P.~van Dokkum, C.~Gilhuly, A.~Bonaca, A.~Merritt, S.~Danieli, D.~Lokhorst, R.~Abraham, C.~Conroy, J.~P.~Greco,
    [arXiv:1906.11260 [astro-ph]].
    
    \bibitem{Kaneda:2007uw}
    H.~Kaneda, T.~Suzuki, T.~Onaka, Y.~Doi, M.~Kawada, B.~C.~Koo, S.~Makiuti, T.~Nakagawa, Y.~Okada and S.~Serjeant, \textit{et al.}
    Publ. Astron. Soc. Jap. \textbf{59} (2007), 463
    [arXiv:0706.0068 [astro-ph]].
    
    \bibitem{Afanasiev:1998xi}
    V.~L.~Afanasiev and O.~K.~Sil'chenko,
    Astron. J. \textbf{117} (1999), 1725
    [arXiv:astro-ph/9812390].
    
    \bibitem{Chudnovsky:1986hc}
    E.~M.~Chudnovsky, G.~B.~Field, D.~N.~Spergel and A.~Vilenkin,
    Phys. Rev. D \textbf{34} (1986), 944-950.
    
    \bibitem{Vilenkin:2018zol}
    A.~Vilenkin, Y.~Levin and A.~Gruzinov,
    JCAP \textbf{11} (2018) 008
    [arXiv:1808.00670 [astro-ph.CO]].
    
    \bibitem{Blasi:2020mfx}
    S.~Blasi, V.~Brdar and K.~Schmitz,
    [arXiv:2009.06607 [astro-ph.CO]].
    
    \bibitem{Ellis:2020ena}
J.~Ellis and M.~Lewicki,
Phys. Rev. Lett. \textbf{126} (2021) no.4, 041304
[arXiv:2009.06555 [astro-ph.CO]].
    
    \bibitem{Arzoumanian:2020vkk}
    Z.~Arzoumanian \textit{et al.} [NANOGrav],
    [arXiv:2009.04496 [astro-ph.HE]].
    
    \bibitem{morris}
    M.R.~Morris, J.-H.~Zhao and W.M.~Goss, Astrophys. J. Lett. \textbf{850} (2017) L23
    [arXiv:1711.04190 [astro-ph.GA]].
    
    \bibitem{YusefZadeh:2003qx}
    F.~Yusef-Zadeh,
    Astrophys. J. \textbf{598} (2003), 325-333
    [arXiv:astro-ph/0308008].
	
    \bibitem{RojasNino:2011xh}
    A.~Rojas-Nino, O.~Valenzuela, B.~Pichardo and L.~A.~Aguilar,
    Astrophys. J. Lett. \textbf{757} (2012) L28
    [arXiv:1112.4510 [astro-ph.GA]].
    
    \bibitem{Rojas-Nino:2015qna}
    A.~Rojas-Niño, L.~A.~Martínez-Medina, B.~Pichardo and O.~Valenzuela,
    Astrophys. J. \textbf{805} (2015) 29
    [arXiv:1503.05861 [astro-ph.GA]].
    
    \bibitem{Laureijs:2011gra}
    R.~Laureijs \textit{et al.} [EUCLID],
    [arXiv:1110.3193 [astro-ph.CO]].
	
\end{thebibliography}
\end{document}